\documentclass[11pt,lot,lof]{puthesis}
\newcommand{\proquestmode}{}

\title{Fluctuation Dynamo in Collisionless and Weakly Collisional Magnetized Plasmas}

\submitted{September 2019}  %
\copyrightyear{2019}  %
\author{Denis Andr\'e St-Onge}
\adviser{Professor Matthew W. Kunz}  %
\department{Astrophysical Sciences \\ Program in Plasma Physics}

\usepackage{amsmath,amssymb, amsfonts}
\usepackage{lmodern}
\usepackage{upgreek}

\usepackage{color}
\usepackage{cancel}
\usepackage[compress,round]{natbib}
\usepackage{bibentry}

\usepackage{macs}
\usepackage{graphicx}
\usepackage{subcaption}
\usepackage{empheq}
\usepackage{verbatim}

\usepackage{multirow}
\usepackage{longtable}

\usepackage{booktabs}

\ifdefined\printmode

\usepackage{url}

\else

\ifdefined\proquestmode
\usepackage{hyperref}
\hypersetup{bookmarksnumbered}

\makeatletter
\hypersetup{pdftitle=\@title,pdfauthor=\@author}
\makeatother

\else
\usepackage{hyperref}
\hypersetup{colorlinks,bookmarksnumbered}

\makeatletter
\hypersetup{pdftitle=\@title,pdfauthor=\@author}
\makeatother

\fi %
\fi %

\ifodd 0
\renewcommand{\maketitlepage}{}
\renewcommand*{\makecopyrightpage}{}
\renewcommand*{\makeabstract}{}

\else

\abstract{

In weakly collisional astrophysical plasmas, such as the intracluster medium of galaxy clusters, the amplification of cosmic magnetic fields by chaotic fluid motions is hampered by the adiabatic production of magnetic-field-aligned pressure anisotropy. This anisotropy drives a viscous stress parallel to the field that inhibits the plasma's ability to stretch magnetic-field lines. We demonstrate through the use of kinetic simulations that in high-$\beta$ plasmas, kinetic ion-Larmor scale instabilities---namely, firehose and mirror---sever the adiabatic link between the thermal and magnetic pressures, reducing this viscous stress and thereby allowing the dynamo to operate. We identify two distinct regimes of the fluctuation dynamo in a magnetized plasma: one in which these instabilities efficiently regulate the pressure anisotropy so that it does not venture much beyond the firehose and mirror instability thresholds, and one in which this regulation is imperfect. Using kinetic and Braginskii-MHD simulations and analytic theory, we elucidate the role of these kinetic instabilities on the plasma viscosity and determine how the fields and flows self-organize to allow the dynamo to operate in the face of parallel viscous stresses. In the case of efficient pressure-anisotropy regulation, the plasma dynamo closely resembles its more traditional ${\rm Pm}\gtrsim{1}$ MHD counterpart. When the regulation is imperfect, the dynamo exhibits characteristics remarkably similar to those found in the saturated state of the MHD dynamo. An analytical model for the latter regime is developed that exploits this similarity. The model predicts that the plasma dynamo ceases to operate if the ratio of field-aligned to field-perpendicular viscosities is too large, a behavior confirmed by numerical simulation. Leveraging these results, we construct a novel set of microphysical closures for fluid simulations that bridges these two regimes---one that exhibits explosive magnetic-field growth caused by a field-strength-dependent viscosity set by the firehose and mirror instabilities.  The dynamo in both collisionless and weakly collisional plasmas are then closely compared to each other, revealing substantial differences in how sub-parallel viscous motions behave. The former (collisionless) scenario experiences a cascade of stretching motions to sub-Larmor scales that lead to increasingly fast dynamo as the magnetic Reynolds number is increased.

}

\acknowledgements{

First and foremost, I would like to thank my advisor Matt Kunz for taking me on as his first Ph.D. student. The plasma dynamo problem was far more difficult than I had envisioned, but Matt, with his guidance and seemingly endless knowledge on turbulence, kinetic theory and astrophysics, was always able to help me through the rough patches. Thank you for your patience while I worked on zonal flows, advice on things both related and unrelated to physics, and for countless enjoyable discussions over the years. I couldn't have asked for a better advisor.

I would also like to thank my collaborators Alexander Schekochihin and Jonathan Squire, with whom Matt and I worked on much of the material in chapter~\ref{ch:brag}. Their invaluable insight helped to make sense of many of the subtleties we encountered when dealing with the Braginskii-MHD system. I look forward to collaborating with you both in the future.

I would very much like to thank the readers of my thesis, Francois Rincon and James Stone. Their feedback helped strengthen the narrative of this work, turning it into something that will hopefully make a difference in the field.  I really can't thank you two enough.

I am also very much indebted to my second-year advisor John Krommes, with whom I worked on who on zonal flows.\footnote{which can be read about in \citet{Stonge_zono} and \citet{Stonge_dimits}.} Thank you for helping me become a better theorist and, more importantly, showing me the kind of care and honesty that is needed to be a successful one. 

I would like to thank the faculty of the program, especially Greg Hammett, who put a great deal of effort into crafting thoughtful responses to all of the various questions I would ask, and to my academic advisor Amitava Bhattacharjee, who always made time to hear me out.

I would also like to thank all my fellow graduate students and post docs here at PPPL. To Lee and Kelsey, who were my first point of contact in Princeton, thank you for welcoming me and Kelly with open arms. To my fellow classmates, Jeff L, Eugene, Ge, Scott, and Chen, it was great working on prelims and generals problems together. Jeff P, thanks for all the help getting started with numerical methods and zonal flows. Jacob, thanks for all your advice on scientific visualization, which has been a great boon for all my presentations. Charles, thanks for keeping the discussions lively and interesting, to say the least. Daniel, your words of encouragement (read: cordial taunts) kept me on my toes as a student and a scientist. Peter B, thanks for giving us the greatest story ever told on the subject of chess, and Peter J, for handing me what is perhaps the greatest solution to an assignment problem while I was a TA for asymptotics. Andy, thanks for the lively debates on the relative merits of fried cheese curds versus poutine (you're wrong, by the way). Deepen and Celia, my family and I are honored to have been part of your marriage. To Noah, thanks for your help in all things Landau-fluid related and giving my family a good reason to visit Jersey City. Hongxuan, it was a pleasure to discuss the intricacies of zonal flows with you. Alex G, I was always happy to hear about your musings on math, physics, and life in general. Jake and Yichen, it was a pleasure sharing an office with you. To Suying, Eric, Ian, Oak, Alex L, Vadim, Mike, Jack, Eli, Eduardo, Ben, Valentin, Nick, and Laura, thanks for being excellent company at the party office, plasma coffee, DBar and APS. I would also like to thank Lev and Silvio over at Peyton for their help, as well as their camaraderie.  And to those I may have missed, thanks for being part in making this experience memorable.

I would also like to deeply thank the Jameses and the Kraufmans.
From dinners to game nights and park outings, you really made Princeton feel like home.  

It would be amiss of me not to give a shout-out to Queenship of Mary, whose parish family helped my own get through some truly difficult times.  I'm not quite sure what we would have done without you. As my wife always says, ``it takes a parish''. 

Finally, I would like to thank my family from the depths of my soul. I thank my parents, Rosie and Andr\'e, and my in-laws, John and Sylvia, for all their love and support. Kelly, you've put up with several relocations, numerous poor decisions and months on end where I was there but not present. Somehow, through all of this, you've decided that this is all still worth fighting for. I owe everything to you. 

And to Adria, my pint-sized muse: without you, I would still be stuck working on zonal flows. A father couldn't possibly ask for a better daughter.

}

\dedication{To Kelly.}

\fi  %

\def\apjs{Astrophys.~J.~Supp.~Ser.}

\def\apjl{Astrophys.~J.~Lett.}
\def\aap{Astron.~Astrophys.}

\newcommand{\rmDelta}{{\Delta}}
\newcommand{\upi}{{\uppi}}
\newcommand{\Pm}{\mr{Pm}}

\newcommand{\pD}[2]{\frac{\partial #2}{\partial #1}}

\newcommand{\D}[2]{\frac{{\rm d} #2}{{\rm d} #1}}

\newcommand{\od}{\operatorname{d}\!}
\newcommand{\oD}{\operatorname{D}\!}
\newcommand\bs[1]{{\boldsymbol{#1}}}
\newcommand\bb[1]{{\boldsymbol{#1}}}
\newcommand\grad{\bb{\nabla}}
\newcommand\bcdot{\,\bb{\cdot}\,}
\newcommand{\bscdot}{\,\bs{\cdot}\,}
\newcommand{\btimes}{\,\bb{\times}\,}
\newcommand{\bstimes}{\bs{\times}}
\newcommand{\mc}[1]{\mathcal{#1}}
\newcommand{\mr}[1]{\mathrm{#1}}

\newcommand{\msb}[1]{\bb{\mathsf{#1}}}

\newcommand{\const}{{\rm const}}
\newcommand{\imag}{{\rm i}}
\newcommand{\rmd}{{\rm d}}
\newcommand{\rme}{{\rm e}}
\newcommand{\eb}{\bb{\hat{b}}}
\newcommand{\ez}{\bb{\hat{z}}}
\newcommand{\ex}{\bb{\hat{x}}}

\newcommand{\eig}{\bb{\hat{e}}}
\newcommand{\unitDyadic}{\mathsfbi{I}}

\newcommand{\ethree}{\boldsymbol{\hat{e}}_3}
\newcommand{\etwo}{\boldsymbol{\hat{e}}_2}
\newcommand{\eone}{\boldsymbol{\hat{e}}_1}

\newcommand{\vth}[1]{v_{{\rm th {#1}}}}
\newcommand{\ROS}{\eb\eb\, \bb{:}\, \grad \bb{u}}
\newcommand{\drive}{\widetilde{\bb{f}}}
\newcommand{\Reeff}{\mathrm{Re}_{\parallel \mathrm{eff}}}
\newcommand{\Reprl}{\mathrm{Re}_\parallel}
\newcommand{\nusl}{\nu_\mathrm{eff}^\mathrm{SL}}

\newcommand{\Deltap}{{\Updelta p}}
\newcommand{\tcorrf}{{t_\mathrm{corr,f}}}
\newcommand{\visc}{\nu}
\newcommand{\kforce}{k_{\rm f}}

\newcommand{\rndb}{\tilde{\hat{b}}}
\newcommand{\nrndb}{\hat{b}}
\newcommand{\rndB}{\tilde{B}}
\newcommand{\rndu}{\tilde{u}}
\newcommand{\rnds}{\tilde{\sigma}}
\newcommand{\rndk}{\tilde{k}}
\newcommand{\ea}[1]{\overline{#1}}

\newcommand{\ii}{\ensuremath{{\mathrm{i}}}}
\newcommand{\ee}{\ensuremath{{\mathrm{e}}}}
\newcommand{\force}{\ensuremath{\widetilde{f}}}
\newcommand{\vforce}{\ensuremath{\widetilde{\bb{f}}}}

\newcommand{\ba}[1]{\left< #1 \right>}
\newcommand{\bas}[1]{\langle #1 \rangle}

\newcommand{\urms}{\ensuremath{u_\mr{rms}}}
\newcommand{\Brms}{\ensuremath{B_\mr{rms}}}
\newcommand{\nueff}{\ensuremath{\nu_\mr{eff}}}
\newcommand{\rhoi}{\ensuremath{\rho_\mr{i}}}
\newcommand{\Omegai}{\ensuremath{\mathit{\Omega}_\mr{i}}}
\newcommand{\vthi}{\ensuremath{v_\mr{thi}}}
\newcommand{\Deltai}{\ensuremath{\Delta_\mr{i}}}
\newcommand{\va}{\ensuremath{v_\mr{A}}}
\newcommand{\rhoio}{\ensuremath{\rho_\mr{i0}}}
\newcommand{\Omegaio}{\ensuremath{\mathit{\Omega}_\mr{i0}}}
\newcommand{\Omegae}{{\mathit{\Omega}_\mathrm{e}}}

\newcommand{\vao}{\ensuremath{v_\mr{A0}}}
\newcommand{\dio}{\ensuremath{d_\mr{i0}}}

\newcommand{\betai}{\beta_{\mathrm{i}}}
\newcommand{\betaio}{\beta_{\mathrm{i}0}}
\newcommand{\betapar}{\beta_{\parallel\mathrm{i}}}

\newcommand{\pperp}{p_{\perp\mathrm{i}}}
\newcommand{\ppar}{p_{\parallel\mathrm{i}}}
\newcommand{\bdbldot}{\,\bb{:}\,}

\DeclareMathAlphabet{\mathsfbi}{OT1}{\sfdefault}{bx}{sl}
\DeclareMathVersion{sfletters}
\SetSymbolFont{letters}{sfletters}{OML}{ntxsfmi}{m}{it}

\begin{document}

\makefrontmatter

\nobibliography*

\chapter{Introduction}\label{ch:intro}

\section{Astrophysical motivation}\label{ch1:motivation}

\begin{figure}
\centering
\includegraphics[scale=0.45]{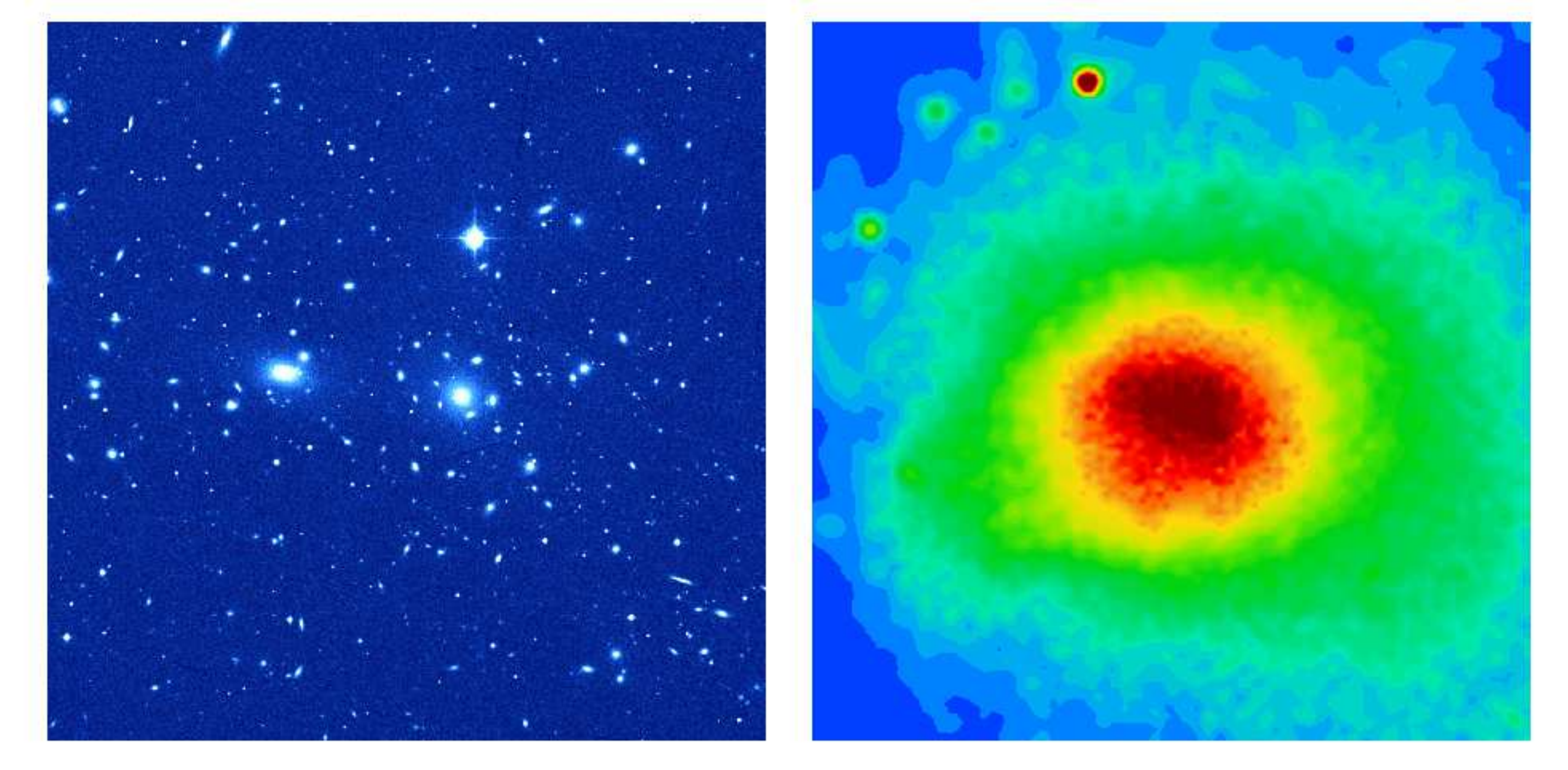}
\caption[Optical and X-ray emission of the Coma cluster.]{\label{intro-coma}Optical (left, UK Schmidt) and X-ray (right, ROSAT) emission of the Coma cluster.}

\end{figure}

Plasmas are ubiquitous in the Universe: they are the most abundant state of matter  apart from dark matter. As an example, consider clusters of galaxies,
 the largest gravitationally bound objects in the Universe. A typical example, such as the Coma cluster, has a mass of  ${\sim}10^{15}$ $M_\odot$ and spans roughly a megaparsec ($1 \textrm{ pc} \approx 3.08 \times 10^{18}\textrm{ cm}$). Most of the mass in this cluster (${\approx}84\%$) is in the form of dark matter, which establishes the gravitational potential well,  while galaxies only comprise ${\approx} 1\%$ by mass. The remainder of the cluster mass (${\approx}15$\%) is made up of a hot, diffuse plasma --- the intracluster medium (ICM) --- which fills the regions between the galaxies. Because these systems are virialized, the ICM tends to be quite hot (ion temperature $T_\mathrm{i}\sim 1 \textrm{--}10$ keV).  These plasmas are also diffuse (density $n \sim 10^{-4} \textrm{--}10^{-2}$ cm$^{-3}$) and, as a result, are weakly coupled ($\Lambda \doteq  4\upi n \lambda_\mathrm{D}^3\sim 10^{14}\textrm{--}10^{17} \ggg 1$, where $\lambda_\mathrm{D} = (T_\mathrm{i}/4\upi n e^2)^{1/2}$ is the Debye length and $e$ is the positive electron charge).  Optical and X-ray images of the Coma cluster are displayed in figure~\ref{intro-coma}. While the optical image shows a sparse collection of galaxies, the hot Bremsstrahlung-emitting plasma that permeates the entire cluster can be seen in  X-ray emission, and is concentrated near the center of the cluster where two supergiant elliptical galaxies reside.
 
\begin{figure}
\centering
\includegraphics[scale=0.91]{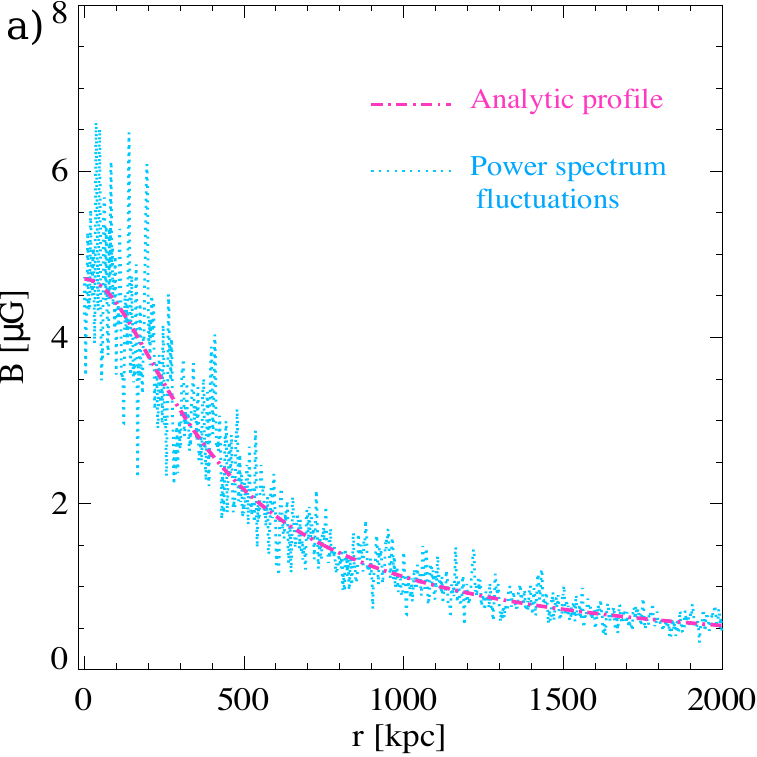}
\hspace{0.5cm}
\includegraphics[scale=0.91]{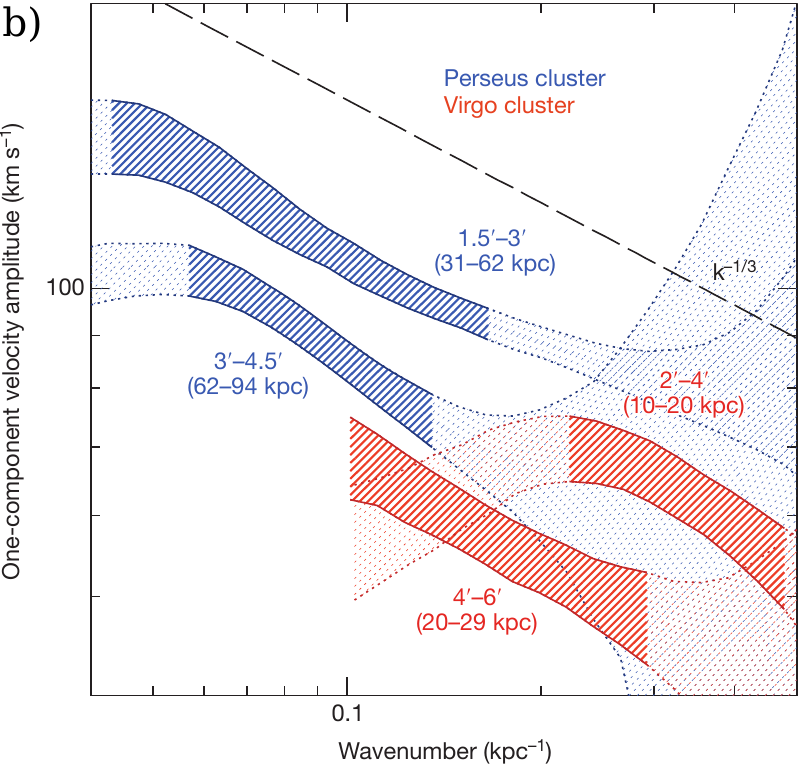}
\caption[Magnetic field and velocity profiles of the Coma cluster.]{\label{intro-coma2} (a) Radial profile of the magnetic-field strength in the Coma cluster obtained by Faraday rotation measurements. From~\citet{bonafede}. (b) One-component velocity amplitude as a function of wavenumber inferred from gas density fluctuations measured for two different annuli in  the Perseus (blue) and Virgo (red) clusters. From~\cite{Zhuravleva}.}
\end{figure}

 The Universe is also magnetized, which is known through observations of Faraday rotation, Zeeman splitting, synchrotron radiation, and dust polarization (e.g.,~\citealt{Carilli02,bonafede,Beck96,Beck15}).  Consider again the Coma cluster, whose magnetic-field strength $B$ is observed to be ${\sim}\upmu$G throughout [see figure~\ref{intro-coma2}(a)].
 Perhaps what is most remarkable is that, for typical ICM plasmas and the interstellar medium of our 
 Galaxy, magnetic-field strengths of
\begin{equation}
    B \sim 2.5\times 10^{-19}\,\biggl(\frac{n}{10^{-3}~\mr{cm}^{-3}}\biggr)\biggl(\frac{T}{5~\mr{keV}}\biggr)^{-3/2}~\mr{G}
\end{equation}
are all that are needed to magnetize the intracluster medium.  This number is obtained by asking for what magnetic-field strength $B$ is the ion cyclotron frequency $\Omegai \doteq eB / m_\mathrm{i}c$ on the order of the ion collision frequency~\citep{Braginskii}
 \begin{equation}
 \nu_{\mathrm{i}} = \frac{4\upi^{1/2}  Z^4 e^4 n_\mathrm{i} \ln \Lambda }{3 m_\mathrm{i}^{1/2} T_\mathrm{i}^{3/2}}.
 \end{equation}
Here, $m_\mr{i}$ and $n_\mathrm{i}$ are the ion mass and density, $c$ is the speed of light, and $Z$ is the ion charge state. 
This $B$ also ensures $\rho_\mr{i}\lesssim\lambda_\mr{mfp}$, where $\lambda_\mathrm{mfp} = v_\mathrm{thi} \tau_\mathrm{i}$ is the collisional mean free path,  $\vthi\equiv(2T_\mr{i}/m_\mr{i})^{1/2}$ is the ion thermal speed, and the inverse ion collision time is $\tau_\mathrm{i} = \nu_\mathrm{i}^{-1}$.
 In the $T_\mr{i}\sim{0.5}~\mr{eV}$ interstellar medium, the same $B$ ensures $\rhoi\lesssim{0.01L}$ for $L\sim{1}~\mr{kpc}$. Typical physical parameters  of ICM plasmas are recorded in table~\ref{ch1:tab_ICM}.

  \begin{table}[b]
  \centering
  \begin{tabular}{r|l c r|l c r|l}
\hline \rule{0pt}{2.4ex}
  	L &  1 Mpc & $\quad \quad$ 			&  $n$  & $10^{-4}$ --$10^{-1}$ cm$^{-3}$         & $\quad \quad$ & $B$ & ${\sim}\upmu$G \\
	$\ell_0$    &  100 kpc  &  			        &    $T_\mathrm{i}$ & 1--10 keV          & & $\rho_{\mathrm{i}}$          & 10$^5$ km  \\
	$\lambda_\mathrm{mfp}$  &  0.1--10 kpc  & & $v_\mathrm{thi}$ & ${\sim}1000$ km/s &  &  $\Omegai $ & 0.2 s$^{-1}$  \\
	$\ell_\visc$          &   10 kpc &			&   $M$ & $0.1$ -- 0.3 & &  $\beta_{\mathrm{i}}$ & 100\\
	$\ell_\eta$          &  $10^4$ km & 		& $L/u_\mathrm{rms}$ & ${\sim}10^9$ years  & &			$\mathrm{Re}$ & 1--100 \\
	$d_\mathrm{i}$              &  5000 km  &  & 	 				Turnover time & ${\sim}10^8$ years				& &   $\mathrm{Rm}$ & $10^{29\textrm{--}31}  $  \\
	 $\lambda_\mathrm{D}$ & 10 km  &	& $\tau_\mathrm{i}$ & ${\sim}10^{6\textrm{--}7}$ years  && $\mathrm{Pm}$ & $10^{27\textrm{--}31}$	\\
   \end{tabular}
   \caption[Typical present-day ICM parameters]{\label{ch1:tab_ICM} Typical present-day parameters of a hot ICM plasma. Included are the ion inertial length $d_\mathrm{i} \doteq (m_\mathrm{i}c^2/4\upi n_\mathrm{i}Z^2 e^2)^{1/2} $, the Mach number $M\doteq u_\mathrm{rms}/v_\mathrm{thi}$, the Reynolds number $\mathrm{Re} \doteq u_0 \ell_0 / \visc $, the magnetic Reynolds number $\mathrm{Rm} \doteq u_0\ell_0 / \eta$, and the magnetic Prandtl number $\mathrm{Pm} \doteq \mathrm{Rm}/\mathrm{Re}$. Here, $\ell_0$ is the outer scale of the fluid motions, $u_0$ is the typical velocity at the scale $\ell_0$, $\visc$ is the kinematic viscosity and $\eta$ is the magnetic diffusivity. The viscous and resistive scales $\ell_\visc$ and $\ell_\eta$ are calculated using the Coulomb collisionality $\tau_\mathrm{i}^{-1} $ and the Spitzer value of the resistivity $\eta$. Both the Reynolds number and magnetic Reynolds number are also calculated using these values.}
  \end{table}

That these systems are not content with hosting weaker fields is surprising, at least until one realizes that the energy density of a ${\sim}\upmu\mr{G}$ field is comparable to that of the observed turbulent motions; e.g., the {\em Hitomi}-observed velocity dispersion ${\approx}160~\mr{km~s}^{-1}$ in the ICM of Perseus \citep{Hitomi1} matches the Alfv\'{e}n speed $\va\equiv{B}/\sqrt{{4\upi}m_\mr{i}n_\mr{i}}$ for the observed number density $n\approx{0.02}~\mr{cm}^{-3}$ if $B\approx{10}~\upmu\mr{G}$. It is then natural to attribute the amplification and sustenance of (at least the random component of) the interstellar and intracluster magnetic fields to the fluctuation (or `turbulent') dynamo \citep{Batchelor,Zeldovich,ChildressGilbert}, by which a succession of random velocity shears stretches the field and leads on the average to its growth to dynamical strengths.

The ability of and efficiency by which the fluctuation dynamo amplifies magnetic fields crucially relies on the material properties of the host plasma. This especially concerns the viscosity, which controls the rate of strain of the field-amplifying motions and thus directly controls the growth rate of the magnetic energy~\citep[e.g.~see the review by][]{Rincon_review}. In this context, it is interesting to note that, short of magnetization effects, the ICM is rather viscous: if we calculate the Reynolds number $\mathrm{Re} = u_0 \ell_0 / \visc$ (where $\ell_0$ is the outer scale of the fluid motions, $u_0$ is the typical speed at that scale, and $\visc$ is the viscosity) using the Coulomb collision frequency, then typical values are $\mathrm{Re} \sim \textrm{1--100}$. On the other hand, the ICM is an excellent conductor: using the Spitzer resistivity,
\begin{equation}
\eta =  \frac{m_\mathrm{e}^{1/2}c^2 Z e^2 \ln \Lambda}{4T_\mathrm{i}^{3/2}},
\end{equation}
the typical value of the magnetic Reynolds number is $\mathrm{Rm} \doteq u_0 \ell_0 / \eta  \sim 10^{29\textrm{--}31}$ for the ICM, and so the magnetic the Prandtl number $\mathrm{Pm} \doteq \mathrm{Rm}/\mathrm{Re} \sim 10^{27\textrm{--}31}$. Thus, at least based on Coulomb collisions, we are interested in the problem of the turbulent dynamo in a $\mathrm{Re} \sim 1\textrm{--}100$ and $\mathrm{\Pm} \ggg1$ system.

As such, much of this thesis addresses the ways in which the transport properties of the plasma impact its ability to amplify magnetic fields and how these transport properties change as the plasma becomes more magnetized. This requires a deep and careful treatment of the underlying equations that describe the interplay between the plasma and electromagnetic fields.  In the next section and starting with the kinetic equation, we derive various descriptions of a plasma for varying levels of magnetization and collisionality.  In the remainder of this chapter, we then describe how the nature of the dynamo changes in each of these systems.

\section{Descriptions of collisional and weakly collisional plasmas}

In order to properly frame the problem of how the dynamo works in a conductive fluid (or plasma), we first need to obtain a system of equations that appropriately describes the interaction between the plasma and the electromagnetic fields.  In this section, a rundown is given of all of the relevant systems of equations that we study in this thesis; equations of particular importance are boxed for emphasis.

\subsection{Kinetics}

 We  begin with the kinetic Vlasov--Landau equation for the probability distribution function $f_s$ of particle species $s$: 
 \begin{equation}\label{eq:int:vlasov}
 \frac{\partial f_s}{\partial t} + \bb{v}\bcdot \grad f_s + \frac{q_s}{m_s}\left(\bb{E} + \frac{\bb{v}}{c}\btimes\bb{B}\right)\bcdot \frac{\partial f_s }{\partial \bb{v}} = C_s[f_s],
 \end{equation}
along with Maxwell's equations of electrodynamics,
\begin{subequations}\label{eq:vlaslan}
\begin{align}
\grad \bcdot \bb{E} &= 4 \upi \rho_\mathrm{c}, && \textrm{(Gauss's Law)} \label{eq:gausslaw}\\
\grad \bcdot \bb{B} &= 0,  && \textrm{(Solenoidality)} \\
 \grad \btimes \bb{E} &= -\frac{1}{c}\frac{\partial \bb{B}}{\partial t}, && \textrm{(Faraday's Law of induction)} \\
\grad \btimes \bb{B} &=  \frac{4\upi}{c} \bb{J} + \frac{1}{c}\frac{\partial \bb{E}}{\partial t}. && \textrm{(Amp\`ere's Law)}
\end{align}
\end{subequations}
 Here,  $q_s$ and $m_s$ are the charge and mass of species $s$, respectively; $C_s[f_s]$ is the collision operator; $\bb{E}$ and $\bb{B}$ are the electric and magnetic fields; and the  charge and current densities are given by
 \begin{subequations} \label{curden}
 \begin{align}
 \rho_\mathrm{c} &= q_s \int \od^3 \bb{v}\, f_s ,\\
  \bb{J} &= q_s \int \od^3 \bb{v}\, \bb{v}f_s,
 \end{align}
 \end{subequations}
 respectively.
 
While equations~\eqref{eq:int:vlasov},~(\ref{eq:vlaslan}a--d) and ~(\ref{curden}a--b) constitute a closed set that can be used to study the dynamo, in practice they are enormously complicated: not only is the distribution function $f_s$ seven dimensional (6 phase-space dimensions, plus time), it also describes time and length scales that are not relevant to the problem at hand. This first simplification we can make here is to assume the plasma is non-relativistic, and so the displacement current in Amp\`ere's law, $(1/c) \partial \bb{E}/ \partial t$, can be neglected.  We also assume that we have a single ion species with charge state $Z=1$, so that $q_\mathrm{i} = -q_\mathrm{e} = e$.

\subsection{Hybrid kinetics}\label{int:hybridkin}

The next step we take is the simplification of the electron dynamics. This can be done by considering the evolution of~\emph{moments} of the distribution function, rather than of the distribution function itself. In particular, the first three moments of the distribution function are the density, fluid velocity, and pressure tensor:
 \begin{subequations}
  \begin{align}
 n_s &= \int \od^3 \bb{v} f_s,\\
n_s\bb{u}_s &= \int \od^3 \bb{v}\, \bb{v}f_s, \\
\mathsfbi{P}_s &=  m_s\int \od^3 \bb{v}\, (\bb{v}-\bb{u}_s)(\bb{v}-\bb{u}_s)f_s,
 \end{align}
 \end{subequations}
 respectively.  We now proceed by taking the first two moments of~\eqref{eq:int:vlasov} for the electron species:
 \begin{subequations}
  \begin{gather}
\frac{\partial n_\mathrm{e}}{\partial t}  + \grad \bcdot n_\mathrm{e}\bb{u}_\mathrm{e} = 0, \\
m_\mathrm{e} \frac{\partial n_\mathrm{e} \bb{u}_\mathrm{e}}{\partial t}  + \grad \bcdot (\mathsfbi{P}_\mathrm{e} + m_\mathrm{e} n_\mathrm{e}\bb{u}_\mathrm{e}\bb{u}_\mathrm{e})  + en_\mathrm{e} \left(\bb{E} + \frac{\bb{u}_\mathrm{e}}{c}\btimes \bb{B}\right) = 0.
\end{gather}
\end{subequations}
 By combining these two equations, one obtains the electron momentum equation
 \begin{equation}\label{eq:int:elecmom}
 m_\mathrm{e}n_\mathrm{e}\left(\frac{\partial \bb{u}_\mathrm{e}}{\partial t}  + \bb{u}_\mathrm{e}\bcdot \grad \bb{u}_\mathrm{e} \right)= - \grad \bcdot \mathsfbi{P}_\mathrm{e}  - e n_\mathrm{e}\left(\bb{E} + \frac{\bb{u}_\mathrm{e}}{c}\btimes \bb{B}\right).
 \end{equation}
 We now take the smallness of the electron mass as a small parameter ($m_\mathrm{e}/m_\mathrm{i}\ll 1$) and assume  $\omega \ll \Omegae$, where $\omega$ is a characteristic frequency of interest (such as the growth rate of the magnetic energy or inverse eddy turnover time) and $\Omegae$ is the electron gyrofrequency. This results in an electron fluid that is Maxwellian, gyrotropic, and isothermal along a field line~(see, for instance, Appendix A1 of~\citealt{Rosin_2011}). Additionally, the frequency ordering renders the electron inertia subdominant to the other terms in the momentum equation.
 As a further approximation, we take the electron fluid to be uniformly isothermal with electron temperature $T_\mathrm{e}$. This approximation may be justified if the magnetic-field lines are chaotic and volume-filling (which, as we shall see in \S\ref{sec:STF}, is a natural outcome of the fluctuation dynamo). 
 This leads to an Ohm's law that is used to determine the electric field $\bb{E}$:
  \begin{equation}
  e n_\mathrm{e} \bb{E} +\frac{ Ze n_\mathrm{i}}{c}\bb{u}_\mathrm{i}\btimes \bb{B}  - \frac{1}{c}\bb{J}\btimes \bb{B}=  -T_\mathrm{e}\grad n_\mathrm{e},
 \end{equation}
 where we have used $\bb{J} = -e n_\mathrm{e}\bb{u}_\mathrm{e} + Ze n_\mathrm{i} \bb{u}_\mathrm{i}$. 
  We also restrict our attention to scales larger than the Debye length, thus replacing Gauss's law ~\eqref{eq:gausslaw} with quasineutrality,
  \begin{equation}\label{eq:quasineutrality}
   \sum_s q_s n_s = e(n_\mathrm{i} - n_\mathrm{e}) =0,
   \end{equation}
   so that $n\doteq n_\mathrm{i} = n_\mathrm{e}$.
   With these approximations, we arrive at a new closed system of equations, the so-called `hybrid-kinetic' system (see equations (1)--(4) and (10) in \citealt{Pegasus}):
\begin{subequations}\label{intro:hybkin}
 \begin{empheq}[box=\fbox]{gather}
  \frac{\partial f_\mathrm{i}}{\partial t} + \bb{v}\bcdot \grad f_\mathrm{i} + \frac{e}{m_\mathrm{i}} \left( \bb{E} + \frac{\bb{v}}{c}\btimes\bb{B}\right)\bcdot \frac{\partial f_\mathrm{i} }{\partial \bb{v}} = 0, \\
  \grad \btimes \bb{B} = \frac{4\upi }{c}\bb{J},\\
  \frac{\partial \bb{B}}{\partial t} = -c\grad \btimes \bb{E}, \label{eq:sim:ind} \\
  \bb{E} + \frac{1}{c}\bb{u}_\mathrm{i}\btimes \bb{B} - \frac{\eta}{c}\grad \btimes \bb{B}  = -\frac{T_\mathrm{e} \grad n}{e n} +  \frac{1}{e n c}\bb{J}\btimes \bb{B}. \label{eq:int:ohms}
 \end{empheq}
 \end{subequations}

Equations~(\ref{intro:hybkin}a--d) are still seven dimensional and are thus difficult to solve both analytically and numerically.  While we do so in chapter~\ref{ch:simulation}, we can also perform the much simpler task of investigating the dynamo in certain asymptotic regimes of these equations.  Specifically, we consider two asymptotic regimes based upon the size of the characteristic dynamical frequency $\omega$ relative to the cyclotron and collision frequencies. One involves strong collisionality and weak magnetization ($\nu_\mathrm{i} \gg \Omegai \gg \omega$) --- the so-called magnetohydrodynamic (MHD) limit; and the other involves weak collisionality and strong magnetization ($\Omegai \gg \nu_\mathrm{i} \gg \omega$), which results in a magnetized, weakly collisional fluid.\footnote{One may ask whether the ordering $\nu_\mathrm{i} \gg \omega$ belies the ``weakly collisional'' moniker. This regime is alternatively called the \emph{dilute} magnetized plasma by~\citet{Balbus01}. Setting this issue of semantics aside, we use the term ``weakly collisional magnetized plasma'' to describe a $\Reprl \lesssim 1$ plasma [viz. equation~\eqref{eqn:Reynolds}] whose collision frequency is much less than the gyro-frequency.}

\subsection{Collisional plasmas (MHD)}

The highly collisional regime $\nu_\mathrm{i} \gg \Omegai$ is derived first, which is done by assuming the ion collisions now serve to render the ion pressure tensor isotropic. The resulting system of equations is known as the magnetohydrodynamic system of equations, or more simply magnetohydrodynamics (MHD):
\begin{subequations}\label{eq:int:mhd}
\begin{empheq}[box=\fbox]{gather}
\frac{\od \rho}{\od t} = -\rho \grad \bcdot \bb{u}, \\
\rho\frac{\od \bb{u}}{\od t} =- \frac{T_\mathrm{e}\grad \rho}{e\rho}  -\grad p + \frac{1}{ c}\bb{J}\btimes \bb{B}, \\
  \grad \btimes \bb{B} = \frac{4\upi }{c}\bb{J},\\
  \frac{\partial \bb{B}}{\partial t} = \grad \btimes (\bb{u}\btimes \bb{B}) + \eta \nabla^2 \bb{B},\label{eq:MHD:ind}
\end{empheq}
\end{subequations}
where $\od / \od t \doteq \partial / \partial t + \bb{u}\bcdot \grad$ is the convective derivative, $\rho \doteq m_\mathrm{i}n$ is the mass density, and  we have defined the center-of-mass velocity $\bb{u} \doteq (m_\mathrm{i}\bb{u}_\mathrm{i} + m_\mathrm{e}\bb{u}_\mathrm{e})/(m_\mathrm{i} + m_\mathrm{e}) = \bb{u}_\mathrm{i}$. This is equal to the mean ion velocity, as we have taken our electrons to be massless.\footnote{A more rigorous derivation of the MHD equations with viscous contributions is done in~\citet{Braginskii}} 
 MHD serves as the starting point of dynamo theory~\citep{Batchelor}, and we will use it to learn the fundamentals on how the dynamo operates to amplify the magnetic energy.
 
Note that we have neglected the last term of the Ohm's law~\eqref{eq:int:ohms}, known as the Hall term, in our induction equation~\eqref{eq:MHD:ind}. For the astrophysical systems we consider in this thesis, this term can be shown to be small:
 \begin{align}
 \frac{\bb{J}\btimes \bb{B}}{q_\mathrm{i}n_\mathrm{i}\bb{u}_\mathrm{i}\btimes \bb{B}} &\sim \frac{B}{\ell_{B}}\frac{1}{Mv_\mathrm{thi}} \frac{c }{4\upi q_\mathrm{i} n_\mathrm{i}} \sim  \frac{1}{M\sqrt{\betai}}\frac{d_\mathrm{i}}{\ell_B},  
 \end{align}
 where $\ell_B$ and $\ell_{n_\mathrm{i}}$ are the gradient scale lengths of the magnetic field and density, $d_\mathrm{i}$ is the ion inertial length, $M \doteq u_\mathrm{rms}/\vthi$ is the Mach number, and $\betai \doteq 8\upi p_\mathrm{i}/B^2$  is the ion plasma beta. The values in table~\ref{ch1:tab_ICM} indicate that, assuming $\ell_\mathrm{B} \sim \ell_\eta$, this ratio is at most ${\sim}10^{-2}$ for present day parameters of the ICM, and should be vanishingly small in the early stages of the dynamo when $\betai \gg 1$, and so we neglect this term for the remainder of this work. Note also that our electron pressure is isobaric (depends only on the density) and thus does not appear in the induction equation (as the curl of a gradient is zero). As a result, we can define a simplified Ohm's law appropriate for standard visco-resistive MHD,
 \begin{equation}
   \bb{E} + \frac{1}{c}\bb{u}_\mathrm{i}\btimes \bb{B} = \frac{\eta}{c}\grad \btimes \bb{B}. \label{int:ohms_simp}
 \end{equation}
 This form of the Ohm's law will be called upon at various points in this chapter.

 \subsection{Collisionless and weakly collisional plasmas}

\subsubsection{drift-kinetic equation} \label{sec:DKEder}

For the magnetized regime, we make an asymptotic expansion in $\omega/\Omegai \ll 1$ or, equivalently, $ \rhoi/\ell \ll 1$, where  $\ell$ is a  typical  length scale of the problem.  In this limit, the distribution function becomes independent of the gyrophase and is thus gyrotropic. Expanding the distribution function in powers of $\rhoi/\ell$, i.e. $f = f_0  + f_1 + \ldots$ and neglecting the collision term, one can derive an equation for the lowest order contribution~\citep{KulsrudMHD}, resulting in the drift-kinetic system of equations:
\begin{subequations}\label{eq:DKEs}
\begin{empheq}[box=\fbox]{gather}\label{eq:DKE}
\frac{\oD f_0}{\oD t} + \frac{\oD\,\ln B}{\oD t} \frac{w_\perp}{2}\frac{\partial f_0}{\partial w_\perp} + \left( \frac{e E_\parallel}{m_\mathrm{i}}  + \frac{w_\perp^2}{2}\grad\bcdot\bb{\hat{b}} - \frac{\oD \bb{u}_{\perp}}{\oD t}\bcdot\bb{\hat{b}}\right)\frac{\partial f_0}{\partial v_\parallel}= 0, \\
\frac{\od n}{\od t} = -n \grad \bcdot \bb{u}, \\
m_\mathrm{i}n \frac{\od \bb{u}}{\od t} =- \frac{T_\mathrm{e}\grad n}{en} - \grad \bcdot [p_\perp (\unitDyadic -\eb\eb) + p_\parallel \eb \eb] + \frac{1}{ c}\bb{J}\btimes \bb{B}, \label{eq:DKE:mom}\\
  \grad \btimes \bb{B} = \frac{4\upi }{c}\bb{J},\\
  \frac{\partial \bb{B}}{\partial t} = -c\grad \btimes \bb{E}, \\ 
  \bb{E} +\frac{1}{c}\bb{u}\btimes \bb{B} =   -\frac{T_\mathrm{e} \grad n}{e n}.
\end{empheq}
\end{subequations}
Here, $ \oD / \oD t \doteq \partial / \partial_t + \bb{u}_{\perp} \bcdot\grad + v_\parallel \bb{\hat{b}}\bcdot\grad$,  $E_\parallel \doteq \eb \bcdot \bb{E}$,  $\bb{w}_\perp = \bb{v}_\perp - \bb{u}_{\perp}$, and perpendicular and parallel pressures, $p_\perp$ and $p_\parallel$, are given by
\begin{subequations}
\begin{align}
p_\perp &\doteq  m_\mathrm{i} \int \od^3 \bb{v}\, f_\mathrm{0} \frac{w_\perp^2}{2},\\
p_\parallel &\doteq  m_\mathrm{i} \int \od^3 \bb{v}\, f_\mathrm{0} (v_\parallel - \eb\bcdot\bb{U} )^2,
\end{align}
\end{subequations} 
respectively. The perpendicular part of the fluid velocity, $\bb{u}_\perp$, is species independent and  simply the $\bb{E}\btimes \bb{B}$ velocity, i.e.  $\bb{u}_\perp \approx c \bb{E}\btimes \bb{B}/B^2$. This implies that the field-perpendicular part of equation~\eqref{eq:DKE:mom} is an evolution equation for $\bb{E}_\perp \doteq -\eb \btimes (\eb\btimes \bb{E})$. This constitutes a closed system of equations. By doing this asymptotic expansion, we have greatly simplified our problem: the system is now six dimensional ($\bb{x}$, $v_\perp$, $v_\parallel$, and $t$) and we have ordered out all the time and length scales associated with the Larmor gyration. We will use equations~(\ref{eq:DKEs}a--f) in appendix~\ref{ap:forcing}, where we determine how a collisionless magnetized plasma accepts energy from a random source.

The physical content of the drift-kinetic equation~\eqref{eq:DKE}, or DKE, can be better appreciated by taking its pressure moments~\citep{CGL}:
\begin{subequations}\label{eq:DKE:pre}
\begin{align}
nB \frac{\od  }{\od t} \left(\frac{p_\perp}{n B}\right) &= - \grad \bcdot (q_\perp \eb) - q_\perp \grad \bcdot \eb, \\
\frac{n^3}{B^2}\frac{\od}{\od t}\left(\frac{p_\parallel B^2}{n^3}\right) &= -\grad \bcdot(q_\parallel \eb) + 2 q_\perp \grad \bcdot \eb,
\end{align}
\end{subequations}
where the perpendicular and parallel heat fluxes are given by
\begin{subequations}
\begin{align}
q_\perp &\doteq m_\mathrm{i} \int \od^3 \bb{v}\, f_0 (v_\parallel-\eb \bcdot \bb{U})\frac{w^2_\perp}{2},\\
q_\parallel &\doteq m_\mathrm{i} \int \od^3 \bb{v}\, f_0 (v_\parallel - \eb \bcdot \bb{U})^3,
\end{align}
\end{subequations}
respectively.  If we make the rather extreme simplification of omitting the heat fluxes entirely, we arrive at the so-called `double-adiabatic' closure~\citep{CGL},
\begin{subequations}\label{eq:CGL}
\begin{align}
 \frac{\od  }{\od t} \left(\frac{p_\perp}{n B}\right) &= 0, \label{eq:CGL1} \\
\frac{\od}{\od t}\left(\frac{p_\parallel B^2}{n^3}\right) &=0. \label{eq:CGL2}
\end{align}
\end{subequations}
The physical meaning of these two equations is immediately apparent: in a collisionless magnetized plasma, the first and second adiabatic invariants of each particle is conserved.  To be more precise, the first adiabatic invariant $\mu$ of a single particle is defined as 
\begin{equation}
\mu \doteq \frac{w_\perp^2}{B}.
\end{equation} 
This comes about via a conservation of the flux enclosed in a gyro-orbit: 
\begin{equation}
B\rho_\mathrm{i}^2 = B \left(\frac{cm_\mathrm{i}w_\perp}{e B}\right)^2= \frac{c^2m_\mathrm{i}^2w_\perp^2}{e^2B}  = \mathrm{const}.
\end{equation}
If $\mu$ is conserved for a single particle, then its average, which is proportional to $p_\perp/nB$, is also conserved, leading to equation~\eqref{eq:CGL1}.  
Equation~\eqref{eq:CGL2} is due to conservation of the second adiabatic invariant, which is associated with the periodic motion of a particle bouncing between two reflection points along an inhomogeneous magnetic-field line. To derive it, one can consider a particle bouncing back and forth in a flux tube of width $L$ and cross-section $A$. Conservation of the flux $BA$, particle number $nAL$ and action $v_\parallel L$ lead immediately to equation~\eqref{eq:CGL2}. Thus, the physics contained in the DKE results primarily from the conservation of these invariants. Additionally, the DKE also captures phase mixing along the magnetic field, magnetic pumping~\citep{Barnes66}, and the collisionless analogues of Alfv\'en waves. Note that equations~\eqref{eq:CGL} directly tie the magnetic field strength to the perpendicular and parallel pressures; this point will be revisited in \S\ref{ch:int:magnetized}.

\subsubsection{Braginskii-MHD}

\label{sec:eq:brag}
We can reintroduce collisions by either choosing a collision operator (such as the Krook  collision operator, $C_s[f_s] = -\nu_s (f_s - f_{\mathrm{M},s})$, where $f_{\mathrm{M},s}$ is a Maxwellian for particle species $s$), or by heuristically positing that their role will be to isotropize the distribution, i.e. to render $p_\perp = p_\parallel$. This results in:
\begin{subequations}
\begin{align}
nB \frac{\od  }{\od t} \left(\frac{p_\perp}{n B}\right) &= - \grad \bcdot (q_\perp \eb) - q_\perp \grad \bcdot \eb - \frac{1}{3}\nu_\mathrm{i}(p_\perp - p_\parallel), \\
\frac{n^3}{B^2}\frac{\od}{\od t}\left(\frac{p_\parallel B^2}{n^3}\right) &= -\grad \bcdot(q_\parallel \eb) + 2 q_\perp \grad \bcdot \eb - \frac{2}{3}\nu_\mathrm{i}(p_\parallel - p_\perp).
\end{align}
\end{subequations}
The asymptotic limit $\Omegai \gg \nu_\mathrm{i} \gg \omega$ is now considered. This renders the heat fluxes and  time derivatives of the pressure  subdominant to the collisional isotropization and $ \od\, \ln B /\od t$. This results in the equation for the pressure anisotropy
\begin{equation}\label{ch:int:dp}
\Delta p = \frac{3 p}{\nu_\mathrm{i}} \left(\frac{\od \, \ln B}{\od t} - \frac{2}{3} \frac{\od \, \ln n}{\od t}\right),
\end{equation}
where $p \doteq (p_\parallel + 2 p_\perp)/3$ is the isotropic pressure, $\Delta p \doteq p_\perp - p_\parallel$ is the pressure anisotropy and  $\Delta p \ll p$ as a result of the ordering $\nu_\mathrm{i} \gg \omega$.\footnote{Note that equation~\eqref{ch:int:dp} implies that if an electron pressure anisotropy were to  be present in the weakly collisional regime, it would be a factor of $\sqrt{m_\mathrm{e}/m_\mathrm{i}}$ smaller than the ion pressure anisotropy.}  This leads us to the Braginskii-MHD system of equations in the limit $\Omegai/\nu_\mathrm{i} \rightarrow \infty$~\citep{Braginskii}:
\begin{subequations}\label{eq:int:brag}
\begin{empheq}[box=\fbox]{gather}
\frac{\od n}{\od t} = -n \grad \bcdot \bb{u}, \\
m_\mathrm{i}n\frac{\od \bb{u}}{\od t}=- \frac{T_\mathrm{e}\grad n}{en}  -\grad p + \grad \bcdot \left[\Delta p \left(\eb\eb - \frac{1}{3}\unitDyadic \right)\right] + \frac{1}{ c}\bb{J}\btimes \bb{B}, \label{ch1:mom_brag} \\
  \grad \btimes \bb{B} = \frac{4\upi }{c}\bb{J},\\
  \frac{\partial \bb{B}}{\partial t} = \grad \btimes (\bb{u}\btimes \bb{B}) + \eta \nabla^2 \bb{B}.
\end{empheq}
\end{subequations}
(Recall that the isotropic and isothermal electron pressure is an \emph{assumption}, rather than a consequence of an asymptotic ordering.)
This system of equations is remarkably similar to that of MHD [equations~(\ref{eq:int:mhd}a--d)], save for the introduction of an anisotropic viscous stress $\grad \bcdot [\Delta p(\eb\eb - \unitDyadic /3)]$ that results from the pressure anisotropy $\Delta p$. The pressure anisotropy itself results from a balance between adiabatic production and collisional isotropization of the pressure tensor, and thus faster growth (decay) of the magnetic field results in larger positive (negative) viscous stress.
 This new stress has a profound effect on the dynamo by primarily damping the motions that lead to changes in the magnetic field strength. We shall study this in more detail in \S\ref{ch:int:magnetized}. First, let us learn how the dynamo works in a highly collisional environment.

\section[The dynamo: what it is and how it works]{The dynamo: what it is and how it works}

In this section, we learn about how the dynamo amplifies the magnetic energy via electromotive forces. We give an idea of the types of magnetic fields that are generated by the dynamo, and as a result we identify systems that can and cannot host a viable dynamo. 

\subsection{Basic features of the dynamo}

The dynamo is the process by which an electrically conducting fluid generates and sustains magnetic field through the electromotive forces brought about by the fluid's underlying motions. The magnetic field $\bb{B}$ evolves according to Faraday's law of induction,
\begin{equation}\label{eq:far1}
\frac{\partial \bb{B}}{\partial t} = - c\grad \btimes \bb{E},
\end{equation}
where $\bb{E}$ is the electric field. As a first step, we consider the dynamo in the MHD system given by~\eqref{eq:int:mhd}. This system has been the foundation upon which much of our knowledge of the dynamo has been built, and so to give a proper context, this is where we begin.
In particular, the induction equation~\eqref{eq:MHD:ind} can be written as
\begin{align}
\frac{\partial \bb{B}}{\partial t} &=  (\bb{B}\bcdot \grad )\bb{u}- (\bb{u}\bcdot \grad)\bb{B}  - \bb{B}(\grad \bcdot \bb{u})+ \eta \nabla^2 \bb{B}, \label{ch1:induction2}
\end{align}
which can be recast as
\begin{equation}
\frac{\od \bb{B}}{\od t} = (\bb{B}\bcdot \grad )\bb{u} + \eta \nabla^2 \bb{B}. \label{ch1:eqn_ind}
\end{equation}
 In equation~\eqref{ch1:eqn_ind}, we have assumed incompressibility ($\grad \bcdot \bb{u} = 0$) for simplicity. For many astrophysical systems, this is a fairly good approximation as the Mach numbers tend to be small (see table~\ref{ch1:tab_ICM}). As such, we will continue to make use of this approximation whenever convenient. The significance of individual terms now becomes apparent: The left-hand side of equation~\eqref{ch1:eqn_ind} describes advection of the magnetic field by the fluid motion, while the second term on the right-hand side represents the  Ohmic dissipation that ultimately converts magnetic energy into heat.  The remaining term (the first one on the right-hand side) signifies compression or stretching of the magnetic field by velocity gradients that lie parallel to the field itself; it is this term that is ultimately responsible for the dynamo~\citep{Batchelor}. To quantify this statement, we define the volume-averaged magnetic energy
\begin{equation}
W \doteq \frac{1}{V} \int \od^3 \bb{x} \, \frac{B^2}{8\upi} \doteq \frac{1}{8\upi}\ba{B^2},
\end{equation}
where $V$ is the  volume under consideration and $\ba{\,\cdot\,}$ defines a spatial average over that volume. Performing the dot product of equation~\eqref{ch1:eqn_ind} with $\bb{B}$ and taking the boundaries to infinity  leads to an evolution equation for the magnetic energy, 
\begin{equation}
\frac{\od  W }{\od t} = 2W (\ROS)  - \frac{1}{\sigma} \ba{J^2} = -\frac{1}{c}\ba{\bb{u}\bcdot \bb{J}\btimes \bb{B}} - \frac{1}{\sigma} \ba{J^2},\label{ch1:mag_en}
\end{equation}
where $\eb \doteq \bb{B}/B$  is the magnetic-field unit vector, $\ROS \doteq \eb \bcdot (\eb \bcdot \grad)\bb{u}$ is the component of the rate of strain oriented parallel to the magnetic field (hereafter, the \emph{parallel rate of strain}), and $\sigma \doteq c^2/4\upi \eta$ is the electrical conductivity. In equation~\eqref{ch1:mag_en}, the final equality was obtained using equation~\eqref{eq:MHD:ind}. Equation~\eqref{ch1:mag_en} demonstrates that the field-aligned component of the symmetrized rate of strain tensor $(\grad \bb{u} + \grad \bb{u}^\mathrm{T})/2$ is responsible for growing the magnetic field.  

The dynamo process results in the transfer of energy from the fluid motions to the magnetic field, which can be seen from the magnetohydrodynamic (MHD) momentum equation
\begin{subequations}
\begin{align}
\rho \frac{\od \bb{u}}{\od t} &=  \frac{\bb{J}\btimes \bb{B}}{c} - \grad p + \rho\visc \nabla^2 \bb{u}, \label{ch1:eq_mom1}
\\ &=  \frac{1}{4\upi} \bb{B}\bcdot \grad \bb{B}  - \grad\left(p + \frac{B^2}{8\upi}\right) + \rho\visc \nabla^2 \bb{u}, \label{ch1:eq_mom2}
\end{align}
\end{subequations}
where $p$ is the scalar pressure and $\visc$ is the isotropic kinematic viscosity. Here we have split up the Lorentz force into a magnetic tension term $\bb{B}\bcdot\grad \bb{B} / 4\upi$ and a magnetic pressure term $B^2 / 8\upi$. By taking the dot product of this with $\bb{u}$, defining the volume-averaged kinetic energy $E\doteq \rho \ba{u^2}/ 2$ and again assuming incompressibility, one arrives at
\begin{equation}
\frac{\od E}{\od t} = \frac{1}{c}\ba{\bb{u}\bcdot \bb{J}\btimes \bb{B}}.
\end{equation}
Comparing this equation with~\eqref{ch1:mag_en}, we find that energy is transferred through the term $\bb{\bb{u}\bcdot \bb{J}\btimes \bb{B}}/c$. For magnetic fields that are dynamically weak ($v_\mathrm{A} \ll u$), the impact of the magnetic tension  and pressure terms on the dynamics can be neglected and, provided that the pressure tensor remains a scalar, the fluid velocity evolves without any influence from the magnetic field. This regime is known as the \emph{kinematic induction} regime (or more commonly the kinematic  regime of the dynamo), and the evolution equation of the magnetic field ~\eqref{ch1:eqn_ind} becomes a linear equation (though \emph{stochastically} nonlinear if $\bb{u}$ is a random variable). The problem of the kinematic dynamo then is whether any appreciable growth of the magnetic field can occur in this regime, and whether the magnetic field can saturate at dynamically important levels.
Nonlinear effects become important when $\rho \bb{u}\bcdot \grad \bb{u} \sim \bb{B}\bcdot \grad\bb{B} /4\upi$, at which point the magnetic field begins to counteract the stretching motions of the underlying fluid.

\subsection{Fast and slow dynamo}\label{sec:STF}

 An important distinction can already be made between two classes of dynamos described by equation~\eqref{ch1:eqn_ind}:  slow and fast dynamos~\citep{vainshtein1972}.  In the former, the dynamo process fundamentally relies on non-ideal effects in Ohm's law; the magnetic-field lines must be able to `slip' from the fluid elements to access free energy (much like the well-known interchange instability).  However, this will result in a growth rate of the magnetic energy that depends on some positive power of the magnetic diffusivity.  This is a problem for astrophysical plasmas, as their resistivities tend to be vanishingly small (i.e. $\mathrm{Rm} \ggg 1$, where the magnetic Reynolds number $\mathrm{Rm} \doteq u_0\ell_0/\eta$,  $\ell_0$ is the outer scale and $u_0$ is the typical outer-scale velocity).  Thus the slow dynamo is defined as a dynamo whose growth rate vanishes as $\eta \rightarrow 0$.

\begin{figure}
\centering
\includegraphics[scale=0.7]{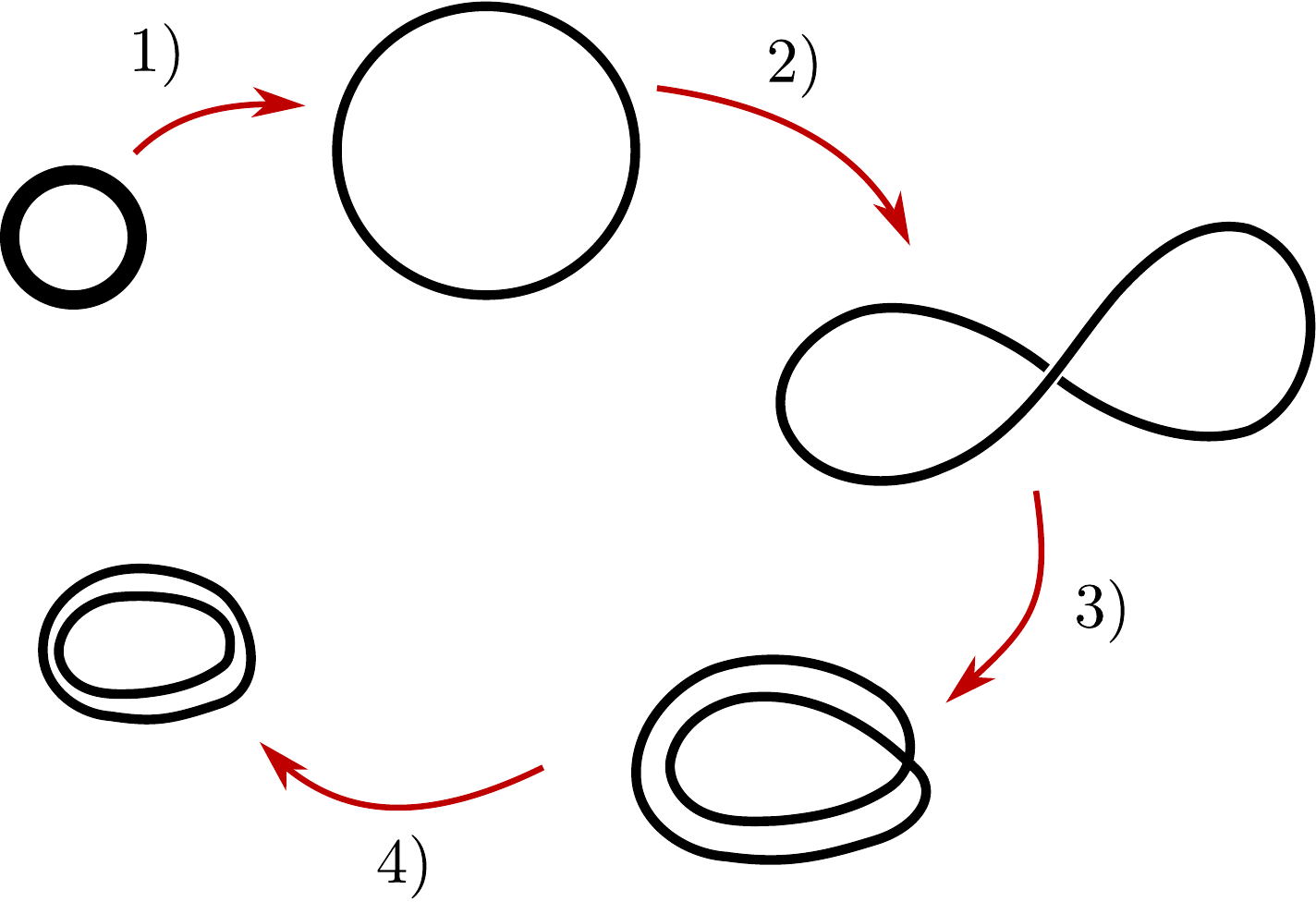}
\caption[Visualization of the `stretch-twist-fold' dynamo mechanism.]{\label{ch1:fig_STF} A visualization of Vainshtein and Zel'dovich's `stretch-twist-fold' dynamo mechanism, which leads to an exponentiation of magnetic energy. First, a magnetic flux tube  is stretched to approximately twice its length. It is then twisted and folded in on itself, resulting in a doubling of the total magnetic energy. Provided with some finite dissipation, the field lines reconnect leaving two flux tubes.   }
\end{figure}

The fast dynamo, on the other hand, exhibits a finite growth rate in the $\eta\rightarrow 0$ limit, and persists in ideal MHD. This type of dynamo relies on the freezing-in of magnetic-field lines to fluid elements. As the fluid carries the magnetic field, the field lines can become stretched, resulting in the amplification of magnetic energy on the average.  This idea lends itself to a phenomenological model of the dynamo called the `stretch-twist-fold' dynamo~\citep{vainshtein1972}. Here, a magnetic flux tube is first stretched, then twisted and folded in on itself, resulting in twice the original amount of magnetic energy.  This process is visualized in figure~\ref{ch1:fig_STF}. While this can result in an arbitrarily strong magnetic field, the field lines themselves  develop arbitrarily fine spatial scales and quickly become non-integrable. A small amount of resistive dissipation is then needed to smooth-out such fine-scale irregularities.

\subsection{Anti-dynamo theorems}\label{ch1:antidyn}

It is emphasized that equation~\eqref{ch1:eqn_ind} contains two nonlinearities: the convective nonlinearity $(\bb{u}\bcdot \grad) \bb{B}$ and the compressive nonlinearity $(\bb{B} \bcdot \grad)\bb{u}$. While the former does not lead to magnetic-field amplification in any volume-averaged sense, in a turbulent fluid it does serve to twist and fold magnetic-field lines, creating ever smaller structures whose scale is only limited by resistive dissipation. As the characteristic scales of the magnetic field continue to shrink and resistive dissipation becomes more important, the dynamo runs into the danger of become resistively dominated, thus leading to net \emph{decay} of the magnetic energy.  The dynamo is thus generically a problem of growing the magnetic energy by stretching field lines in the face of ever-decreasing magnetic-field length scales and thus ever-increasing resistive dissipation of magnetic energy. In many types of systems the resistive dissipation eventually always dominates over magnetic field stretching. This leads to the concept of \emph{anti}-dynamo theorems, two of which are discussed in this section. 

\subsubsection{Zel'dovich}\label{ch1:zel2d}

Perhaps the simplest anti-dynamo theorem was put forth by~\citet{Zeldovich57}, who considered a purely two-dimensional system in planar geometry, i.e. $u_z = B_z = 0$ and $\partial_z \rightarrow 0$. To derive this theorem, we express Ohm's law~\eqref{int:ohms_simp} in terms of the scalar potential $\phi$ and vector potential $\bb{A}$ (\emph{viz.} $\bb{E} = -\partial_t \bb{A} - \grad \phi $ and $\bb{B} = \grad \btimes \bb{A}$): 
\begin{subequations}
\begin{align}
\frac{\partial \bb{A}}{\partial t}  &= \bb{u}\btimes \grad \btimes \bb{A} + \eta \nabla^2 \bb{A} - \grad \phi \\
 &=\grad(\bb{A}\bcdot\bb{u}) -\bb{A}\btimes \bb{\omega} - \bb{A}\bcdot \grad \bb{u} - \bb{u}\bcdot \grad \bb{A}+ \eta \nabla^2 \bb{A} - \grad \phi,
\end{align}
\end{subequations}
where $\bb{\omega} \doteq \grad \btimes \bb{u}$ is the flow vorticity and the Coulomb gauge $\grad \bcdot \bb{A} = 0$ has been assumed.
Due to the planar geometry, $\bb{A} = A_z\ez$, $\bb{\omega} = \omega_z \ez$, and $\partial_z \phi = 0$, leading to the $z$-component
\begin{align}\label{eq:zel2d1}
\frac{\od A_z}{\od t}  = \eta \nabla^2 A_z.
\end{align}
While this is a pure diffusive equation, the nonlinearity $\bb{u}\bcdot \grad A_z$ can lead to transient amplification of the magnetic field by simply creating small-scale structures in the vector potential, i.e. $B \sim A/ \ell$ increases when the length scale $\ell$ decreases. However, this generation of small-scale structure is checked by resistive dissipation, which occurs  over roughly one turnover time $\tau_\mr{c} \sim |\grad \bb{u}|$ of the fluid motion. If we assume~\citet{Kolmogorov1941} phenomenology where the dominant stretching is done by the viscous-scale eddies, the magnitude of this transient amplification can be estimated by balancing the resistive and nonlinear times $(\eta \nabla^2 \sim |\grad \bb{u}|  \rightarrow \eta \ell_\eta^{-2} \sim \mathrm{Re}^{1/2} u_0/\ell_0$, where $\mathrm{Re} = u_0\ell_0/\visc$ is the Reynolds number). Assuming an initial scale of $\bb{A}_0$ to be the outer scale, this leads to
\begin{align}
\frac{B_\mathrm{f}}{B_0} = \frac{\ell}{\ell_\eta} = \mathrm{Re}^{1/4}\left(\frac{u\ell}{\eta}\right)^{1/2} = \mathrm{Re}^{1/4}\mathrm{Rm}^{1/2} = \mathrm{Re}^{3/4}\mathrm{Pm}^{1/2},
\end{align}
where  $B_0$ and $B_\mathrm{f}$ are the initial and final magnetic energies and $\mathrm{Pm}\doteq \mathrm{Rm}/\mathrm{Re} = \visc / \eta$ is the magnetic Prandtl number.
Once this occurs, the scale refinement of the magnetic field ends and the magnetic energy can only be damped, leading to the magnetic field dying off in the long-time limit. This example highlights the struggle of amplifying magnetic field in the face of resistive dissipation.

\subsubsection{Cowling}

An even earlier theorem due to~\citet{Cowling1933} was derived in the context of sources of magnetic fields found around sun spots. This theorem rules out the possibility that an axisymmetric field can be maintained through dynamo action alone.

Consider an axisymmetric system in cylindrical coordinates ($r$,$\varphi$,$z$). The magnetic field and velocity will then have the form 
\begin{equation}
\bb{B} = B_\varphi \bb{\hat{\varphi}} + \bb{B}_\mathrm{p}(r,z), \qquad \bb{u} = \bb{u}_\varphi+ \bb{u}_\mathrm{p},
\end{equation}
where $\mathrm{p}$ denotes the poloidal component (which has both radial and vertical extent). We also define the poloidal flux function $A$ such that $\bb{B}_\mathrm{p} = \grad \btimes (A \bb{\hat{\varphi}})$.  The poloidal and azimuthal components of the induction equation~\eqref{ch1:eqn_ind} are then 
\begin{subequations}
\begin{align}
\frac{\partial A}{\partial t} + \frac{1}{r}\bb{u}_\mathrm{p}\bcdot \grad(rA) &= \eta\left(\nabla^2 - \frac{1}{r^2}\right)A,\label{eq:cowling1}\\
\frac{\partial B_\varphi}{\partial t} + r \bb{u}_\mathrm{p}\bcdot \grad \frac{B_\varphi}{r} &= \eta\left(\nabla^2 - \frac{1}{r^2}\right)A +  r \bb{B}_\mathrm{p}\bcdot \grad \frac{u_\varphi}{r}. \label{eq:cowling2}
\end{align}
\end{subequations}
These equations are the axisymmetric analog to equation~\eqref{eq:zel2d1}, though now a source appears as the last term in equation~\eqref{eq:cowling2}, by which a toroidal field is generated by shearing a poloidal one. However, notice that this source is  finite only if $\bb{B}_\mathrm{p}$ is also finite. Unfortunately, there is no source term in equation~\eqref{eq:cowling1}, and so $\bb{B}_\mathrm{p}$ must damp away by an argument analogous to Zel'dovich's anti-dynamo theorem. With no poloidal field, the source term in equation~\eqref{eq:cowling2} vanishes, and so too must the toroidal field. Thus, neither poloidal nor toroidal magnetic fields can be sustained purely by dynamo action in an axisymmetric system.

\subsection{The Moffatt--Saffman--Zel'dovich 3D model of the dynamo}\label{ch1:zeldo_lin}

In light of the previous two theorems, one might wonder which classes of systems, if any, can support the growth of magnetic energy through the dynamo in the face of resistive dissipation. While the stretch-twist-fold phenomenology from section~\ref{sec:STF} certainly suggests that a viable dynamo is possible, what is needed is a workable quantitative example that gives the dynamo some firm theoretical footing.
\citet{MoffattSaffman64}  put forward a simple example of a workable dynamo in a constant linear shear that illustrated the need for full three-dimensional geometry. \citet{Zeldovich} generalized the result for the case of random linear shear.

This analysis goes as follows: consider the kinematic stage of the dynamo starting with a magnetic field embedded in an incompressible, linear velocity field of infinite extent with the form $u_i = \sigma_{ij}x_j$. Here, $\sigma_{ij}$ is the rate-of-strain tensor $\partial_{x_j} u_i$. As the velocity field is assumed to be incompressible, $\sigma_{ij}$ is also traceless. The eigenvalues $\lambda_i$ of $\sigma_{ij}$ denote the strain rates of the velocity field and are ordered as $\lambda_1 \ge \lambda_2 \ge \lambda_3$; their corresponding eigenvectors are $\eig_i$. To simplify this discussion, we only deal with eigenvalues that are distinct, resulting in a rate-of-strain tensor that is diagonalizable; the general case is dealt with in~\citet{Zeldovich} using Jordan normal forms.
The eigenvectors $\eig_1$, $\eig_2$, $\eig_3$ correspond respectively to the stretching, `null', and compression directions of the incompressible velocity shear, with the `null' direction corresponding either to  additional stretching ($\lambda_2 > 0$), compression ($\lambda_2 < 0$), or true neutrality ($\lambda_2 = 0$). Typically, in turbulent systems, $\lambda_2 >0$ with $\lambda_1$, $| \lambda_3| \gg \lambda_2$.

 To proceed, we adopt the {\it Ansatz}
\begin{equation}\label{ch1:zel_ansatz}
    \bb{B}(t,\bb{x}) = \int \od^3\bb{k}_0 \, \widetilde{\bb{B}}(t,\bb{k}_0) \rme^{\imag \bs{x}\bscdot \widetilde{\bs{k}}(t,\bs{k}_0)} ,
\end{equation}
which states that magnetic field is composed of plane waves with initial conditions $\widetilde{\bb{B}}(t=0,\bb{k}_0) = \bb{B}_0$ and whose wavenumbers depend on time and evolve from $\bb{k}_0$.  It then follows from the induction equation~\eqref{ch1:eqn_ind} that
\begin{equation}
 \frac{\partial \widetilde{\bb{B}}}{\partial t} +  \imag \widetilde{\bb{B}}\left(  \bb{x}\bcdot  \frac{\partial \widetilde{\bb{k}}}{\partial t}\right) + \imag \widetilde{\bb{B}}(\bb{u}\bcdot \widetilde{\bb{k}} ) = \widetilde{\bb{B}} \bcdot \grad \bb{u} + \eta \widetilde{k}^2 \widetilde{\bb{B}}.
\end{equation}
Assuming statistical homogeneity, the above equation must be satisfied at every point $\bb{x}$. This leads to  two separate evolution equations for $\widetilde{\bb{B}}$ and $\widetilde{\bb{k}}$:
\begin{subequations}
\begin{align}
\frac{\partial \widetilde{B}_i}{\partial t} &= \sigma_{ij} \widetilde{B}_j - \eta \widetilde{k}^2 \widetilde{B}_{i}, \\
\frac{\partial \widetilde{k}_i}{\partial t} &= - \sigma_{ji}\widetilde{k}_j.
\end{align}
\end{subequations}
Notice that the above equations imply $\widetilde{\bb{k}}\bcdot \widetilde{\bb{B}} = 0 $ if $\bb{k}_0\bcdot \bb{B}_0 = 0$. With an appropriate change of coordinates, the solution to these equations in the longtime limit is
\begin{subequations}
\begin{align}
\widetilde{\bb{B}}(t,\bb{k}_0) &= \left( \rme^{\lambda_1 t}{B}_{01}\eig_1 +  \rme^{\lambda_2 t}{B}_{02} \eig_2  + e^{-(\lambda_1+\lambda_2)t }{B}_{03}\eig_3 \right) \exp \left( - \eta \int_0^t \od t' \widetilde{k}^2(t',\bb{k}_0)\right), \\
{\widetilde{\bb{k}}}(t,\bb{k}_0) &= \rme^{-\lambda_1 t} k_{01} \eig_1 + \rme^{-\lambda_2 t}k_{02} \eig_2 + \rme^{(\lambda_1 + \lambda_2)t}k_{03}\eig_3,
\end{align}
\end{subequations}
where ${B}_{0i} \doteq {\bb{B}}_0 \bcdot \eig_i$ and $k_{0i} \doteq \bb{k}_0 \bcdot \eig_i$.
This leads to 
\begin{subequations}
\begin{align}
|\widetilde{B}|^2(t,\bb{k}_0) &= \left( \rme^{2\lambda_1 t}|{B}_{01}|^2 +  \rme^{2\lambda_2 t}|{B}_{02}|^2  + e^{-2(\lambda_1+\lambda_2)t }|{B}_{03}|^2 \right) \rme^{ -2 \eta \int_0^t \od t' \widetilde{k}^2(t',\bb{k}_0)}, \label{ch1:zel_en} \\
\widetilde{k}^2(t,\bb{k}_0) &= \rme^{-2\lambda_1 t}k^2_{01} + \rme^{-2\lambda_2 t}k^2_{02}  + \rme^{2(\lambda_1 + \lambda_2)t}k_{03}^2.
\end{align}
\end{subequations}
This indicates that, since $\lambda_1$ is the largest eigenvalue, the magnetic field aligns itself in the direction of $\eig_1$. However, the characteristic scale of the field, which aligns itself with $\eig_3$, decreases at an exponential rate $-(\lambda_1 + \lambda_2)$ and thus the resistive term in equation~\eqref{ch1:zel_en} has the possibility of growing \emph{super}-exponentially. In order for the dynamo to be viable then, that term  must not become too large. At time $t$, the condition for the magnetic field with initial wavenumber  $\bb{k}_0$ to be growing is 
\begin{align}
\frac{\eta}{t \lambda_1}\int_0^t \od t'& \, \widetilde{k}^2 (t',\bb{k}_0) \nonumber \\ 
 &\approx \frac{\eta}{2t\lambda_1^2}k^2_{01} + \frac{\eta}{2t\lambda_1\lambda_2}\left(1-\rme^{-2\lambda_2 t}\right)k^2_{02}  + \frac{\eta} {2t\lambda_1(\lambda_1+ \lambda_2)}\left(\rme^{2(\lambda_1 + \lambda_2)t} -1\right)k_{03}^2  \nonumber \\ &< 1.
\end{align}
This defines the boundary of an ellipsoid in $k_0$-space with volume
\begin{equation}
\sim \lambda_1^2(\lambda_2  |\lambda_3|)^{1/2}\left(\frac{t}{\eta}\right)^{3/2}\rme^{-2(\lambda_1+\lambda_2)t}(1 - \rme^{-2\lambda_2 t})^{-1/2}
\end{equation}
that contains all modes ${\bb{B}}_0(\bb{k}_0)$ that exhibit growth at time $t$. It is clear that this volume  contracts exponentially fast in the $\eig_3$ direction (as $\lambda_1 + \lambda_2 > 0$), resulting in either a thin tube when $\lambda_2 < 0$, or a flat pancake when $\lambda_2 > 0$. If we consider the case $\lambda_2 > 0$, the volume decays similarly. Then the magnetic field at any point is, using \eqref{ch1:zel_ansatz},
\begin{equation}
\bb{B}(t,\bb{x}) \sim \underbrace{\rme^{\lambda_1 t}}_{\mathrm{stretching}} \underbrace{\rme^{-(\lambda_1 + \lambda_2)t}}_{k_0\textrm{ volume}} \sim \rme^{-|\lambda_2| t}.
\end{equation}
However, the \emph{total} magnetic energy, $\ba{B^2} = (2\upi)^{-3}\int \od^3 \bb{k}_0 \,|\bb{B}(t,\bb{k}_0)|^2$ by Parseval's theorem, grows:
\begin{equation}
\bas{B^2} \sim \underbrace{\rme^{2\lambda_1 t}}_{\mathrm{stretching}} \underbrace{\rme^{-(\lambda_1 + \lambda_2)t}}_{k_0\textrm{ volume}} \sim \rme^{(\lambda_1 -\lambda_2) t}.
\end{equation}
As $\lambda_1 > \lambda_2$, this leads to net growth of the magnetic energy.

The case with $\lambda_2 < 0$ is similar, but an extra consideration must be taken into account: here, the volume in $k_0$-space of the initial modes that grow at time $t$ now shrinks at the rate $\rme^{-(|\lambda_2| + |\lambda_3|)t}$. In order to enforce solenoidality of the magnetic field, the magnitude of the original Fourier modes must not exceed a certain threshold, viz.
\begin{equation}\label{ch1:zeldo2}
k_{01}{B}_{01} \approx k_{02}{B}_{02} \quad \Longrightarrow \quad {B}_{10}   < \frac{k_{10}}{k_{02}}{B}_{02} \sim \rme^{-|\lambda_2|t}.
\end{equation}
(Here we have neglected  $k_{03}{B}_{03}$ as it decays exponentially faster than $k_{02}{B}_{02}$.) Equation~\eqref{ch1:zeldo2}  must be taken as another threshold that the initial Fourier modes must satisfy to exhibit growth at time $t$. Thus  
\begin{equation}
\bb{B}(t,\bb{x}) \sim \rme^{(\lambda_1 -|\lambda_2|)t} \rme^{-(|\lambda_2| +|\lambda_3| )t} \sim \rme^{-|\lambda_2| t},
\end{equation}
as, in this case, $|\lambda_3| = \lambda_1 - |\lambda_2|$. Thus, the magnetic field grows similarly as in the case with $\lambda_2 > 0$, though now the magnetic energy grows as
\begin{equation}
\bas{B^2} \sim \rme^{2(\lambda_1 -|\lambda_2|)t} \rme^{-(|\lambda_2| +|\lambda_3| )t} \sim \rme^{(|\lambda_3|-|\lambda_2|) t}.
\end{equation}
Again, with $|\lambda_3| > |\lambda_2| $, this leads to net growth. The analysis for this case holds in the 2D case as well; setting $\lambda_3 = 0$ and $\lambda_1 > \lambda_2$ leads to exponential decay of both the magnetic flux and magnetic energy, consistent with~\citet{Zeldovich57} (see \S\ref{ch1:zel2d}).

The above analysis features the somewhat counter-intuitive result that, while the point-wise magnetic field \emph{decreases} in time, the total energy of the magnetic field \emph{increases}. This example was generalized to a linear but \emph{random} series of velocity shears in \citet{Zeldovich}. Their main result is that, for a given realization of the velocity field, there exists a basis to which the magnetic field and wavevectors converge  in $1/t$ time with eigenvalues that represent the finite-time Lyapunov exponents of the flow~\citep{Goldhirsch87}.  The problem in the long-time limit then becomes isomorphic to the case with constant velocity shear and the results presented inthis section continue to hold.

\begin{figure}
\centering
\includegraphics[width=0.8\textwidth]{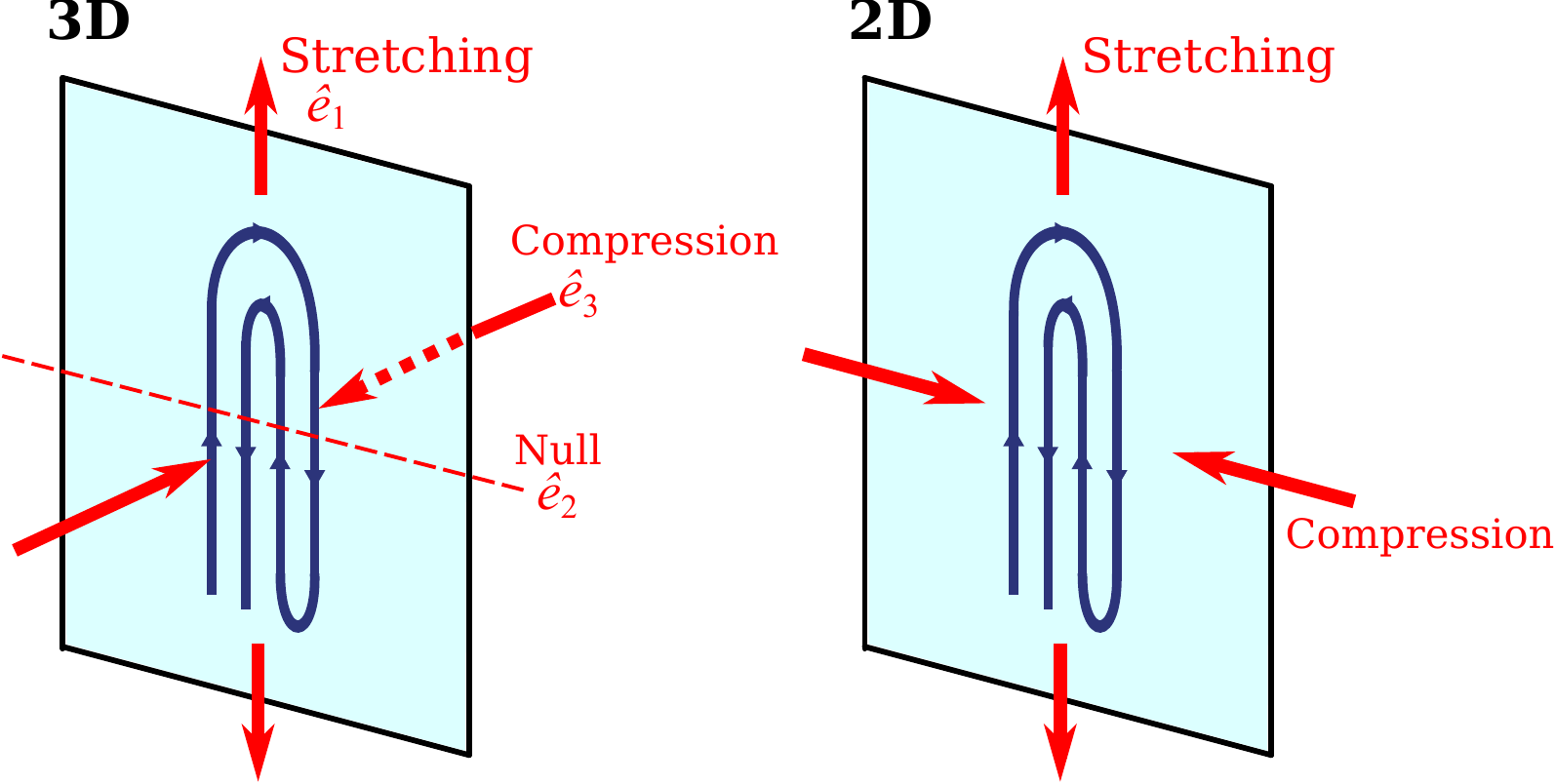}
\caption[Visual explanation of the 3D dynamo and 2D anti-dynamo theorems.]{\label{ch1:compression} A visual explanation of why the dynamo operates in three dimensions but not two. In the former case, the magnetic field develops fine-scale variations (field reversals) in the null direction $\eig_2$. The field is allowed to get compressed in the $\eig_3$ direction without incurring any resistive penalties, as the characteristic variations along that direction are small.  In two dimensions, the fine-scale variations must develop in the compression direction, and thus resistive losses will always overcome energy growth via stretching. (Figure taken from~\citealt{Scheko_sim}.) }
\end{figure}

An intuitive picture, visualized in figure~\ref{ch1:compression}, on why the dynamo works in three-dimensions, but not two, is now manifest. The fluctuation dynamo process necessarily produces magnetic fields with small variations along some direction.  In three dimensions, the fields that survive are those that orient their small variation along the `null' direction, which does not experience significant stretching or compression.  Then, resistive damping is minimized. In two dimensions, however, these variations are always compressed, and resistive dissipation overcomes any growth the magnetic field may experience via stretching.

\section{The fluctuation dynamo in a collisional system}

We have learned in the previous section how the dynamo works in a generic way by stretching and folding magnetic fields lines, and that this processes fundamentally requires three dimensional geometry in order for this stretching to overcome resistive annihilation of the magnetic field.  Let us now move onto the specific problem of the \emph{fluctuation} dynamo. First, let us look at the historical developments that have led to our current understanding of the fluctuation dynamo, which has been primarily been studied within the framework of MHD.  

\subsection{A qualitative picture}\label{ch1:flucdyn}

The fluctuation dynamo is the process that amplifies an initial seed magnetic field via a series of random shears by a background  velocity field. This results in the generation of magnetic energy ($\ba{B^2}$), but not magnetic flux ($\ba{\bb{B}}$, which is generated by the \emph{mean-field} dynamo).
As this process is typically seen in systems exhibiting fluid turbulence, it is sometimes called the `turbulent dynamo', though this is somewhat of a misnomer: all that is needed is a velocity field which exhibits random shearing. A smooth but chaotic single-scale `Stokes' flow ($\mathrm{Re} \lesssim 1$) is sufficient for the dynamo to take place.  

\begin{figure}
\centering
\includegraphics{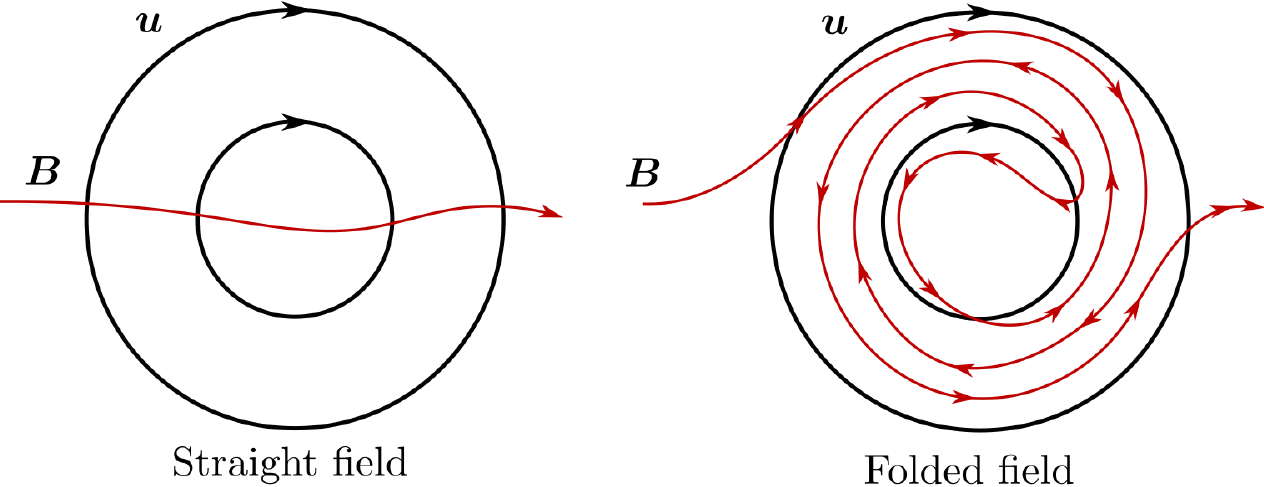}
\caption[The eddy-rotation picture of the fluctuation dynamo.]{\label{ch1:eddy}  A qualitative illustration of the fluctuation dynamo in a turbulent fluid: a magnetic flux tube initially embedded in a turbulent eddy will be rolled-up into elongated folded structures. The length of these folds is comparable to the size of the eddy, while the fold separation is limited by magnetic diffusivity and is comparable to the resistive scale. From~\citet{Scheko_sim}.}
\end{figure}

The fluctuation dynamo is also sometimes called the `small-scale' dynamo to distinguish it from the mean-field or `large-scale' dynamo.  This is because the fluctuation dynamo typically results in the creation of small-scale magnetic-field fluctuations. To see why, we  consider the following qualitative picture illustrated in figure~\ref{ch1:eddy}: imagine a magnetic flux tube embedded in some large-scale turbulent eddy. As the eddy rotates, it carries along the flux tube with it. This will initially shear the tube, eventually causing it to roll up on itself.  As this rotation continues, the tube begins to develop an elongated folded structure, where the length of the fold is comparable to the size of the eddy while the fold separation exponentially decreases until it becomes limited by magnetic diffusivity (i.e. reaches the resistive scale $\ell_\eta$).  In~\citet{Kolmogorov1941} turbulence, the smallest eddies have the shortest correlation time $\tau_\mathrm{eddy} = (\ell_0/u_\mathrm{rms})(\ell_\mathrm{eddy}/\ell_0)^{2/3} $, where $\ell_0$ is the forcing scale and $\ell_\mathrm{eddy}$ is the characteristic size of an eddy. Thus, the growth rate of the turbulent dynamo in the kinematic regime is mainly controlled by the smallest-scale eddies, which are those residing at the viscous scale $\ell_\visc = \mathrm{Re}^{-3/4}\ell_0$. The scale refinement of the magnetic field also occurs exponentially, being controlled by the viscous eddy turnover time. The time it takes for resistivity to become important is then
\begin{equation}
t \sim \frac{\ell}{u_\mathrm{rms}} \mathrm{Re}^{-1/2} \ln (\ell_\visc/\ell_\eta) \sim \gamma^{-1} \ln \mathrm{Pm}^{1/2},
\end{equation}
assuming the initial scale of the magnetic field to be  comparable to the viscous scale.

For systems with $\mathrm{Pm} > 1$, Nonlinear effects in the collisional MHD dynamo come into effect when the magnetic tension becomes comparable to the Reynolds stress in equation \eqref{ch1:eq_mom2}: $\rho\bb{u}\bcdot \grad \bb{u} \sim \bb{B}\bcdot \grad \bb{B} /4\upi$.  Care must be taken into consideration, however; while the magnetic field develops fine-scale structures that are limited by resistivity, it is clear from figure~\ref{ch1:eddy} that the variation along the magnetic field itself is set by the size of the eddy. Thus
\begin{equation}
\bb{B}\bcdot \grad \bb{B} \sim \frac{B^2}{\ell_\visc} \sim \frac{u_\visc^2}{\ell_\visc} \Rightarrow B^2 \sim (4\upi \rho) u_\visc^2.
\end{equation}
Therefore, nonlinear effects become important when the magnetic energy becomes comparable to the energy of the viscous-scale eddies.

\subsection{The Kazantsev-Kraichnan model of the kinematic dynamo}

When the magnetic field is weak, it exerts no dynamical influence on the velocity field and the dynamo becomes a problem that is linear in $\bb{B}$ (though non-linear in the random fields). Perhaps the most important contribution in the study of the fluctuation dynamo is due to~\citet{Kazantsev} and \citet{Kraichnan_mag} whose results form the foundation of the continuing body of research on the fluctuation dynamo, even today. In this work, analytical progress was made by assuming a delta-correlated Gaussian model of the underlying velocity field:
\begin{equation}
\ea{u^i(t,\bb{x})} = 0, \quad\quad \ea{u^i(t,\bb{x})u^j(t',\bb{x}')} = \delta(t-t')\kappa^{ij}(r),
\end{equation} 
where  $\bb{r} = \bb{x} - \bb{x}'$, $\kappa_{ij}$ is the velocity field correlator and $\ea{\vphantom{|}\cdot\vphantom{|}}$  denotes the ensemble average.  
This model of the velocity field is called the `Kraichnan model' of passive advection, which is based on work
performed concurrently to that of Kazantsev~\citep{Kraichnan_passive}, and has been used as an analytical starting point for many problems in turbulence.\footnote{\citet{Kraichnan_mag} also derived a similar result to~\citet{Kazantsev} one year earlier, though in the former work the  important $k_\eta \gg k_\nu$ limit was not considered, and thus equation~\eqref{FK} was ultimately absent.} As the velocity field is random, it serves to twist and stretch the magnetic field, creating the folded structure as described above.  
The problem of solving the induction equation~\eqref{ch1:eqn_ind} can be greatly simplified if one considers the $\mathrm{Pm} \gg 1$ limit, which exhibits a separation between the magnetic-field scales and the viscous scales.  In this limit, the viscous eddies appear as a random and smooth linear shear, much like the approach used in~\S\ref{ch1:zeldo_lin}. Then the velocity correlator can be Taylor expanded assuming incompressibility:
\begin{equation}
\kappa^{ij}(r) = \kappa_0 \delta^{ij} - \kappa_2 \frac{r^2}{2} \left(\delta^{ij} - \frac{1}{2}\frac{r^ir^j}{r^2}\right) + \ldots.
\end{equation}
Using this in~\eqref{ch1:eqn_ind} leads to an equation for the magnetic spectral energy density $M(t,k) \doteq \frac{1}{2}\int \od \Omega_\bb{k}\langle |\bb{B}(t,\bb{k})|^2 \rangle$:
\begin{equation}\label{FK}
\frac{\partial M}{\partial t} = \frac{\overline{\gamma}}{5}\left(k^2 \frac{\partial^2 M}{\partial k^2} - 2 k \frac{\partial M}{\partial k} + 6 M\right) - 2\eta k^2 M,
\end{equation}
where $\overline{\gamma} \doteq -(1/6)[\nabla^2 \kappa^{ii}(\bb{x})]_{\bb{x}=0} = \kappa_2$  and $\int \od \Omega_\bb{k}$ is the integral over the solid angle for each wavenumber $\bb{k}$. This equation has been derived in a variety of publications~\citep{Kazantsev, Kulsrud1992}, and its derivation in the context of a more general model that includes velocity statistics that are anisotropic with respect to the magnetic field direction is given in appendix~\ref{ap:kazantsev}.  This equation can also be transformed into a Fokker-Planck equation:
\begin{equation}\label{FK2}
\frac{\partial M}{\partial t} = \frac{\overline{\gamma}}{5} \frac{\partial}{\partial k}\left(k^2 \frac{\partial M}{\partial k} - 4 k M\right) + 2\overline{\gamma} M- 2\eta k^2 M,
\end{equation}
which highlights the diffusion of magnetic energy through $k$-space (the term in parentheses), along with growth via stretching and resistive diffusion (second and third terms, respectively). 

 If we consider the diffusion-free regime (where the $\eta$ term is negligible) then equation~\eqref{FK} has an exact solution. This can be quickly found by making the substitution $z = \ln k$, leading to a diffusion equation with constant coefficients that can be solved via Fourier transform. This results in the solution
\begin{equation}
M(t,k) = \rme^{(3/4)\overline{\gamma} t} \int_0^\infty \frac{\od k'}{k'} \,M_0(k') \left(\frac{k}{k'}\right)^{3/2}\frac{1}{\sqrt{(4/5 \upi \overline{\gamma} t)}}\exp \left(-\frac{[\ln(k/k')]^2}{(4/5)\overline{\gamma} t}\right),
\end{equation}
where $M_0(k')$ is the initial spectrum. This solution shows that
\begin{enumerate}
\item the width of the spectrum grows exponentially at the rate (4/5)$\overline{\gamma}$;
\item every individual mode grows exponentially at the rate (3/4)$\overline{\gamma}$;
\item the peak, or bulk of the magnetic energy, moves toward larger $k$ leaving a power spectrum of $k^{3/2}$.
\end{enumerate}
All of these effects combine to give a growth rate of $2\overline{\gamma}$ for the total magnetic energy.
It is clear that no magnetic energy can move past the resistive scale, and so once the bulk of the energy reaches that scale, the refinement of the fluctuating magnetic field comes to an end.  The power law it leaves behind, $k^{3/2}$, is referred to as the `Kazantsev spectrum', Simulations of the fluctuation dynamo  in collisional MHD typically obey this scaling in the kinematic regime~\citep[see, for instance,][]{Scheko_sim, Haugen04}.

\begin{figure}
\centering
\includegraphics[width=\textwidth]{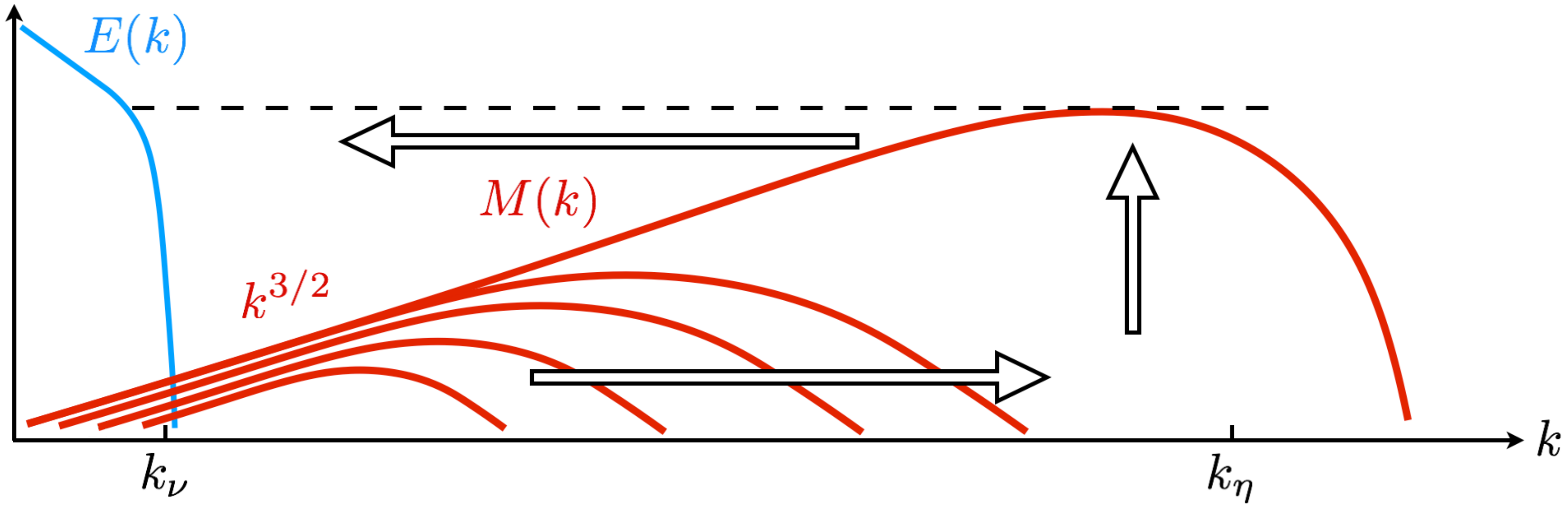}
\caption[The progression of the kinematic $\mathrm{Pm} \gg1$ fluctuation dynamo.]{\label{ch1:kinematic_prog}  An illustration of the $\mathrm{Pm}\gg1$ fluctuation dynamo in the kinematic regime, leading into the nonlinear regime. First, the peak of the magnetic energy migrates to smaller scales, eventually being limited by resistivity.  Second, the entire envelope of the spectrum grows exponentially at the rate $\lambda \overline{\gamma}$ given by~\eqref{ch1:growthrate}. Lastly, the peak of the magnetic energy migrates to larger scales as the stretching motions are disabled, leading to selective decay of the smallest scale modes.}
\end{figure}

If the magnetic energy reaches the resistive scale before nonlinear effects become important, then the dynamo enters a kinematic regime in which diffusion now plays a role.  Equation~\eqref{FK} can be solved  in the asymptotic limit $\eta \rightarrow 0$ as an eigenvalue problem. By imposing a zero-flux boundary condition at some $k_*$ that satisfies $k_\visc \ll  k_* \ll k_\eta$, one gets an approximate expression for the magnetic-energy spectrum for $k \gg k_*$~\citep{Kulsrud1992,Scheko_theory}: 
\begin{equation}\label{ch1:eigensol}
M(k) \approx k^{3/2}\, \rme^{\lambda \overline{\gamma} t} K_0\left(k\sqrt{10\eta/\overline{\gamma}}\right),
\end{equation}
where $K_0$ is the Macdonald function and 
\begin{equation}\label{ch1:growthrate}
\lambda \simeq \frac{3}{4} - \frac{\upi^2}{5[\ln(k_*/2k_\eta)]^2}.
\end{equation}
The second term is of order $1/\ln^2 (\mathrm{Pm^{1/2}})$ and so, in the limit $\eta\rightarrow 0$, the growth-rate becomes independent of the magnetic diffusivity --- a fast dynamo. Notice that $\lambda$ only converges to $3/4$ square-logarithmically  in the magnetic Prandtl number.  The structure of the magnetic field as dictated by \eqref{ch1:eigensol} is an envelope with a scaling of $k^{3/2}$ that is cut-off exponentially at the resistive scale. As time progresses, the entire envelope grows at a rate given by $\lambda \overline{\gamma}$. This evolution is illustrated in figure~\ref{ch1:kinematic_prog}.

\subsection{Nonlinear regime and saturation}\label{ch1:saturation}

Once the magnetic energy becomes comparable with the turbulent energy at any given scale, nonlinear effects become important and the dynamo ceases to be kinematic. For $\mathrm{Pm} \gtrsim 1 $ plasmas, this occurs when the magnetic energy reaches the energy of the smallest turbulent eddies. In this nonlinear stage, the Lorentz force becomes important ($\rho\bb{u}\bcdot \grad \bb{u} \sim \bb{B}\bcdot \grad \bb{B} /4\upi$) and the velocity must be self-consistently determined through the momentum equation~\eqref{ch1:eq_mom2}. As the magnetic field begins to counteract the motions of these viscous scale eddies,  their ability to stretch in the direction parallel to $\eb$ becomes suppressed and the smallest-scale eddies cease to grow the magnetic field (see the discussion in \S\ref{ch1:flucdyn}). 

Various semi-quantitative theories have been put forward to explain what happens next.  One such scenario, proposed by ~\citet{Scheko_theory}, is that the \emph{next} smallest eddies at scale $\ell_\mathrm{s} >  \ell_\visc$ now dominate stretching of the field, and thus the growth of the magnetic energy in this stage is mediated by progressively larger eddies. Here,  $\ell_\mathrm{s}$ denotes the scale of the smallest eddies whose stretching motions have yet to be suppressed.  An estimate of the growth rate in this regime may be obtained by positing that the Lorentz force disables all eddies with energy less than the total magnetic energy. If one assumes Kolmogorov scalings with eddies at scale $\ell_\mathrm{s}$ having energy $E(\ell_\mathrm{s}) \propto (\ell_\mathrm{s}/\ell_0)^{2/3}$, then the scale of the smallest eddies that appreciably stretch and grow the field is given by $E(\ell_\mathrm{s}) \sim \langle B^2\rangle/2 \Rightarrow \ell_\mathrm{s}/\ell_0 \propto \langle B^2 \rangle^{3/2}$. This leads to secular evolution:
    \begin{align}
       \frac{1}{2}\D{t}{\langle B^2\rangle} = \langle B^2\ROS \rangle \sim \frac{\langle B^2 \rangle u_0}{\ell_0}\left(\frac{\ell_0}{\ell_\mathrm{s}}\right)^{2/3} \sim \const.
    \end{align}
  Therefore, as the field-stretching scale shifts, exponential growth gives way to linear-in-time growth of magnetic energy, which has been observed in numerical simulation~\citep{Scheko_theory,Maron04,cho09}. As this happens, the resistive scale, and thus the bulk of the magnetic energy, also begins to migrate to larger scales. This process is termed `selective decay'~\citep{Scheko_theory}, and is a direct result of balancing the magnetic-field growth rate with the resistive damping time, $\gamma \sim \eta \ell_{\visc}^{-2} $.  As the stretching component of the smallest-scale eddies is progressively disabled, $\gamma$ decreases, causing an increase in the resistive scale given by
  \begin{equation}\label{ch1:selective}
  \frac{\ell_\eta}{\ell_{\eta 0 }} \sim \left(\frac{\ell_\mathrm{s}}{\ell_\visc}\right)^{1/3},
  \end{equation}
  where $\ell_{\eta 0}$ is the resistive scale in the kinematic regime.
  Saturation of the dynamo occurs when the suppression of field-aligned stretching ceases  and the scale containing the smallest eddies most responsible for the growth of the magnetic field, $\ell_\mathrm{s}$, stops increasing.  At this point the system has reached a steady state and the magnetic energy ceases to grow any further~\citep{Scheko_theory}. 

The above theory is non-local in nature in the sense that large-scale fluid motions interact directly with magnetic fields at the resistive scale. An alternative scenario was put forward~\citet{Beresnyak12} for $\mathrm{Pm} \sim 1$ systems, which posits that the important interactions are local. Here, an Alfv\'enic cascade is set up with scale-by-scale equation between the magnetic and kinetic energies at a scale $\ell^*$ such that the sum of the kinetic energy below this scale equals the total magnetic energy.  The peak of the magnetic energy, then, is located at the scale whose eddies contribute the strongest stretching. This scenario also results in linear-in-time growth of the magnetic energy.

\subsection{Open questions in the MHD fluctuation dynamo}

 One of the main outstanding questions on the fluctuation dynamo in a collisional system is how far up the inertial range can the stretching motion of eddies be suppressed, and which of the two scenarios presented in the previous section pertain to $\mathrm{Pm} \gg 1$ systems.
 In the Stokes flow regime where  $\mathrm{Re} \sim 1$ and $\ell_0 \sim \ell_\visc$, the system comes into saturation precisely when the nonlinearities become important on the outer scale and the final magnetic energy is indeed of the same order as the kinetic energy. This question is far more difficult to answer in the case $\mathrm{Re} \gg 1$, where a large inertial range is expected. While the stretching component of the smallest eddies is progressively suppressed by the Lorentz force, the mixing component is allowed to remain, and so resistive dissipation may put an upper limit on $\ba{v_\mathrm{A}^2}/\ba{u^2}$~\citep{Scheko_saturated}. In the extreme case where the mixing remains efficient,  $\ell_\mathrm{s} \sim \ell_\visc$. In this case, the magnetic energy saturates with a value that is smaller than the kinetic energy by a factor of $\mathrm{Re}^{1/2}$~\citep{Batchelor}. On the other hand, if $\ell_\mathrm{s}$ is allowed to get as large as $\ell_0$, then the magnetic energy will be of the same order as the kinetic energy.  Resolving both  $\mathrm{Re} \gg 1$ and $\mathrm{Pm} \gg 1$ is currently infeasible even with current computational resources, and so the best we can say at present is that $\ell_\mathrm{s}$ seems to saturate somewhere in between $\ell_0$ and $\ell_\visc$~\citep{Scheko_sim,Haugen04,Maron04}.

\begin{figure}
\centering
\includegraphics[width=0.7\textwidth]{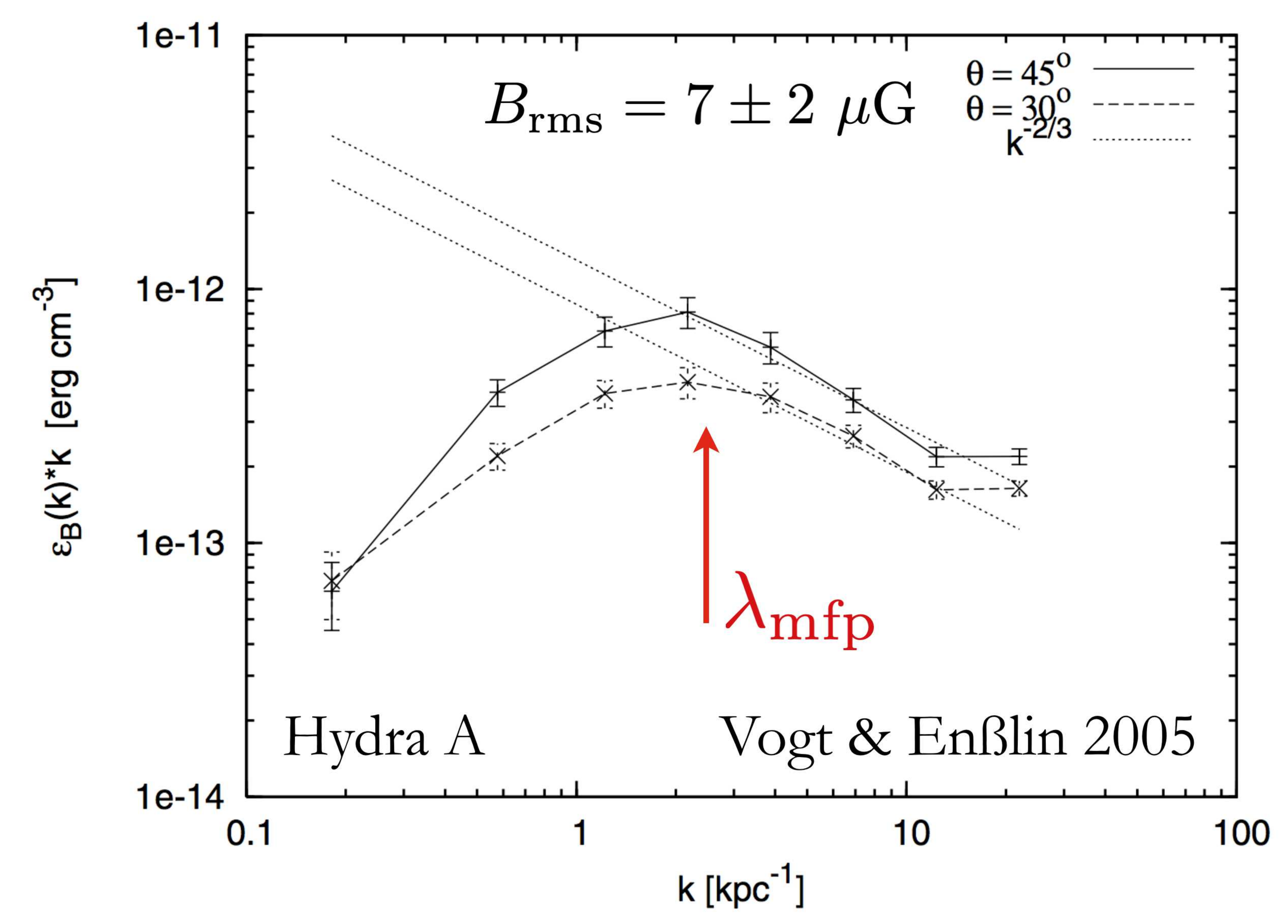}
\caption{\label{ch1:Vogt}Magnetic energy spectra of the Hydra A cluster.}
\end{figure}

The other important remaining question left unanswered on fluctuation dynamo in a collisional system is at what scale the bulk of the magnetic energy spectra ultimately resides, and whether the saturated state yields a scale-by-scale equipartition between the kinetic and magnetic energies~\citep{Biermann1951}.  This is not just an academic question, as  observations of magnetic fields in galaxy clusters indicates that the intracluster magnetic field eventually saturates with the bulk of its energy on scales comparable to the  collisional mean free path (see figure~\ref{ch1:Vogt}). 
This, of course, poses a significant challenge: in the kinematic regime, the characteristic scale of the magnetic field is roughly the resistive scale $\ell_\eta$, which is a factor of $\mathrm{Pm}^{1/2}$ smaller than the viscous scale. For typical ICM parameters assuming Spitzer resistivity and Coulomb collisions, this can be a factor of ${\sim}10^{15}$ (see table~\ref{ch1:tab_ICM}), and so somehow the spectral peak of the magnetic energy would have to migrate upwards by orders of magnitude to reach the scales needed to match observations. 
While this scale does migrate by simply deceasing the stretching rate (\S\ref{ch1:saturation}), if the stretching motion of every eddy up to the outer scale were suppressed, this would only result in an increase of the resistive scale by a factor of $\mathrm{Re}^{1/4}$, see equation~\eqref{ch1:selective}. The current collection of numerical evidence seems to be contradictory, with some pointing to resistive-scale magnetic fields~\citep{Maron04,Scheko_sim} while others claiming scales approaching the viscous-scale eddies and beyond~\citep{Haugen04,ChoDynamo,Beresnyak12}.  However, we have mentioned that this saturation process is strongly dependent on the material properties of the plasma, namely the Reynolds number.  While for the ICM, $\mathrm{Re}\sim 10$ based on the Coloumb scattering rate, suggesting that the ICM is reasonably well approximated by the Stokes flow regime, we shall see in the next section that this situation is not nearly as clean-cut as it first appears.

\section{The plasma dynamo} \label{ch:int:magnetized}

\subsection{Anisotropic viscous stress}

\label{int:anivisc}
Recall from \S\ref{ch1:motivation} the fact that, for typically  parameters, an ICM plasma is not rigorously a magnetohydrodynamic (MHD) fluid~\citep{Scheko_2005,KulsrudZweibel}. First, the ion--ion Coulomb collision frequency
\begin{equation}
    \nu_\mr{i} \approx 0.2 \left(\frac{n}{10^{-3}~{\rm cm}^{-3}}\right) \left(\frac{T}{5~{\rm keV}}\right)^{-3/2}~{\rm Myr}^{-1} 
\end{equation} 
is only a factor of ${\sim}100$ larger than the inverse dynamical time of the turbulent fluid motions at the largest scales,\footnote{Here we have normalized $n$ and $T$ to typical cluster values, for which the Coulomb logarithm ${\approx}40$.  The representative values given for the outer scale $\ell_0$ and its characteristic turbulent velocity $u_0$ are motivated by a variety of observational constraints on gas motions in nearby clusters \citep[e.g.][]{Hitomi1,Zhuravleva2018,Simionescu2019}.}
\begin{equation}
    t^{-1}_{\rm dyn} \approx 0.002 \left(\frac{u_0}{200~{\rm km~s}^{-1}}\right) \left(\frac{\ell_0}{100~{\rm kpc}}\right)^{-1}~{\rm Myr}^{-1} .
\end{equation}
Thus, ${\sim}1\%$ deviations from local thermodynamic equilibrium are to be expected -- i.e., the ICM plasma is {\em weakly collisional}. That the energy density of these deviations is comparable to that stored in the observed turbulent motions and magnetic-field fluctuations indicates that ${\sim}1\%$, while small, is nevertheless enough to be dynamically important. Second, even magnetic-field strengths as small as
\begin{equation}\label{eqn:Bmin}
    B \sim 10^{-18} \left(\frac{n}{10^{-3}\textrm{ cm}^{-3}}\right)\left(\frac{T}{5\textrm{ keV}}\right)^{-3/2}~{\rm G}
\end{equation}
are sufficient to ensure that the ICM plasma is {\em magnetized}, i.e., that the ion gyrofrequency $\Omegai \doteq eB/m_\mr{i}c$ is larger than $\nu_\mr{i}$. As seed magnetic fields are thought to be produced by various processes in the era preceding galaxy formation with magnitudes ${\sim}10^{-22}$--$10^{-19}~{\rm G}$~\citep[e.g.][]{Biermann1950,KulsrudZweibel},
the amplification of the intracluster magnetic field via the fluctuation dynamo occurs almost exclusively in the weakly collisional, magnetized regime, and is thus not appropriately described~\emph{a priori} by MHD with isotropic transport. Thus we must study the dynamo in a weakly collisional regime, hereafter referred to as the \emph{plasma dynamo}.

At magnetic-field strengths larger than that given by (\ref{eqn:Bmin}), departures of the plasma from local thermodynamic equilibrium are biased with respect to the magnetic-field direction~\citep{CGL}, and the transfer of momentum and energy across magnetic-field lines becomes stifled by the smallness of the particles' Larmor radii. In weakly collisional plasmas like the ICM, this system can be described using the Braginskii-MHD system~[see equations (\ref{eq:int:brag}a--d)]. We are now in a position to recast the pressure anisotropy given by equation~\eqref{ch:int:dp} using equation~\eqref{ch1:induction2}
\begin{equation}\label{eqn:bragvisc}
    \rmDelta p = 3 \visc_\mathrm{B} \left(\eb\eb - \frac{1}{3}\unitDyadic \right)\!\boldsymbol{:}\!\grad \bb{u},
\end{equation}
where $\visc_\mathrm{B} = p/\nu_\mathrm{i}$ is the field-aligned Braginskii viscosity~\citep{Braginskii}.\footnote{Factors of order  unity as derived by~\citet{Braginskii}, which vary amongst different collision operators, are subsumed into the definition of the collision frequency.} Notice that this pressure anisotropy, as well as the resulting viscous stress in equation~\eqref{ch1:mom_brag}, imply a significant departure from the kinematic regime of the dynamo in a \emph{collisional} plasma: while $\bb{B}$ may be small, $\eb$ is always $\mathcal{O}(1)$, and so the fluid velocity always has knowledge of the magnetic field through the parallel viscous stress. As a result, the `kinematic' regime of the plasma dynamo is fundamentally nonlinear. Note that it is the parallel rate of strain that appears in equation~\eqref{eqn:bragvisc}, and so it is worthwhile to reiterate the point made in \S\ref{sec:eq:brag} that faster growth/decay of the magnetic field results in larger amounts of parallel viscosity, and so in Braginskii-MHD the parallel viscosity attempts to damp any motion that changes the magnetic field strength, \emph{greatly hindering the dynamo!}

 The extreme limit of $\nu_\mathrm{i} /\Omegai \ll 1$ is even more disastrous to the dynamo: equation~\eqref{eq:CGL1} states that any increase in the strength of the magnetic field must be accompanied adiabatically by a commensurate increase in the pressure perpendicular to the field. To increase the magnetic-field strength tenfold (let alone by $10^{10}$), the thermal pressure must also increase tenfold, which would require an enormous amount of free energy.  Indeed, simulations of the dynamo in the double-adiabatic regime with energy injection comparable to $u_0^3/\ell_0$ result in \emph{no appreciable growth of the magnetic energy whatsoever}~\citep{SantosLima,Helander_constraints}. Thus the fluctuation dynamo cannot exist in a purely collisionless system without some mechanism to break the conservation of adiabatic invariants. Even in a weakly collisional environment, equation~\eqref{eqn:bragvisc} taken at face value would suggest a parallel Reynolds number $\mathrm{Re}_\parallel$ set by Coulomb collisions.  In the ICM, this would lead to only modest values of the parallel Reynolds number (${\sim}10$), and would place significant limits on how fast the dynamo would be allowed to operate.  This puts constraints on the viability of some dynamo models to explain the observed magnitude of cosmic magnetic fields: as the lifetime of the Universe is finite, too sluggish a dynamo may not amplify a seed field sufficiently fast to be consistent with observations.  However, we shall see in the next section that this constraint will be swiftly alleviated.

\subsection{Larmor-scale kinetic instabilities}

In practice, the Braginskii parallel viscous stress
\begin{equation}\label{eq:bragten}
    \msb{\Uppi}_\parallel \doteq -\left(\eb\eb - \frac{1}{3} \unitDyadic \right) \rmDelta p, 
\end{equation}
with the pressure anisotropy $\rmDelta p$ specified by equation \eqref{eqn:bragvisc}, is only suitable for plasmas with small to order-unity values of $\betai \doteq 8 \upi p_\mathrm{i}/ B^2$, the ratio of the ion thermal and magnetic pressures. The reason is that plasmas are susceptible to several kinetic instabilities when $|\rmDelta p/p| \gtrsim 1/\betai$, such as the firehose~\citep{Rosenbluth56,Parker58,Chandrasekhar58,HellingerMatsumoto00} and mirror~\citep{Barnes66,Hasegawa69,SouthwoodKivelson93,Hellinger07} instabilities. Before we can understand their (crucial) role in the operation of the dynamo in weakly collisional and collisionless plasmas, let us first get an overview of their mechanism for instability and their effect on the electromagnetic fields and plasma.

\subsubsection{firehose instability}

\begin{figure}
\centering
\includegraphics[width=\textwidth]{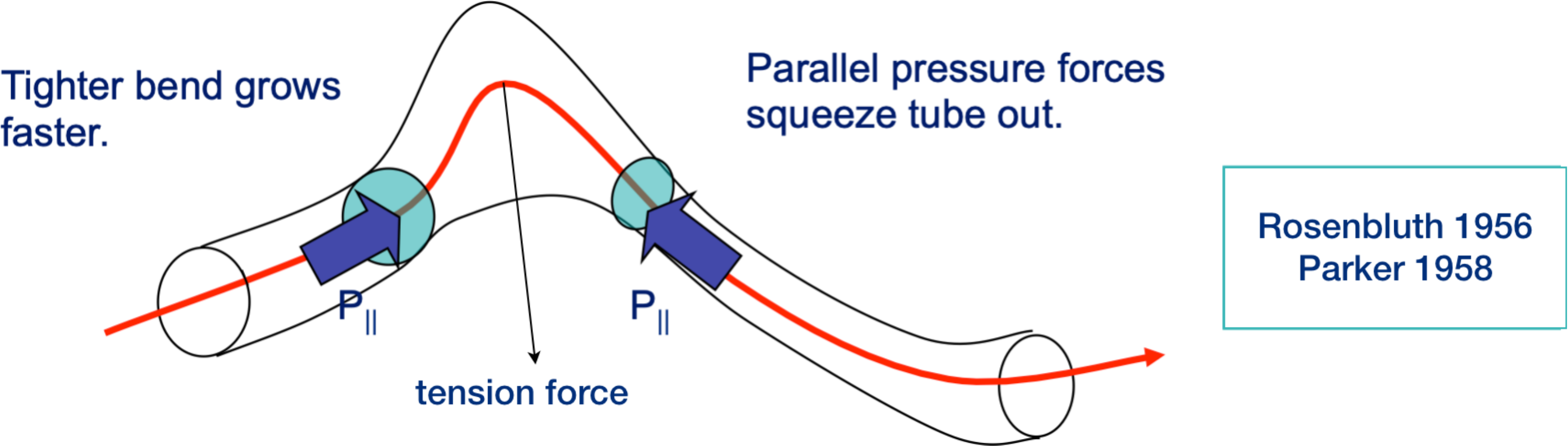}
\caption[The firehose instability.]{\label{ch1:fig_firehose} An illustration of the firehose instability, whereby an excess of thermal pressure parallel to the magnetic field causes any kink along the field line to buckle, leading to instability. See~\citet{Rosenbluth56,Parker58}.}
\end{figure}

We consider the (parallel) firehose instability first, which is captured by an MHD system supplemented with the Braginskii viscosity (equation~\ref{ch1:mom_brag}). One can see from equation~\eqref{ch1:mom_brag} that, if the parallel pressure satisfies
\begin{equation}
p_\parallel > p_\perp +\frac{B^2}{4\upi},
\end{equation}
then the magnetic tension force becomes unable to undo any bending of the field and the system becomes unstable (akin to a high-pressure firehose). This scenario is illustrated in figure~\ref{ch1:fig_firehose}. We can rewrite this stability requirement by defining the (ion) anisotropy parameter $\Deltai$ (distinct from $\rmDelta p$) as
\begin{equation}
\Deltai \doteq \frac{p_\perp - p_\parallel}{p_\parallel}.
\end{equation}
Then the requirement for firehose instability becomes
\begin{equation}
\Deltai < - \frac{2}{\betai}.
\end{equation}
 (This analysis is performed in detail in appendix~\ref{ap:linear}.)
 Once the firehose instability grows, sharp structures in the magnetic field develop on the order of the ion gyroradius. This gives rise to fluctuations in both the magnetic and velocity fields that modify the rate of strain in such a way as to introduce \emph{positive} pressure anisotropy, which attempts to cancel out the initial negative pressure anisotropy driving the instability~\citep{Rosin_2011}. 
 When kinetic effects are considered, an additional branch of the firehose instability appears called the oblique firehose~\citep{HellingerMatsumoto00}. This branch is not only faster than the parallel branch, but also has the ability to scatter particles and thus break adiabatic invariance~\citep{Kunz_kin}.

\begin{figure}
\centering
\includegraphics[width=\textwidth]{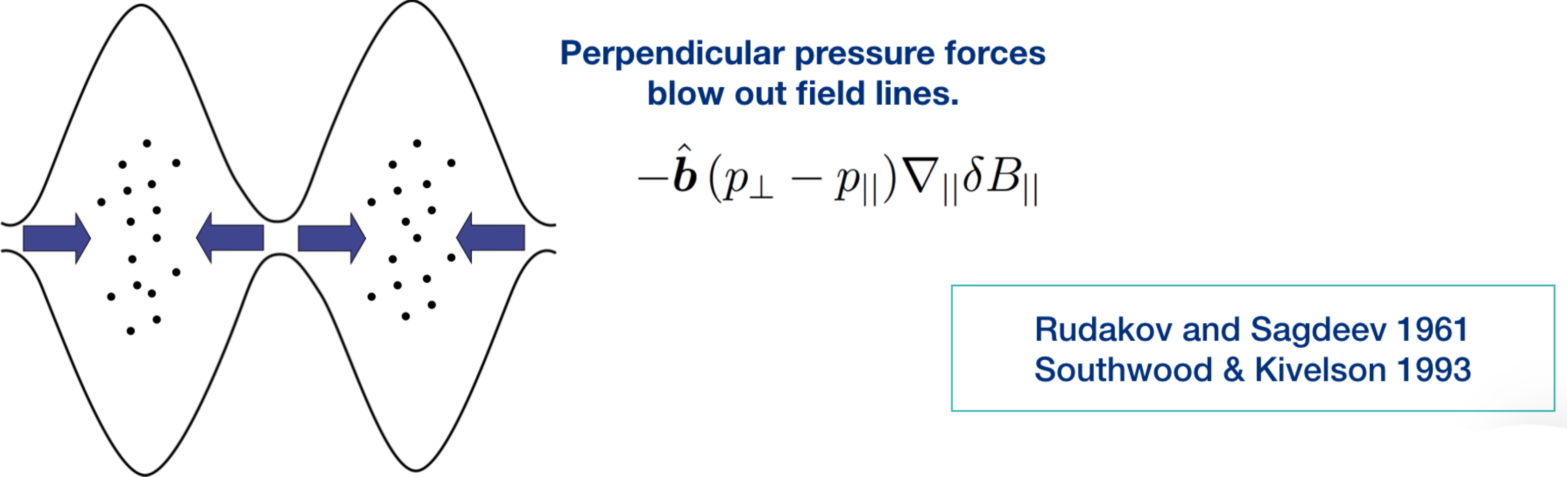}
\caption[The mirror instability.]{\label{ch1:fig_mirror} An illustration of the mirror instability, whereby particles interact in `bubble'-like magnetic mirror structures. A resonant particles whose parallel velocity relative to wave is small experience Barnes damping, giving energy to the magnetic field and increasing the pressure in the troughs. If the initial perpendicular pressure exceeds the parallel one, the increase in pressure is more than can be balanced by the magnetic pressure, and the magnetic field strength at the cusps (troughs) get stronger (weaker), leading to stronger Barnes damping and thus instability, see~\citet{SouthwoodKivelson93}.}
\end{figure}

\subsubsection{mirror instability}

Unlike the firehose instability, the mirror instability has no true fluid analog and must be treated using  kinetic theory. Consider a slow-mode perturbation to an otherwise uniform magnetic field, resulting in regions of high and low field strength. Particles that are resonant with this mode ($v_\parallel \sim 0 $ in the wave frame) exchange energy with the wave via the mirror force~\citep{Barnes66}. For an initial distribution function having a majority of large-pitch-angle particles, the proportional increase of these particles in the magnetic troughs ($\delta B_\parallel < 0$) inflates the field lines (in order to maintain perpendicular pressure balance). If the concentration of these particles leads to more perpendicular pressure than can be stably balanced by the magnetic pressure, the troughs must grow deeper to compensate, strengthening the mirror force, and thus leading to instability~\citep{SouthwoodKivelson93, Kunz_2015}. It is shown in appendix~\ref{ap:linear} that the process becomes unstable when the anisotropy parameter satisfies~\citep{SouthwoodKivelson93}
\begin{equation}
\Deltai \gtrsim  \frac{1}{\betai}.
\end{equation}
As the magnetic-field lines balloon outwards, they eventually develop sharp bends on length scales comparable to the  ion gyroradius~\citep{Kunz_kin}. These bends serve as scattering centers for particles, breaking adiabatic invariance and saturating the instability.

\subsection{Kinetic instabilities, pressure anisotropy and the dynamo}

\label{sec:pressureDynamo}

\begin{figure}
\centering
\includegraphics[width=1\textwidth]{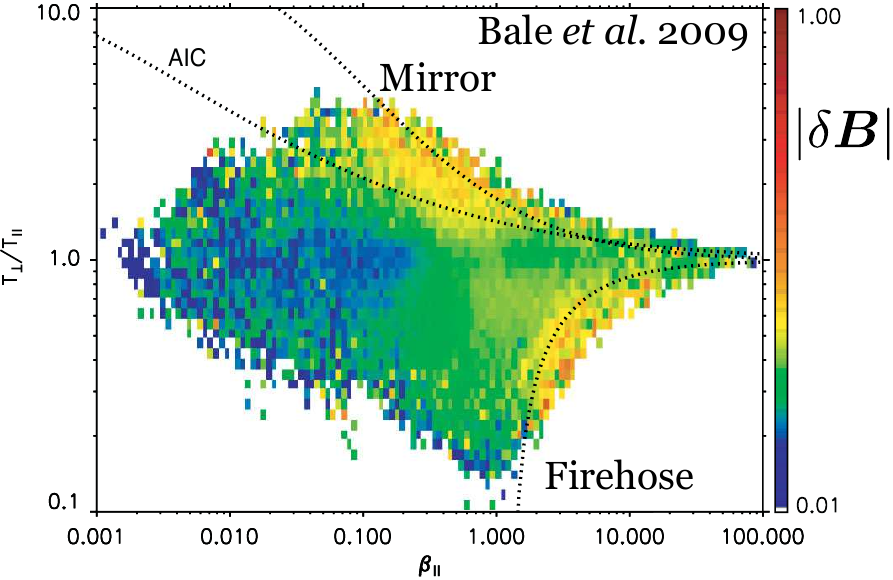}
\caption[PDF of temperature anisotropy and magnetic fluctuations in the solar wind.]{\label{ch1:bale}Joint PDF of the temperature anisotropy and the magnitude of the magnetic field fluctuations in the solar wind, which clearly indicates regulation of the pressure anisotropy by the firehose and mirror instabilities.  From~\citet{Bale}.}
\end{figure}

As these instabilities develop and progress beyond their linear stages, they begin to limit the pressure anisotropy by scattering and trapping particles in such a way as to isotropize the distribution. In particular, if the effective collisionality $\nueff$ of the system satisfies
\begin{equation}\label{eqn:nu_eff}
\nueff \gtrsim \betai |\ROS|,
\end{equation}
then the instabilities shape the particle distribution function such that the pressure anisotropy 
 \begin{equation}\label{stab_bound}
 \Deltai \in \left[-\frac{2}{\betai}, \frac{1}{\betai}\right],
 \end{equation}
 where the lower (upper) threshold is determined by the firehose (mirror) instability.
This  effect has been diagnosed in various kinetic particle-in-cell simulations \citep{Kunz_kin,Riquelme_2015,Hellinger_2015,Melville} and directly observed using \emph{in situ} measurements of particle distribution functions and magnetic fluctuations in the solar wind~\citep[][see also figure~\ref{ch1:bale}]{Kasper02,hellinger06,Bale,Chen16}. 
Thus, as the magnitudes of pressure anisotropy (and thus the parallel viscosity) specified by equation \eqref{eqn:bragvisc} are often unphysically large in high-$\betai$ plasmas, any fluid model of the fluctuation dynamo must adopt some form of microphysical closure to account for this otherwise absent regulation of the pressure anisotropy.

 One may ask, then, how exactly do these instabilities operate in the typical folded fields produced by the fluctuation dynamo? We know that firehose and mirror instabilities reside in regions of negative and positive pressure anisotropy, respectively, or equivalently from equations~(\ref{eq:CGL}a--b), regions of decreasing and increasing magnetic-field strength. Thus in regions where the magnetic field is growing, the mirror instability should occur. On the other hand, regions of magnetic field decay should be populated by the firehose instability. Through the appearances of these Larmor-scale instabilities, one can hope to alleviate the concerns raised in \S\ref{int:anivisc}.

\subsection{The three regimes of the plasma dynamo} \label{sec:paniso}

We must learn how efficiently these instabilities can regulate the pressure anisotropy in high-$\betai$ plasmas. In this subsection, we argue that there are three relevant operational regimes of the plasma dynamo, and give approximate magnetic field amplitudes for when they occur in the ICM.

We do this by considering when in the evolution of the dynamo can equation~\eqref{eqn:nu_eff}, which is the criterion needed for perfect regulation of the pressure anisotropy, be satisfied.
Following the reasoning presented in \S\,4.2.2 of \citet{Melville}, we estimate the effective parallel viscosity $\visc_{\parallel{\rm eff}} \doteq \vth{i}^2/\nu_\mathrm{eff}$, and thus the effective parallel-viscous Reynolds number $\Reprl$ associated with the enhanced collisionality \eqref{eqn:nu_eff}, as follows: The~\citet{Kolmogorov1941} scaling $u_{\ell_\parallel} \propto \ell_\parallel^{1/3}$ for the field-stretching turbulent velocity at parallel scale $\ell_\parallel$ implies that the magnitude of the field-parallel rate of strain $|\ROS| \sim u_{\ell_\parallel}/{\ell_\parallel} \propto {\ell_\parallel}^{-2/3}$ is largest at the effective parallel-viscous scale $\ell_{\visc_\parallel}$, where such motions are dissipated. The value of $\Reeff$ corresponding to \eqref{eqn:nu_eff} is then
\begin{align}\label{eqn:Re_eff}
\Reeff \doteq \frac{u_0\ell_0}{\visc_{\parallel{\rm eff}}} \sim u_0 \ell_0 \frac{\nu_\mathrm{eff}}{v_\mathrm{thi}^2}\sim \betai M^2 \Reeff^{1/2} \quad \Longrightarrow\quad \Reeff \sim \betai^2 M^4 ,
\end{align}
where we have used $|\ROS| \sim (u_0/\ell_0) \Reeff^{1/2}$. (Recall that $u_0$ is the characteristic speed of the outer-scale bulk fluid motions, $\ell_0$ is the energy injection (outer) scale, and $M \doteq u_0/v_\mathrm{thi}$ is the Mach number.) This implies a smaller viscous cutoff,
\begin{equation}\label{eqn:ellvisc}
    \ell_{\visc_\parallel} \sim \ell_0 \, \Reeff^{-3/4} \sim %
    \ell_0 \, \betai^{-3/2} M^{-3} ,
\end{equation}
than the cutoff effected by Coulomb collisions 
and therefore a larger maximal shear rate,
\begin{equation}\label{eqn:Svisc}
    S_{\visc_\parallel} \sim \frac{u_0}{\ell_0} \, \Reeff^{1/2} \sim \frac{u_0}{\ell_0}\, \betai M^2 .
\end{equation}
Note that $\ell_{\visc_\parallel}\propto B^3$ increases and $S_{\visc_\parallel} \propto B^{-2}$ decreases with increasing magnetic-field strength. %

We now use the value of $S_{\visc_\parallel}$ given by~\eqref{eqn:Svisc} in \eqref{eqn:nu_eff}. For $\betai \gtrsim (\ell_0/d_{\rm i})^{2/5} M^{-6/5}$, where $d_{\rm i}$ is the (field-strength-independent) ion inertial scale, or
\begin{equation}\label{eqn:ultrahighbeta}
B \lesssim 6 \left(\frac{n}{10^{-3}\textrm{ cm}^{-3}}\right)^{2/5}\left(\frac{T}{5\textrm{ keV}}\right)^{1/2}\left(\frac{M}{0.2}\right)^{3/5}\left(\frac{\ell_0}{100\textrm{ kpc}}\right)^{-1/5}~\mathrm{nG},
\end{equation}
the collision frequency needed to pin the pressure anisotropy to the marginal-stability threshold is greater than the ion gyrofrequency $\Omegai$. (This is called the `ultra-high-$\betai$' limit in \citealp{Melville}.) If the effective collision frequency is comparable to the maximum growth rates of the firehose and mirror instabilities, which are smaller than the gyrofrequency by factors of ${\sim}\sqrt{|\rmDelta p| / p}$ and ${\sim}\rmDelta p/p$, respectively \citep[e.g.][]{HellingerMatsumoto00,Hellinger07,Rosin_2011}, the pressure anisotropy cannot in this case be efficiently regulated to be bounded by the instability thresholds.

The conditions \eqref{eqn:Bmin} and \eqref{eqn:ultrahighbeta} suggest three distinct regimes: 
\vspace{1ex}
\begin{enumerate}
    \item the {\em unmagnetized regime}, when $B \lesssim 10^{-18}$ G; 
    \item the {\em magnetized `kinetic' regime} (ultra-high-$\betai$), when $10^{-18}\textrm{ G} \lesssim B \lesssim 6 \textrm{ nG}$ and for which the regulation of the pressure anisotropy by kinetic instabilities is inefficient; and 
    \item the {\em magnetized `fluid' regime}, when $B \gtrsim 6\textrm{ nG}$ and for which the pressure anisotropy can be well regulated by the instabilities (i.e.~$\nu_{\rm eff} \lesssim \Omegai$). 
\end{enumerate}
\vspace{1ex}
The saturated state of the dynamo, in which the magnetic and kinetic energies are comparable, would be obtained when
\begin{equation}
B_{\rm sat} \sim 3 \left(\frac{n}{10^{-3}\textrm{ cm}^{-3}}\right)^{1/2}\left(\frac{T}{5\textrm{ keV}}\right)^{1/2}\left(\frac{M}{0.2}\right) ~\upmu\mr{G} 
\end{equation}
and thus occurs in the magnetized regime. Perhaps coincidentally, this value is close to the field strength at which the effective scattering due to Coulomb collisions is sufficient to satisfy~\eqref{eqn:nu_eff}:
\begin{equation}\label{eqn:collisional}
B \gtrsim 2 \left(\frac{n}{10^{-3}\textrm{ cm}^{-3}}\right)^{1/4}\left(\frac{T}{5\textrm{ keV}}\right)\left(\frac{M}{0.2}\right)^{3/4}\left(\frac{\ell_0}{100\textrm{ kpc}}\right)^{-1/4}~\upmu\mathrm{G}.
\end{equation}

 \subsubsection{explosive growth}

An intriguing feature of these regimes is that, while the Reynolds number in the unmagnetized regime is set by Coulomb collisions, resulting in $\Reprl \sim 1 \textrm{--} 100$~\citep{Scheko_2005}, at the transition from the second (magnetized kinetic) regime to the third (magnetized fluid) regime we find $\Reeff \sim \betai^2 M^4 \gg 1$. This suggests that $\Reeff$ must experience a large increase at some time between these two epochs. Since the viscous-scale rate of strain increases as $\Reeff^{1/2}$, the dynamo in this intermediate second regime should be self-accelerating, with the field-stretching eddies becoming smaller and faster as the magnetic field is amplified.  This can potentially lead to explosive growth of the magnetic energy. One could imagine a scenario where the scattering rate is controlled by the firehose instability with growth rate ${\sim}(\Delta p /p)^{1/2} \Omegai$, resulting in $\nueff \sim B^\alpha$ and $\alpha$ is a positive exponent. Then
\begin{align}
\frac{\od\, \ln B}{\od t} \propto \Reeff^{1/2} \propto B^{\alpha/2},
\end{align}
and so $B(t) \sim B_0/ ( 1-t/ t_\mathrm{c} )^{2/\alpha}$, where $t_\mathrm{c}$ is a constant dependent on the specifics of the scattering.  This exhibits explosive growth in finite time $t = t_\mathrm{c}$; similar scenarios have previously formed the basis for theories of explosive dynamo in collisionless plasmas~\citep{SchekoCowley06a,SchekoCowley06b,Melville,Mogavero}.

\subsection{Previous results and current status of the plasma dynamo}

The first two regimes have been previously studied through the use of hybrid-kinetic numerical simulations by~\citet{Rincon_2016}, who observed the generation of firehose and mirror instabilities as the dynamo entered the magnetized regime.  However, due to computational constraints, they were not able to go much further than the initial diffusion-free regime for simulations that were initially magnetized ($L/\rhoi  > 1$).  \citet{SantosLima} have recently studied the effects of pressure-anisotropy regulation during the fluctuation dynamo using a collisionless double-adiabatic closure to evolve $p_\perp$ and $p_\parallel$~\citep{CGL} supplemented by a non-zero collision frequency $\nu_\mr{eff}$ that is activated in spatial regions of kinetic instability. It was found that simulations with instantaneous pressure-anisotropy relaxation exhibited magnetic-field growth rates similar to those in isotropic MHD, with some minor details differing in the saturated state. As the effective collision frequency was lowered, the dynamo growth rates and the final value of the saturated magnetic energy decreased. In the entirely collisionless case of $\nu_\mr{eff}=0$, the pressure anisotropy was allowed to grow arbitrarily large, and no growth of the magnetic energy was observed. This is consistent with recent theoretical considerations of magnetic-field amplification occurring under adiabatic invariance in collisionless plasmas~\citep{Helander_constraints}, which found that dynamo action always requires collisions or some small-scale kinetic mechanism for breaking the adiabatic invariance of the magnetic moment.

 At the moment, the third (magnetized fluid) regime is prohibitively expensive to investigate using kinetic simulation in any regime except near the saturated state. To appreciate this difficulty, let us imagine that one wishes to resolve two decades of magnetic-energy growth (equivalently, one decade in growth of the magnetic-field strength) in this regime in a single simulation. The constraints on the initial plasma beta $\betaio$ required to simulate this regime are, in terms of the controllable simulation parameters,
\begin{equation}
    \frac{A}{M^2} \lesssim \betaio \lesssim \frac{1}{M^{3/2}}  \left(\frac{\ell_0}{\rhoio}\right)^{1/2}, %
\end{equation}
where $A$ is the desired magnetic-energy amplification factor and $\rhoio$ is the ion Larmor radius at the transition from the second to this third regime. The first inequality follows from the dynamo  not yet being saturated (i.e.~$\betai M^{-2} \lesssim 1$), while the second inequality follows from the requirement $\nu_{\rm eff} \lesssim\Omegaio$. For such a range of $\betaio$ to exist, 
\begin{align*}
    A \lesssim M^{1/2} \left(\frac{\ell_0}{\rhoio}\right)^{1/2} .
\end{align*}
To allow for an appreciable range of field amplification in this regime, we must then maximize $M$ and $\ell_0/\rhoio$. If we demand $M \lesssim 0.2$ (so as to avoid the possibility of shocks occurring at larger Mach numbers and to maintain relevance to the sub-sonic turbulence observed in the ICM), then resolving two decades of energy growth ($A \sim 10^2$) requires $\ell_0/\rhoio \gtrsim 50,000$. If we then wish to resolve the ion-Larmor radius at the end of this growth by a minimum of two cells, then the number of cells needed in each spatial direction is ${\sim}10^6$! For a problem that is intrinsically three-dimensional~\citep{Cowling1933,Zeldovich57}, this requirement is well beyond current computational capabilities.

\section{This thesis}

\subsection{Our goals}

It is clear from the previous sections that the plasma dynamo is a rich and complex problem that must incorporate numerous aspects of plasma physics, from transport theory to Larmor-scale kinetic physics, in order to successfully describe how a collisionless plasma self-consistently allows itself to amplify magnetic fields.

\begin{figure}
\centering
\includegraphics[width=\textwidth]{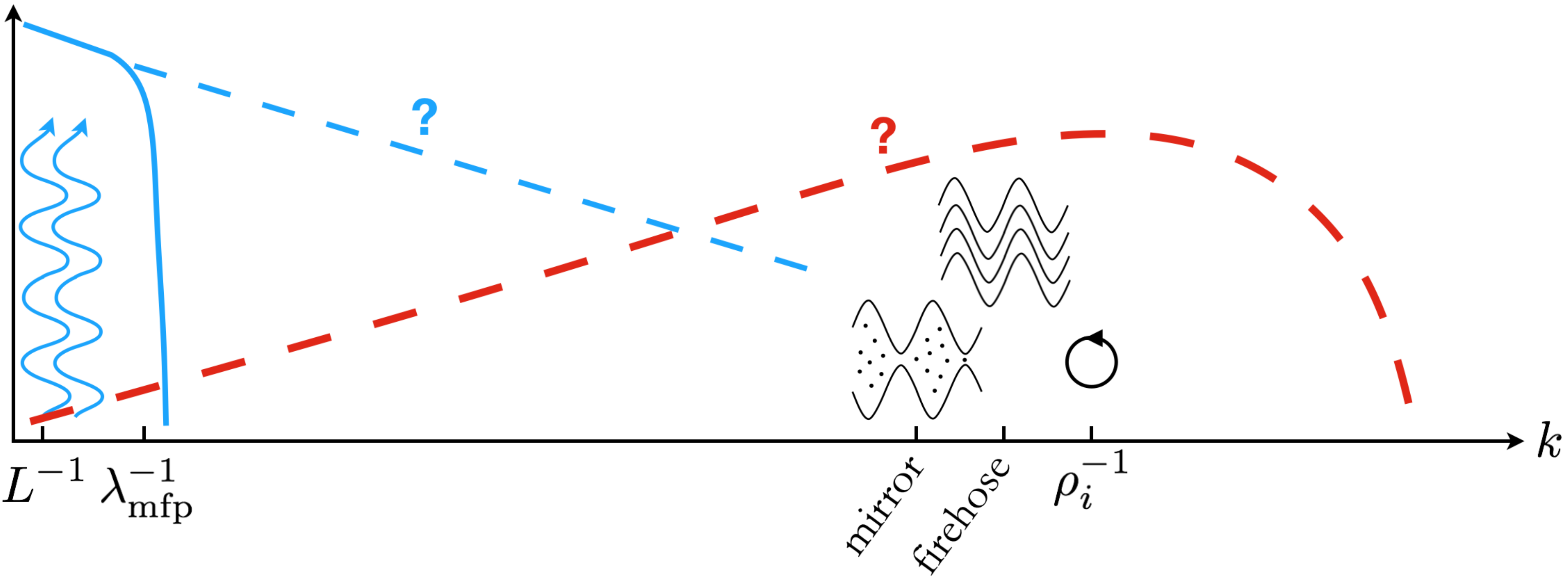}
\caption[Kinetic consideration embedded on the fluctuation dynamo. ]{\label{ch1:kinematic_kinetic} The length scales of interest in the plasma dynamo.  Along with the viscous and resistive scales, we also need to consider the ion gyroradius and the characteristic scales of the mirror and firehose instabilities. The latter may lead to changes in the plasma viscosity, which may result in an extended energy cascade. }
\end{figure}

Thus, one component of this thesis is to discover how the firehose and mirror instabilities influence the natural progression of the dynamo. This will be done by exploring the second plasma dynamo regime using the hybrid-kinetic, particle-in-cell code \textsc{Pegasus} \citep{Pegasus}, which is discussed in \S\ref{ch:method}. 
The picture we shall keep in mind as we do so is illustrated in figure~\ref{ch1:kinematic_kinetic}. Previously with the fluctuation dynamo in collisional MHD, small-scale eddies (denoted by the blue lines) give rise to a Kazantsev magnetic spectrum in the kinematic regime, leading to a magnetic field with a spectral peak at the resistive scale. Moving into the weakly collisional regime, we must now place on this diagram an ion Larmor radius somewhere between the viscous eddies and resistive scale.  Along with this gyroradius are the Larmor-scale instabilities, namely firehose and mirror, that will affect the particle dynamics, leading to changes in the plasma viscosity.  These changes may lead to an extension of the kinetic energy cascade, potentially resulting in faster stretching and thus faster growth of the magnetic energy.

\begin{figure}
\centering
\begin{subfigure}[b]{0.45\textwidth}
\includegraphics[width=\textwidth]{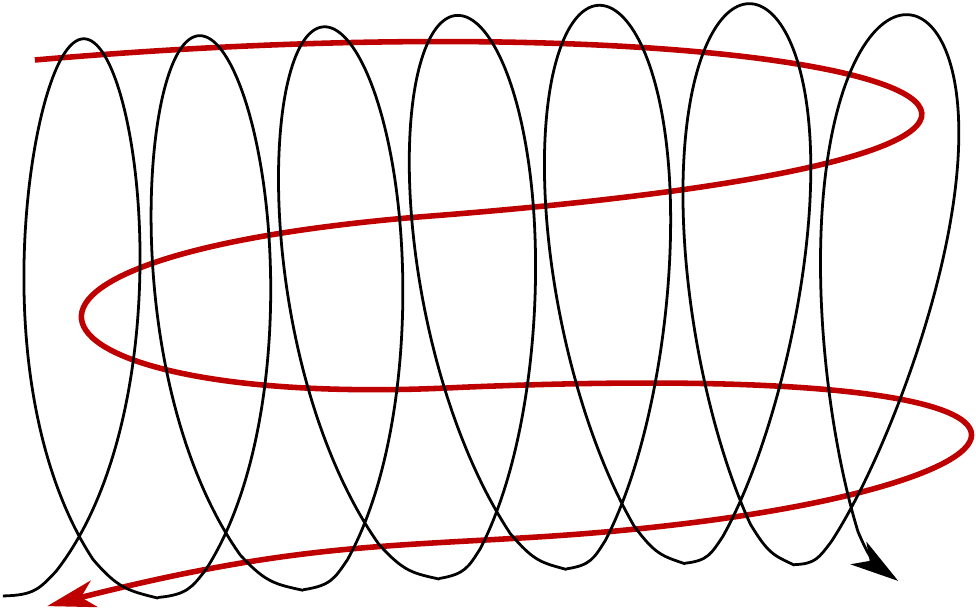}
\caption{$k_{\scriptscriptstyle \bs{B}\bstimes\bs{J}}\rhoi \gg 1$}
\end{subfigure}
\begin{subfigure}[b]{0.45\textwidth}
\includegraphics[width=\textwidth]{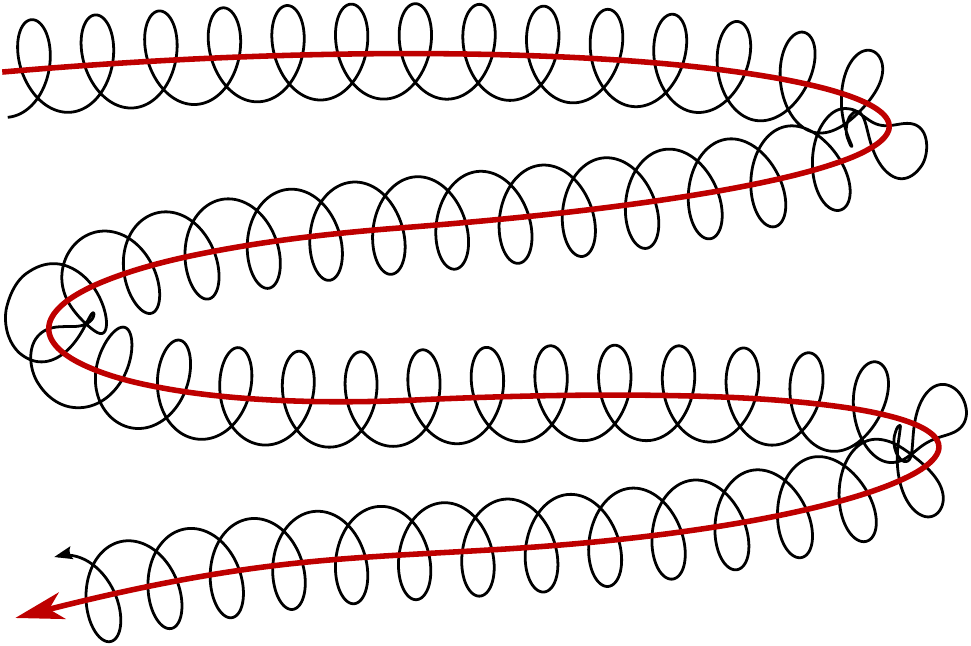}
\caption{$k_{\scriptscriptstyle \bb{B}\bstimes\bb{J}}\rhoi \ll 1$}
\end{subfigure}
\caption[Relative size of magnetic field folds and ion gyroradius.]{\label{fig:rhoi_fold}Two potential scenarios of the dynamo arise depending on the relative size of the ion gyroradius $\rhoi$ and the magnetic field fold separation ($k_{\scriptscriptstyle \bb{B}\bstimes\bb{J}}$, see equation~\ref{char-wavenumbers}). \emph{Left:} the ion samples several different magnetic fields, and in a sense becomes unmagnetized. \emph{Right:} the ion travels along the length of the fold, potentially being scattered by firehose near the fold bends or getting trapped by mirrors in the straight regions (see \S\ref{sim:sec:kinematic}).}
\end{figure}

 We must also be cognizant of finite ion-Larmor-radius effects when considering the weakly collisional and collisionless regimes. In particular, if the system is collisionless and the effective resistive scale is comparable to the electron gyroradius or skin depth, then the fold separation of the resulting magnetic structures is \emph{always} smaller than the ion gyroradius, which is illustrated in figure~\ref{fig:rhoi_fold}(a). In this scenario, an ion can sample several  magnetic fields in opposing directions which may lead them to lose their sense of magnetization. As this happens during a single gyro-orbit, these particles may undergo Bohm-like diffusion $D_\mathrm{B}$ through the magnetic field, where a particle undergoes a `collision' essentially once a gyro-orbit:
 \begin{equation}
 D_\mathrm{B} \sim \frac{(\textrm{step size})^2}{\textrm{time between steps}} \sim \frac{\rho_\mathrm{i}^2}{\Omegai^{-1}} \sim \frac{cT_\mathrm{i}}{eB}.
 \end{equation}
 On the other hand, if some other heretofore unknown process limits the fold separation in such a way that it becomes larger than the gyroradius, or if a particle is insensitive to the details of sub-Larmor-scale magnetic fields, then an ion can travel along the length of a magnetic fold [figure \ref{fig:rhoi_fold}(b)], possibly becoming trapped in mirror instabilities or scattering by firehoses.  How these Larmor-scale instabilities operate in the former regime is an interesting question in its own right;  the analytical theory built for these instabilities has so-far been derived using guide fields which possess net flux and a well-defined Larmor radius. However, the fluctuation dynamo deals with system \emph{without} net flux and naturally results in magnetic fields with fine-scale structures and regions of both small and large magnetic energies. In this case, the size of a Larmor radius depends on the particular location of the particle.

We also study the plasma dynamo in the third regime using a reduced set of fluid equations that takes the effects of kinetic instabilities into account through the use of microphysical closures that limit the pressure anisotropy. For this, we use the incompressible MHD equations including the parallel component of the Braginskii viscosity tensor, the latter being limited by hand to kinetically stable values (see (\ref{stability}) and \S\,\ref{method:braginskii}). In addition, in order to understand better the effects of this regulation and the nature of the dynamo in the (second) magnetized `kinetic' regime, we also perform a number of Braginskii-MHD simulations without any microphysical pressure-anisotropy regulation.  We also ascertain whether certain aspects of the hybrid-kinetic simulations can be faithfully reproduced by these fluid simulations with microphysical closures. Finally, we devise a new set of closures that aim to capture the explosive scenario discussed in this chapter, a scenario which has yet to be observed in simulations of the fluctuation dynamo employing full geometry.

\subsection{The thesis layout}

This Thesis is laid out as follows: In chapter~\ref{ch:method}, we explain the numerical codebases, statistical procedures and diagnostics we use to perform simulations of the plasma dynamo and study their results. In~\ref{ch:simulation}, we perform \emph{ab initio} hybrid-kinetic simulations of the plasma dynamo in the magnetized regime using the particle-in-cell code \textsc{Pegasus} \citep{Pegasus}.  In chapter~\ref{ch:brag}, we perform simulations of the plasma in a weakly collisional, Braginskii-MHD plasma, both with and without pressure-anisotropy limiters.  By doing so, we can investigate the effects of anisotropic viscosity on the dynamo in a controlled environment and gain insight on how pressure anisotropy limiters change the character of the dynamo. With the knowledge gained in these two chapters, we revisit a result found chapter~\ref{ch:simulation}, which illustrates that the nature of the dynamo in a collisionless plasma exhibiting features of both $\mathrm{Re}\sim 1$ and $\mathrm{Re} \gg 1$ dynamos, and thus does not fit neatly into either of the categories of limited or unlimited Braginskii-MHD. Finally, motivated by the discussion in \S\ref{sec:paniso}, we formulate a novel set of pressure anisotropy limiters in order to access the explosive growth scenario proposed by~\citet{SchekoCowley06a,SchekoCowley06b}.  This thesis is then summarized in chapter~\ref{ch:conclusion}, and various avenues of future research are proposed.

\chapter{Methods of Solution and Diagnostics}\label{ch:method}

In this section we list the two numerical codebases that we utilize in our simulations and the equations they solve.  We also record several different  Reynolds numbers and present various diagnostics used to the analyze the output of simulation data.

\section{Numerical codes}

\subsection{Hybrid-kinetics using \textsc{Pegasus}}\label{method:hybrid}

 \textsc{Pegasus} \citep{Pegasus} solves the hybrid-kinetic system of equations, which treat ions kinetically and electrons as a fluid. This system is derived in \S\ref{int:hybridkin}, and the resulting  equations are [\emph{cf.} equations~\eqref{intro:hybkin}]
  \begin{subequations}
 \begin{gather}
  \frac{\partial f_\mathrm{i}}{\partial t} + \bb{v}\bcdot \grad f_\mathrm{i} + \frac{1}{m_\mathrm{i}}\left[ \tilde{\bb{f}} +Ze\left( \bb{E} + \frac{\bb{v}}{c}\btimes\bb{B}\right)\right]\bcdot \frac{\partial f_\mathrm{i} }{\partial \bb{v}} = 0, \\
  \grad \btimes \bb{B} = \frac{4\upi }{c}\bb{J},\\
  \frac{\partial \bb{B}}{\partial t} = -c\grad \btimes \bb{E},\\
  \bb{E} + \frac{1}{c}\bb{u}_\mathrm{i}\btimes \bb{B} - \frac{\eta}{c}\grad \btimes \bb{B} +   \frac{\eta_\mathrm{H}}{c}\grad \btimes \nabla^2 \bb{B}  = -\frac{T_e \grad n_\mathrm{i}}{e n_\mathrm{i}} +  \frac{1}{Ze n_\mathrm{i}c}\bb{J}\btimes \bb{B}. \label{eq:sim:ohms}
 \end{gather}
 \end{subequations}
 The Ohm's law~\eqref{eq:sim:ohms} that we utilize in our hybrid-kinetic system includes the Hall electric field (last term on the RHS of eq.~\ref{eq:sim:ohms}) as well as the thermo-electric field driven by pressure gradients in the massless electron fluid (first term on the RHS of eq.~\ref{eq:sim:ohms}). 
  It also contains magnetic diffusivities in the form of an Ohmic resistivity $\eta$ and hyper-resistivity $\eta_\mathrm{H}$; as the Ohmic resistivity formally vanishes in the limit $m_\mathrm{i}\rightarrow 0$, these terms are meant to be a sink of magnetic energy and serve as a microphysical closure for dissipative electron dynamics that are not captured by hybrid kinetics. We also include a random forcing $\tilde{\bb{f}}$ in the Vlasov equation that serves as a source of free energy.

 In Pegasus, the \emph{full} distribution function $f_\mathrm{i}$ is sampled using macroparticles of finite extent.  Second-order--accurate triangle-shaped stencils are used for interpolating the electromagnetic and forcing fields to the particle positions, as well as for depositing moments of $f_\mathrm{i}$ onto the grid. These macroparticles are evolved using the Boris method~\citep{Boris}, which has been shown to conserve phase-space volume~\citep{HongBoris}.  The electromagnetic fields are evolved using the constrained transport method~\citep{Evans_CT}. Here, the components of the magnetic field $B_i$ are evaluated at the center of their respective grid-cell face, while the components of the electric field $E_i$ are evaluated at the center of their respective grid-cell vertex, a configuration known as the `Yee lattice'~\citep{Yee}. This method ensures that a magnetic field that is initially solenoidal will remain so with machine precision for all time.  A three-point, low-pass filter is applied to these moments twice per time step to mitigate small-scale discrete-particle noise in the computed $\bb{E}$ and $\bb{u}$.

 Time integration is performed using a predictor-predictor-corrector approach that is second-order accurate and has the advantage of stably propagating Whistler waves.  This is done by first predicting the evolved values of $\bb{B}$ and $\bb{E}$ using the current positions and velocities of the particles (first predictor step). The particles are then evolved using the average of the current and predicted values of $\bb{B}$ and $\bb{E}$ which place the electromagnetic fields at the half-time-step needed for the Boris algorithm. These new positions and velocities are then used to re-evaluate the predicted $\bb{B}$ and $\bb{E}$ (the second predictor step).  Using these new predicted values, the fields and particles are again evolved to yield their new values for the next time step (corrector step). This algorithm is described in detail in~\citet{Pegasus}.
 
\subsection{Braginskii-MHD using \textsc{Snoopy}}\label{method:braginskii}

We use a version of the pseudospectral incompressible-MHD code \texttt{Snoopy}~\citep{Lesur2007} to solve the incompressible Braginskii-MHD equations. This system was derived in \S\ref{sec:eq:brag}, and the resulting equations are [\emph{cf.}  equations~\eqref{eq:int:brag}]
\begin{subequations}\label{brag_MHD}
\begin{align}\label{brag_MHD:mom}
\D{t}{\bb{u}} &\doteq \left(\pD{t}{} + \bb{u}\bcdot\grad\right)\bb{u} =  \bb{B}\bcdot\grad\bb{B} - \grad p + \grad \bcdot(\eb\eb\rmDelta p) - (-1)^h \mu_h \nabla^{2h}\bb{u}+ \drive, \\*
\D{t}{\bb{B}} &\doteq \left(\pD{t}{} + \bb{u}\bcdot\grad\right)\bb{B} = \bb{B}\bcdot \grad \bb{u} - (-1)^h \eta_h\nabla^{2h} \bb{B} ,\label{brag_MHD:ind}
\end{align}
\end{subequations}
where the magnetic field $\bb{B}$ is expressed in Alfv\'{e}nic units and the mass density has been scaled out. Here, the isotropic pressure $p$ incorporates  the isotropic component of the ion and electron pressures, the magnetic pressure, and the isotropic component of the parallel viscous stress ($\Delta p \unitDyadic/3$). The last term on the right-hand side of \eqref{brag_MHD:mom}, $\drive$, is a random driving body force (see \S\,\ref{sec:forcing}). The additional diffusive terms in \eqref{brag_MHD}, featuring $\visc_h$ and $\eta_h$, are Laplacian ($h=1$) or hyper ($h=2$) viscosity and diffusivity, respectively; these are introduced to truncate the cascades of kinetic and magnetic energy near the smallest wavelengths captured in our simulations. The pressure anisotropy is given by~\eqref{eqn:bragvisc} except when limited by heuristic micro-instability limiters, equations \eqref{eqn:mirror_hard} and \eqref{eqn:firehose_hard}. Equations \eqref{brag_MHD} have four free parameters: the two isotropic diffusivities $\visc_h$ and $\eta_h$, the anisotropic Braginskii viscosity $\visc_\mathrm{B}$, and the specifics of the random forcing $\tilde{\bb{f}}$, described in \S\,\ref{sec:forcing}.

 \texttt{Snoopy} takes the pseudospectral approach to solving equations~\ref{brag_MHD}. Rather than solving the primitive variables in real space, evolution equations for their complex Fourier  amplitudes are solved instead. By doing so, the spatial derivatives can be calculated exactly, becoming one-point quantities rather than operators that couple adjacent grid cells. However, the non-linear terms, which are \emph{local} in real space (as derivatives couple nearby points), become \emph{global} in Fourier space, requiring a convolution over all Fourier modes. In one dimension, this operation is $\mathcal{O}(N_\mathrm{cell}^2)$, which becomes prohibitively expensive for large grids.
 This issue can be avoided by first calculating the non-linear terms in \emph{real space}, and then transforming back to Fourier space (thus giving the pseudo in \emph{pseudo}spectral).  This operation is only $\mathcal{O}(N_\mathrm{cell}\ln N_\mathrm{cell})$, thus making it more affordable. 
 
 This approach necessitates an explicit calculation of the non-linear term and also introduces an effect called `aliasing'.  To see how this works, consider the product of two cosines:
 \begin{equation}
 \cos(a)\cos(b) = \frac{1}{2}[\cos(a+b) + \cos(a-b)],
 \end{equation}
 and so a quadratic nonlinearity that multiplies two Fourier modes with wavenumbers $k_1$ and $k_2$ will result in two modes with wavenumbers $k_1 \pm k_2$. However, the Nyquist criterion states that a grid of length $L$ with $N_\mathrm{cell}$ collocation points can only resolve modes with wavenumbers $k$ satisfying 
 \begin{equation}\label{eq:nyquist}
 |k | \le  \frac{\upi N_\mathrm{cell}}{L} \doteq k_\mathrm{Nyquist},
 \end{equation}
  with modes with larger wavenumber magnitudes being downsampled by $2 \upi /L$ until they satisfy equation~\eqref{eq:nyquist}. This can potentially result in energy being transferred unphysically to large-scale modes from unresolved small-scale ones, which is called aliasing. To avoid this unphysical behavior, simulations are typically `de-aliased' by zeroing out modes beyond a certain wavenumber $k_\mathrm{max}$ at every step, and thus modes that would cause aliasing do not have an opportunity to do so.  The threshold $k_\mathrm{max}$ that optimizes usage of the grid can be calculated by considering which threshold will result in two modes with wavenumber $k_\mathrm{max}$ being downsampled to a mode with wavenumber $-k_\mathrm{max}$.  Thus any mode with  $|k| < k_\mathrm{max}$ will not have a large enough wavenumber to be downsampled into the range of modes that are evolved. To wit,
 \begin{equation}
 2k_\mathrm{max} - \frac{2\upi}{L}= - k_\mathrm{max},
 \end{equation}
 or
 \begin{equation}
 k_\mathrm{max} = \frac{2\upi}{3L} = \frac{2}{3}k_\mathrm{Nyquist}.
 \end{equation}
 This is then known as the `2/3' rule, with $(2k_\mathrm{Nyquist}/3)^{-1}$ known as the de-aliasing scale, which is effectively the smallest resolved scale in a pseudospectral simulation employing the 2/3 rule.
 
  All simulations are run on a triply periodic grid with $2/3$ de-aliasing. The parallel Braginskii stress involves a quintic nonlinearity and, in a spectral code like \texttt{Snoopy}, formally introduces aliasing due to division by $B^2$. However, numerical tests show that, if such aliasing effects are present, they produce no quantifiable difference in magnetic-energy growth rates, field statistics, or turbulent spectra between simulations with $1/3$ de-aliasing and a grid of size $448^3$ and those with $2/3$ de-aliasing and a grid of size $224^3$. 
   \texttt{Snoopy} also calculates the isotropic diffusivities semi-analytically using an operator-split approach by first calculating the non-linear terms and anisotropic diffusivities, then multiplying the  resulting velocity fields (magnetic fields) by $\rme^{\imag \visc k^2 \Delta t}$ ($\rme^{\imag \eta k^2 \Delta t}$), where $\Delta t$ is the simulation time step. Time integration is performed using the third-order Runge-Kutta method.
   
By adopting the incompressibility assumption, the thermal velocity $v_\mr{thi}$, and thus $\betai$, are eliminated from the equations. Accordingly, we formulate the stability thresholds~\eqref{stability} in terms of $\rmDelta p$ and $B^2$ directly and subsume $p$ and $\nu_\mr{i}$ into the definition of $\visc_\mr{B}$. Code units are based on a box size $L=1$ and energy injection rate $\varepsilon =1$. This leads to a saturated turbulence amplitude of order unity ($u_\mathrm{rms} \sim 1$).

\subsubsection{Pressure anisotropy limiters}\label{sec:limiters}

In chapter \ref{ch:brag}, we perform simulations of Braginskii-MHD that incorporate microphysical closures to capture the regulation of pressure anisotropy by kinetic instabilities as described in~\S\ref{sec:pressureDynamo}.
Here, we adopt a popular closure used in fluid simulations of weakly collisional, high-$\betai$ plasmas that limits, by hand, the pressure anisotropy to remain within the firehose and mirror instability thresholds:
\begin{equation}
\label{stability}    
-\frac{B^2}{4\upi}\lesssim \rmDelta p \lesssim \frac{B^2}{8\upi}.
\end{equation}
The resulting `hard-wall' limiters, which have their origin in pioneering work on the kinetic magnetorotational instability by \citet{Sharma06} and have since been used in Braginskii-MHD simulations of magnetothermal and magnetorotational turbulence by \citet{Kunz_2012} and \citet{Kempski_2019}, respectively,   take the form
\begin{align}\label{eqn:mirror_hard}
    \rmDelta p = \mr{min}\left(\frac{B^2}{8\upi},\, 3\visc_\mr{B}\ROS\right)
\end{align}
on the mirror ($\rmDelta p>0$) side and 
\begin{align}\label{eqn:firehose_hard}
    \rmDelta p = \mr{max}\left(-\frac{B^2}{4\upi},\, 3\visc_\mr{B}\ROS\right)
\end{align}
on the firehose ($\rmDelta p<0)$ side (again, assuming incompressibility). Similarly effective limiters, in the form of a large anomalous collision frequency enacted in regions of firehose/mirror instability, were employed by \citet{SantosLima} in simulations of turbulent dynamo using the double-adiabatic \citet{CGL} equations~\eqref{eq:CGL}.  

We also develop a novel set of microphysical closures that are more suitable for the `kinetic' magnetized regime of the dynamo, as discussed in \S\ref{sec:paniso}. The description of these closures and the simulations that employ them are presented in chapter~\ref{ch:explosive}.

\section{Forcing prodedure}\label{sec:forcing}

Nearly incompressible turbulence is driven in all simulations by applying a random, solenoidal, zero-net-helicity body force $\tilde{\bb{f}}(t,\bb{r})$ to the ions on the largest scales, $\kforce L/2\upi\in[1,2]$, and whose power is distributed evenly across Fourier modes. This forcing procedure is consistent with the simulations of non-helical turbulent dynamo as performed by~\citet{Meneguzzi}, and has been adopted by many others~\citep[e.g.,][]{Scheko_sim, Maron04}. 
 In this work we choose our forcing to be time-correlated using an Ornstein-Uhlenbeck process:
\begin{align*}
\drive(t+\Delta t) = \drive(t) \,\rme^{-\Delta t/\tcorrf} + \left[\frac{\varepsilon}{\tcorrf}(1-\rme^{-2 \rmDelta t /t_\mr{corr}})\right]^{1/2}\bb{\widetilde{g}}, 
\end{align*}
where $\Delta t$ is the simulation time step, $\tcorrf$ is the correlation time of the forcing,  $\bb{\widetilde{g}}$ is Gaussian noise at wavenumber $\kforce$ generated at every time step, and $\varepsilon$ controls the magnitude of the forcing~\citep{gillespie}. Using a time-correlated forcing is a more physically realistic approach to driving turbulence when compared to white-noise forcing, and has the advantage of avoiding spurious particle acceleration due to resonances with high-frequency power in kinetic simulations~\citep{Lynn12}. However, it has the disadvantage of necessarily injecting a small amount of net momentum in each step: $\int \od V\, n_\mathrm{i}(t)\tilde{\bb{g}}(t)$ can me made to be zero for a given time step, but as $n_\mathrm{i}$ evolves and a portion of $\tilde{\bb{g}}(t)$ is carried over to the next time step, $\int \od V \, n_\mathrm{i}(t+\Delta t)\tilde{\bb{g}}(t)$ cannot be guaranteed to be zero, and in general is not. Likewise, time-correlated forcing also injects a finite amount of net helicity every time step as well, unless the cross-phase of every forced mode is chosen as to not contain net helicity and are kept constant between time steps. Such an approach is rather pathological, however, and may be even less physically relevant than white noise.  Symmetry considerations dictate that the ensemble average of these injected quantities be zero, though they may grow unbounded in any given realization. In practice, we do not find any difficulty with net momentum in the box (i.e. $V^{-1} \int \od V n_i \bb{u}_\mathrm{i} \ll n_\mathrm{i}u_\mathrm{rms}$, where $\urms$ is the rms ion flow speed), and that while the simulations do  exhibit some net helicity, its mean value hovers around zero throughout the entire runtime.

The correlation time $\tcorrf$ is chosen as  $\tcorrf\approx (k_\mathrm{f}u_\mathrm{rms})^{-1}$, which corresponds to the inverse decorrelation rate at the outer scale for $\mathrm{Re} \ge 1$ turbulence.  The initial state of the forcing in all simulations is zero, i.e. $\bb{f}(t=0) = \bb{0}$.

 \section{Reynolds numbers}

 For our analysis, it is useful to define the following Reynolds and Prandtl numbers:
\begin{equation}\label{eqn:Reynolds}
\mathrm{Re} \doteq \frac{u_0 \ell_0}{\visc}, \quad
\mathrm{Re}_\parallel \doteq \frac{u_0 \ell_0}{\visc_\mathrm{B}}, \quad
\mathrm{Rm} \doteq  \frac{u_0 \ell_0}{\eta}, \quad
\mathrm{Pm} \doteq \frac{\mathrm{Rm}}{\mathrm{Re}},
\end{equation}
where $u_0$ is the typical velocity at the outer scale, $\ell_0 \doteq 2 \upi / L \sim k_\mr{f}^{-1}$ is the outer scale, $L$ is the size of the simulation domain, and $k_\mr{f}$ is the forcing wavenumber. In the unlimited case, the effective Reynolds number $\Reeff = \mathrm{Re}_\parallel$; in the limited case, it satisfies $\mathrm{Re}_\parallel \le \Reeff\le \mathrm{Re}$. The definitions \eqref{eqn:Reynolds} are suitable for Laplacian dissipation, but generalized Reynolds numbers can be formulated for higher-order dissipation. To do so, we  make the substitutions $\visc\rightarrow \ell_\visc^{-2(h-1)}\visc_h$ and $\eta \rightarrow \ell_\eta^{-2(h-1)}\eta_\mathrm{h}$ in the standard definitions of the Reynolds numbers \eqref{eqn:Reynolds}. This replacement is done so that, for a given value of $\mathrm{Re}_h$ ($\mathrm{Rm}_h$), $\ell_\visc/\ell_0$ ($\ell_\eta/\ell_0$) is held fixed for all values of $h$. This allows one to compare two systems by directly comparing their generalized Reynolds number. Assuming Kolmogorov scalings ({\it viz.}~$|\grad \bb{u}| \sim \mathrm{Re}_h^{2/3}$) to compute the dissipation scales $k_\visc$ and $k_\eta$, we have
\begin{align*}
    \mathrm{Re}_h &\doteq \frac{u_0 \ell_0}{\ell_\visc^{-2(h-1)} \visc_{h}} =\left(\frac{u_0 \ell_0^{2h-1}}{\visc_h}\right)^{2/(3h-1)},\\
    \mathrm{Rm}_h &\doteq \frac{u_0\ell_0}{\ell_\eta^{-2(h-1)} \eta_{h}} = \left(\frac{ u_0 \ell_0^{2h-1}}{\eta_h \Reeff^{(h-1)/2}}\right)^{1/h}.
\end{align*}
While more general numbers can be defined without assuming Kolmogorov scalings, they will generically depend on some characteristic of the underlying fields that must be determined \emph{a posteriori}.

\section{Averaging procedures}

In the analysis of our simulation data, we make use of volume and time averages. These are denoted as  $\langle\,\cdot\,\rangle$ and $\langle \,\cdot\, \rangle_t $, respectively.  Additionally in chapter~\ref{ch:simulation} where particle-in-cell simulations are performed, we also make use of averaging over simulation particles, denoted as $\langle\,\cdot\,\rangle_\mr{p}$.  The root-mean-square (rms) value of a quantity $A$ is given by $A_\mr{rms} \doteq \langle A^2 \rangle^{1/2}$.

\section{Diagnostics}\label{sec:diagnostics}

Before presenting our results, we define various diagnostics that are used to study the structure and statistics of the turbulent velocity and magnetic fields.

\subsubsection{Characteristic wavenumbers}

A useful diagnostic for characterizing the structure of the magnetic field is the following assortment of characteristic wavenumbers~\citep[following][]{Scheko_sim}:
\begin{subequations}\label{char-wavenumbers}
\begin{gather}
k_\parallel \doteq \left( \frac{\left\langle|\bb{B}\bcdot\grad\bb{B}|^2\right\rangle}{\langle B^4\rangle}\right)^{1/2} ,  \quad 
k_{\bs{B}\bstimes\bs{J}} \doteq \left( \frac{\left\langle|\bb{B}\btimes\bb{J}|^2\right\rangle}{\langle B^4\rangle}\right)^{1/2}, \tag{\theequation {\it a,b}}\\*
k_{\bs{B}\bscdot\bs{J}} \doteq \left( \frac{\left\langle|\bb{B}\bcdot\bb{J}|^2\right\rangle}{\langle B^4\rangle} \right)^{1/2}, \quad
k_{\mr{rms}} \doteq \left( \frac{\left\langle|\grad\bb{B}|^2\right\rangle}{\langle B^2\rangle} \right)^{1/2} . \tag{\theequation {\it c,d}}
\end{gather}
\end{subequations}
These quantities have simple interpretations: $k_\parallel$ measures the variation of the magnetic field along itself, and is typically set by the smallest-scale field-aligned stretching motions; 
$k_{\bs{B}\bstimes\bs{J}}$ measures the variation of the magnetic field across itself, and corresponds to the field reversals in folds, which are ultimately limited by resistive dissipation; $k_{\bs{B}\bscdot\bs{J}}$ measures the variation of the field in the direction both orthogonal to $\bb{B}$ and $\bb{B}\btimes \bb{J}$, which tends to orient itself along the direction of greatest compression~\citep{Zeldovich}; and $k_{\mr{rms}}$ provides a general measure of the overall variation of the magnetic field. For magnetic fields that are arranged in folded sheets -- a typical realization during the kinematic stage of the ${\rm Pm}\gg{1}$ MHD dynamo -- the relative ordering of these wavenumbers is $k_\parallel \lesssim k_{\bs{B}\bscdot \bs{J}} \ll k_{\bs{B}\bstimes\bs{J}} \sim k_\mr{rms} \sim k_\eta$, where $k_\eta$ is the spectral cut-off due to resistivity. For a magnetic field arranged in folded ribbons -- a typical realization during the saturated state of the ${\rm Pm}\gg{1}$ MHD dynamo -- $k_\parallel \ll  k_{\bs{B}\bscdot \bs{J}} \lesssim k_{\bs{B}\bstimes\bs{J}}\approx k_\mr{rms} \sim k_\eta$.

\subsubsection{Transfer functions}

We calculate the shell-filtered kinetic-energy transfer function $\mathcal{T}_k[\bb{A}]$ of an arbitrary vector field $\bb{A}$, defined as
\begin{equation}\label{eq:shell_trans}
    \mathcal{T}_k[\bb{A}] \doteq \frac{2}{u_\mathrm{rms}^{2}} \sum_{q \in (2^{-1/4}k, 2^{1/4}k]} \bb{u}_\bb{q}^* \bcdot \bb{A}_\bb{q},
\end{equation}
where the star denotes the complex conjugate, $\bb{A}_\bb{q}$ denotes the Fourier amplitude of $\bb{A}$ with wavenumber $\bb{q}$, and a round (square) bracket in the wavenumber range indicates exclusivity (inclusivity). This quantity has units of inverse-seconds when its argument is a term from the momentum equation \eqref{brag_MHD:mom}, and represents the rate of kinetic energy flowing due to $\bb{A}$ into the kinetic energy of the Fourier shell at $k$ of width $2^{1/2}k$. For example, $\mathcal{T}_k[\bb{B}\bcdot \grad \bb{B}]$ denotes the energy flowing into the velocity field at shell $k$ due to the Lorentz force. This diagnostic can be used to probe the energy balance in $k$-space between each term in \eqref{brag_MHD}. We also define the root-mean-square shell-filtered kinetic-energy transfer function,
\begin{equation}\label{eq:shell_trans_rms}
    \mathcal{T}^{\mathrm{rms}}_k[\bb{A}] \doteq \frac{2}{\langle u^4\rangle^{1/2}} \left[\sum_{q \in (2^{-1/4}k, 2^{1/4}k]}\left(\frac{1}{2}\Re( \bb{u}_\bb{q}^* \bcdot \bb{A}_\bb{q})\right)^2\right]^{1/2}.
\end{equation}
This diagnostic serves as an alternative to the fourth-order spectra previously used by~\citet{Scheko_sim} and features the added benefit of enabling quantitative comparison between the nonlinear terms in the momentum equation.

To supplement these diagnostics, we also define a shell-filtering procedure on vector field $\bb{A}$ as
\begin{equation}\label{eq:shell_u}
    \bb{A}^{[\mathrm{range}]} \doteq \int_{k \in \mathrm{range}} \frac{\od^3 \bb{k}}{(2\upi)^3} \,\bb{A}_\bb{k} \rme^{\imag \bb{k}\bcdot \bb{x}},
\end{equation}
where the integration is taken over a specified range in $k$-space. We utilize three ranges: $[k]$ denotes modes in the shell of width $\sqrt{2}$ with range $(2^{-1/4}k,2^{1/4}k]$; $[{<}k]$ denotes all modes with wavenumber magnitude less than $2^{-1/4}k$; and $[{>}k]$ denotes all modes with wavenumber magnitude greater than $2^{1/4}k$. The shell-filtered quantity $\bb{A}^{[\mathrm{range}]}$ can be used to determine the amount of energy transfer from one region of $k$-space to another. For instance, the quantity $\mathcal{T}_k[\bb{u}\bcdot \grad \bb{u}^{[<K]}]$ denotes the net transfer of kinetic energy from all modes with wavenumber magnitudes ${<}2^{1/4}K$ to modes in the shell $k \in (2^{-1/4}K,2^{1/4}K]$. Such shell-filtered quantities have been used in analyses of spectral energy transfer in MHD guide-field turbulence~\citep[e.g.][]{Alexakis2005,Grete2017} and have also been used alongside H\"older's inequality to establish constraints on non-local transport in the fluctuation dynamo~\citep{Beresnyak12}. In this paper, we use the shell-filtered quantity $\bb{u}^{[\mathrm{range}]}$ to compare the relative strengths of the hydrodynamic nonlinearity and the viscous stresses (figures~\ref{fig:trans_unlim} and~\ref{fig:stokes_trans}), as well as to determine how the  motions at scale $k$ affect the growth of the magnetic energy (figure~\ref{fig:unlim_trans}).

\subsubsection{Structure functions}

To probe the structure of the turbulent velocity field, it is useful to introduce structure functions, which relay information about the scale-by-scale structure and spatial anisotropy with respect to the \emph{local} magnetic-field direction. In particular, we employ three-point, second-order structure functions for increment $\bb{\ell}$, defined by
\begin{align}\label{eqn:strf}
    \mathrm{SF}_2[\bb{u}](\bb{\ell}) &\doteq \langle | \bb{u}(\bb{x}+\bb{\ell}) - 2\bb{u}(\bb{x}) + \bb{u}(\bb{x}-\bb{\ell}) |^2\rangle.
\end{align}
The structure functions can  be used to extract information about variations of a given field along and across the local magnetic-field direction by conditioning the box average on the alignment of the point-separation vector $\bb{\ell}$ with the local magnetic field, defined by $\bb{B}_\bs{\ell} \doteq [\bb{B}(\bb{x}+\bb{\ell})+\bb{B}(\bb{x})+\bb{B}(\bb{x}-\bb{\ell})]/3$. This conditioning yields the parallel and perpendicular structure functions
\begin{subequations}\label{eqn:strf_prl_prp}
\begin{align}
    \mathrm{SF}_2[\bb{u}](\ell_\parallel) &\doteq \langle | \bb{u}(\bb{x}+\bb{\ell}) - 2\bb{u}(\bb{x}) + \bb{u}(\bb{x}-\bb{\ell}) |^2 ; \:0 \le \theta < \upi/18 \rangle, \\
    \mathrm{SF}_2[\bb{u}](\ell_\perp) &\doteq \langle | \bb{u}(\bb{x}+\bb{\ell}) - 2\bb{u}(\bb{x}) + \bb{u}(\bb{x}-\bb{\ell}) |^2 ;\: 8\upi/18 < \theta \le \upi/2\rangle,
\end{align}
\end{subequations}
respectively, where $\theta \doteq \arccos | \bb{B}_\bs{\ell}\bcdot\bb{\ell}/B_\ell \ell|$ is the angle between the point separation vector and the local magnetic field. For the purposes of computing one-dimensional, field-biased structure functions, increments whose angles lie within $10^\circ$ of $0^\circ$ or $90^\circ$ are considered to be `parallel' or `perpendicular', respectively \citep[cf.][]{chen12}. 

 The parallel and perpendicular structure functions may be combined to calculate the scale-dependent anisotropy of the fluctuations. For example, equating the two,
\begin{equation}\label{eqn:scale_aniso}
   \mathrm{SF}_2[\bb{u}](\ell_\parallel) =  \mathrm{SF}_2[\bb{u}](\ell_\perp),
\end{equation}
provides $\ell_{\parallel,\bs{u}}$ as a function $\ell_{\perp,\bs{u}}$, or vice versa.

Note that power laws with exponent $\alpha$ that appear in second-order structure functions translate to Fourier spectra with spectral index $-\alpha-1$. The use of a three-point stencil allows one to resolve spectral indices that are less steep than $-5$, or a power law appearing in a structure function with exponent less than $4$~\citep{lazarianPogosyan}.\footnote{A five-point stencil can do even better, resolving spectral indices larger than $-9$~\citep{cho09}, though such steep spectra are not encountered here.}

To compute the structure functions from our simulation data, we first choose 5000 random spatial locations $\bb{x}_1$ on our grid.  For every one of our randomly chosen points, we perform a loop over all possible increments $\bb{\ell} = (i\Delta x , j\Delta x, k \Delta x)$, where $0\le i$, $j$, $k\le L/2\Delta x$ are integers, $\Delta x$ is the grid scale and $L$ is the box size of the simulation.  The structure functions are then binned, and an average is performed for every bin by dividing by the number of additions into that bin.

\chapter{Numerical Simulation of the Collisionless Plasma Dynamo }\label{ch:simulation}

\section{Overview}

In this chapter we perform numerical simulations of the plasma dynamo using the second-order--accurate, hybrid-kinetic, particle-in-cell code \textsc{Pegasus} \citep{Pegasus}.  Our attention is primarily concentrated on two runs: The first focuses on the early production of pressure anisotropy, its regulation by kinetic instabilities, the consequent generation of an effective collisionality, and the impact of these processes on magnetic-field amplification in the ``kinematic'' phase. The second focuses on the the ``non-linear'' regime  and how kinetic instabilities affect the plasma and magnetic field in saturation.\footnote{These results have been presented in~\bibentry{St-Onge-dynamo}.} In addition, we present the results of a number of parameter scans which reveal how our findings depend on various parameters, such as the magnetic Reynolds number $\mathrm{Rm}$ and initial magnetization $L/\rhoio$, as well as perform a series of convergence tests on the number of particles per cell and resolution.

One may question whether modeling the plasma dynamo using \emph{kinetic} ions and \emph{fluid} electrons constitutes a reasonable and worth-while effort.  Indeed, in a truly collisionless system properties of the bulk flow (at ion scales) and magnetic diffusion (at electron scales) should fundamentally rely on kinetic effects, at least in an \emph{a priori} sense. Thus, considering kinetic physics occurring at one scale is only half the problem. While one would hope to model the dynamo in a fully kinetic fashion, the scale separation needed in order to achieve sufficiently high $\mathrm{Rm}$ and $\mathrm{Re}$ required for the dynamo to operate in the magnetized regime would preclude any effort to simulate the problem with even modest mass ratios.\footnote{If one assumes a mass ratio of $m_\mathrm{i}/ m_\mathrm{e} = 100 $, then to perform an initially magnetized simulation ($L/\rho_\mathrm{i0}\sim10$) for a 100 fold amplification of the magnetic energy would require a grid size of $N_\mathrm{cell} > 8000^3$ assuming the final electron gyroradius  is twice the electron skin depth $\lambda_\mathrm{ie}$  and that one desires $\lambda_\mathrm{ie} / \Delta x  > 4$. Assuming further that one uses 32 particles per cell per species, this would require ${\sim}1$ petabyte of memory to contain the particle phase-space-coordinate data of 6 double-precision floats. } However, the important properties of the fluctuation dynamo (such as the growth rate of magnetic energy) are strongly dependent on the characteristics of the underlying turbulent fluid (due to ions), while in the $\mathrm{Pm} \gg 1$ limit are only weakly dependent on the specifics of small-scale magnetic diffusion~\citep{Scheko_theory}. As the problem of determining the plasma viscosity in a turbulent fluid threaded by a chaotic small-scale magnetic field is an interesting problem in its own right, we believe that the hybrid-kinetic approach is a valid and important first step in understanding how the dynamo might behave in a collisionless environment.

\section{Numerical set-up}

Both simulations are initialized with a stationary, spatially uniform, Maxwellian, ion-electron plasma in a triply periodic box of size $L^3$, threaded by a random, zero-net-flux magnetic field $\bb{B}_0$ with power at wavenumbers $kL/2\upi\in[1,2]$. The electrons are assumed isothermal with temperature $T_\mr{e}=T_\mr{i0}$, where $T_\mr{i0}$ is the initial ion temperature.   The amplitude of  $\varepsilon$ is chosen such that the steady-state Mach number $M\equiv\urms/\vthi\sim{0.1}$.  This amplitude is fixed; the amount of energy accepted by the plasma varies as its impedance self-consistently evolves. The correlation time is chosen as  $\tcorrf\approx (\kforce u_\mathrm{rms})^{-1}$, which corresponds to the inverse decorrelation rate at the outer scale for $\mathrm{Re} \ge 1$ turbulence.

\begin{table}
\centering
\begin{tabular}{ccccc cccc}
\hline 
\hline 
Run & $\betaio$ & $L$ & $L/\rho_\mathrm{i0}$ & $\eta$ & $\eta_\mathrm{H}$ & $\tcorrf$ &$N_\mathrm{cell}$ & $N_\mathrm{ppc}$   \\[0.1em] 
\hline
1 & $10^6$ & 16000 & 16 & 12.8 & 32800 & 16 & $504^3$ & 216 \\
2 & $10^4$ & 1000 & 10 & 0 & 6 & 10 & $252^3$ & 216 \\
\hline 
\end{tabular}
\caption{\label{tab:sim:fiducial}Parameters for the two hybrid-kinetic simulations discussed in section~\ref{sim:sec:results}.}
\end{table}

The first simulation has $\betaio=10^6$ and $L/\rhoio=16$,  It uses $504^3$ cells, $N_\mr{ppc}=216$ particles per cell, $\Omegaio\tcorrf=16$, $\eta_\mr{Ohm}/\vao\dio=12.7$, and $\eta_\mr{H}/\vao\dio^3=32800$.  The latter two parameters correspond to $\mr{Rm}_2\approx 32,000$ and $\mr{Rm}_4\approx 9000$, assuming $\Reeff \sim 1$.  The second run uses $\betaio=10^4$, $L/\rhoio=10$, $252^3$ cells, and $N_\mr{ppc}=216$. These parameters ensure that $\rhoi$ is well resolved in the second run, even in the saturated state in which $\betai{M^2}\sim{1}$ is anticipated. To maximize scale separation, only hyper-resistivity is used in this run, with $\eta_\mr{H}/\vao\dio^3=6$ ($\mr{Rm}_4\approx 4000$). In both runs, the plasma starts well magnetized. For reference, the simulation parameters are recorded in table~\ref{tab:sim:fiducial}.

 As a precaution to the issue of momentum injection discussed in \S\ref{sec:forcing}, the simulation $\betaio = 10^4$ has momentum zeroed out at every timestep, though simulations with identical parameters and no momentum zeroing do not seem to differ in any significant way.

\section{Results}\label{sim:sec:results}

\begin{figure}[t]
    \centering
\includegraphics[width=1.00\textwidth]{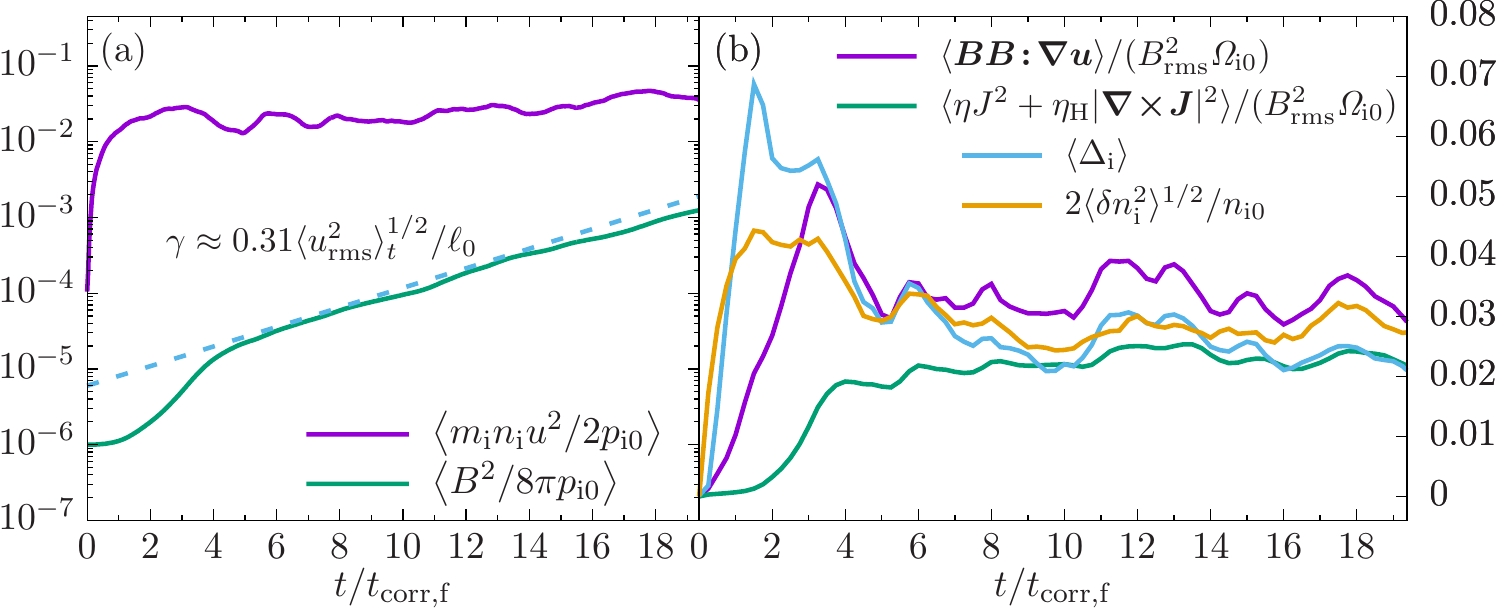}
    \caption[ Energies, rate-of-strain and pressure anisotropy for $\betaio = 10^6$. ]{\label{sim:beta6_energy} (a) Kinetic and magnetic energies; (b) parallel rate of strain, total magnetic dissipation, and pressure anisotropy; both for $\betaio=10^6$}
\end{figure}

The plasma dynamo can be characterized by four distinct stages: (1) an initial period of fast, diffusion-free growth, during which ion-Larmor-scale firehose/mirror instabilities are excited; (2) a reduction in growth rate, leading to steady exponential growth similar to the kinematic regime of MHD dynamo; (3) a non-linear regime, in which the magnetic field becomes strong enough to influence the bulk plasma motion via the Lorentz force; and (4) the saturated state, in which the magnetic and kinetic energies become comparable. Results from both runs are used to elucidate each stage.

\begin{figure}[t]
    \centering
\includegraphics[width=1.00\textwidth]{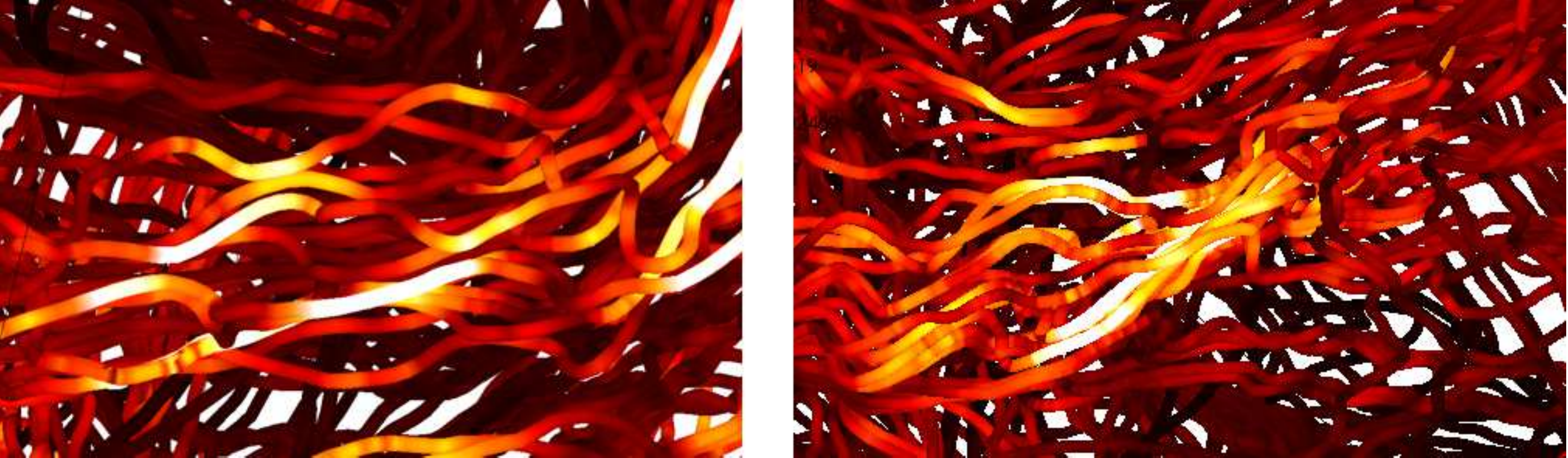}  
\caption[Mirror modes in the magnetic field for $\betaio=10^6$. ]{\label{sim:mirror_modes} Visualization of the magnetic-field lines at two locations and $t/\tcorrf = 1.25$ in the $\betaio = 10^6$ simulation, demonstrating the existence of mirror modes. Field lines are color-coded based on magnetic field strength, with brighter regions indicating stronger fields.}
\hfill
\includegraphics[width=1.00\textwidth]{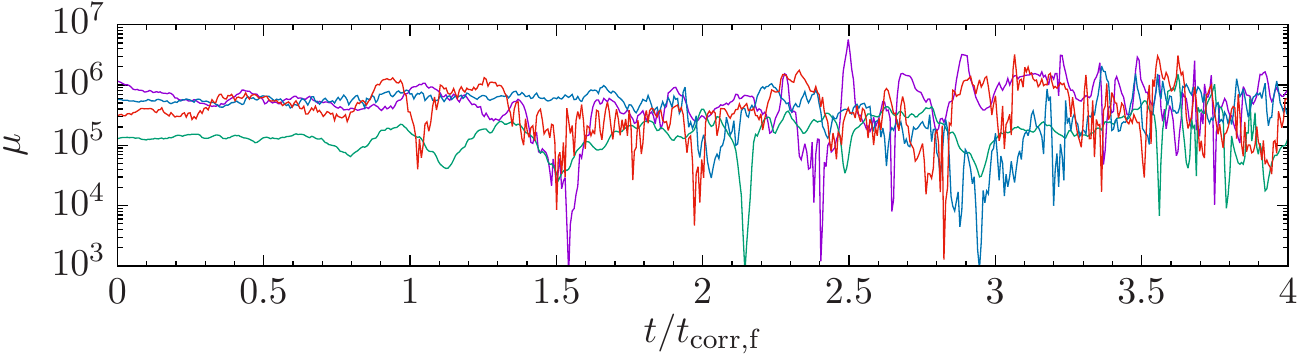}
\caption[Time trace for randomly chosen particles in the $\betaio=10^6$ simulation. ]{\label{sim:particle_track} The time evolution of the first adiabatic invariant $\mu$ for four randomly chosen particles from the $\betaio=10^6$ simulation. }
\end{figure}

\begin{figure}[t]
    \centering
\includegraphics[width=1.00\textwidth]{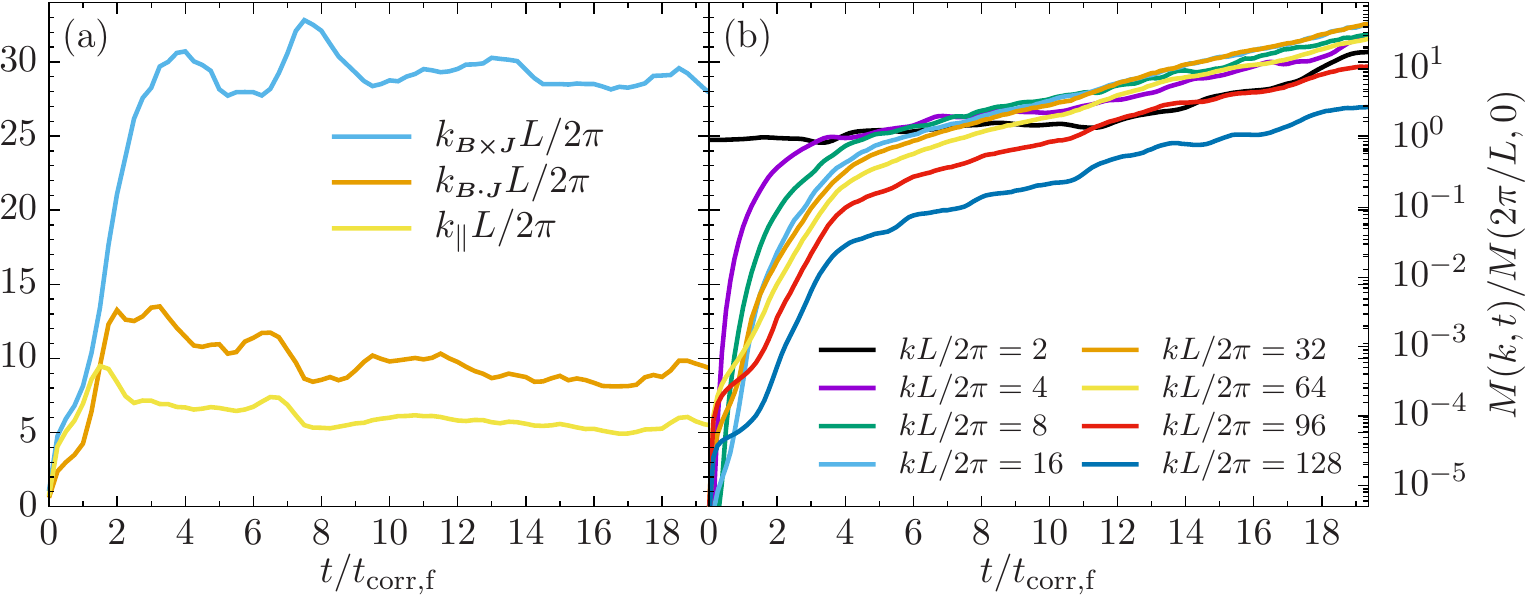}
    \caption[Evolution of characteristic wavenumbers of modal growth for $\betaio=10^6$.]{\label{sim:beta6_wavenumber} Time evolution of (a) characteristic parallel and perpendicular wavenumbers;  and (b) magnetic energy spectrum for select wavenumbers; both for $\betaio=10^6$.}
\end{figure}

\subsection{Initial rapid-growth phase}%
\begin{figure*}
    \centering
    \includegraphics[width=0.99\textwidth]{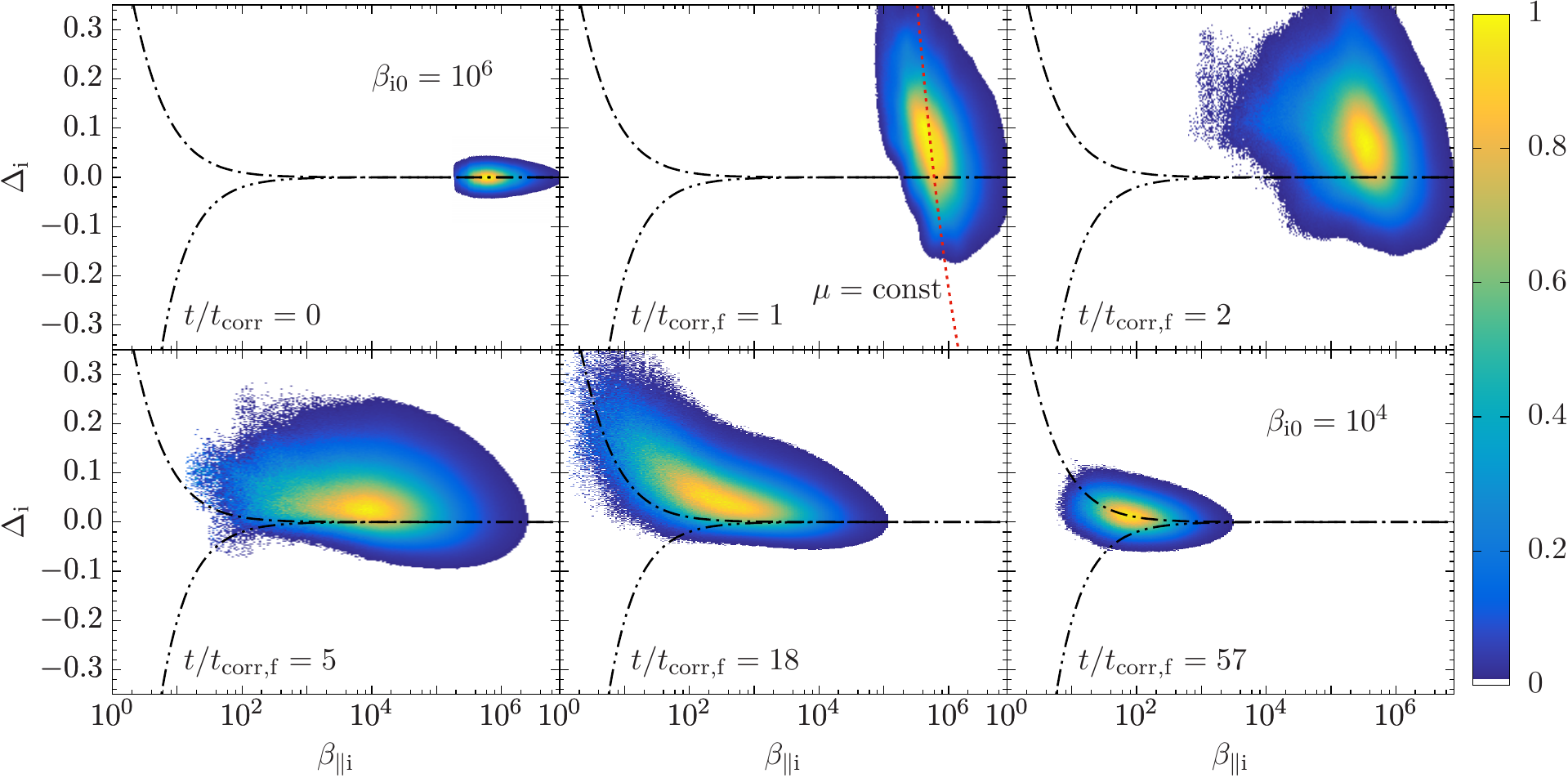}
    \caption[Joint PDF (Brazil plot) pressure anisotropy and  $\betapar$. ]{\label{sim:aniso}Distribution of pressure anisotropy versus $\betapar$ in the rapid-growth ($t/\tcorrf=0,1,2$) and kinematic ($t/\tcorrf=5,18$) phases for $\betaio=10^6$, and in the saturated state ($t/\tcorrf=57$) for $\betaio=10^4$. Dot-dashed (dot-dot-dashed) lines denote approximate mirror (firehose) instability thresholds. The red dotted line traces $\pperp/\ppar\propto\betapar^{-2}$, corresponding to evolution with $\mu=\const$.}
\end{figure*}

\label{sim:fastgrowthphase}
Figure~\ref{sim:beta6_energy}(a) displays the box-averaged kinetic and magnetic energies versus time for the $\betaio=10^6$ run. The kinetic energy saturates within $t\approx\tcorrf$ and a large-scale smooth flow is established. On the average, this flow amplifies the large-scale seed magnetic field, and rapid growth of magnetic energy occurs at $k\rhoi\approx{0.5}$--$1$ ($kL/2\upi\approx{4}$--$8$), adiabatically driving $\ba{\Deltai}>0$ (Figure~\ref{sim:beta6_energy}(b); see also Figure~\ref{sim:aniso}, $t/\tcorrf=1$).
Because $\betaio\gg{1}$, mirror instabilities are readily excited. Such modes can be observed by looking for `bubble'-like structures in the magnetic field. Figure~\ref{sim:particle_track} displays two mirror modes that appear in the magnetic field at $t/\tcorrf = 1.25$. The `bubble'-like mirror structures, with strong magnetic field at the cusps and weaker field in the central region, trap particles in the central region.  As the magnetic field grows, these particles gain an excess of perpendicular energy due to adiabatic invariance, resulting in a blowing out of the field lines, reinforcing the field strength at the cusps, thereby trapping more particles.
From the standpoint of these mirror modes, the initial seed field ($kL/2\upi=1,2$) behaves as a local `mean' field on which they grow with oblique polarization $k_{\bs{B}\bstimes\bs{J}}>k_\parallel>k_{\bs{B}\bscdot\bs{J}}$ (Figure~\ref{sim:beta6_wavenumber}(a), $t/\tcorrf\lesssim{1.5}$). Firehose-unstable modes are also triggered on ion-Larmor scales in regions of locally decreasing field and, in concert with mirror-unstable modes, ultimately generate sharp features in the magnetic field that trap and pitch-angle scatter particles. The latter produces an effective collisionality $\nueff$, which drives $\Deltai$ towards marginal stability (Figure~\ref{sim:aniso}, $t/\tcorrf={2,5}$) and ties the pressure anisotropy to the parallel rate of strain (Figure~\ref{sim:beta6_energy}(b), $t/\tcorrf\gtrsim{3}$). This leads to a Braginskii-like relation, $\Deltai\approx{3}\ROS/\nueff$, in which a balance obtains between adiabatic production and collisional relaxation, with $\nueff\lesssim\Omegai$. This type of scattering can be seen in the particle tracks themselves, four of which are shown in figure~\ref{sim:particle_track}; while their adiabatic invariant $\mu$ is well conserved for the initial correlation time, an effective collisionality is quickly established, and $\mu$ is no longer well conserved after $t/\tcorrf > 1$.

\begin{figure*}
    \centering
        \includegraphics[height=1.2\textwidth]{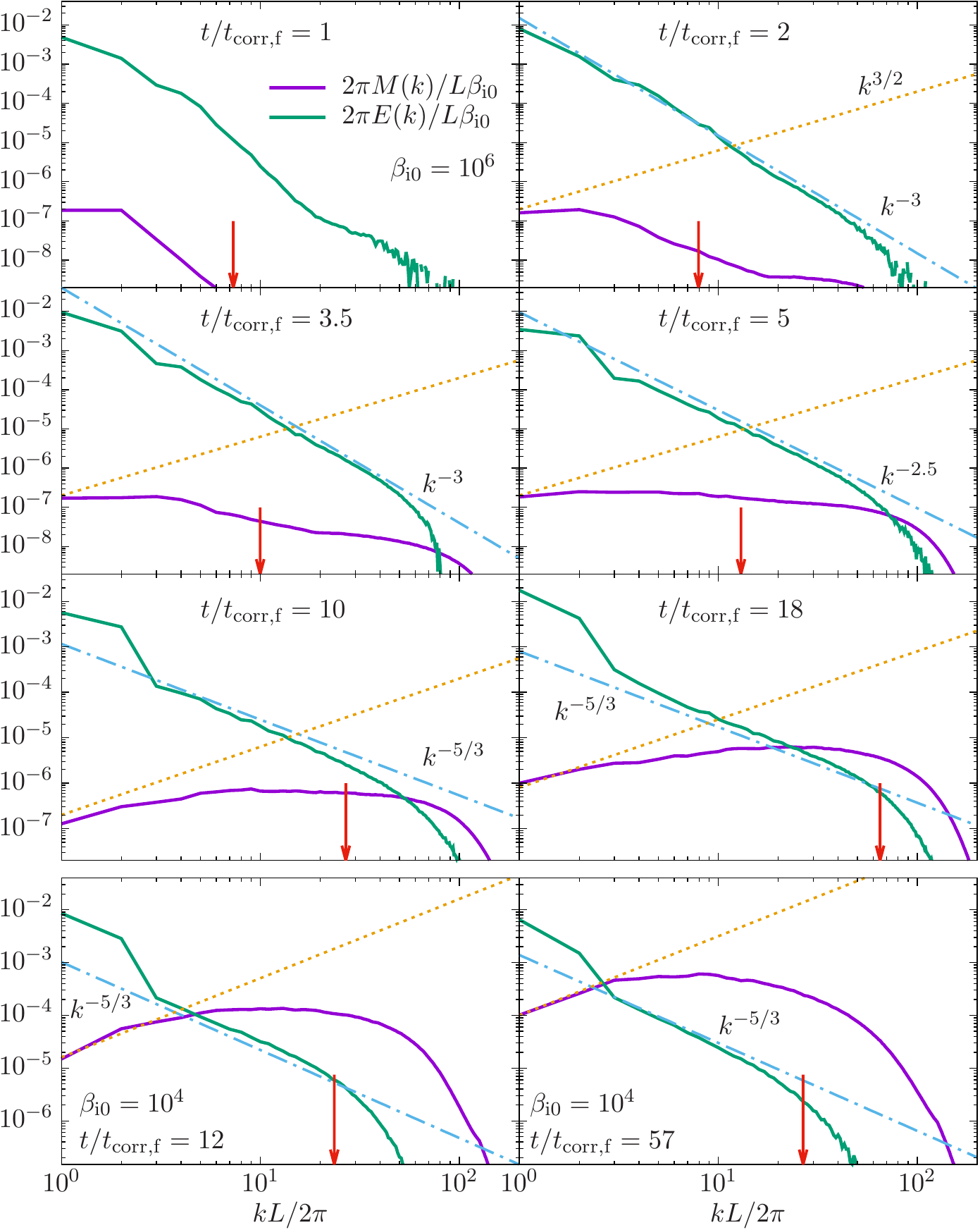}
    \caption[Magnetic and kinetic energy spectra of the hybrid-kinetic simulations.]{\label{sim:spectra} Magnetic- and kinetic-energy spectra for $\betaio=10^6$ ($t/\tcorrf=1$, $2$, $3.5$, $5$, $10$, $18$) and $\betaio=10^4$ ($t/\tcorrf= 12$, $57$). Red arrows denote the wavenumber $\upi/\rho_\mr{median}$, where $\rho_\mr{median}$ is the median value of $v_{\mathrm{thi}\perp}/\Omegai$.}
\end{figure*}

\begin{figure*}
    \centering
    \includegraphics[height=1.3\textwidth]{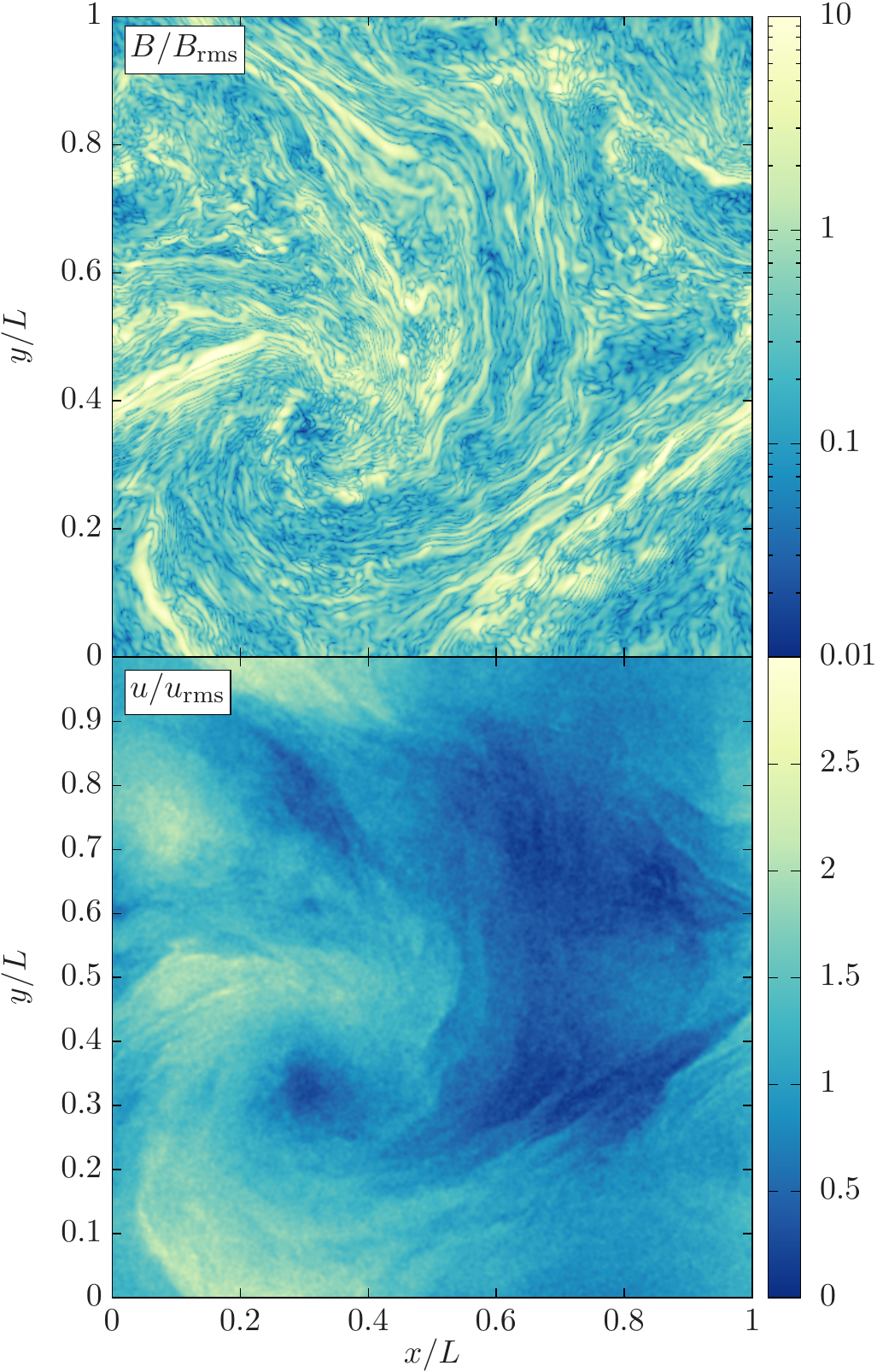}
    \caption[Pseudo-color images of $B/\Brms$ and $u/\urms$.]{\label{sim:slices} Pseudo-color images of $B/\Brms$ and $u/\urms$ in a 2D cross-section during the kinematic phase at $t/\tcorrf=15$ for $\betaio=10^6$.}
\end{figure*}

At the same time that the firehose and mirror instabilities saturate at $k\rhoi\lesssim{1}$ with $\delta{B}/B_0\sim{1}$, the magnetic field acquires energy at sub-ion-Larmor scales due to field-line stretching and folding by the large-scale flow (Figure~\ref{sim:beta6_energy}(d), $t/\tcorrf\gtrsim{5}$). The result is a much flatter angle-integrated magnetic-energy spectrum, $M(k)\equiv\frac{1}{2}\int\rmd\Omega_{\bs{k}}\,k^2\langle|\bb{B}(\bb{k})|^2\rangle$ (Figure~\ref{sim:spectra}, $t/\tcorrf=5$), than is seen in corresponding $\Pm\gtrsim{1}$ MHD simulations, with mirror and firehose fluctuations at $k \rhoi  < 1$ and fold reversals at $k\rhoi > 1$. A change in the dominant magnetic-field topology accompanies this growth, with $k_{\bs{B}\bstimes\bs{J}}>k_{\bs{B}\bcdot\bs{J}}>k_\parallel$ indicating a folded geometry in which the field varies quickly (slowly) across (along) itself (Figure~\ref{sim:beta6_wavenumber}(a), $t/\tcorrf\gtrsim{2}$).\footnote{The steady-state value of $k_\parallel$ in Figure~\ref{sim:beta6_wavenumber}(a) is an overestimate of the inverse fold length, being biased towards larger $k_\parallel$ by ion-Larmor-scale firehose/mirror fluctuations.}

\subsection{``Kinematic'' phase}%
\label{sim:sec:kinematic}

Eventually, this period of rapid growth ends. Figure~\ref{sim:beta6_energy}(b) indicates that the reduction in growth rate is due to two effects. First, an appreciable fraction of the magnetic energy migrates to resistive scales, and magnetic diffusion becomes important. Secondly, the energy-weighted rate-of-strain $\boldsymbol{BB\!:\! \nabla u}/B^2_\mathrm{rms}$ is sharply reduced between $t/\tcorrf\approx{3}$--$5$, a feature we attribute to feedback from firehose/mirror fluctuations  \citep[e.g.,][]{Scheko_2008,Rosin_2011,Rincon_2015}.  This is concurrent with a $\approx$30\% reduction in the rms value of density fluctuations during this time, signifying particles that were trapped in magnetic troughs have now begun to scatter. 
Also concurrent is the development of an angle-integrated kinetic-energy spectrum, $E(k)\equiv\frac{1}{2}\int\rmd\Omega_{\bs{k}}\,k^2\langle|\bb{u}(\bb{k})|^2\rangle$ (Figure~\ref{sim:spectra}, $t/\tcorrf=5,18$), that is \citet{Kolmogorov1941} (i.e., ${\propto}k^{-5/3}$).

Thereafter, $\ba{B^2}$ grows exponentially (Figure~\ref{sim:beta6_energy}(a), $t/\tcorrf\gtrsim{5}$), much as in the kinematic-diffusive stage of the large-$\Pm$ MHD dynamo \citep[e.g.,][]{Scheko_theory}, with a growth rate $\gamma\doteq\rmd\ln\ba{B^2}/\rmd{t}= 0.31 \langle u^2_\mathrm{rms}\rangle_t^{1/2}/\ell_0$ that becomes comparable at all scales (Figure~\ref{sim:beta6_wavenumber}(b), $t/\tcorrf\gtrsim{5}$). The folded magnetic-field geometry previously established persists [Figure~\ref{sim:beta6_wavenumber}(a)], and $M(k)$ develops a \citet{Kazantsev} $k^{3/2}$ scaling with a peak near the resistive scale (Figure~\ref{sim:spectra}, $t/\tcorrf=18$). Such folded structure, accompanied by ion-Larmor-scale irregularities driven by firehose/mirror, is evident in the rightmost panels of Figure~\ref{sim:slices}, which display pseudo-color images of $B/\Brms$ and $u/\urms$ in a representative 2D cross-section. Anisotropization of the plasma viscosity is also apparent; while the turbulent velocity field is primarily large-scale, filamentary structures of near-constant $u$ develop along magnetic lines of force. Thus, there is a dynamical feedback of the magnetic field on the large-scale flow, even in the absence of a dynamically important fluid-scale Lorentz force, belying the ``kinematic'' moniker.

Because of the continuous energy injection and consequent magnetic-field amplification, along with insufficient scale separation between $L$ and $\rhoi$, exact marginal firehose/mirror stability cannot be maintained and a residual $\ba{\Deltai}\approx(0.02-0.03)\gg{1}/\betai$ persists for $t/\tcorrf\gtrsim{5}$ [Figure~\ref{sim:beta6_energy}(b)], with the bulk of the plasma approximately following the mirror threshold as $\betai$ decreases (Figure~\ref{sim:aniso}, $t/\tcorrf=18$). The regulation of $\Deltai$ is imperfect since, in order for saturated firehose/mirror instabilities to tightly regulate the pressure anisotropy near marginal stability, $\nueff\sim{S}\betai$ \citep{Kunz_kin,Melville}, where $S$ is the parallel rate of strain at the viscous scale (where it is largest). However, at $t/\tcorrf=5$, $S/\Omegai\sim{10^{-2}}$ and $\betai\sim{10^5}$, thus requiring $\nueff\sim{10^3}\,\Omegai$ (!) Instead, $\nueff\ll\Omegai$ in the kinematic phase in both simulations, a point we have confirmed both indirectly, by comparing $\ROS$ and $\Deltai$ to infer 
\begin{equation}\label{eq:sim:bragmes}
\nueff\approx{3}\ba{\bb{BB}\bdbldot\grad\bb{u}}/\ba{B^2\Deltai},
\end{equation}
 and directly, by calculating the mean time over which $\mu$ changes for individually tracked particles (using the method described in \citealt{Kunz_kin} and \citealt{SquireKunz17}). The result is shown in Figure~\ref{sim:beta4_energy}(d) for $\betaio=10^4$; qualitatively identical behavior is observed for $\betaio=10^6$.

 \begin{figure}
\centering
\includegraphics[width=1.0\textwidth]{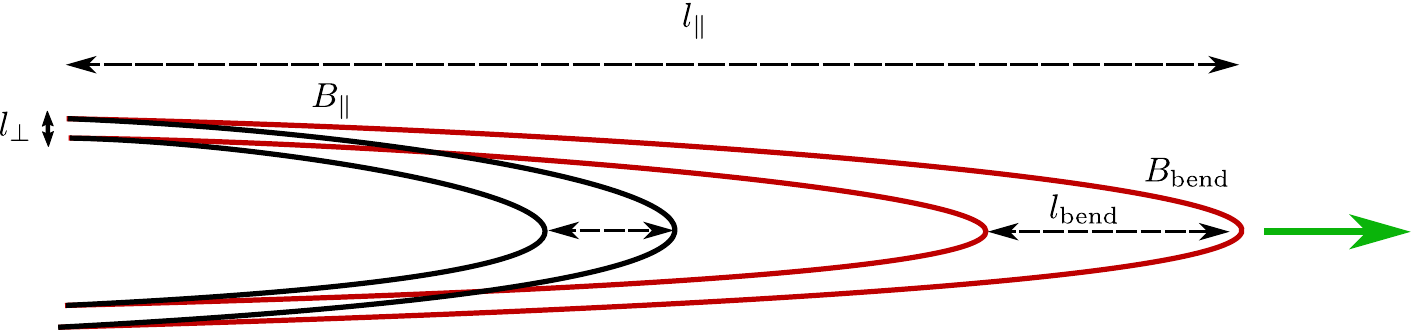}
\caption[Illustration of a fold undergoing stretching by a linear shear.]{\label{ch1:curvature} A visualization of a fold undergoing a linear shear, where $l_\parallel$  denotes  the length of the fold  and $l_\perp$ ($l_\mathrm{bend}$) denotes the fold separation in the straight (bent) region.  The magnitudes of the magnetic field in the straight and bent regions are denoted by $B_\parallel$ and $B_\mathrm{bend}$, respectively. }
\end{figure}

Two processes that may in principle contribute to $\nueff$, depending upon whether the majority of the particles' gyroradii is above or below the reversal scale of the magnetic field. In the former case, those particles sample several field-reversing folds during their gyromotion and thus undergo Bohm-like diffusion with $\nueff\sim\Omegai$. On the other hand, if the majority of particles have gyroradii below the field-reversal scale and remain well magnetized, or if a particle is \emph{indifferent} to the fact that it samples several different magnetic fields, then it can travel along the length of a fold, scattering on sharp magnetic field structures that can appear in regions of firehose and mirror activity.  As the firehose instabilities tend to scatter particles more efficiently than the mirror instability,\footnote{The mirror instability only weakly scatters particles throughout much of its nonlinear evolution \citep{Kunz_kin,Melville}. Moreover, in turbulence where $S$ is a fluctuating quantity, the mirror instability is suppressed when $\betai>\Omegai/S$ due to residual firehose fluctuations; see fig.~13 of \citet{Melville}.}  we expect particles to scatter at firehose sites, and so $\nueff$ is determined mainly by pitch-angle scattering off of firehose fluctuations, which populate regions of weak magnetic field where $\Deltai<0$.
To understand where these regions appear along a typical folded field, we re-derive here a result of the fluctuation dynamo that these folds exhibit an anti-correlation between the magnetic-field strength $B$ and the magnitude of the magnetic-field curvature $\bb{K} \doteq \eb \bcdot \grad \eb$.  To see this, we imagine our fold visualized in figure~\ref{ch1:curvature} undergoes stretching by a linear shear. Here, the length of the fold is $l_\parallel$, the fold separation distance is $l_\perp$, the separation at the bend is $l_\mathrm{bend}$, and the magnetic-field magnitude at the straight and bent regions are $B_\parallel$ and $B_\mathrm{bend}$, respectively.  Flux conservation dictates that $l_\perp B_\parallel \sim l_\mathrm{bend} B_\mathrm{bend} \propto l_\parallel B_\mathrm{bend}$.
 As the length of the fold is stretched by a factor $s$, $l_\parallel \rightarrow s l_\parallel$ while $l_\perp$ remains constant.  By volume and flux conservation, $B_\parallel \rightarrow s B_\parallel$, and so $B_\mathrm{bend}$ remains unchanged as well. 
  \begin{figure}
\centering
\includegraphics[width=0.8\textwidth]{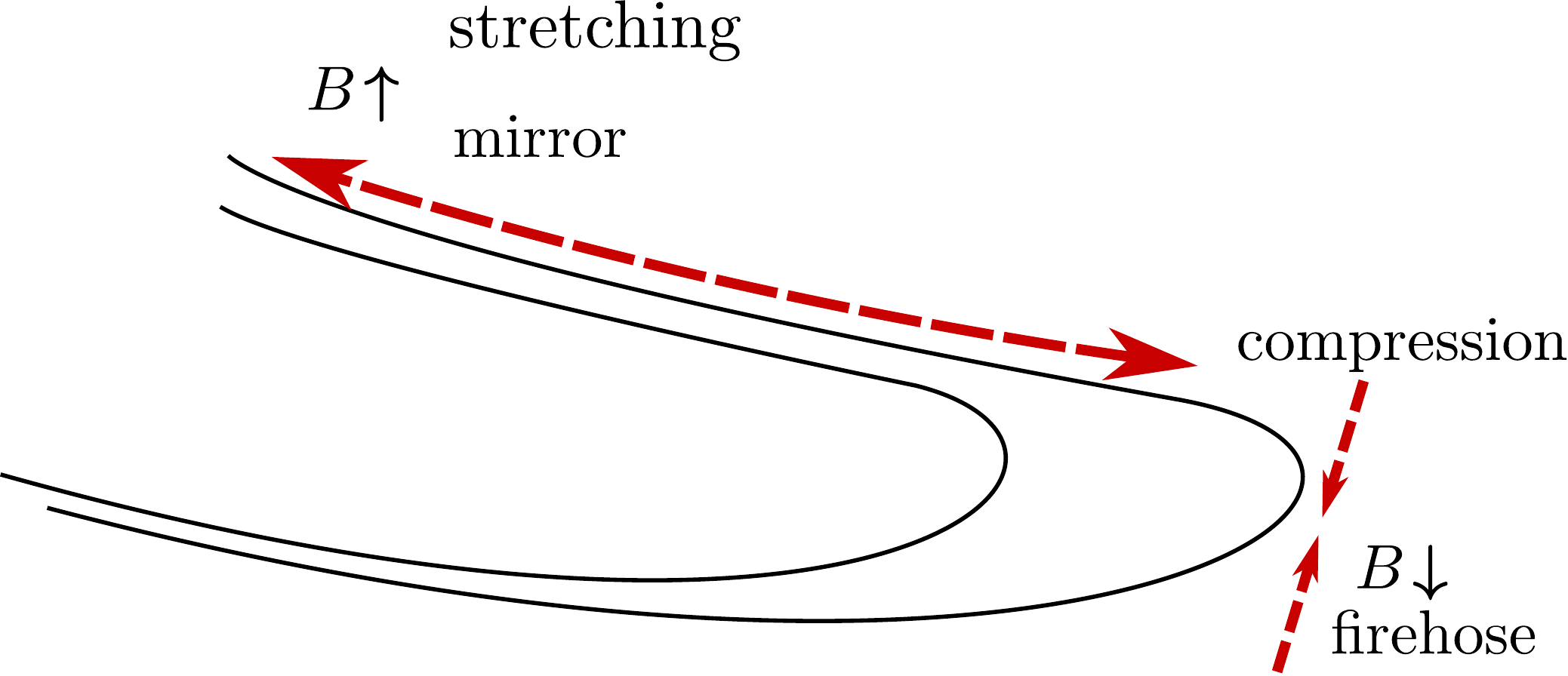}
\caption[Firehose and mirrors embedded on a dynamo-generated magnetic fold.]{\label{ch1:TS_rotated} A illustration indicating the regions of a dynamo-generated magnetic field fold in which firehose and mirror instabilities should occur.}
\end{figure}
 As this process is akin to stretching an ellipse,\footnote{A somewhat different argument was originally put forth by~\citet{Scheko_sim2}, where it was posited that $K_\mathrm{bend}\sim l_\perp^{-1} $, $K_\parallel \sim l_\parallel^{-1}$. This leads to $BK \sim \mathrm{const}$ throughout the fold. This result, however, is not borne out in simulation.} we posit that the field curvatures at the straight and bent regions are given by $K_\parallel \sim l_\perp / l_\parallel^2$ and  $K_\mathrm{bend} \sim l_\parallel / l_\perp^2$, respectively. Notice that the curvature of the bent region then increases by a factor of $\sqrt{s}$ during the stretching, while the curvature in the straight region decreases by $s^{-2}$,  leading to anti-correlation between $B$ and $K$. To get a prediction of how this anti-correlation should behave,  consider in the straight region the product $B_\parallel K_\parallel^{1/2} \sim l_\perp^{1/2} (B_\parallel / l_\parallel) \sim \mathrm{const}$.  As the fluctuation dynamo progresses, straight regions eventually become volume filling while regions with high curvature become intermittent, and thus $BK^{1/2} \sim \mathrm{const}$ should be followed throughout most of the physical domain. This is indeed witnessed in simulation~\citep{Scheko_sim}. 
 Therefore,  as illustrated in figure~\ref{ch1:TS_rotated}, we should expect to find the mirror instability occurring along the straight region of a magnetic fold while the firehose instability should appear in the curved region near the end of a fold which is aligned in the `null' direction of the stretching. Of course, in actual turbulence this region will typically be also stretched, but there will always be regions in space which this direction is compressive; it is in those regions that the firehose instability should occur. 
 The collision frequency that results from a particle scattering off of firehose instabilities at the bends of folds is then $\nueff\sim{k}_\parallel\vthi$, the inverse timescale for a thermal particle to traverse the length of a fold. 
 Both the contribution from Bohm-like diffusion and firehose instability scattering may be important, depending upon the structure of the magnetic field and the local magnetization of the plasma. In our runs, we witness only a brief moment in the evolution with $\nueff\sim\Omegai$. This may, however, be coincidental and not related to Bohm diffusion. Eventually, $\nueff\sim{k}_\parallel\vthi\ll\Omegai$ in the kinematic phase. It is only once $k_\parallel\vthi\sim{S}\betai$ that efficient regulation of $\Deltai$ is possible (\S\ref{sec:saturation}).

One consequence of $\nueff\ll\Omegai$ is an anisotropic viscosity, with Reynolds numbers $\mr{Re}\equiv u_0/(\kforce\visc)$ differing in the parallel and perpendicular directions: $\mr{Re}_\parallel\ll\mr{Re}_\perp$ \citep{Braginskii}. While the magnetic-field growth is controlled by $\mr{Re}_\parallel$ (since $\rmd\ln{B}/\rmd{t}\simeq\ROS\sim(\urms/\ell_0)\mr{Re}^{1/2}_\parallel$), the viscous cutoff $\ell_\visc$ seen in Figure~\ref{sim:spectra} is arguably determined by $\mr{Re}_\perp$ through the Kolmogorov relation $\ell_\visc\sim \ell_0\mr{Re}_\perp^{-3/4}$. Using classical transport theory to estimate the effective perpendicular ion viscosity $\visc_\perp \sim{0.1}\rhoi^2\nueff$, we find $\ell_0/\ell_\visc\sim(ML\Omegai/\rhoi\nueff)^{3/4}$. Taking $M$, $\Omegai$, $\rhoi$, and $\nueff$ from the run, we calculate a minimum value of $\ell/\ell_\visc\sim{10}$ at $t/\tcorrf\approx{5}$, which grows exponentially to $\ell/\ell_\visc \sim{100}$ at $t/\tcorrf\approx{18}$. This roughly agrees with the evolution shown in Figure~\ref{sim:spectra}. Likewise, $\mr{Re}_\parallel$ can be calculated using the parallel viscosity for a magnetized plasma, $\visc_\mathrm{B} \sim\vthi^2/\nueff$. Once $\nueff\sim{k}_\parallel\vthi$, $\mr{Re}_\parallel\sim{M}(k_\parallel/\kforce)\sim{1}$, suggestive of a $\mr{Pm}\gg{1}$ dynamo and consistent with adrop over an order-of-magnitude in $E(k)$ at $kL/2\upi \approx 2$.\footnote{Such drops directly beyond the forcing range are not uncommon in forced isotropic simulations of hydrodynamic turbulence, although drops larger than an order of magnitude are atypical.} The Braginskii-MHD dynamo simulations presented in chapter~\ref{ch:brag} with $1\sim\mr{Re}_\parallel\ll\mr{Re}_\perp$ and $-2/\betai\le\Deltai\le{1}/\betai$ enforced (e.g., following \citealt{Sharma06} and \citealt{Kunz_2012}) exhibit similar spectra and field-anisotropic flow to those presented here.

\begin{figure}
    \centering
        \includegraphics[width=1.00\textwidth]{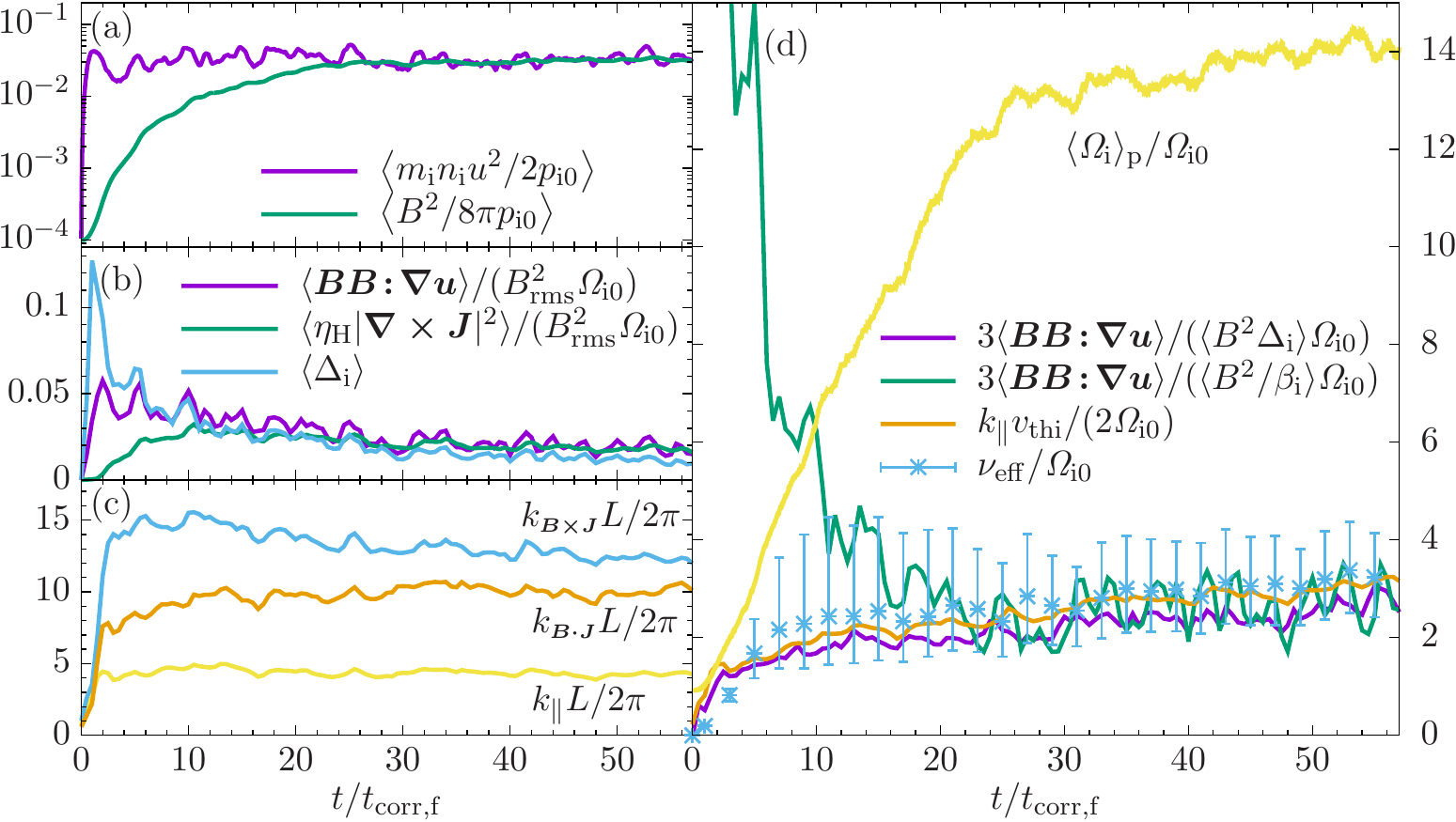}
    \caption[Energies and scattering frequencies of the kinetic $\betaio=10^{-4}$ run.]{\label{sim:beta4_energy} Time evolution of (a) kinetic and magnetic energies, (b) energy-weighted parallel rate-of-strain, resistive dissipation and pressure anisotropy, and (c) characteristic wavenumbers for $\betaio=10^4$. (d) Effective collision frequency (blue), compared to a ``Braginskii'' collision frequency (purple), the collision frequency required to maintain marginal firehose/mirror stability (green), a parallel-streaming frequency (orange), and the particle-averaged $\Omegai$ (yellow).}
\end{figure}

\subsection{Nonlinear regime and saturation}%
\label{sec:saturation}

Figure~\ref{sim:beta4_energy}(a) shows the evolution of kinetic and magnetic energies for the $\betaio=10^4$ run. After evolving through the rapid-growth phase and a brief exponential kinematic phase, saturation is reached with $\ba{B^2/4\upi}\sim\ba{m_\mr{i}nu^2}$ via a reduction of $\ROS$ (Figure~\ref{sim:beta4_energy}(b), $t/\tcorrf\gtrsim{25}$; \citealt{Scheko_sim}). The ordering $k_{\bs{B}\bstimes\bs{J}}>k_{\bs{B}\bscdot\bs{J}}>k_\parallel$ established in the kinematic phase is preserved [Figure~\ref{sim:beta4_energy}(c)], but the two perpendicular scales become closer to one another in saturation; i.e., the folded sheets evolve towards a ribbon-like structure, as seen in the $\Pm\gtrsim 1$ MHD dynamo~\citep{Scheko_sim}. Despite the box-averaged equipartition between kinetic and magnetic energies, this balance is not scale-by-scale (Figure~\ref{sim:spectra}, $t/\tcorrf = 57$). Rather, there is an excess of the former at the forcing scales (since $E(k)\propto{k^{-5/3}}$) and an excess of the latter at smaller scales, where the $k^{3/2}$ scaling is approximately maintained. It is because the folds exhibit spatial coherence at the flow scale that allows them to exert a back-reaction on the flow via the Lorentz force.

As in the $\betaio=10^6$ run, the pressure anisotropy becomes Braginskii-like, with $\ba{\Deltai}\propto\ba{\bb{BB}\bdbldot\grad\bb{u}}/\ba{B^2}>0$ (Figure~\ref{sim:beta4_energy}(b), $t/\tcorrf\gtrsim{5}$) and $\nueff\sim{k}_\parallel\vthi$ (Figure~\ref{sim:beta4_energy}(d), $t/\tcorrf\gtrsim{5}$). However, once $\betai$ decreases to ${\sim}50$ ($t/\tcorrf\gtrsim{20}$), $\nueff\sim{S}\betai$ and $\Deltai$ is regulated close to the firehose/mirror thresholds (Figure~\ref{sim:aniso}, $t/\tcorrf=57$).

\begin{figure}
    \centering
        \includegraphics[width=1.0\textwidth]{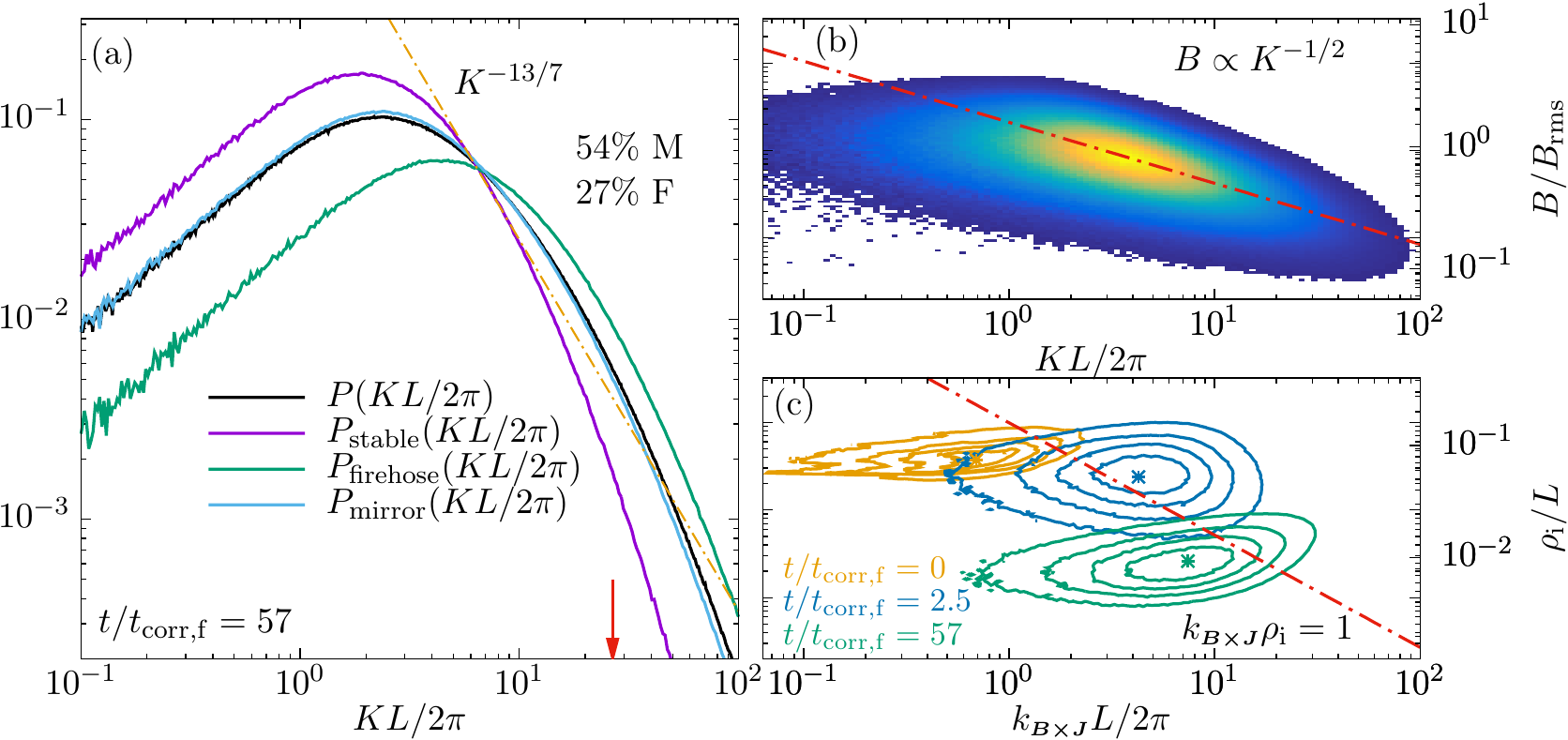}
    \caption[Curvature and joint PDF of $\rhoi$, $k_{\bb{B}\btimes \bb{J}}$. ]{\label{sim:curvature}(a) PDF of field-line curvature $K$ in saturation ($\betaio=10^4$, $t/\tcorrf=57$) for firehose-unstable (green), mirror-unstable (blue), firehose/mirror-stable (purple), and all (black) regions. The predicted $K^{-13/7}$ scaling \citep{Scheko_theory2} is shown for comparison. The red arrow denotes the wavenumber $\upi/\rho_\mr{median}$. (b) Distribution of $K$ and $B$ in saturation. (c) Distribution of locally computed $\rhoi$ and $k_{\bs{B}\bstimes\bs{J}}$ for $\betaio=10^4$; contours are evenly spaced between $0.2$ and $1$.}
\end{figure}

Figure~\ref{sim:curvature}(a) shows the probability distribution function $P(K)$ of the magnetic curvature $K\equiv|\eb\bcdot\grad\eb|$. In the MHD case, the tail of $P(K)$ relaxes to a $K^{-13/7}$ scaling \citep{Scheko_theory2} throughout the kinematic and saturated phases, depending only weakly on $\mr{Pm}$ (see fig.~25 of~\citealt{Scheko_sim}). While $P(K)$ in the plasma dynamo is peaked at similar values as those found in~\citet{Scheko_sim} ($KL/2\upi\approx{2}$), it is generally broader, and is dependent upon whether the host plasma is mirror unstable (blue; 54\% by volume), firehose unstable (green; 27\%), or stable (purple). Regions that are firehose unstable tend to have the largest curvature, for two reasons. First, $\Delta<0$ is generically produced in the stretched bends of the field lines, where $\rmd\ln{B}/\rmd{t}<0$ and $K$ is large. The reduction in effective field-line tension by $\Delta<0$ reinforces this trend. Secondly, firehose grows fastest at $k\rhoi\sim{1}$ and generates sharp kinks in the field lines on these scales. $K$ in mirror-unstable regions is also enhanced by the generation of mirror-shaped field lines. Despite this difference, there remains a strong anti-correlation between $B$ and $K$ in saturation [Figure~\ref{sim:curvature}(b)], with $B\propto{K}^{-1/2}$ similar to the MHD case (cf.~fig.~17 of \citealt{Scheko_sim}).

Finally, Figure~\ref{sim:curvature}(c) displays the joint distribution of $\rhoi$ and $k_{\bs{B}\bstimes\bs{J}}$, each computed cell by cell, initially (orange), at the start of the kinematic phase (blue), and in saturation (green). Points rightward (leftward) of the dot-dashed line exhibit perpendicular magnetic structure on scales ${\lesssim}\rhoi$ (${\gtrsim}\rhoi$). At early times, this structure is driven by kinetic instabilities and the emergent folded-field geometry, with an appreciable fraction of the plasma having $\rhoi$ larger than the field-reversal scale. As $B$ increases, the mode of the distribution crosses into the magnetized region at $t/\tcorrf\approx{5}$ and settles when the dynamo saturates ($t/\tcorrf\approx{25}$). As this happens, the bulk of the plasma becomes well magnetized on the folding scale.   

On reason why Bohm-like diffusion may be subdominant to firehose scattering in our simulations is that they are never truly in the $k_{\bs{B}\bstimes\bs{J}} \rhoi \gg 1$ regime: rather, figure~\ref{sim:curvature}(c) indicates that after a few correlation times, $k_{\bs{B}\bstimes\bs{J}} \rhoi \sim 1$. In this scenario, a particle could potentially travel along the interface of two opposing folds in the region of low magnetic energy, much like how a particle will travel along the null line of a neutral current sheet with an orbit width of ${\sim}(k_{\bs{B}\bstimes\bs{J}}^{-1}\rhoi)^{1/2}$~\citep{Sonnerup1971,Chen1984}. As particles travel along this interface, they also experience a $\bb{B}\bcdot \grad B$ drift in the direction of $\bb{J}$, and thus their  correlation time may be as long as $(k_{\bs{B}\bcdot\bs{J}} \vthi)^{-1}$.  In our simulations, $k_{\bs{B}\bcdot\bs{J}}  \sim k_\parallel$, and so this effect may be difficult to discern from scattering off of firehose fluctuations.

\subsection{Dependence on physical and numerical parameters}
\label{sec:param_scan}

\begin{figure}
    \centering
        \includegraphics[width=0.7\textwidth]{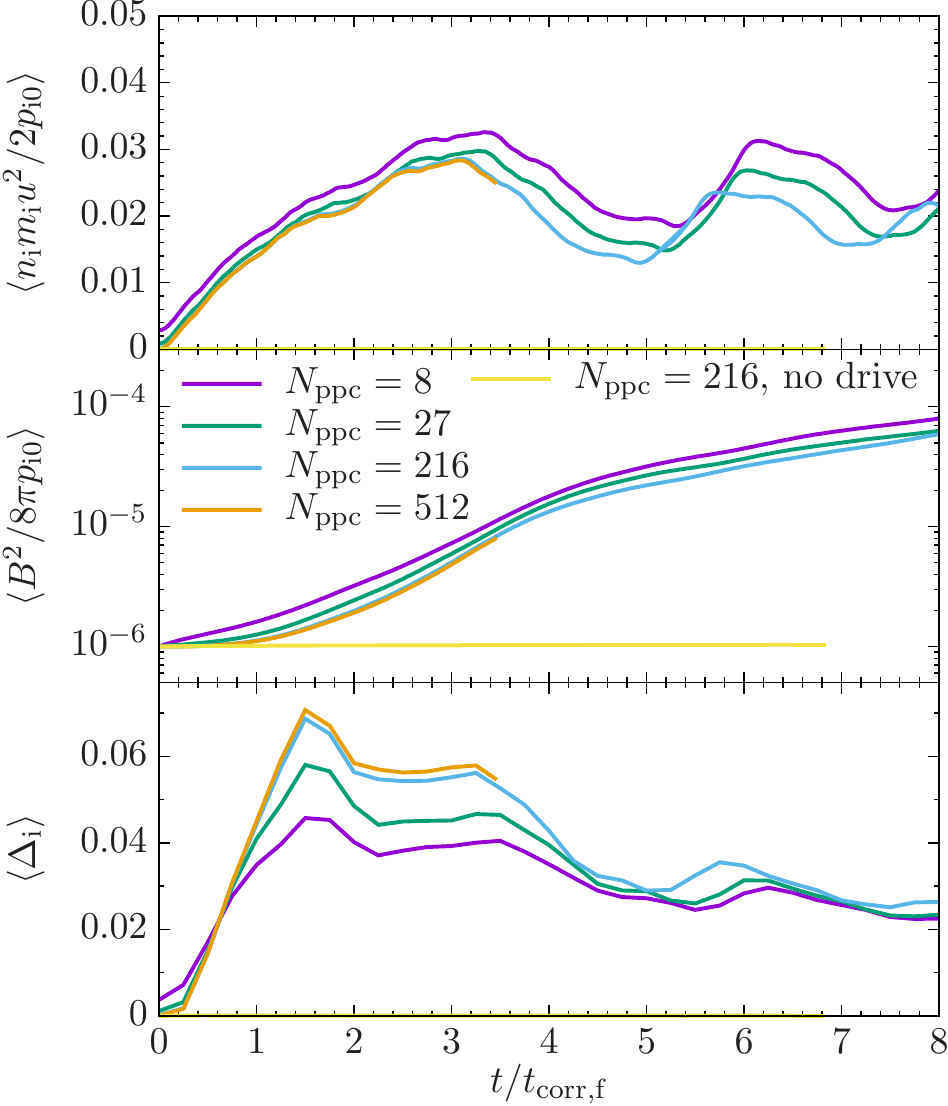}
    \caption[Evolution of energy vs. particle per cell for the kinetic simulations. ]{\label{sim:dynamo_PPC} Time evolution of the kinetic (top) and magnetic (bottom) energies as a function of particles per cell for $\betaio =10^{6}$ ,  $L/\rho_\mathrm{i0} = 16$, and $N_\mathrm{cell}=504^3$.}
\end{figure}

We present results from a series of parameter scans to ascertain how the results of the previous section change along with parameters both physical (e.g. magnetic Reynolds number) and numerical (e.g.  resolution, particles per cell). 

\subsubsection{discrete particle noise}

First, we consider the effect of discrete particle noise on the evolution of the magnetic and kinetic energies, as well as the pressure anisotropy $\Delta_\mathrm{i}$. This noise is a result of the limited and often rather diminutive number of particles one must use in order to make a simulation computationally feasible, leading to grid-scale fluctuations in all moments of the distribution function.  This also has the effect of polluting quantities that immediately depend on these moments, such as $\bb{E}\approx \bb{u}_\mathrm{i}\btimes \bb{B}$. For a given set of parameters, the noise level can quantified, and is done so for the velocity fluctuations in appendix~\ref{ch:noise}.  Figure~\ref{sim:dynamo_PPC}  displays the evolution of the kinetic and magnetic energies as well as the pressure anisotropy for simulations $\betaio = 10^6$, $L/\rho_\mathrm{i0} = 16$, $N_\mathrm{cell}=504^3$, $\eta = 12.7$, $\eta_\mathrm{H} = 32800$, and varying the number of particles per cell. In addition, a single simulation with 216 particles per cell and no random forcing (i.e. $\varepsilon = 0$) is also included to demonstrate that particle noise is not sufficient to appreciably grow the magnetic energy. From this figure, we can conclude that the simulation using 216 particles per cell is converged, and that little is gained from going to 512 particles per cell. Surprisingly, even with only 8 particles per cell the quantitative difference in the evolution of both the kinetic and magnetic energy is only somewhat different than the higher resolution cases. The pressure anisotropy, on the other hand, does change significantly in the lower resolution simulations, as this feature of the distribution function gets overcome by discrete particle noise.

\subsubsection{$L/\rhoi$ scan}

\begin{figure}
    \centering
        \includegraphics[width=0.7\textwidth]{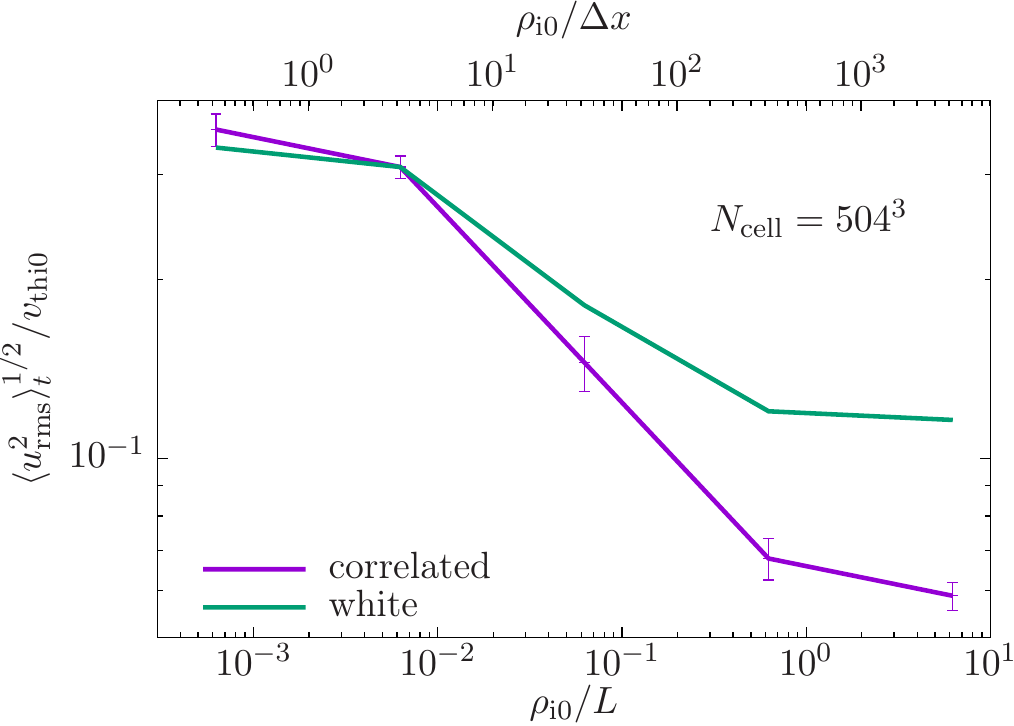}
    \caption[Saturated $u_\mathrm{rms}$ as a function of $L/\rho_\mathrm{i0}$ ]{\label{sim:scan_lrhoi} Time-averaged value of the saturated  $u_\mathrm{rms}$ as a function of $\rho_\mathrm{i0}/L$ at fixed $\varepsilon$. Purple (green) lines indicate systems with time-correlated (delta-correlated) driving. As the magnetization of the system is increased ($\rho_\mathrm{i0}/L$ is decreased), the system readily accepts more energy.  The effect is less pronounced in the case of white-noise forcing.    }
\end{figure}

We also consider the effect of magnetization (parameterized by $L/\mathrm{\rhoi}$) on the ability for the plasma to accept energy from the random driving. A theoretical analysis of this effect is performed in appendix~\ref{ap:forcing} for the asymptotic cases of no magnetization ($\bb{B} = 0$)  and full magnetization ($\rhoi \rightarrow  0$) using the drift-kinetic equation. It was found that the unmagnetized regime is extremely viscous, while certain motions in the drift-kinetic-equation go undamped, and can thus grow with impunity in the linear regime.  We perform a series of simulations to support these conclusions. This is done by varying the initial plasma beta while keeping all else fixed. The results are plotted in figure~\ref{sim:scan_lrhoi}, which demonstrates a smooth transition between the unmagnetized ($L/\rho_\mathrm{i0}  < 1$) and magnetized ($L/\rho_\mathrm{i0}  > 1$)  regimes. This is due to the fact that as the magnetization is increased, particles lose their ability to travel across the magnetic field, and so the perpendicular viscosity drops accordingly. This effect is somewhat less pronounced when using white-noise driving, but it present nonetheless. The finding of increasing $\gamma$ with decreasing $\betaio$ by \citet{Rincon_2016} (${\propto}(\kforce\rhoio)^{2}$ in their set-up), then, might partly be due to the role of $\kforce\rhoio$ in setting $M$ for a given energy-injection rate and in facilitating initially rapid magnetic-field amplification by kinetic instabilities. Additionally, the inset of their figure 3 may  overestimate the growth rates for the simulations with $\betaio \le 10^7$ as they are biased towards the fast initial growth phase caused by kinetic instabilities (as discussed in \S\ref{sim:fastgrowthphase}). Further comparison between the dynamo in unmagnetized and magnetized plasmas will be a topic of future inquiry. 

\subsubsection{$\mathrm{Rm}$ scan, $\mathrm{Re}$ fixed}

\begin{figure}
    \centering
        \includegraphics[width=1.0\textwidth]{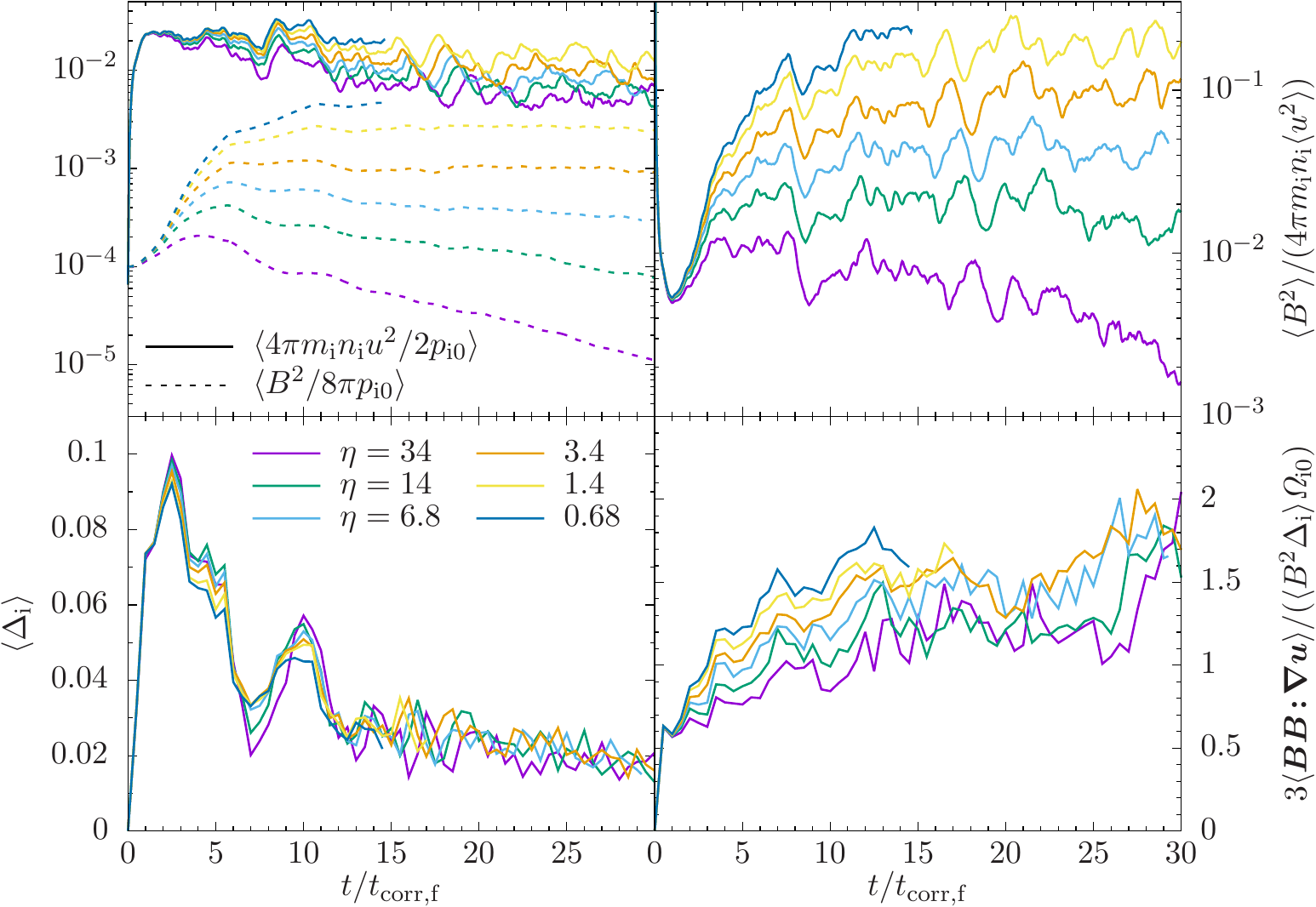}
    \caption[Dependence of various quantities vs. Rm in the kinetic simulations. ]{\label{sim:scan_energy} Evolution of the (\emph{top left}) magnetic and kinetic energies; (\emph{top right}) ratio of the magnetic and kinetic energies; (\emph{bottom left}) box-averaged pressure anisotropy; and (\emph{bottom right}) effective scattering frequency as functions of the magnetic Reynolds number Rm.  Parameters are $\betaio = 10^{4}$, $L/\rho_\mathrm{i0} = 10$, $N_\mathrm{cell}=252^3$ and $N_\mathrm{ppc} = 343$.     }
\end{figure}

We also perform a scan of the magnetic Reynolds number by varying the Ohmic resistivity $\eta$. To do so, we use the same parameters as our $\betaio=10^4$ simulation above, except for a somewhat different resolution ($N_\mathrm{cell}=256^3$) and a higher particle-per-cell count ($N_\mathrm{ppc}=343$).  As only the resistivity is changed, this is akin to keeping the Reynolds number---whatever it may be---fixed.
 This campaign employs Ohmic resistivity, though a small amount of hyper-resistivity $\eta_\mathrm{H}= 1$ is added to ensure sufficient resistive dissipation in the simulation employing $\eta = 0.68$ to avoid numerical issues;  it is checked \emph{a posteriori} that these simulations rely primarily on the Ohmic dissipation. For reference, the parameters for these simulations are displayed in table~\ref{tab:sim:RmScan}.
\begin{table}
\centering
\begin{tabular}{cc}
\hline 
\hline 
$L/d_\mathrm{i0}$ & 1000 \\
$\betaio$ & $10^4$ \\
$L/\rho_\mathrm{i0}$ &  $10$ \\
$\eta_\mathrm{H}$  	& $1$ \\
$\tcorrf$ & 10\\
$N_\mathrm{cell}$ & $256^3$ \\
$N_\mathrm{ppc}$ & 343 \\
\hline
\end{tabular}
\hspace{2cm}
\begin{tabular}{ccc}
\hline 
\hline 
$\eta$ & $u_\mathrm{rms}/v_\mathrm{A0} $ & $\mathrm{Rm}$\\
\hline 
34  &12.8  & 60 \\
14 & 13.6  & 155\\
6.8 & 14.2 & 330\\
3.4 & 14.5  & 680 \\ 
1.4 & 15.0  &1700 \\
0.68 & 15.3 &3600\\
\hline 
\end{tabular}
\caption[Parameters for scan in $\mathrm{Rm}$]{\label{tab:sim:RmScan} Parameters and quantities of interest for simulations scanning a range of $\mathrm{Rm}$. Left table lists parameters fixed across simulation, while the right table records the chosen values of $\eta$, along with some averaged quantities from each simulation.}
\end{table}

Figure~\ref{sim:scan_energy} displays the evolution of the kinetic and magnetic energies, the pressure anisotropy, as well as the Braginskii measure of the effective scattering frequency given by equation~\eqref{eq:sim:bragmes}, for this suite of simulations. While the magnetic energy decays for $\mathrm{Rm} \le 250$, one must also take into consideration the slow decay of the kinetic energy as the plasma heats up (see appendix~\ref{ap:forcing}). If one instead considers the ratio of the magnetic energy to the kinetic energy, then a steady state is reached for $\mathrm{Rm} \ge 125$. This is in line with previous measurements of the critical magnetic Reynolds number seen in simulations of $\mathrm{Re}\sim 1$ MHD dynamo~\citep{Scheko_sim,Rincon_2016}.  The exact value of this ratio in saturation is strongly dependent  on $\mathrm{Rm}$, though it should asymptote to ${\sim}1$ in the limit $\mathrm{Re}\rightarrow 1$ and $\mathrm{Pm}\rightarrow \infty$ (see the discussion in \S\ref{ch1:saturation}); this was also noted in~\citet{Scheko_sim}. On the other hand, figure~\ref{sim:scan_energy} indicates that the average pressure anisotropy appears to be independent of $\mathrm{Rm}$, and that the scattering frequency increases no more than a factor of two as the magnetic Reynolds is increased by a factor of 60. This is consistent with the idea that the inverse scattering frequency is the time it takes for  a particle to travel the length of a fold, and hence to visit different firehose sites. The typical fold length of a magnetic field is independent of $\mathrm{Rm}$ in the isotropic MHD dynamo~\citep{Scheko_sim}, and if this still holds in the kinetic regime, then $\nu_\mathrm{eff}$ should be approximately constant as well.

\begin{figure}[t]
    \centering
        \includegraphics[width=0.85\textwidth]{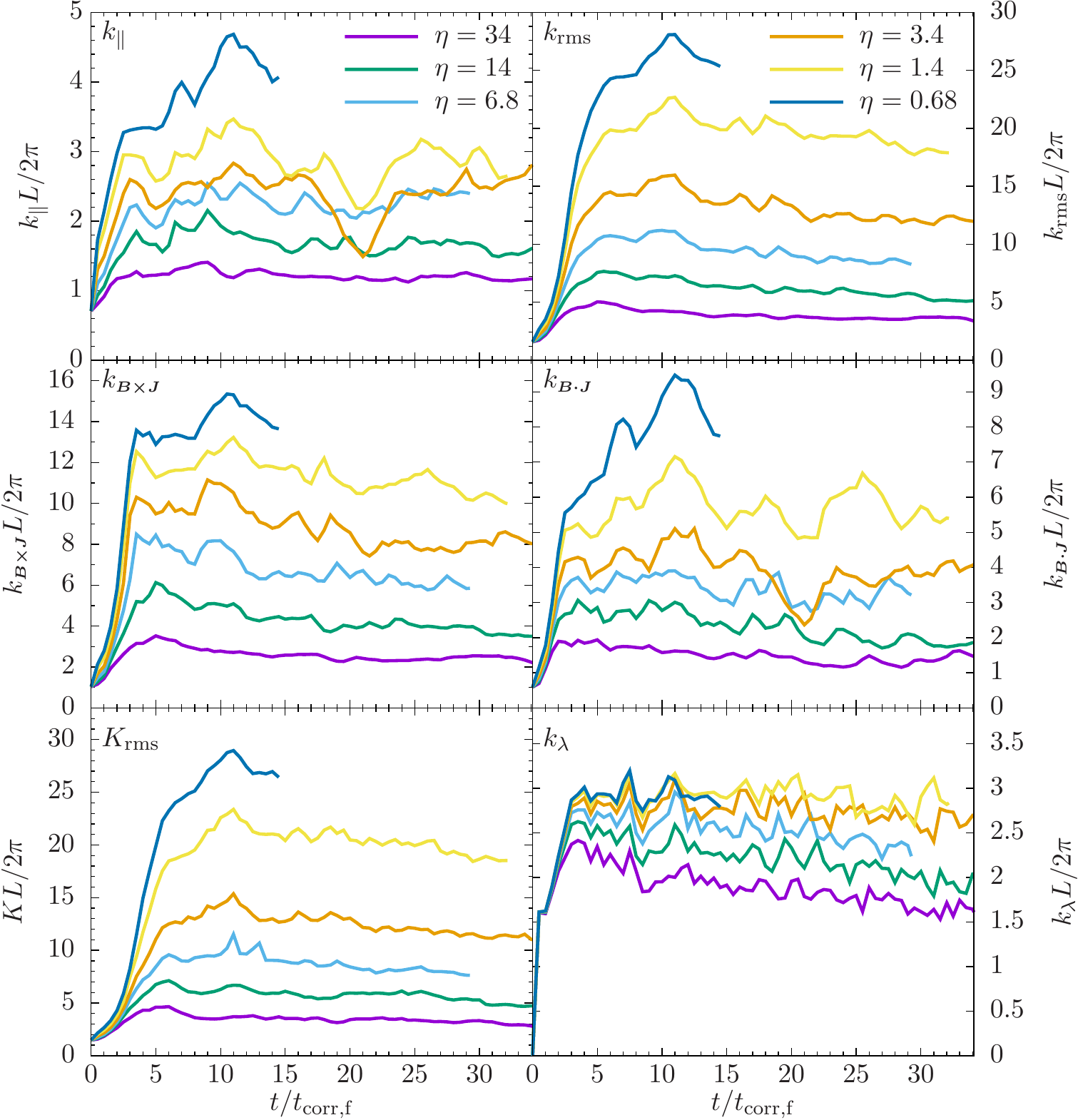}
    \caption[Characteristic wavenumbers vs. Rm for kinetic simulations.]{\label{sim:scan_wave} Time evolution of the characteristic wavenumbers given by~\eqref{char-wavenumbers}, as well as the magnetic field curvature $K$ and the rms wavenumber of the turbulence $k_\lambda \doteq \ba{|\grad \bb{u}|^2}^{1/2}/u_\mathrm{rms}$, for simulations scanning a range of $\mathrm{Rm}$.}
\end{figure}

Figure~\ref{sim:scan_wave} displays the evolution of the characteristic wavenumbers~\eqref{char-wavenumbers}, as well as the curvature and rms  wavenumber of the velocity field $k_\lambda \doteq \ba{|\grad \bb{u}|^2}^{1/2}/u_\mathrm{rms}$.  All the magnetic field length scales decreases as a function of $\mathrm{Rm}$, with $K_\mathrm{rms}$ and $k_{\scriptscriptstyle \bb{B\times  J}}$ experiencing the strongest change. While this is line with the MHD dynamo for most of the magnetic field quantities, $k_\parallel$ also  increases appreciably here, whereas for $\mathrm{Re}\sim 1$ MHD this should remain constant, with $k_\parallel \sim k_\nu$ being set by the viscous scale of the turbulence. This increase may be due to an over-estimation of the fold-length when the simulation is permeated by small-scale firehose and mirror fluctuations.  On the other hand, the size of the gyroradius, which sets the scale of the fastest growing mirror and firehose modes, is not affected by changing $\mathrm{Rm}$, and it is hard to imagine that these small-scale modes are solely responsible for the four-fold increase of $k_\parallel$ witnessed here. This point will be revisited in \S\ref{ch:brag}, where a greater understanding of how the anisotropic plasma viscosity affects the structure of the magnetic field.  Finally, the rms wavenumber of the turbulence asymptotes to a finite value as $\mathrm{Rm}\rightarrow \infty$, which indicates that the viscosity itself is sensitive to the characteristic scales of the magnetic field.

\begin{figure}[t]
    \centering
        \includegraphics[width=1.0\textwidth]{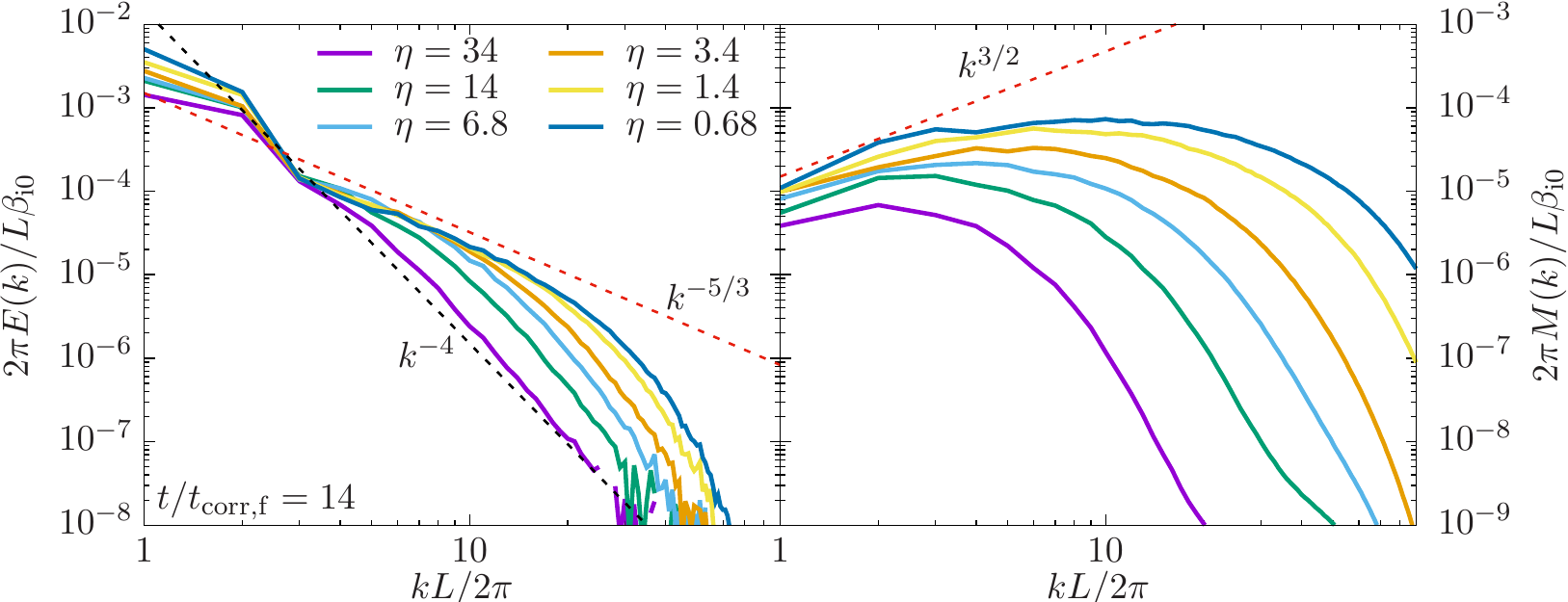}
    \caption[Kinetic and magnetic energy spectra vs. Rm for kinetic simulations.]{\label{sim:scan_spec} Kinetic (left) and magnetic (right) energy spectra for simulations scanning a range of $\mathrm{Rm}$.}
\end{figure}

This last point is reinforced in figure~\ref{sim:scan_spec}, which displays the kinetic and magnetic energies for the parameter scan in $\mathrm{Rm}$.  The most striking feature here is the strong influence the magnetic Prandtl number has on the turbulent cascade of energy beyond the forcing (or parallel viscous) scales.  Note that the systems at $t/\tcorrf=14$  have varying degrees of $L/\rhoi$, and so this may be more a function of magnetization than $\mathrm{Rm}$. The magnetic energy spectra displayed on the right panel of fig.~\ref{sim:scan_spec} exhibit spectral indices that are  shallower than the Kazantsev $k^{3/2}$, an indication that magnetic energy is being transferred to smaller scales quicker than in isotropic MHD, either by kinetic instabilities or interchange motions of the turbulence. A modified version of the Kazantsev model presented in~\citet{Scheko_saturated} that takes into account the anisotropization of the underlying turbulence does indeed predict this behavior; this model will be presented in \S~\ref{ch:brag}.

\begin{figure}
    \centering
        \includegraphics[width=0.7\textwidth]{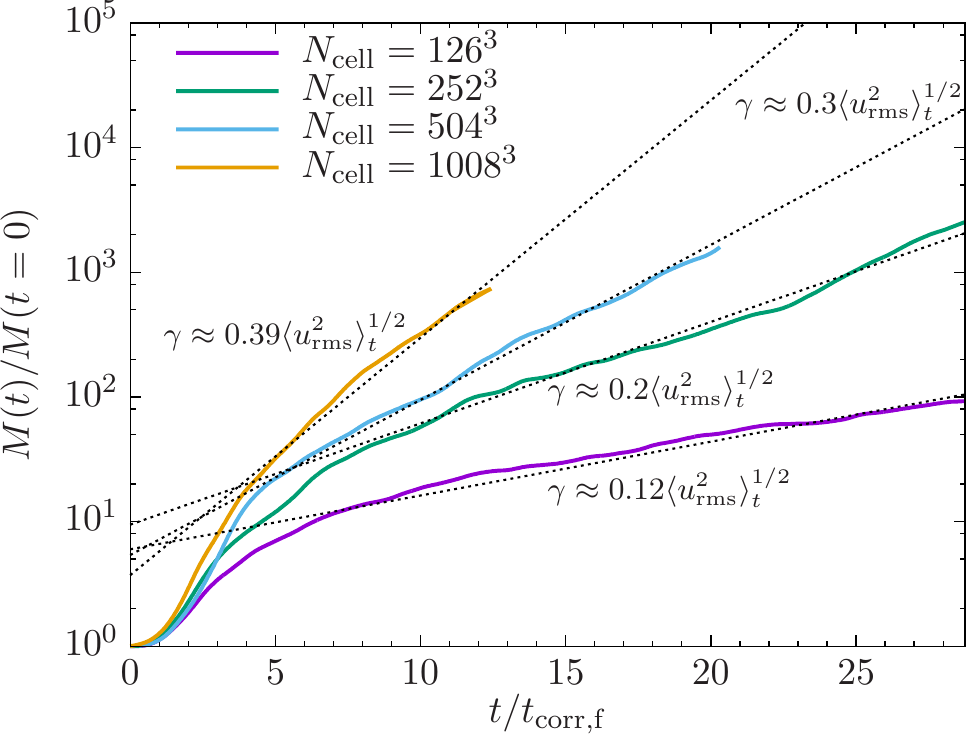}
    \caption[Evolution of magnetic energy versus grid resolution.]{\label{sim:ROS_blowup} The evolution of magnetic energy for various simulations with varying grid resolution. For these runs, the magnetic diffusivity is adjusted so as to have the resistive scale near the grid. }
\end{figure}

\subsubsection{$\mathrm{Rm}$ scan, $\mathrm{Pm}$ fixed}
\label{sec:sim:rmpm}

\begin{table}
\centering
\begin{tabular}{cc}
\hline 
\hline 
$L$ & 16000 \\
$\betaio$ & $10^6$ \\
$L/\rho_\mathrm{i0}$ &  $16$ \\
$\tcorrf$ & 16\\
$N_\mathrm{ppc}$ & 216 \\
\hline
\end{tabular}
\hspace{0.5cm}
\begin{tabular}{cccccccc}
\hline 
\hline 
$N_\mathrm{cell}$ & $N_\mathrm{ppc}$ & $\eta$ & $\eta_\mathrm{H}$ & $\langle u_\mathrm{rms}^2 \rangle_t^{1/2}$ & $\mathrm{Rm}$ &$\mathrm{Rm}_2$ & $k_\eta L/2\upi$ \\
\hline 
126 & 216 &50.8  & 1099200 & 133 &6700 & 1400 &20 \\
252 &216 &25.2 & 262400 & 147&15000 & 3000 &35 \\
504 &216 &12.8 & 32800 & 144&29000&8500 &60 \\
1008 & 27&6.35  & 4100 &152 & 61000  & 25000& 100 \\ 
\hline 
\end{tabular}
\caption[Parameters for scan in $N_\mathrm{cell}$]{\label{tab:sim:GRScan} Parameters and quantities of interest for simulations scanning a range of $N_\mathrm{cell}$. Left table lists parameters fixed across simulation, while the right table records the chosen values of $\eta$, along with some averaged quantities from each simulation. The resistive scale $k_\eta$ is found from the simulation data by finding the maximum of $k^2 M(k)$, the magnetic spectra weighted by $k^2$. The hyper-resistive magnetic Reynolds numbers $\mathrm{Rm}_2$ are calculated assuming $\Reeff = 1$.}
\end{table}

As a final parameter scan, we perform a series of simulations to probe the limit $\mathrm{Rm}\rightarrow \infty$ and $N_\mathrm{cell} \rightarrow \infty$ by varying the grid resolution while changing the magnetic diffusivity in such a way to have the resistive scale near the grid. As the kinetic energy cascade is ultimately terminated by the three-point-digital filter (provided it can cascade to the grid scale), this is akin to performing a scan in $\mathrm{Rm}$ while keeping the (perpendicular) Prandtl number  $\mathrm{Pm} \sim 1$.
 The parameters for these runs are recorded in table~\ref{tab:sim:GRScan}. In the $\mathrm{Re}\sim 1$ isotropic MHD dynamo, this limit should lead to magnetic growth rate that asymptotes to a fixed value $\gamma \sim u_\mathrm{rms}/L$ as $\mathrm{Rm} \rightarrow \infty$~\citep{Scheko_theory}.  If the plasma dynamo is truly a $\mathrm{Re}\sim1$ dynamo, then this behavior should also manifest. Figure~\ref{sim:ROS_blowup} displays the time evolution of the magnetic energy as a function of grid resolution. Surprisingly, the magnetic field growth rate seems to keep increasing without any indication of reaching an asymptotic value as $N_\mathrm{cell}$ is increased. \emph{How can this be?} Before we answer this question, we will first study the dynamo in the weakly collisional regime in chapter~\ref{ch:brag} to get a better understanding on the effects of anisotropic viscosity.  These results shall be used as a basis for comparison to the phenomenon shown in figure~\ref{sim:ROS_blowup}, which will be studied in more detail in chaper \ref{ch:structure}.

\section{Summary}

The initiation and sustenance of the plasma dynamo rely heavily on the production and saturation of kinetic Larmor-scale instabilities, which effectively render the plasma weakly collisional by pitch-angle scattering particles. This scattering causes much of the overall evolution of the plasma dynamo to resemble the large-Pm MHD dynamo, including an analogous ``kinematic'' phase during which the magnetic energy experiences steady exponential growth across several orders of magnitude. %
There are also several differences, such as ion-Larmor-scale structure driven by firehose/mirror instabilities, a Kolmogorov-like cascade of perpendicular kinetic energy, and a field-biased anisotropization of the velocity field.

There is only one other publication to date using kinetic simulations to investigate the plasma dynamo \citep{Rincon_2016}.\footnote{A hybrid-kinetic study of dynamo in collisionless magnetorotational turbulence was presented in \citet{Kunz_mri}.} Those authors focused on the transition from the unmagnetized ($L/\rhoi\ll{1}$) to the magnetized ($L/\rhoi\gg{1}$) regime, with a parameter study conducted to obtain the critical $\mr{Rm}$ at which the dynamo operates. Where our results overlap with theirs, we find broad agreement.
However, computational expense prevented those authors from reaching saturation in simulations starting in the unmagnetized regime ($L/\rhoio < 1$), while also preventing them from proceeding
 far beyond the initial rapid-growth phase driven primarily by kinetic instabilities for their initially-magnetized simulations  ($L/\rhoio >2$). Our finding that this rapid growth eventually gives way to a more prolonged and leisurely exponential growth casts doubt upon their suggestion that the plasma dynamo is self-accelerating, with $\gamma$ increasing as $B$ grows.  The finite resolution of our simulations preclude any determination on the viability of the explosive dynamo, though the results presented in~\S\ref{sec:sim:rmpm} seem to indicate that the plasma dynamo does get faster as the range of accessible small-scale motions increases.  This will be revisited in chapter~\ref{ch:structure}.

\chapter{Dynamo in Weakly Collisional Braginskii MHD}\label{ch:brag}

In this chapter we investigate the fluctuation dynamo in the weakly collisional, magnetized plasma ($\Omegai \gg \nu_\mathrm{i} \gg \omega$) by performing simulations of the Braginskii-MHD system given by equations~(\ref{eq:int:brag}a--d). By using this reduced model, we can study the dynamo in a controlled setting and, in doing so, can gain insight on how a plasma self-organizes to amplify a magnetic field in the face of anisotropic viscous stress. Here we study two of the regimes outlined in \S\ref{sec:paniso}: one of poor regulation of the pressure anisotropy, and one of perfect regulation of the pressure anisotropy using the `hard-wall' limiters described in \S\ref{sec:limiters}.

\section{Outline} 

As this chapter is rather extensive, we give the following brief outline of its structure and results.    We open the results section (\S\,\ref{sec:results}) with a brief overview of the fluctuation dynamo, broken down into its four evolutionary stages (\S\,\ref{sec:overview}). We then present evidence from our simulations in favor of our first conclusion, that the limited Braginskii-MHD dynamo is similar to the isotropic-MHD dynamo in a ${\rm Re}\gg{1}$, ${\rm Pm}\gtrsim{1}$ fluid (\S\,\ref{sec:limited}). Section \ref{sec:unlimited} provides evidence for our second conclusion, that the structure and statistics of the unlimited Braginskii-MHD dynamo imitate those in the saturated state of the more standard ${\rm Pm}\gtrsim{1}$ MHD dynamo (see figures \ref{fig:printout_unlim}--\ref{fig:curvature}). Much of the rest of the chapter is devoted to exploring the consequences of this similarity. This leads us to a treatment of the anisotropization of the fluid flow by the Braginskii viscosity (\S\,\ref{sec:anisotropization}). Some further details related to the interpretation of the latter are relegated to appendix \ref{ap:ROS}. In \S\,\ref{sec:magnetoimmutability}, we consider the relationship of these concepts to that of `magneto-immutability,' previously examined in the context of guide-field Alfv\'enic \citep{Jono_magnetoimmutability} and magnetorotational \citep{Kempski_2019} turbulence. 
Motivated by the results in the previous sections, in \S\,\ref{sec:kazantsev} we formulate an analytic model for the kinematic stage of the unlimited Braginskii dynamo, which is based on an extension of the Kazantsev--Kraichnan model to anisotropic magnetic-field statistics first proposed to describe the saturated MHD dynamo by \citet{Scheko_saturated}. 
 It  provides a reasonable explanation for the behavior of a separate set of simulations in the Braginskii `Stokes-flow' regime ($\Reeff\lesssim 1$), described in \S\,\ref{sec:stokes}, in which we see the dynamo shut off for sufficiently small $\Reeff\lesssim 1$. We summarize these findings in \S\,\ref{brag:sec:discussion}.

\section{Numerical parameters}\label{sec:method}

Simulations are initiated with $\bb{u} =0$ and a random zero-net-flux magnetic field with power at the two largest scales of the box (i.e.~at $k \in [2 \upi/L, 4\upi/L]$). All runs have an initial rms field strength $B_\mr{0,rms} = 10^{-3}$.  The correlation time is chosen as  $\tcorrf=(2\upi)^{-1}$, which corresponds to the inverse decorrelation rate at the outer scale ($\ell_0/u_{\rm rms}$) for $\mathrm{Re} \ge 1$ turbulence.  
 All simulations are run on a triply periodic, $224^3$ grid with $2/3$ de-aliasing (with the exception of the `Stokes-flow' runs discussed in \S\ref{sec:stokes}. For comparison, we have also performed simulations of the dynamo in standard incompressible MHD with isotropic diffusion (i.e.~$\visc_\mr{B}=0$) for Reynolds numbers ranging from order unity to very large values. A list of all the simulations used in this work, along with some parameters of note, is given in table~\ref{tab:runs}. 

\begin{table}
    \centering
    \small
    \begin{tabular}{cccccccccccc}
    \hline 
    \hline
            Run & Res. & $\varepsilon$ &  $\visc_\mr{B}^{-1}$ & $\visc^{-1}$  & $\eta^{-1}$ & $\!\!\langle u_\mathrm{rms}^2 \rangle^{1/2}_t\!\!\!\!$ & $\langle B_0^2\rangle^{1/2}\!\!$ &Re$_\parallel$ & Re & Rm & limiter   \\
    \hline
    MHD1 & $224^3$& 1 & $\infty$ & 20 &  1500  & 0.56 & $10^{-3}$ &  --- & 1.8 & 130 & --- \\
    MHD2 & $224^3$& 1 & $\infty$ &100 &  1500  & 1.07  & $10^{-3}$ &  --- & 17 & 260  & --- \\
    MHD3 & $224^3$ & 1 & $\infty$ & 500 & 1500 & 1.35 & $10^{-3}$&  --- & 110  & 320 & ---\\
    MHD4 & $224^3$ & 1 &  $\infty$ & 1500 &1500 & 1.43 & $10^{-3}$&  --- &340  &  340 & ---\\
    MHDH & $224^3$ & 1 &  $\infty$ &(H) & (H) & 1.50 & $10^{-3}$&  --- & 100  &  100 & ---\\
    \\
    U1 & $224^3$&1 &   20 &1500 & 1500 & 1.21 & $10^{-3}$ & 3.9 & 290 & 290 & unlimited \\
    U2 & $224^3$ &1 &   20 & 600 &1500 & 1.18 & $10^{-3}$& 3.8 & 110 & 280 & unlimited\\
    U3 & $224^3$ &1 &   20 & 240 &1500 & 1.07 & $10^{-3}$ & 3.4 &40 &  255 & unlimited\\
    U4 & $224^3$ & 1 & 20 &96 &  1500 &  0.90& $10^{-3}$ & 2.9   & 14 & 210  & unlimited\\
    U1a & $224^3$& 1 & 100 &1500 &  1500 & 1.38 & $10^{-3}$ & 20 &330 & 330 & unlimited \\
    U1b & $224^3$& 1 & 500 &1500 &  1500 & 1.45 & $10^{-3}$ & 115 & 350 & 350 & unlimited \\
    U1H & $224^3$& 1 &  20 &(H) & (H) & 1.21   & $10^{-3}$ & 3.9 & 100 & 300 & unlimited \\

    \\
    L1 & $224^3$&1 &   20 &1500 & 1500 & 1.47 & $10^{-3}$ & 4.7 & 350 &350 & hard-wall\\
    L2 & $224^3$&1 &  20 &600 &  1500 & 1.43 & $10^{-3}$ &4.6 & 140 & 340 & hard-wall\\
    L3 & $224^3$&1 &   20 &240 & 1500 & 1.33  & $10^{-3}$& 4.2 & 50 & 320 & hard-wall\\
    L4 & $224^3$& 1 &  20 &96 & 1500 & 1.12 & $10^{-3}$ & 3.6 & 17  & 270 & hard-wall\\
    L1m & $224^3$ &1 &  20 &  1500 &1500 & 1.42 & $10^{-3}$& 4.5 &340  &340  & mirror \\
    L1a & $224^3$ & 1 &   100 &1500 & 1500 & 1.43 & $10^{-3}$ &23  & 340 &  340 & hard-wall\\
    L1b & $224^3$ & 1 & 500 & 1500 & 1500 & 1.44 & $10^{-3}$ &  115 & 340 & 340 & hard-wall\\
     L1H & $224^3$& 1 &   20 &(H) & (H) & 1.47 & $10^{-3}$ & 4.7 & 100 & 100 & hard-wall\\
     \\
    MHDSa & $112^3$ & 1 & $\infty$ &  20 & (H)  & 0.60 & $10^{-3}$ &  --- & 2 & 180 & --- \\
    MHDSb & $112^3$ & 20 &  $\infty$ &4 & (H)  & 0.81  & $10^{-3}$ &  --- & 0.5 & 290  & --- \\
    MHDSc & $112^3$ & 500 &  $\infty$ & 0.5 &(H) & 0.67 & $10^{-3}$&  --- &0.05  & 460  & ---\\
    USa & $112^3$ & 1 &  20  &1500 & (H)  & 1.06 & $10^{-3}$ &  3.4 & 250 &  200 & unlimited \\
    USb & $112^3$ & 1 & 10  &1500 &  (H)  & 1.01& $10^{-3}$ &  1.6 & 240 & 240 & unlimited \\
    USc & $112^3$ & 1 &  6  &1500 & (H)  & 0.96 & $10^{-3}$ &  0.9 & 230 & 270 & unlimited \\
    USd & $112^3$ & 2 &  4  &1500 &(H)  & 1.15 & $10^{-3}$ &  0.7 & 275 & 310 & unlimited \\
    USe & $112^3$ & 3 & 2  & 1500 & (H)  & 1.16 & $10^{-3}$ &  0.4 & 275 & 370 & unlimited \\
    USf & $112^3$ & 4 &  1  &1500 & (H)  & 1.11 & $10^{-3}$ &  0.18 & 265 &  440 & unlimited \\            
    USg & $112^3$ & 5 &  0.5 & 1500 &(H)  &  1.07 & $10^{-3}$ &  0.09 & 260 & 515 & unlimited \\
    USg$^*$  & $112^3$ & 50 &  0.5 &4 & (H)  &  0.69 & $10^{-3}$ &  0.05 & 0.4 & 470 & unlimited \\
    \hline
    \end{tabular}
    \caption[Index of runs for the Braginskii-MHD dynamo campaign.]{Index of runs, sorted into those using isotropic MHD (prefix `MHD'), unlimited Braginskii MHD (prefix `U'), and limited Braginskii MHD (prefix `L'). Run names adorned by an `S' employ viscosities approaching and entering the `Stokes-flow' regime (\S\,\ref{sec:stokes}). \emph{Note:}  `Res.' denotes the number of collocation points in the simulation, with the effective resolution reduced by a factor of $(2/3)^3$ due to de-aliasing.  Viscosity and diffusivity values with an (H) indicate simulations with hyper-dissipation with value $\visc_\mathrm{H}$, $\eta_\mathrm{H} =1.8\times 10^7$. `mirror' (`firehose') denotes a simulation with a hard-wall limiter at the mirror (firehose) threshold and no limiter at the firehose (mirror) threshold. Time averages for $\langle u_\mathrm{rms}^2 \rangle^{1/2}_t$ are taken over the kinematic stage.}
    \label{tab:runs}
\end{table}

\section{Results}\label{sec:results}

\subsection{Stages of fluctuation dynamo}\label{sec:overview}

We remind ourselves of the four stages in the typical evolution of the fluctuation dynamo:
\vspace{1ex}
\begin{enumerate}
    \item The {\em diffusion-free regime}, during which the diffusion due to resistivity has yet to become large enough to influence the growth of the magnetic field. This occurs only if the Prandtl number is sufficiently large and the scale of the initial field is much larger than the resistive scale. Once the magnetic field has become sufficiently folded that the bulk of the magnetic energy reaches the resistive scale, magnetic diffusion becomes important and the dynamo enters\dots
    \item ...the {\em kinematic stage}, in which the magnetic energy continues to grow despite the resistivity (if the flow is a dynamo!) but the Lorentz force ($\bb{B}\bcdot \grad \bb{B}$) is still too feeble to exert any dynamical feedback on the field-amplifying turbulence. In MHD, the kinematic dynamo is linear in $\bb{B}$ (though  nonlinear in the random fields), resulting in exponential growth of $B_\mathrm{rms}$~\citep{Kazantsev,Kulsrud1992}. The Braginskii viscosity introduces a dependence in the velocity equation on the unit vector $\eb$. If the parallel viscous stress (which is just the Maxwell stress with $B^2/4\upi$ replaced by $\rmDelta p$) is sufficiently large, the `kinematic' phase is then fundamentally nonlinear: even though the magnetic field is dynamically weak, its structure influences the properties of the flow, which in turn affects induction in a nonlinear way. On the other hand, if the Braginskii viscosity is subject to hard-wall limiters, then its efficacy is reduced to be comparable with that of the Maxwell stress, which is negligible in the kinematic regime. Thus, `hard-wall' limiters again render this stage truly `kinematic'. 
    \item Eventually, the magnetic field becomes strong enough for the Lorentz force to exert a back reaction on the smallest-scale eddies, suppressing their ability to amplify the magnetic field. The dynamo then enters the {\em nonlinear stage}, in which the magnetic-field amplification is driven by progressively larger (and slower) eddies and the dynamo begins to slow down \citep[e.g.][]{Scheko_theory,Maron04,cho09,Scheko_sim,Beresnyak12}, giving way to a linear-in-time growth of magnetic energy (see \S\ref{ch1:saturation}).
    
    \item The fourth and final stage of the dynamo is {\em saturation}, which is achieved when the magnetic and kinetic energies become comparable (though not necessarily scale by scale---see, e.g., \citealp{Scheko_theory, Scheko_sim}).
\end{enumerate}
\vspace{1ex}
In what follows, the evolution and characteristics of the dynamo in each of these stages are examined using results from the hard-wall-limited Braginskii-MHD, unlimited Braginskii-MHD, and isotropic-MHD simulations. We begin with a comparison of the limited Braginskii-MHD and $\mathrm{Pm} = 1$, isotropic-MHD dynamo, which we find to be similar to one another in almost every respect.

\subsection[Limited Braginskii dynamo is similar to high-Re MHD dynamo]{Limited Braginskii-MHD dynamo is similar to ${\rm Re}\gg{1}$, ${\rm Pm}\gtrsim{1}$ MHD dynamo}\label{sec:limited}

The first two stages of the dynamo take place while the magnetic field is dynamically weak. As a result, unless the Braginskii viscosity is negligibly small, a majority (by volume) of the plasma will have pressure anisotropies that  exceed the firehose and mirror instability thresholds, {\em viz.}~$ \visc_\mr{B}|\ROS| \gtrsim B^2$. Applying the hard-wall limiters then effectively disables the Braginskii viscosity in most of the plasma volume, effectively rendering  viscous transport mostly isotropic, at least until the saturated state is reached and  the magnetic field becomes dynamically strong. In what follows, we demonstrate this point both qualitatively and quantitatively through a series of diagnostics.

\subsubsection{Visual appearance of the flow and magnetic field}

\begin{figure}
    \centering
    \includegraphics[scale=0.74]{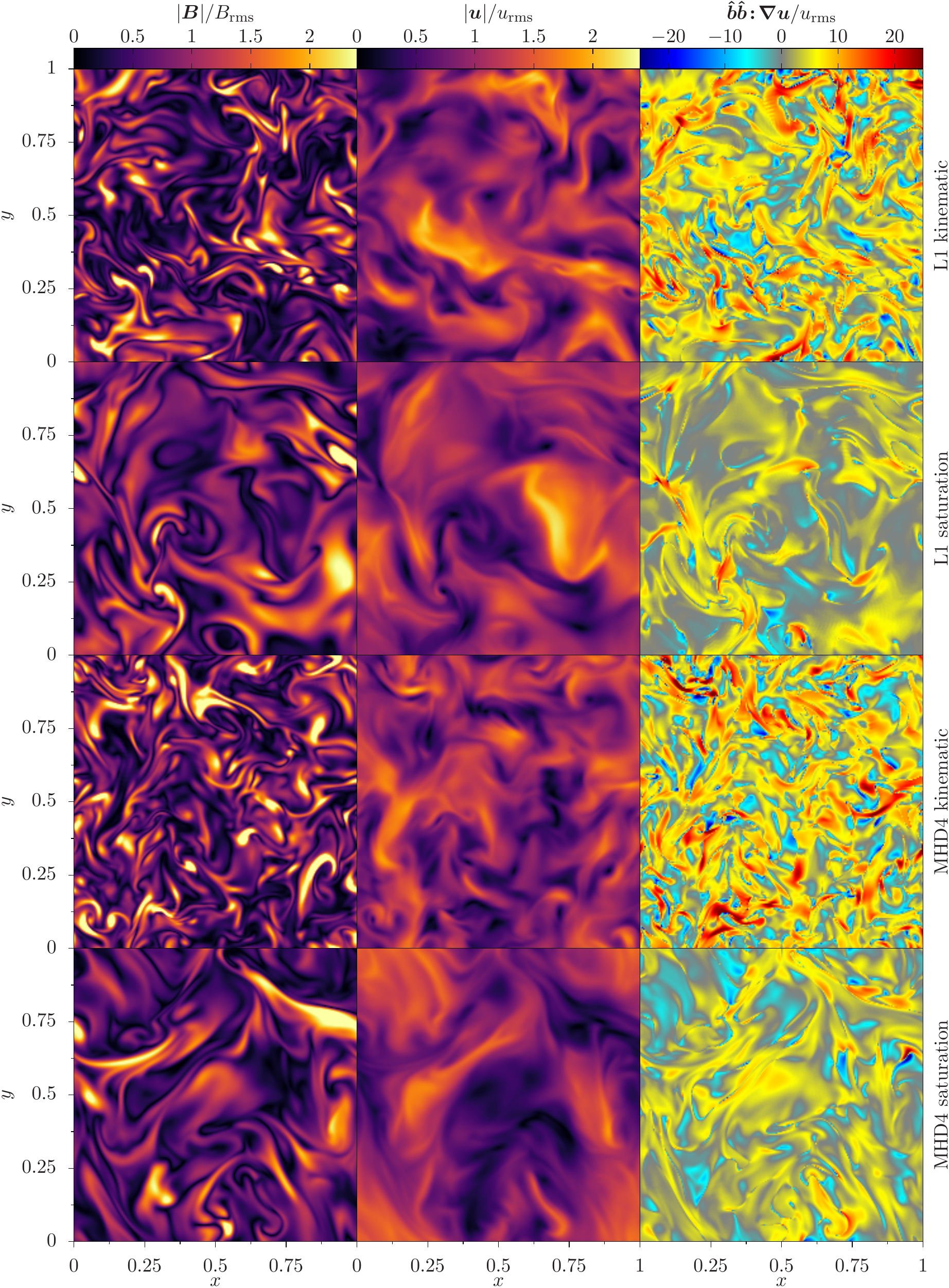}
    \caption[\small Cross-sections from limited Braginskii-MHD dynamo]{A two-dimensional cross-section of the magnetic field and flow in a limited Braginskii-MHD simulation and in a comparable MHD run. Left (center) [right] panels display the magnetic-field strength (velocity magnitude) [parallel rate of strain]. The top two rows display results from the hard-wall-limited simulation L1 (parameters $\visc_\mr{B}^{-1}=20$, $\visc^{-1}=\eta^{-1}=1500$): the first row in the kinematic stage, the second row in the saturated state. The bottom two rows display results from the MHD simulation MHD4 (parameters $\visc^{-1}=\eta^{-1}=1500$): the first row in the kinematic stage, the second row in the saturated state. All plots are on a linear scale, with brighter regions denoting higher magnitudes.}
    \label{fig:printout_lim}
\end{figure} 

We first demonstrate that most of the qualitative features of the Braginskii-MHD dynamo with pressure-anisotropy limiters are similar to those found in isotropic MHD. Figure~\ref{fig:printout_lim} displays two-dimensional cross-sections in the $x$-$y$ plane of the magnetic-field strength, the velocity magnitude, and the parallel component of the rate-of-strain tensor from a limited Braginskii-MHD simulation (run L1) and from a comparable ${\rm Re}\gg{1}$, ${\rm Pm}=1$ simulation using isotropic MHD (run MHD4), both in the kinematic stage and in saturation. The only difference between these simulations is that $\visc^{-1}_{\rm B} = 20$ in the former. Despite this difference, the cross-sections of all displayed quantities are difficult to distinguish between the two systems. In both runs, the magnetic field is dominated by small-scale fluctuations that grow to somewhat larger scales in the saturated state. The classic folded structure of the magnetic field, with direction reversals at the resistive scale and field lines curved at the scale of the flow \citep[e.g.][]{Scheko_sim}, is manifest in both the kinematic and nonlinear regime. Both cases also feature ${\rm Re}\gg{1}$ flow characterized by chaotic structures across multiple scales, with a tendency for the flow to shift to somewhat larger scales in the saturated state when the dynamically important Lorentz force is able to exert an influence on the dynamics. In principle, the Braginskii viscous stress could also influence the flow structure and dynamics, but its regulation to values comparable to the Maxwell stress by the hard-wall limiters merely serves to bolster the Maxwell stress by a factor of order-unity.

\subsubsection{Evolution of magnetic energy}

\begin{figure}
    \centering
    \includegraphics[width=\textwidth]{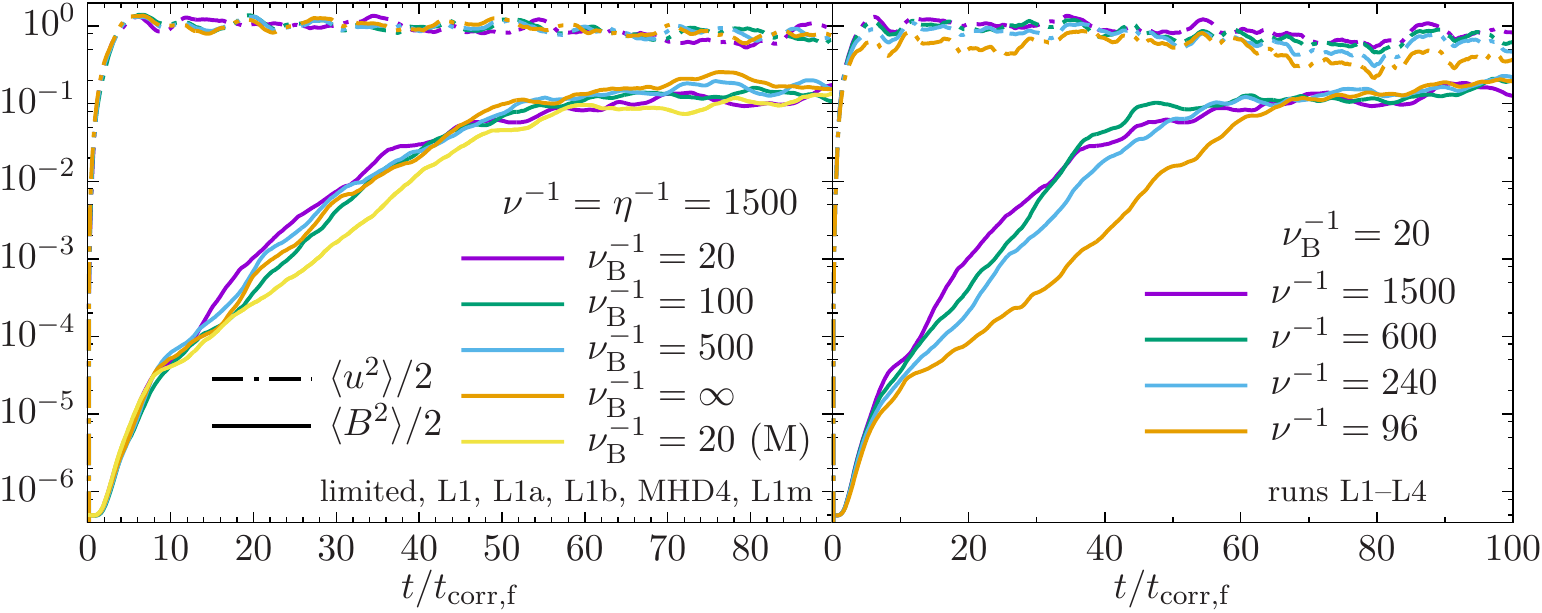}
    \caption[Energy evolution for limited Braginskii-MHD dynamo]{Evolution of the kinetic (dash-dotted lines) and magnetic (solid lines) energies for simulations employing hard-wall pressure-anisotropy limiters with varying Braginskii viscosity  and fixed isotropic viscosity (left) and  varying isotropic viscosity and fixed Braginskii viscosity (right). A Braginskii-MHD simulation employing hard-wall limiters only on the mirror side, marked with an `(M)', is included on the left panel. All simulations use equal levels of resistivity $\eta^{-1} = 1500$.}
    \label{fig:energy_lim} 
\end{figure}

Figure~\ref{fig:energy_lim} displays the evolution of the boxed-averaged magnetic energy $\langle B^2 \rangle /2$ for a series of Braginskii-MHD simulations with hard-wall limiters and isotropic-MHD simulations. In the left panel, various Braginskii viscosities $\visc_\mr{B}$ and fixed Laplacian diffusivities ($\visc=\eta = 5\times10^{-4}$) are used.  For all simulations, the resistivity is fixed ($\eta^{-1}=1500$). When the magnetic field is very weak, the hard-wall limiters affect the majority of the plasma, thus effectively disabling the parallel viscous stress. As a result, the growth rate of the magnetic energy in the kinematic stage is largely independent of $\visc_\mr{B}$ and exhibits magnetic-field growth closely resembling its isotropic MHD counterpart (the orange line). In the right panel, the Braginskii viscosity is fixed at $\visc^{-1}_{\rm B}=20$ while the isotropic viscosity $\visc$ is varied and the magnetic diffusivity is held fixed at $\eta^{-1}=1500$. As in the left panel, the parallel viscosity is effectively disabled at $\langle B^2\rangle\ll 1$ by the hard-wall limiters, and so the scale of the fastest eddies is determined instead by the isotropic viscosity. Accordingly, the growth rate of the magnetic energy in the right panel initially increases as the viscosity is decreased ($\nu^{-1}=96$ to $\nu^{-1} = 240$) and eventually reaches a maximum around $\mathrm{Pm} \sim 1$. If the Reynolds number where to continue to increase, the growth rate would begin to \emph{decrease}, owing to $\mathrm{Pm} < 1$ effects, see~\citet{Vincenzi2002,Iskakov2007}.

\subsubsection{\label{sec:powerspec} Power spectra of the velocity and magnetic fields}

\begin{figure}
    \centering
    \includegraphics[width=\textwidth]{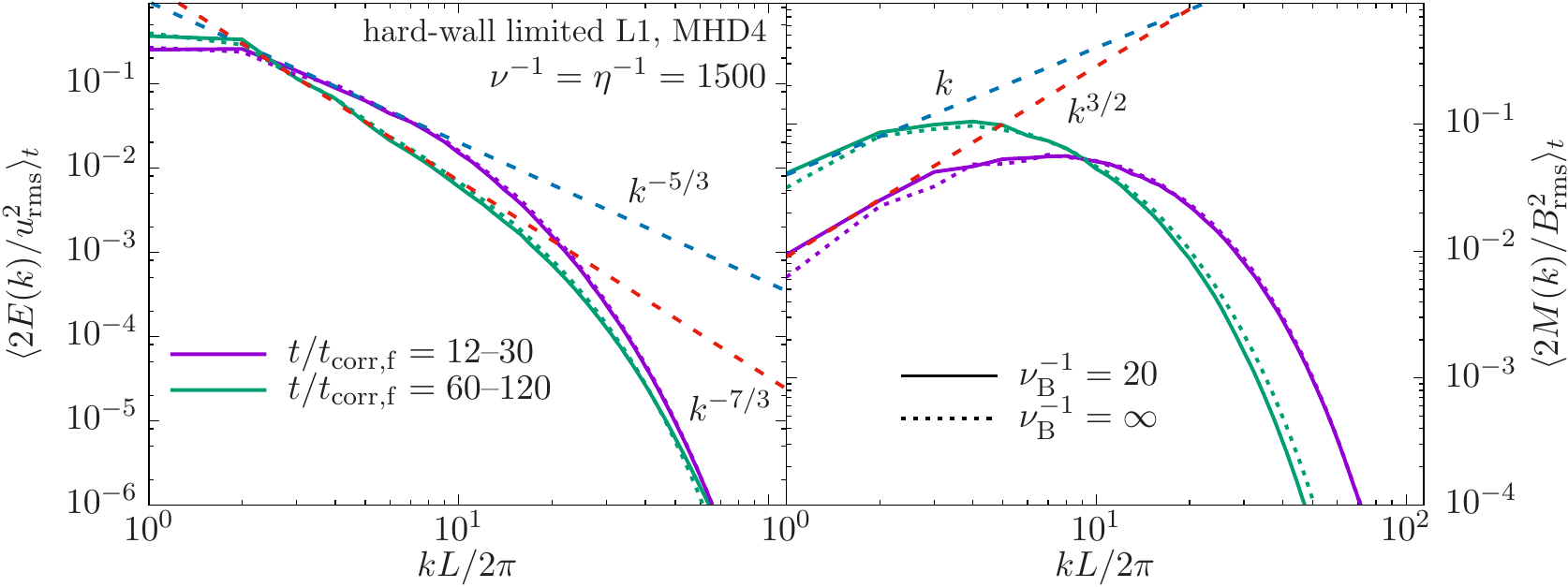}
    \caption[Energy spectra for limited Braginskii-MHD dynamo]{Time-averaged wavenumber spectra of  kinetic energy (left) and  magnetic energy (right) from run L1 and with limited pressure anisotropy (solid lines) and run MHD4 with ${\rm Pm}=1$ (dotted lines) in the exponential phase (purple lines) and in saturation (green lines).}
    \label{fig:lim_spec}
\end{figure}

The similarity between MHD and limited Braginskii-MHD is further illuminated in figure~\ref{fig:lim_spec}, which displays kinetic and magnetic energy spectra for the hard-wall-limited Braginskii simulation L1 and for the $\mathrm{Pm} = 1$ isotropic-MHD simulation MHD4, both in the kinematic stage (purple lines) and in saturation (green lines). The kinetic-energy spectrum has all the characteristics of a high-Re turbulent flow:  a Kolmogorov $-5/3$ spectrum in the kinematic stage and a steeper  $-7/3$ spectrum in the saturated state (as already seen but not explained at these moderate Reynolds numbers by ~\citealp{Scheko_sim}). The magnetic spectrum in the kinematic stage is consistent with the \citet{Kazantsev} $k^{3/2}$ scaling at small wavenumbers, while in saturation it is shallower and closer to being ${\propto}k$. The peak of the magnetic spectra for both simulations also move to larger scales as saturation is reached.  In the isotropic-MHD dynamo, this migration to larger scales has been explained as a consequence of `selective decay' \citep{Scheko_sim}---the increased importance of resistive dissipation on smaller-scale magnetic-field fluctuations as the Lorentz force begins to suppress field-stretching motions. Because limited Braginskii viscosity in regions of magnetic-field growth effectively enhances the magnetic tension by only a factor of 3/2 (since $B^2/4\upi \rightarrow B^2/4\upi + \Delta p = (3/2) B^2/4\upi$ at the mirror threshold), the similarities between the spectra in the saturated state of the limited-Braginskii and isotropic-MHD runs is not particularly surprising. Regions of magnetic-field decay are instead adiabatically pushed towards the firehose threshold, where the limited Braginskii stress exactly nullifies the magnetic tension. The effect of this cancellation on the magnetic spectrum in saturation appears to be minimal, however, likely because the volume-filling factor of regions whose pressure anisotropy lies beyond the firehose threshold is small (see the top panels of figure \ref{fig:brazil}).

\subsubsection{Structure functions of the flow and characteristic scales of the magnetic field}

\begin{figure}
    \centering
    \includegraphics[width=\textwidth]{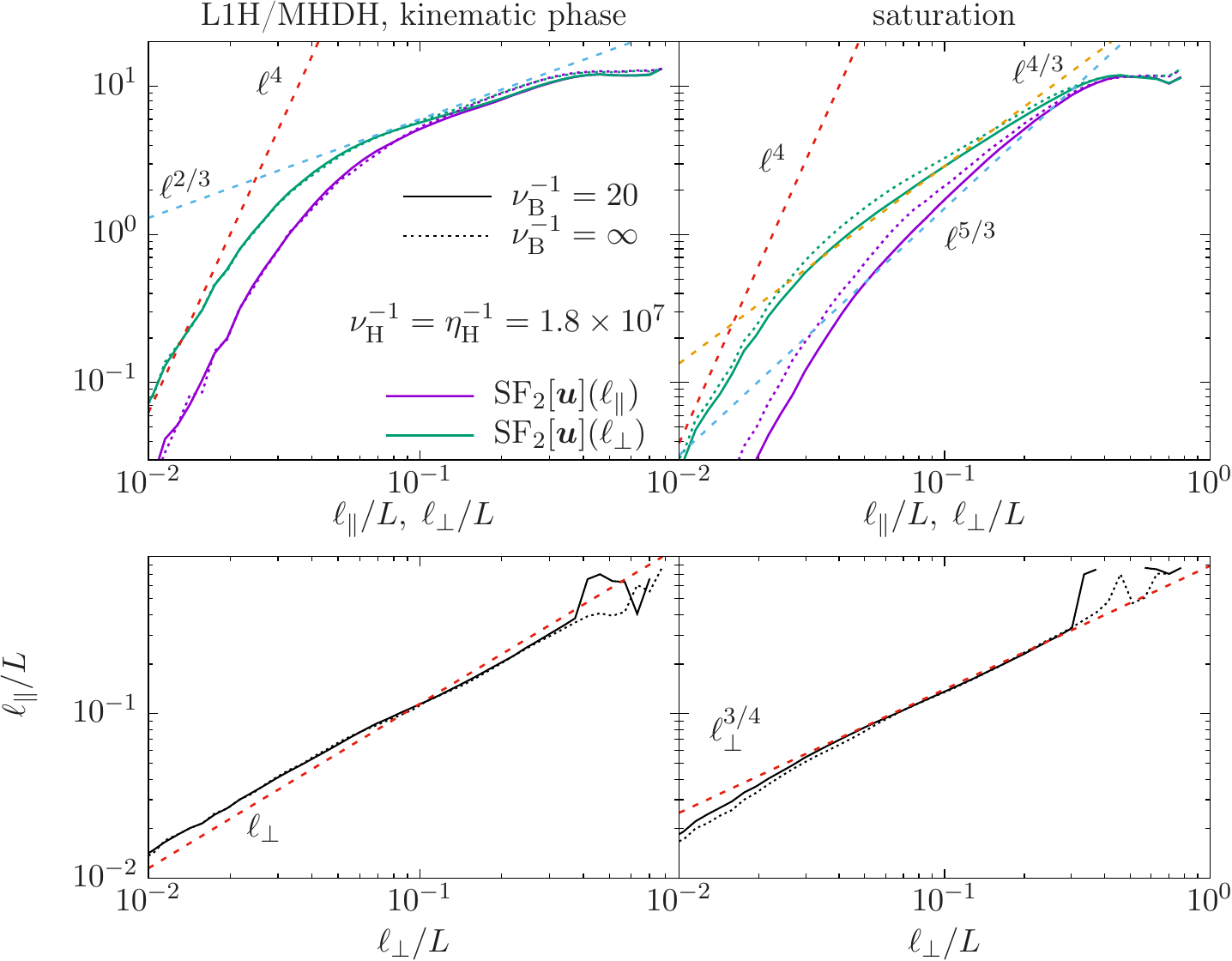}
    \caption[Structure functions for limited Braginskii-MHD dynamo]{Various three-point, second-order structure functions for the Braginskii-MHD system with hard-wall-limited pressure anisotropy in the (left) kinematic and (right) saturation stages. `$\parallel$' (`$\perp$') refers to the direction parallel (perpendicular) to the local scale-dependent magnetic field [see (\ref{eqn:strf_prl_prp})]. Bottom panels show the scale-dependent anisotropy scaling of the parallel variation $\ell_\parallel$, as defined by~\eqref{eqn:scale_aniso}.}
    \label{fig:struct_lim}
\end{figure}

The energy spectra shown in figure \ref{fig:lim_spec} do not provide information about the relative contributions of field-parallel/perpendicular gradients of field-parallel/perpendicular quantities to the overall energetics. To obtain this information, we calculate the structure functions \eqref{eqn:strf} and \eqref{eqn:strf_prl_prp} of the velocity field in the hard-wall-limited, $\visc^{-1}_\mr{B}=20$ Braginskii-MHD simulation L1H, and well as for the MHD simulation employing hyper-diffusion (MHDH).\footnote{Runs L1H and MHDH use hyper-diffusion in order to maximize the inertial range. This facilitates cleaner measurements of scale-dependent anisotropy as compared to runs with only Laplacian dissipation.} The top row of figure~\ref{fig:struct_lim} displays the resulting curves in the kinematic stage (left) and in saturation (right). The structure functions for both systems during their kinematic stage are largely isotropic and remarkably similar, exhibiting a $\ell^{2/3}$ power law (corresponding to a spectral index of $-5/3$) in the inertial range. Beyond the viscous cutoff, all structure functions steepen to a slope close to ${\rm SF}_2 \propto \ell^4$,\footnote{Note that structures functions using a three-point stencil exhibit an $\ell^4$ power law for smooth flows, rather than the $\ell^2$ power law seen using two-point stencils.} which is the maximal slope measurable by structure functions using a three-point stencil. In the saturated state, both runs exhibit an anisotropisation of the turbulence, with the parallel and perpendicular structure functions exhibiting different scalings. The perpendicular structure functions are steeper than Kolmogorov, being roughly proportional to $\ell^{4/3}_\perp$ (corresponding to a $-7/3$ spectral index).  These two scalings were previously observed by~\citet{Yousef2007}, who studied the effects of disparate-scale interactions between turbulence and dynamo-generated magnetic fields on the exact scaling laws of structure functions typically found in isotropic MHD turbulence. The slopes of the parallel structure functions are even steeper, with a scaling of approximately $\ell^{5/3}_\parallel$ (corresponding to a $-8/3$ spectral index). Unlike in the kinematic regime, the structure functions of the limited-Braginskii and isotropic-MHD systems differ in the saturated state, with those from run L1H being slightly steeper than those from run MHDH. This is likely due to addition of the parallel viscous stress to the magnetic tension force, resulting in a stronger magnetic influence on the flow.

The difference in perpendicular and parallel scalings implies a scale-dependent anisotropy in the saturated state of the dynamo, which we quantify using \eqref{eqn:scale_aniso} and display in the bottom-right panel of figure \ref{fig:struct_lim}. Both runs exhibit a scaling close to $\ell_\parallel \sim \ell_\perp^{3/4}$.\footnote{This is to be contrasted with the $\ell_\parallel \sim \ell_\perp^{2/3}$ and $\ell_\parallel \sim \ell_\perp^{1/2}$ scalings predicted for guide-field MHD turbulence respectively without \citep{Goldreich1995} and with \citep{Boldyrev2006,Chandran2015,Mallet2017} scale-dependent dynamic alignment and intermittency corrections.}  By contrast, in the kinematic stage (bottom-left panel) the anisotropy scaling is linear in both systems, indicating isotropic turbulence.

\begin{figure}
    \centering
    \includegraphics[width=0.7\textwidth]{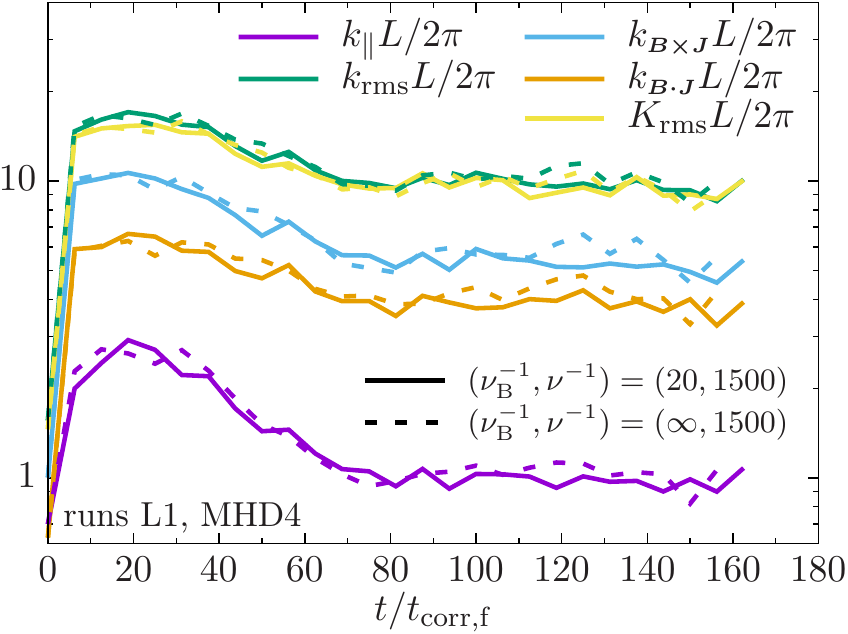}
    \caption[Characteristic wavenumbers for limited Braginskii-MHD dynamo]{Evolution of the characteristic wavenumbers [equations~\eqref{char-wavenumbers}] for isotropic MHD (dashed) and Braginskii MHD with hard-wall pressure-anisotropy limiters (solid lines).}
    \label{fig:wavenumbers_lim}
\end{figure}

\subsubsection{Characteristic scales}

As a final point of contact with results from isotropic MHD, we present in figure \ref{fig:wavenumbers_lim} the characteristic wavenumbers quantifying the geometry of the magnetic field from runs L1 and MHD4. The agreement between the two systems is remarkable, even in the saturated state in which the limited Braginskii viscosity is dynamically important. As the magnetic field is stretched and folded by the flow, it is organized into long thin structures (folds). As a result, the wavenumbers in the kinematic phase satisfy the ordering $k_\parallel < k_{\bs{B\cdot J}} < k_{\bs{B\times J}}$, with each of these wavenumbers decreasing and the latter two scales becoming more comparable in the saturated state as the magnetic folds become more filamentary. Because the magnetic field is only significantly curved in the bends (turning points), the root-mean-square value of the magnetic-field-line curvature $K\doteq|\eb\bcdot\grad\eb|$ is comparable to $k_{\rm rms} \sim k_\eta$. The PDF of $K$ (not shown) is nearly identical to that found in run MHD4 (the blue line in figure \ref{fig:curvature} below), having a peak concentrated near the viscous scale and a power-law tail ${\propto}K^{-13/7}$ \citep[as predicted by][]{Scheko_theory2}. This PDF is representative of a three-dimensional field whose regions of large curvature occupy only a small fraction of the volume.

\subsection[Unlimited Braginskii dynamo is similar to saturated MHD dynamo]{Unlimited Braginskii-MHD dynamo is similar to saturated MHD dynamo}\label{sec:unlimited}

Having established that pressure-anisotropy limiters revert the Braginski-MHD dynamo to its more mundane ${\rm Re}\gg{1}$, ${\rm Pm}\gtrsim{1}$ counterpart, and motivated by the observation of imperfect pressure anisotropy regulation in the hybrid-kinetics simulations seen in chapter~\ref{ch:simulation}, we now turn off those limiters and let the full Braginskii viscous stress act unabated on the flow. In this case, the dynamo takes on a very different character. Left unchecked by limiters, and without a rapidly growing mirror instability in Braginskii MHD to reign it in (see appendix \ref{ap:linear}), the pressure anisotropy will grow proportionally to the parallel rate of strain and, for large values of $\visc_{\rm B}$, spill over both the firehose and mirror thresholds. No longer bound to the relatively meager Lorentz force in the kinematic regime, the large parallel viscous stress exerts a strong dynamical feedback on those fluid motions responsible for amplifying the magnetic field. The result is a strong viscous anisotropization of the fluid flow leading to suppresion of $\ROS$ and, because \eqref{brag_MHD:ind} implies $\rmd\ln B/\rmd t = \ROS$ in the absence of resistivity, a less efficient dynamo. Since the Braginskii viscous stress is similar in form to the magnetic tension force $(\bb{B}\bcdot \grad \bb{B}) = \grad \bcdot (B^2 \eb\eb)$, with the pressure anisotropy playing the role of the magnetic-field strength ({\it viz.}~$B^2 \rightarrow \Deltap \propto \od_t B^2$), one may expect similarities between the unlimited Braginskii-MHD dynamo and the isotropic-MHD dynamo in its saturated state. As in \S\,\ref{sec:limited}, we confirm these expectations by using a variety of diagnostics taken from the unlimited Braginskii-MHD runs and comparing them to our isotropic-MHD runs. Notable differences are also highlighted.

\subsubsection{Visual appearance of the flow and magnetic field}

\begin{figure}
    \centering
    \includegraphics[scale=0.74]{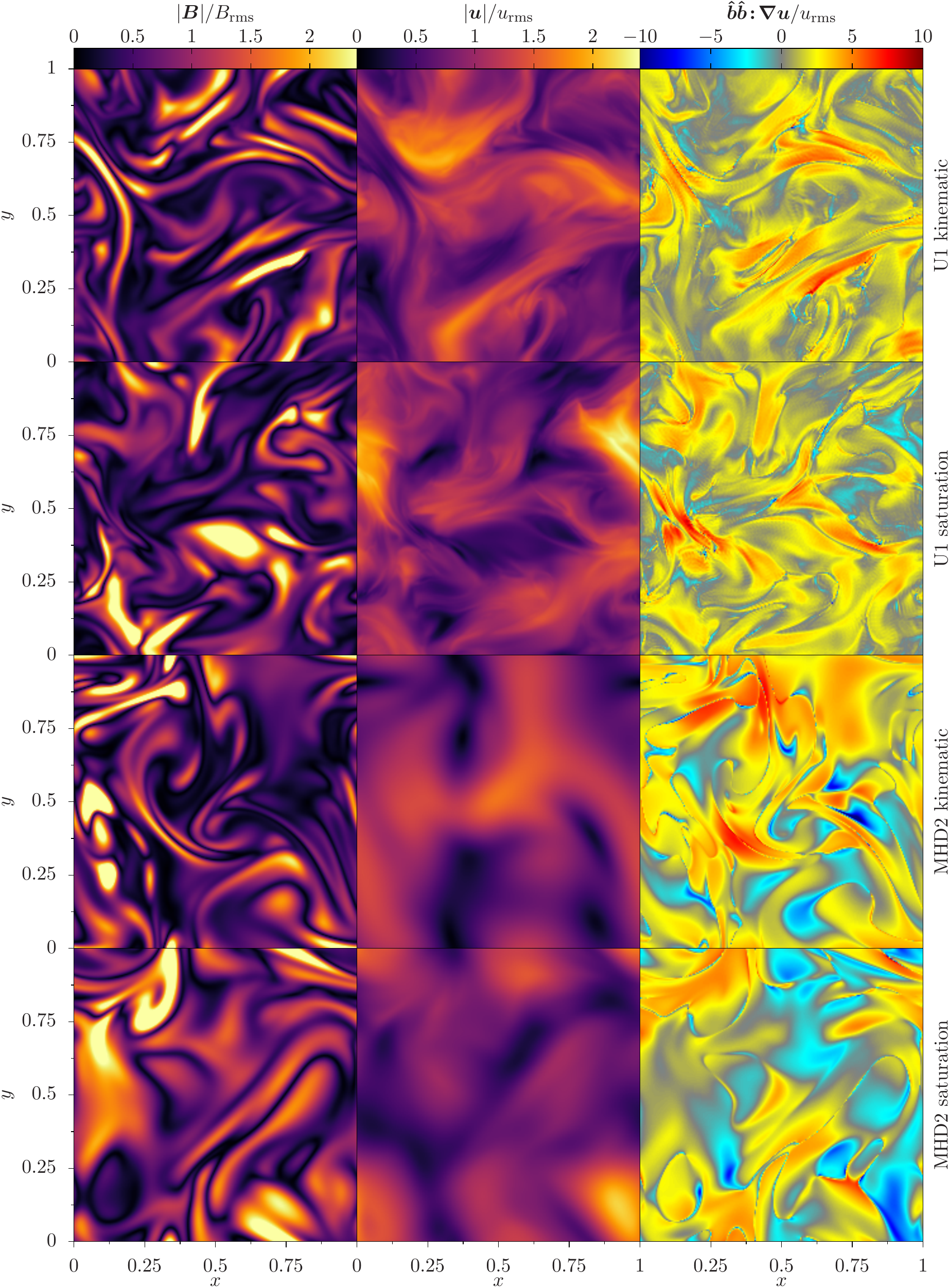}
    \caption[Cross-sections from the unlimited Braginskii-MHD dynamo]{\small Same as figure \ref{fig:printout_lim} but for an unlimited Braginskii-MHD simulation (U1) and a comparable MHD run (MHD2). Left (center) [right] panels display the magnetic-field strength (velocity magnitude) [parallel rate of strain]. The top two rows display results from the unlimited simulation U1 (parameters $\visc^{-1}=1500$, $\visc_\mr{B}^{-1}=20$, $\eta^{-1}=1500$): the first row in the kinematic stage, the second row in the saturated state. The bottom two rows display results from the MHD simulation MHD2 (parameters $\visc^{-1}=100$,  $\eta^{-1}=1500$): the first row in the kinematic stage, the second row in the saturated state. All plots are on a linear scale, with brighter regions denoting higher magnitudes.}
    \label{fig:printout_unlim}
\end{figure}
 
Figure~\ref{fig:printout_unlim} displays the same quantities as in figure~\ref{fig:printout_lim} but now for a Braginskii-MHD simulation without limiters (U1) and an isotropic-MHD simulation with $\visc=\visc_\mathrm{B}/5$ (MHD2). This choice of isotropic viscosity for comparison was advocated by~\citet{Malyshkin} as an effective closure for systems with a magnetic field that is isotropically tangled on sub-viscous scales. Compared to the evolution of the quantities shown in figure \ref{fig:printout_lim}, the magnetic field, velocity, and parallel rate of strain in the unlimited run all exhibit remarkably little change in going from the kinematic stage to the saturated state. This is notable. There are differences from the accompanying MHD panels, but they are relatively minor, being due to the imprint of the anisotropic viscosity on the fluid flow. Indeed, while the MHD system exhibits only large-scale motions typical of order-unity Reynolds numbers, the unlimited Braginskii system features thin striations in the velocity field with sharp gradients across the local magnetic-field direction. Because of this, the Braginskii turbulent state more closely resembles the saturated state of the high-Re MHD system (cf.~bottom centre panel of figure \ref{fig:printout_lim}).

A less subtle difference between runs U1 and MHD2 concerns the parallel rate of strain shown in the rightmost panels of figure \ref{fig:printout_unlim}. The MHD simulation features larger-scale patches of $\ROS$, with more extreme values, than found in the Braginskii-MHD case. This is because, in the unlimited Braginskii system, there is a dynamical feedback whereby the full pressure anisotropy driven by the field-stretching motions ({\it viz.}~$\ROS$) dynamically suppresses those very same motions. There are also fewer regions that exhibit strong negative values of $\ROS$ in the unlimited run, most likely because the act of decreasing the magnetic-field strength with $\ROS<0$ is unstable to the production of firehose fluctuations that grow small-scale magnetic fields (and thus contribute a positive $\ROS$; see \citet{Scheko_2008},  \citet{Rosin_2011} and~\citet{Melville} for further discussion of this point in the context of the parallel firehose instability). We revisit these issues in \S\,\ref{sec:magnetoimmutability}.

\subsubsection{Evolution of magnetic energy}

\begin{figure}
    \centering
    \includegraphics[width=\textwidth]{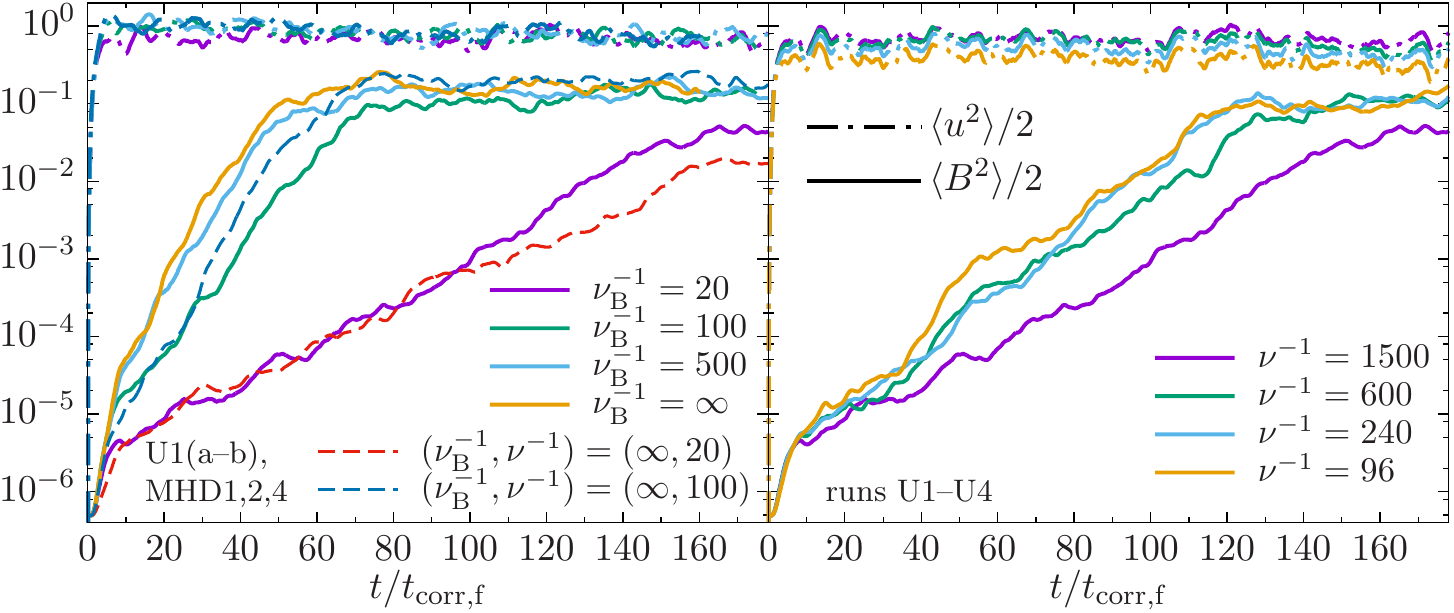}
    \caption[Energy evolution for the unlimited Braginskii-MHD dynamo]{Evolution of the kinetic (dash-dotted lines) and magnetic (solid lines) energies for simulations that do not employ pressure-anisotropy limiters with  varying Braginskii viscosity and fixed isotropic viscosity (left) and varying isotropic viscosity and fixed Braginskii viscosity (right). The evolution of magnetic energy in the MHD1 and MHD2 simulations ($\visc^{-1} = 20$ and $100$, respectively) is included for reference in the left panel (dashed lines). Solid lines in the left (right) panel have $\visc^{-1} = 1500$ ($\visc_\mathrm{B}^{-1} = 20$).}
    \label{fig:energy_unlim}
\end{figure}

The dynamical feedback of the full Braginskii viscosity on the flow affects the time evolution of the magnetic energy, shown in figure \ref{fig:energy_unlim}. As the Braginskii viscosity is increased, the viscous scale of the field-stretching motions becomes larger and the dynamo growth rate decreases accordingly (left panel), indicating that in the unlimited regime the dynamo growth rate is ultimately controlled by the Braginskii viscosity. In the right panel, the Braginskii viscosity is fixed at $\visc^{-1}_{\rm B} = 20$ while the isotropic viscosity is varied. Somewhat counter-intuitively, the growth rate appears to decrease as the isotropic viscosity is decreased. One explanation of this result is that a small isotropic viscosity can allow a cascade of perpendicular (or `interchange'-like) motion to small-scales. These motions, while not responsible for growing the magnetic field, can bring field lines closer together and thereby accelerate the  resistive destruction of the field. The deleterious effect of these mixing motions is a central issue in the $\mathrm{Pm} < 1$ dynamo~\citep{Vincenzi2002,Boldyrev2004,Scheko_Pm,Haugen04,Iskakov2007}, where mixing from all kinetic energy scales promotes resistive annihilation, while only motions larger than the resistive scale can aid in amplification of the magnetic energy via stretching.
This idea is further developed in \S\,\ref{sec:kazantsev}. Another possible explanation is that the rate of strain of the smaller-scale motions allowed by the decreased isotropic viscosity could act to cancel partially the parallel rate of strain driven by the large scales, an effect recently seen in Braginskii-MHD simulations of the magnetorotational instability \citep{Kempski_2019}. This partial cancellation is investigated further in \S\,\ref{sec:magnetoimmutability}. It will be seen that, while both effects are present in our simulations, the latter is of only minor consequence, while the former, that of the efficiency of sub-parallel-viscous mixing, has significant impact on whether or not  unlimited Braginksii-MHD can exhibit a dynamo.

\subsubsection{Power spectra of the velocity and magnetic fields}

\begin{figure}
    \centering
    \includegraphics[width=\textwidth]{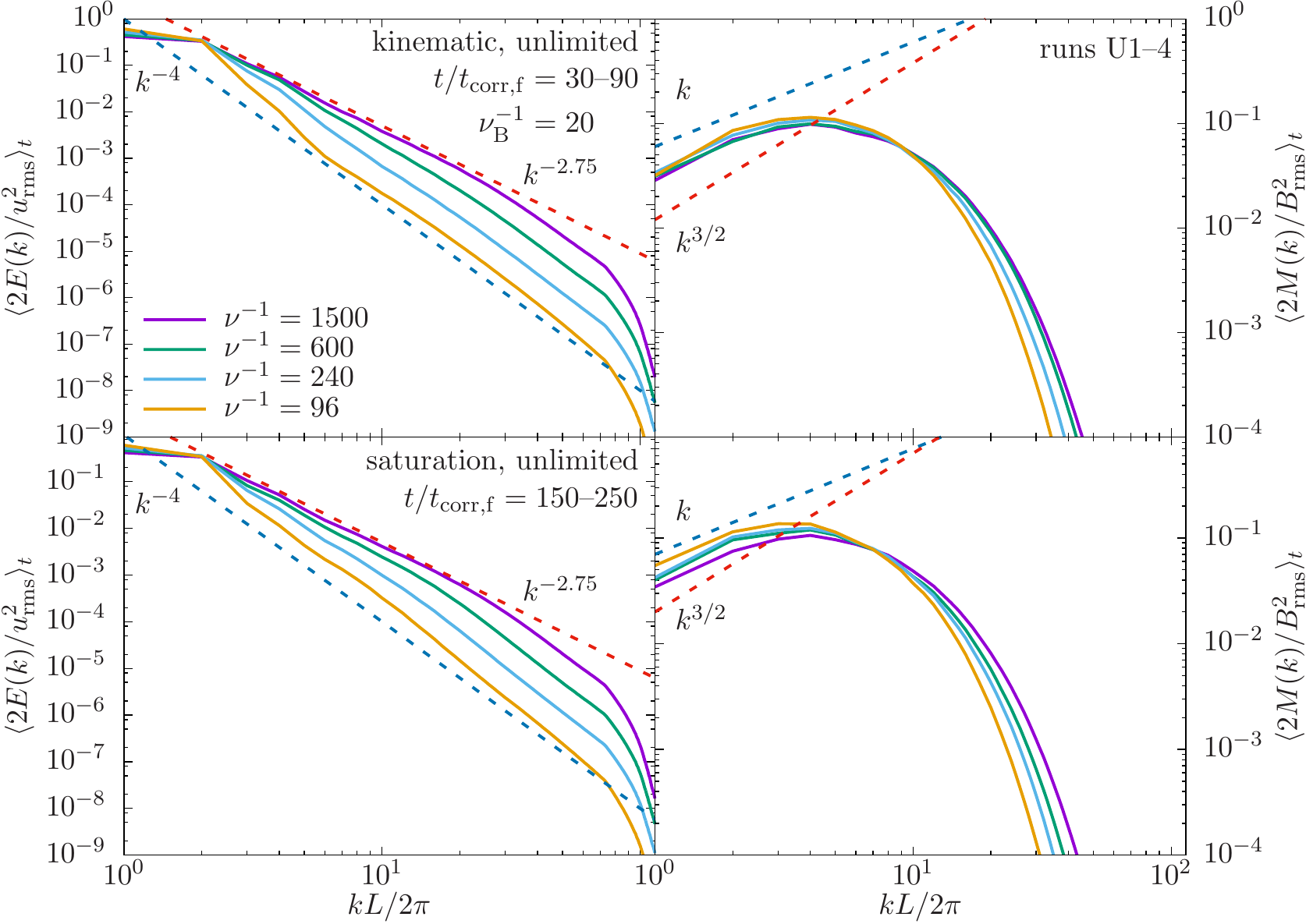}
    \caption[Energy spectra for the unlimited Braginskii-MHD dynamo]{Time-averaged wavenumber spectra of  kinetic energy (left) and  magnetic energy (right) from simulations with unlimited pressure anisotropy for various values of isotropic viscosity $\visc$ in  the kinematic stage (top) and  the saturated state (bottom).}
    \label{fig:unlim_spec}
\end{figure}

\begin{figure}
    \centering
    \includegraphics[width=\textwidth]{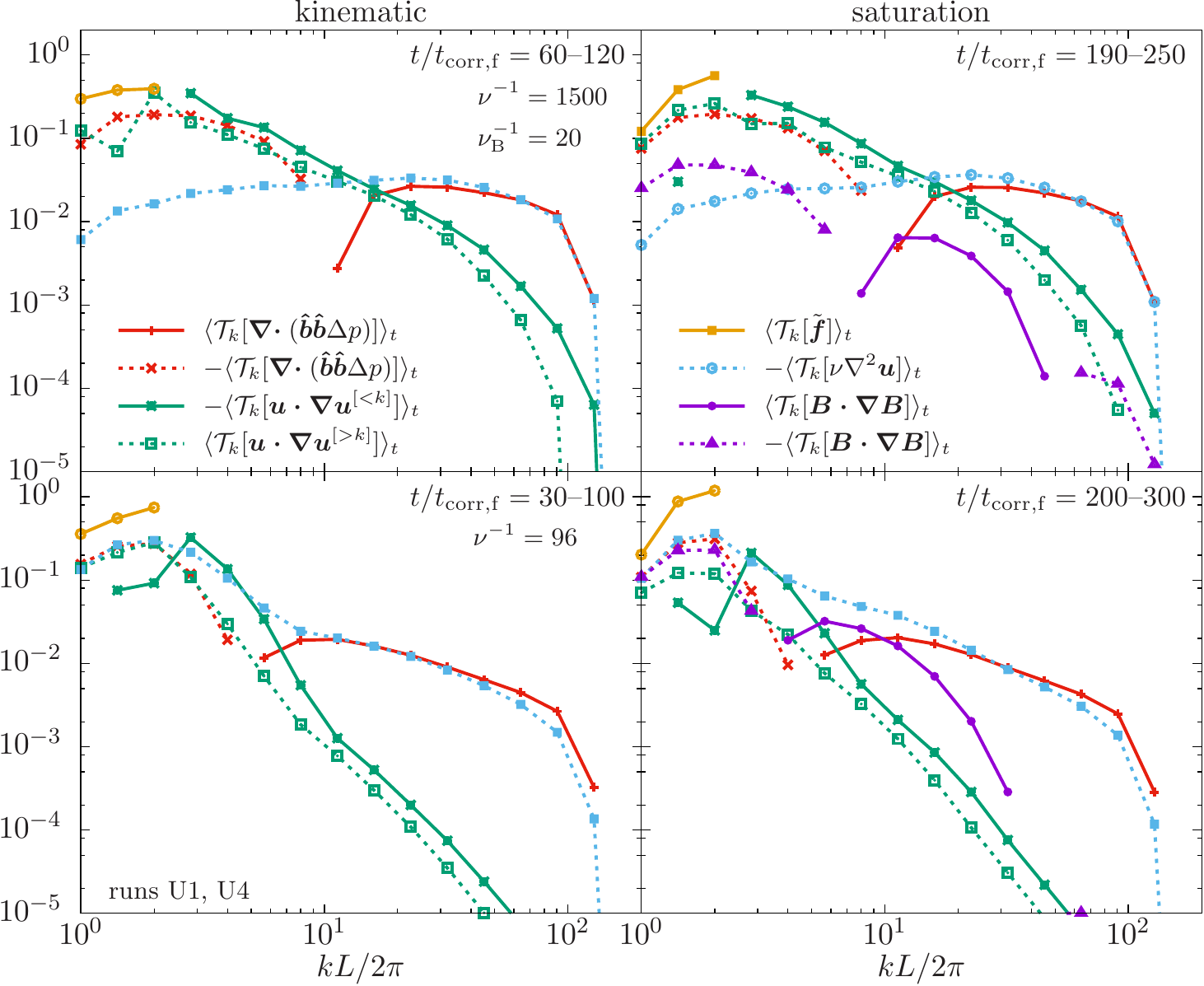}
    \caption[Shell-to-shell transfer for the unlimited Braginskii-MHD dynamo]{Shell-filtered kinetic-energy transfer function $\mathcal{T}_k$~\eqref{eq:shell_trans} for the unlimited Braginskii-MHD system in the (left) kinematic and (right) saturated stages. Top (bottom) row utilizes $\visc^{-1} = 1500$ ($\visc^{-1} = 96$). Solid (dotted) lines denote energy flowing into (out of) the shell centered at mode $k$. }
    \label{fig:trans_unlim}
\end{figure}

Both of these explanations rely on fluid motions getting to sub-parallel-viscous scales. The left panel of figure \ref{fig:unlim_spec} shows that they indeed do, with less isotropic viscosity allowing a shallower spectrum and thus stronger small-scale velocity fluctuations. At the smallest nonzero value of isotropic viscosity (purple line), the unlimited runs exhibit a kinetic-energy spectrum $E(k)\sim k^{-2.75}$, which appears to be asymptotic in $\visc \rightarrow 0$ for fixed $\visc_\mathrm{B}^{-1} = 20$, based on preliminary studies at even higher Re. While steep, this power law  still implies a rate of strain increasing with $k$. Eventually the spectrum experiences a break at $kL/2\upi \approx 30$, whereupon it exhibits a $k^{-4}$ power law down to the grid. This break can be taken as the effective \emph{perpendicular} viscous scale, the slope beyond it being sufficiently steep (spectral index $<{-3}$) that the rate of strain decreases with $k$, i.e., the fastest eddies occur at the largest scales. The run with the next smallest value of isotropic viscosity (green line) also exhibits a similar spectrum, but the spectral break occurs around $kL/2\upi \approx 8$. For $\visc^{-1}=96$ (orange line), almost the entire kinetic-energy spectrum is proportional to $k^{-4}$. Note that a kink exists at the de-aliased grid-scale wavenumber $kL/2\upi = 224/3 \approx 75$ in the kinetic-energy spectra, regardless of the value of the isotropic viscosity. This is a result of small-scale energy injection by the unregulated mirror and firehose instabilities that are present in unlimited Braginskii MHD, the latter of which leads to an ultraviolet catastrophe when $\visc  = \eta = 0$ (see appendix~\ref{ap:linear}). In our simulations, this small-scale energy injection is balanced by the isotropic viscosity, which will be seen in figure~\ref{fig:trans_unlim}.

The kinetic-energy spectra appear to be independent of whether the dynamo is in the kinematic stage (top row) or in saturation (bottom row)---a notable difference from the high-Re MHD dynamo, in which the kinetic-energy spectrum steepens from the Kolmogorovian $k^{-5/3}$ to $k^{-7/3}$ in the saturated state (at least at these limited resolutions; cf. \citealp{Scheko_sim}). The accompanying magnetic-energy spectra, shown in the right panels of figure \ref{fig:unlim_spec}, also show little evolution from the kinematic stage to saturation (consistent with the visualization of the magnetic-field evolution shown in figure \ref{fig:printout_unlim}). For reference, both the Kasantsev $k^{3/2}$ scaling and a $k$ scaling\footnote{This linear scaling results from a calculation of the magnetic spectra for the unlimited Braginskii-MHD dynamo under certain assumptions for the velocity field; see~\citet{Malyshkin}.} are shown. These magnetic spectra are shallower than the Kazantsev 3/2 scaling in both the kinematic and saturated regimes, being much closer to the scaling of the high-Re MHD simulation in the saturated state (see figure \ref{fig:lim_spec}).

The origin of the observed $k^{-2.75}$ slope in the kinetic-energy spectrum  is currently unknown, whereas the $k^{-4}$ spectrum of sub-perpendicular-viscous motions can be understood as  a result of a balance between the pressure-anisotropy stress and the isotropic viscosity. Similar behaviour was measured and explained in \citet{Scheko_sim} by balancing the viscous dissipation $\visc \nabla^2 \bb{u}$ and the magnetic tension $(\bb{B}\bcdot \grad \bb{B})$, resulting in sub-viscous motions that, while initially small, grow along with the magnetic energy. These sub-perpendicular-viscous motions, however, are passive motions that do not exert influence on the evolution of the turbulent motions or the magnetic field. In the unlimited Braginskii-MHD system, the same scenario applies, but now the pressure anisotropy stress $\grad \bcdot (\eb \eb \rmDelta p)$ takes on the role of the magnetic tension $\grad \bcdot (\eb \eb B^2)$. In order for the balance to be valid, it must be the case that $|\bb{u}\bcdot \grad \bb{u}| \ll |\visc \nabla^2 \bb{u}|$, $|\grad \bcdot (\eb \eb \rmDelta p) |$, inequalities that are satisfied in the unlimited Braginskii-MHD simulation beyond the spectral break. Indeed, the shell-filtered energy transfer functions shown in figure~\ref{fig:trans_unlim} confirm that sub-perpendicular-viscous motions arise from a balance between the isotropic viscosity (blue lines) and the Braginskii viscosity (red lines). Surprisingly, in this range the Braginskii viscosity \emph{gives} energy to the velocity field instead of dissipating it; this feature is further discussed in \S~\ref{sec:magnetoimmutability}.

\subsubsection{Structure functions of the flow}

\begin{figure}
    \centering
    \includegraphics[width=\textwidth]{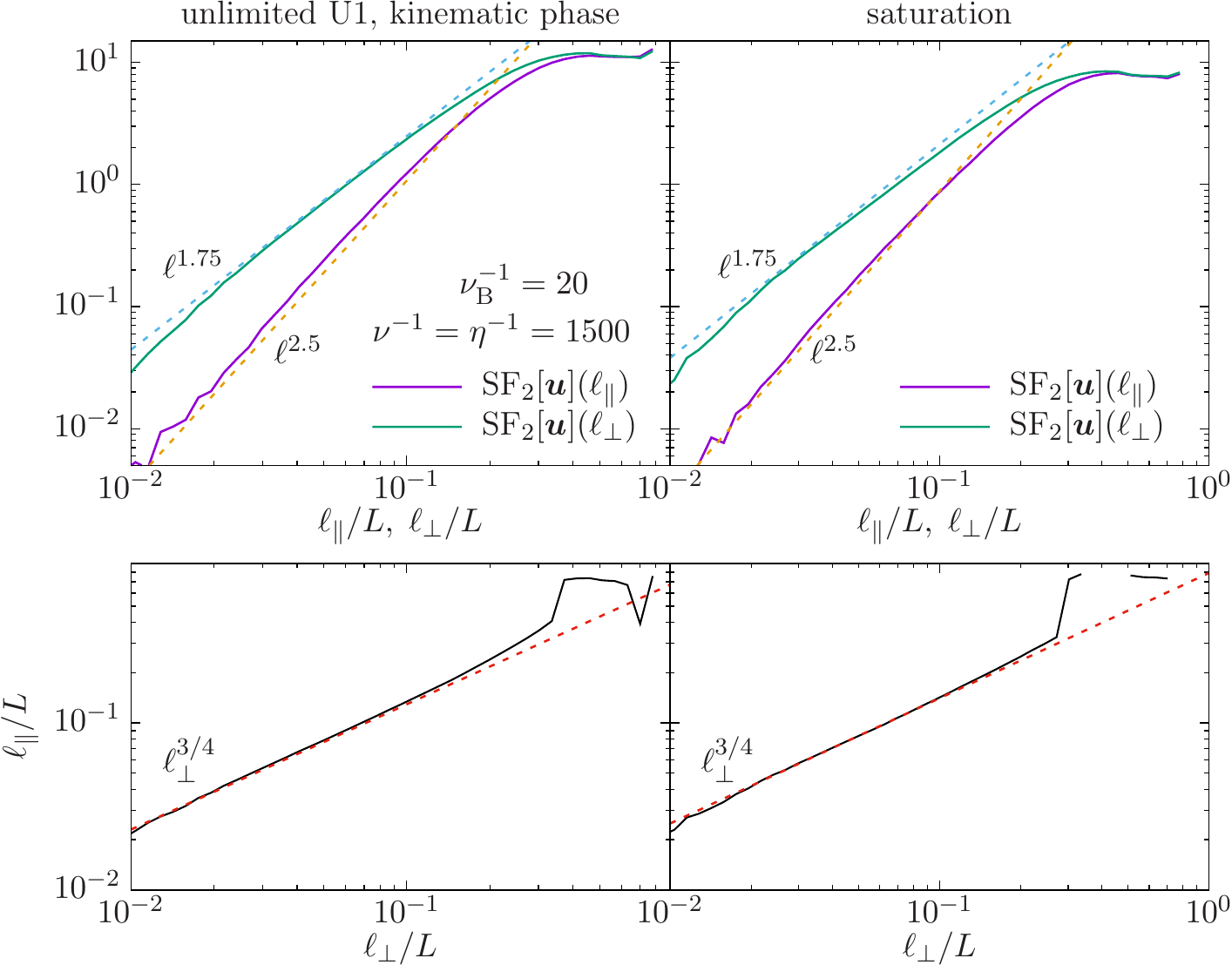}
    \caption[Structure functions for the unlimited Braginskii-MHD dynamo]{Various three-point, second-order structure functions for the unlimited Braginskii-MHD system in the kinematic (left) and saturated (right)stages.}
    \label{fig:struct_unlim}
\end{figure}

We have seen that the presence of an unlimited anisotropic viscous stress strongly biases the properties of the flow with respect to the magnetic-field direction. To quantify this feature better, we calculate the structure functions \eqref{eqn:strf} and \eqref{eqn:strf_prl_prp} of the velocity field in the unlimited $\visc^{-1}_\mr{B}=20$ Braginskii-MHD simulation (run U1) in the kinematic phase and in saturation. The result is shown in figure \ref{fig:struct_unlim}. We find $\mathrm{SF}_2 \propto \ell_\perp^{1.75}$ for the perpendicular structure function (corresponding to the $-2.75$ slope seen in figure \ref{fig:unlim_spec}) and a much steeper $\mathrm{SF}_2 \propto \ell_\parallel^{2.5}$ for the parallel structure function. The steepness of the parallel structure function confirms that small-scale stretching motions play no dynamical role in the dynamo here, and so the largest scales are those primarily responsible for growing the magnetic field. The corresponding spectral anisotropy scaling in both the kinematic and saturated stages is roughly $\ell_\parallel \sim \ell_\perp^{3/4}$, the same as in  the saturated states of the limited Braginskii-MHD and isotropic-MHD dynamos (cf.~figure \ref{fig:struct_lim}). Notably, none of these properties change  from the kinematic stage to saturation.

\subsubsection{Characteristic scales of the magnetic field}

\begin{figure}
    \centering
    \includegraphics[width=\textwidth]{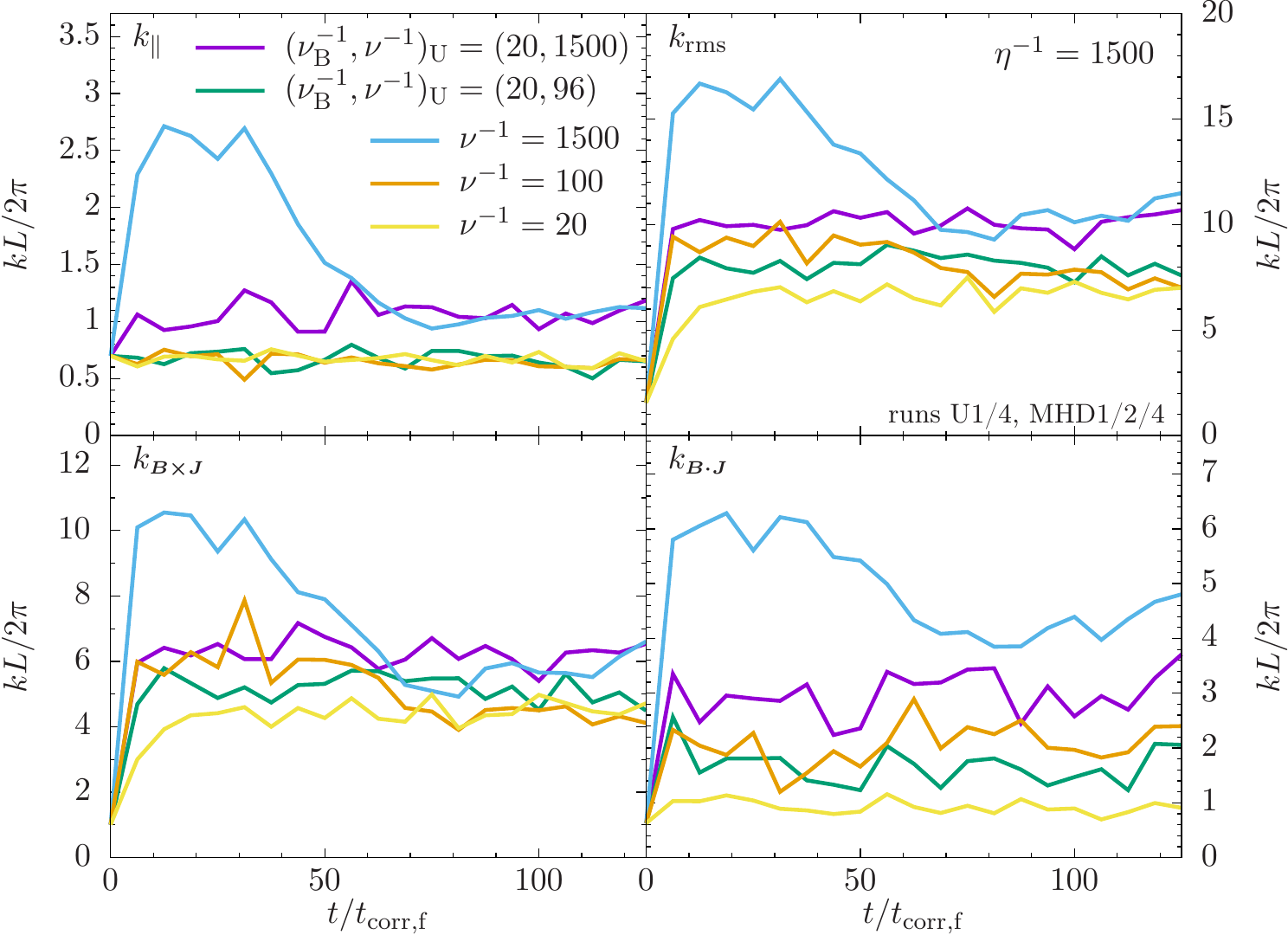}
    \caption[Characteristic wavenumbers for the unlimited Braginskii-MHD dynamo]{Evolution of the characteristic wavenumbers (equations~\eqref{char-wavenumbers}) in run U1 (purple line) and in runs MHD1,2, 4 (yellow, orange and blue lines, respectively).}
    \label{fig:wavenumbers_unlim}
\end{figure}

The geometry of the magnetic field in run U1 does not change much either as the dynamo saturates. Figure~\ref{fig:wavenumbers_unlim} shows the evolution of the characteristic wavenumbers (\ref{char-wavenumbers}) from runs U1 and MHD1, 2 and 3. The wavenumbers in the unlimited Braginskii system (purple lines) are roughly constant in time, holding values remarkably similar to those found in the saturated state of the $\visc^{-1}=1500$ MHD run (blue lines). A somewhat surprising result is that run U1 exhibits a larger $k_\parallel$ than the isotropic-MHD simulations with $\visc^{-1}= \visc_\mr{B}^{-1} = 20$ or $\visc^{-1} = \visc_\mr{B}^{-1}/5 = 100$. This is likely because anisotropic viscosity allows a cascade of perpendicular energy to the perpendicular viscous scale (which is set by $\visc$), and thus small-scale mixing motions are allowed that can bring field lines closer together and promote resistive annihilation. This ultimately leads to characteristically shorter fold lengths. As the isotropic viscosity for the unlimited Braginskii simulations is increased, the saturated value for $k_\parallel$ approaches that of small-Re MHD runs (not shown). 

Figure \ref{fig:wavenumbers_unlim} also makes clear that the unlimited Braginskii viscosity does not result in magnetic fields with larger-scale structure than those produced in isotropic-MHD simulations. It has been argued (for instance by~\citealp{Malyshkin}) that anisotropic viscosity might cause the turbulent dynamo to inverse cascade the saturated magnetic fields from resistive scales to the larger viscous scales (i.e., scales independent of ${\rm Rm}$), by allowing perpendicular motions that are able to unfold the field to cascade to small scales. If true, this could explain the relatively large scale of the observed magnetic fields in the ICM. At least at our modest resolutions, no additional unwinding seems to take place, as the magnetic fields generated by the Braginskii systems exhibit similar $k_{\bs{B}\bstimes \bs{J}}$ and $k_\mr{rms}$ to those found in the saturated state of the isotropic-MHD runs. Efforts to extend our work to larger Rm, and thus larger scale separation, could clarify what sets the peak of the magnetic-energy spectrum in the Braginskii dynamo.

\begin{figure}
    \centering
    \includegraphics[width=\textwidth]{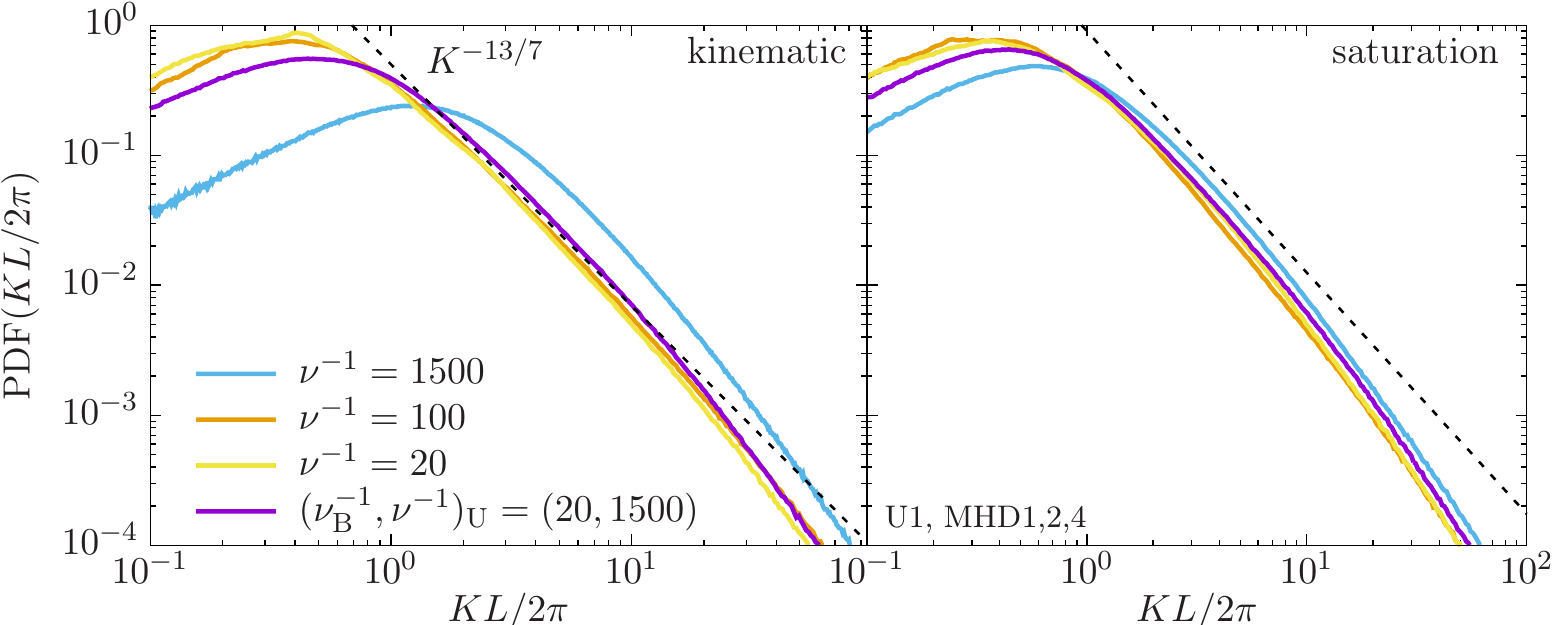}
    \caption[PDF of magnetic-field curvature for Braginski-MHD]{PDF of the magnetic-field curvature $K \doteq |\eb \bcdot \grad \eb|$ in run U1 (purple line) and in runs MHD1, 2 and 4 (yellow, orange and blue lines, respectively) in the kinematic stage and in saturation.}
    \label{fig:curvature}
\end{figure}

An additional useful diagnostic of the magnetic-field geometry is the field-line curvature, defined by $K \doteq |\eb\bcdot \grad \eb|$ and displayed in figure~\ref{fig:curvature} for various runs. The theory of high-Pm MHD dynamo \citep{Scheko_theory2} predicts an asymptotic form of the curvature probability-distribution function (PDF) having a peak concentrated near the viscous scale and a power-law tail of $K^{-13/7}$. This PDF, representative of a three-dimensional field whose regions of large curvature occupy only a small fraction of the volume, is manifest in all of our simulations. The PDFs of the curvature for the unlimited case in both the kinematic stage and saturated state agree closely with those for the MHD systems in saturation. The form of the PDF changes somewhat for the $\mathrm{Pm} = 1$ case (blue line), in which the peak moves to smaller scales in accordance with the higher Reynolds number. By contrast, the unlimited Braginskii-MHD simulation exhibits slightly more curvature than comparable MHD simulations (runs MHD1 and MHD2). This small increase occurs for the same reasons that the characteristic wavenumber $k_\parallel$ increases in the unlimited Braginskii case (see preceding paragraph): small-scale motions that mix field lines are allowed by the Braginskii viscosity (but not by the isotropic viscosity).

\subsection{Viscous anisotropization of the rate of strain}\label{sec:anisotropization}

As motivated at the start of \S\,\ref{sec:unlimited} and supported by the accompanying figures, the unlimited Braginskii-MHD dynamo has many characteristics in common with the saturated state of the ${\rm Pm}\gtrsim{1}$ isotropic-MHD dynamo. This is because the Braginskii viscous stress has a form very similar to that of the magnetic tension, which, in saturation, biases the fluid flow with respect to the magnetic-field direction to reduce the parallel rate of strain $\ROS$. In this section, we explore further this bias, as driven by the Braginskii viscous stress. As part of this discussion, further evidence for the similarity between the unlimited-Braginskii and saturated MHD dynamos is unveiled.

\begin{figure}
    \centering
    \includegraphics[scale=1.0]{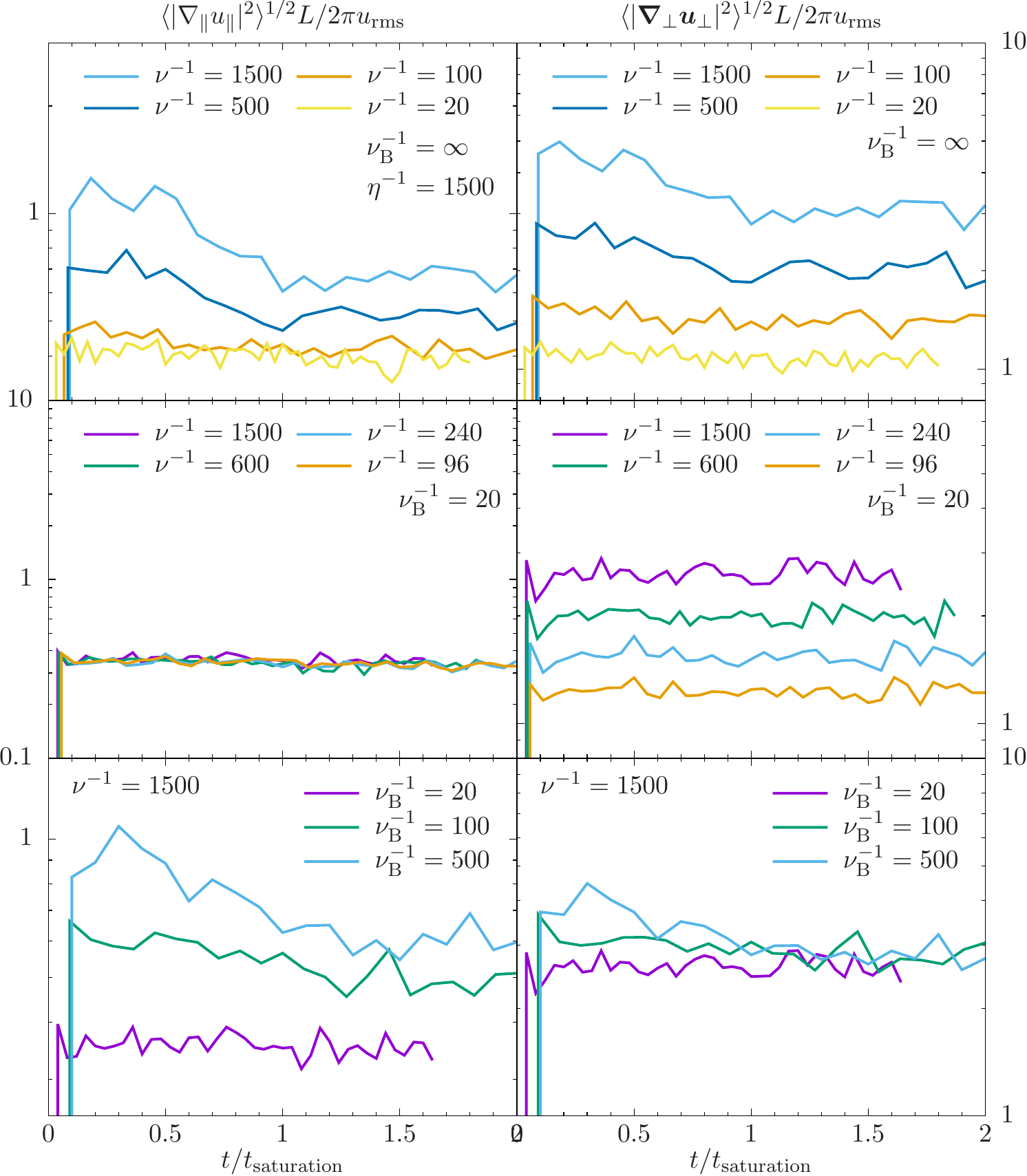}
    \caption[Evolution of the rate of strain for MHD and unlimited Braginskii-MHD]{Evolution of the rms values of the parallel ($\nabla_\parallel u_\parallel \doteq \ROS$) and perpendicular [$\grad_\perp \bb{u}_\perp \doteq ( \mathsfbi{I} - \eb\eb) \bcdot \grad \bb{u} \bcdot ( \mathsfbi{I} - \eb\eb) $] rates of strain for MHD (top) and unlimited Braginskii MHD with varying isotropic viscosity (center) and varying parallel viscosity (bottom). Times are normalized by the time the system takes to reach saturation. All simulations have $\eta^{-1}= 1500$.}
    \label{fig:ROS_all}
\end{figure}

Figure~\ref{fig:ROS_all} displays the time evolution of the rms parallel rate of strain $\ROS$ (left) and the rms value of the perpendicular variations of the perpendicular velocity $\grad_\perp \bb{u}_\perp \doteq (\mathsfbi{I}-\eb\eb)\bcdot \grad \bb{u}\bcdot(\mathsfbi{I}-\eb\eb)$ (right) for various simulations of (top) isotropic MHD and (middle and bottom) unlimited Braginskii MHD. The latter quantity corresponds to motions that mix field lines. For the high-$\mathrm{Re}$ MHD simulations, the saturated state is characterized by a reduction in the magnitude of the parallel rate of strain $|\ROS|=|\nabla_\parallel u_\parallel|$ (see the light and dark blue lines in the upper-left panel), as the magnetic tension force becomes dynamically active. The mixing motions (upper-right panel) are partially suppressed as well. By contrast, the parallel rate of strain in the unlimited Braginskii run with $\visc^{-1}_{\rm B}=20$ is nearly constant in time and independent of isotropic viscosity (middle-left panel). The perpendicular rate of strain is nearly constant as well, but varies with $\visc^{-1}$ in a predictable way: smaller viscosity allows smaller-scale mixing motions, as in isotropic MHD. When the parallel viscosity is decreased at fixed $\visc$ (bottom-left panel), the Braginskii-MHD system converges to the behaviour seen in isotropic MHD, as expected. Interestingly, the overall levels of the perpendicular mixing motions in the unlimited-Braginskii MHD simulations are comparable to those found in the isotropic-MHD systems \emph{in the saturated state}, indicating that these motions are partially suppressed even though the parallel viscosity does not affect them directly. It will be shown in \S\,\ref{sec:kazantsev} that, in order for the dynamo to be viable, the relative sizes of the perpendicular (mixing) motions cannot greatly exceed that of the parallel (stretching) ones.

An alternative way to quantify the anisotropisation of the velocity caused by the Braginskii viscosity is found by examining the relationship between the magnetic-field unit vector $\eb$ and the eigenvectors of the symmetrized rate-of-strain tensor $\mathsfbi{S} \doteq (\grad \bb{u} + \grad \bb{u}^\mathrm{T})/2$. As we are considering only incompressible turbulence, the trace of the tensor $\mathsfbi{S}$, and thus the sum of its eigenvalues $\lambda_i$, are zero. We order these eigenvalues so that $\lambda_1 > \lambda_2 > \lambda_3$. Because $\mathsfbi{S}$ is real and symmetric, its eigenvalues are real and its eigenvectors $\boldsymbol{\hat{e}}_i$ are orthogonal, thus $|\eone\bcdot \eb|^2 + |\etwo\bcdot \eb|^2 + |\ethree\bcdot \eb|^2 = 1$. The eigenvectors $\eig_1$ and $\eig_3$ correspond to the directions of stretching and compression in the incompressible flow, while $\eig_2$ points in the so-called `null' direction (which can either stretch or compress). The alignment angles $\theta_i$ are defined such that $|\eig_i \bcdot \eb| = \cos \theta_i$. For isotropic turbulence, $|\lambda_1| \sim |\lambda_3| \sim 5 |\lambda_2|$, with $\lambda_2 \gtrsim 0$.

For a maximally efficient kinematic dynamo, the magnetic-field unit vector should align itself mainly with the stretching direction, $\eig_1$, in order to maximize $\ROS$, and preferentially away from the compressing direction, $\eig_3$, in order to minimize resistive diffusion~\citep{Scheko_sim}. One way to satisfy these constraints while having a magnetic geometry consisting of field-reversing folds (as we observe in our simulations) is to align the field at its turning point with the null direction, $\eig_2$.\footnote{Similar types of alignment occur in neutral-fluid turbulence, in which it has been observed that the vorticity $\grad \btimes \bb{u}$ aligns itself with $\eig_2$~\citep{ashurst1987}.} 
Because $\lambda_2 > 0$ typically, this configuration also assists in field amplification somewhat.
By contrast, there can be several different ways in which a dynamo can saturate. One such scenario is by having the magnetic field align itself in such a way that $\ROS \sim \lambda_2$, which can sufficiently reduce the stretching motions that grow the magnetic energy. This can occur if $\theta_2 = 0$ or if the magnetic field aligns itself between $\eig_1$ and $\eig_3$ (i.e. $\theta_1 = \theta_3$) with some component along $\eig_2$ (and thus $\ROS \propto \lambda_2$). 
The configuration $\theta_1 = \theta_3 = 45^\circ$, $\theta_2 = 0$ has the added advantage of minimizing mixing motions (i.e.~those with $|\grad_\perp \bb{u}_\perp|$) that promote resistive dissipation of the field, and thus should be more apparent in simulations with large $\mathrm{Re}$.
The two scenarios described above are studied and quantified in appendix \ref{ap:ROS}.

\begin{figure}
    \centering
    \includegraphics[scale=0.95]{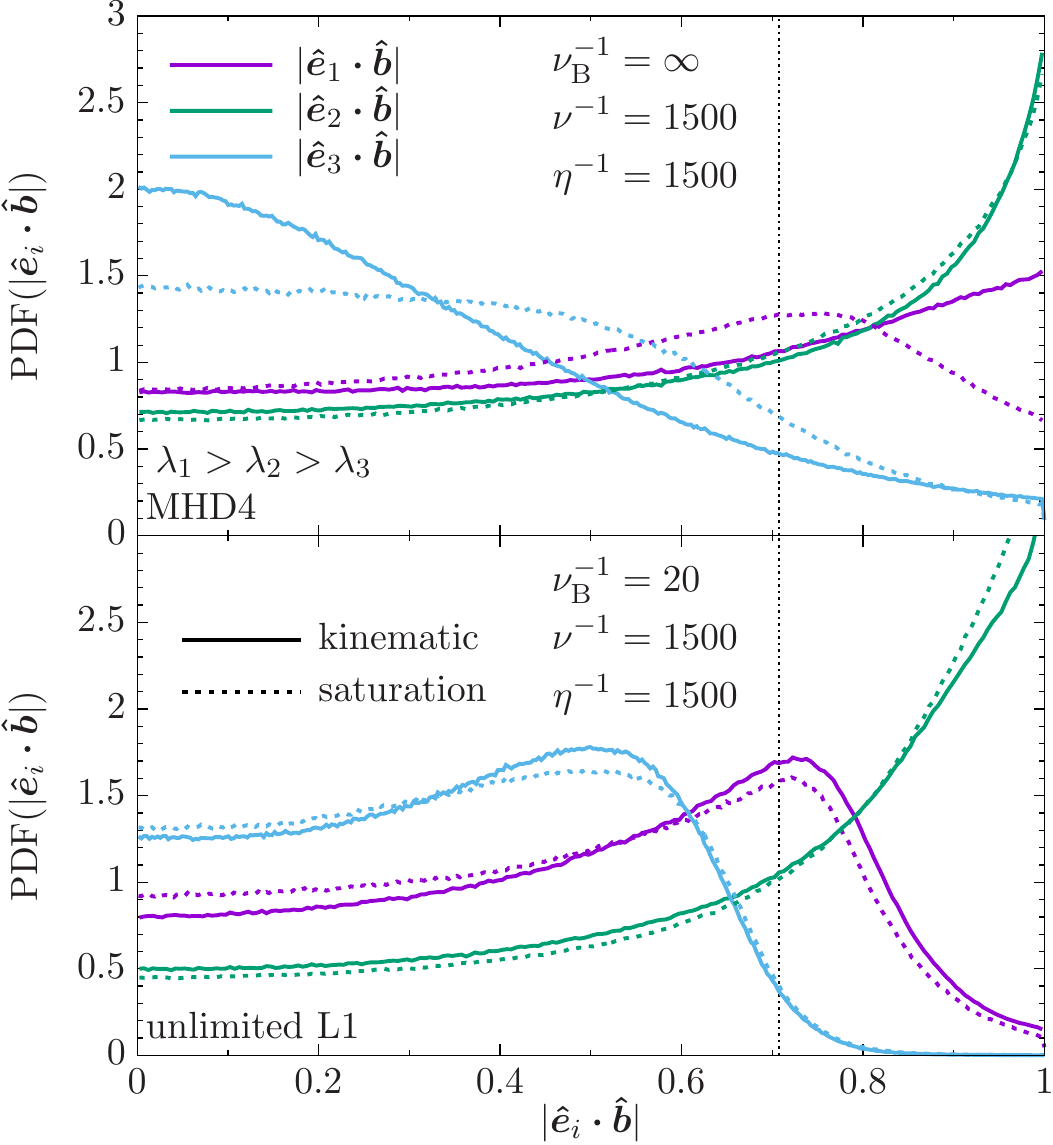}
    \caption[One-dimensional PDF of the magnetic field/rate of strain alignments]{PDF of the symmetric rate-of-strain eigenvector alignments for simulations of (top) isotropic MHD and (bottom) unlimited Braginskii MHD in the kinematic stage (solid line) and saturated state (dotted line). Eigenvalues are ordered such that $\lambda_1 > \lambda_2 > \lambda_3$. }
    \label{fig:angle_some}
\end{figure}
\begin{figure}
    \centering
    \includegraphics[width=0.95\textwidth]{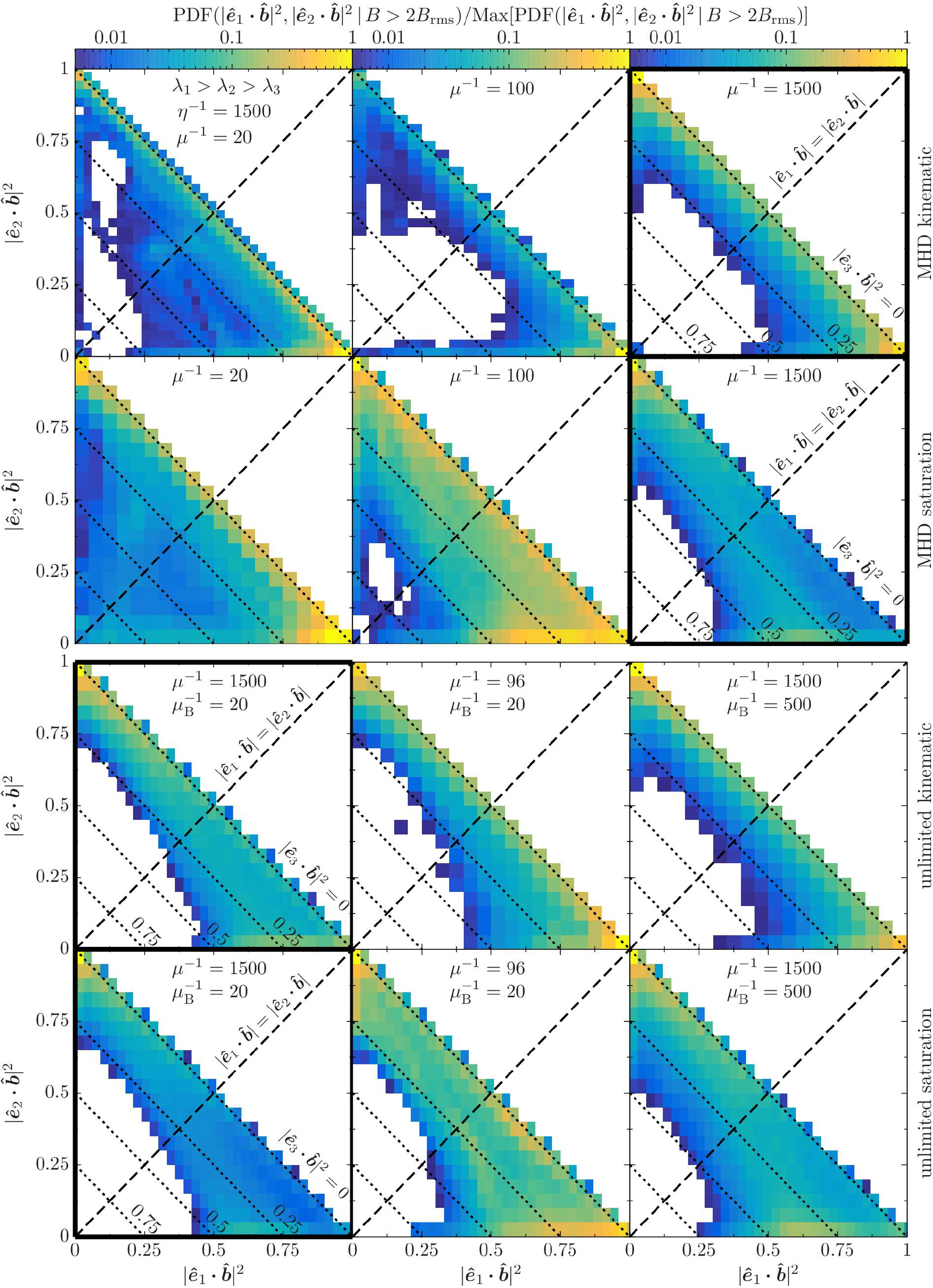}
    \caption[Two-dimensional PDF of the magnetic field/rate of strain alignments ]{\small Two-dimensional PDF of the cosine of the angles between the magnetic field and the first two ordered eigenvectors of the symmetric rate-of-strain tensor for various MHD and unlimited Braginskii-MHD simulations. PDFs are conditioned on regions of high magnetic energy ($B > 2 B_\mathrm{rms}$). Dashed lines corresponds to $|\eone\bcdot \eb| = |\etwo\bcdot \eb|$; dotted lines correspond to solutions of  $|\eone\bcdot \eb|^2 + |\etwo\bcdot \eb|^2+ |\ethree\bcdot \eb|^2 =1$ for fixed $|\eig_3\bcdot \eb|$ (i.e., $|\eig_3\bcdot \eb|$ is fixed along these lines). The simulations featured in figure~\ref{fig:angle_some} are marked by the thicker boxes.}
    \label{fig:angle2d_all}
\end{figure}

Figure~\ref{fig:angle_some} displays the PDFs of the alignments $|\eig_i \bcdot \eb|$ for ${\rm Re}\gg{1}$, $\mathrm{Pm} =1$ MHD (run MHD4; top) and unlimited Braginskii MHD (run U1; bottom) in the kinematic stage (solid line) and saturated state (dashed lined). For MHD the most probable values of alignment for both $\eig_1$ and $\eig_2$ in the kinematic stage are 1, while for $\eig_3$ it is 0. Thus the magnetic field wants to align itself either with $\eig_1$ or $\eig_2$. In saturation, fields that were initially aligned with $\eig_1$ now rotate towards $\eig_3$, consistent with the theoretical arguments made in the prior paragraph and in appendix \ref{ap:ROS}. Unlimited Braginskii MHD, however,  mimics the statistics of the saturated MHD dynamo throughout its evolution.

This behaviour can also be seen in figure~\ref{fig:angle2d_all}, which presents two-dimensional PDFs of the measures of alignment $|\eone\bcdot \eb|^2$ and $|\etwo\bcdot \eb|^2$ both for  simulations of MHD (top two rows) and for unlimited Braginskii MHD (bottom two rows) in the kinematic stage (first row of group) and saturated state (second row of group). In order to highlight the alignment of dynamically important fields, these PDFs are conditioned on regions of high field strength, $B > 2 B_\mathrm{rms}$. As the alignment of $\eb$ with $\eig_3$ can be related to the other two alignments through $|\eone\bcdot \eb|^2 + |\etwo\bcdot \eb|^2 + |\ethree\bcdot \eb|^2 = 1$, these PDFs contain all the information about the field-alignment statistics (e.g.~density appearing in the bottom-left corner of these plots signify magnetic field primarily aligned with $\eig_3$). It is clear that, in the MHD simulations, the magnetic field wants to align itself principally with either $\eig_1$ or $\eig_2$ in the kinematic stage, rather than with some combination of all three eigenvectors. The statistics in saturation vary significantly for different amounts of viscosity; for $\mathrm{Re}\sim 1$, the dynamo saturates by having the magnetic field align itself uniformly between $\eig_1$ and $\eig_2$ along the $|\eig_3 \bcdot \eb| = 0$ contour, while for $\mathrm{Re} \gg 1$ it aligns itself between $\eig_1$ and $\eig_3$ (${\approx}40^\circ$ to $\eig_1$ and ${\approx}50^\circ$, respectively). The simulation with intermediate Reynolds number (center panel) contains regions where both of these situations occur. This behaviour is likely due to the increasing ratio of mixing to stretching with Reynolds number, and so the dynamo becomes progressively more reliant on suppressing these motions in saturation as $\mathrm{Re}$ is increased.

For unlimited Braginskii MHD with large parallel viscosity (left panels of bottom system in figure~\ref{fig:angle2d_all}), the statistics of the field alignment change only slightly between the kinematic stage and saturation, and both look remarkably similar to the panel labeled `MHD saturation', $\visc^{-1}=1500$. This again emphasizes that the role of the parallel Braginskii viscosity is to render the statistics of the magnetic field nearly identical to those found in the saturated state of the MHD dynamo. The slight difference between the kinematic stage and saturation---there being more overall alignment with $\eig_2$ in saturation---does lead to the eventual saturation of the dynamo. Increasing the isotropic viscosity in this simulation leads to an increase of field alignment with $\eig_1$ both in the kinematic stage and in saturation, which may be explained by the elimination of small-scale perturbations that tend to \emph{decrease} the parallel rate of strain when the isotropic viscosity is increased (see \S\,\ref{sec:magnetoimmutability} for more).
Decreasing the parallel viscosity again converges towards the isotropic-MHD result.

\begin{figure}
    \centering
    \includegraphics[width=\textwidth]{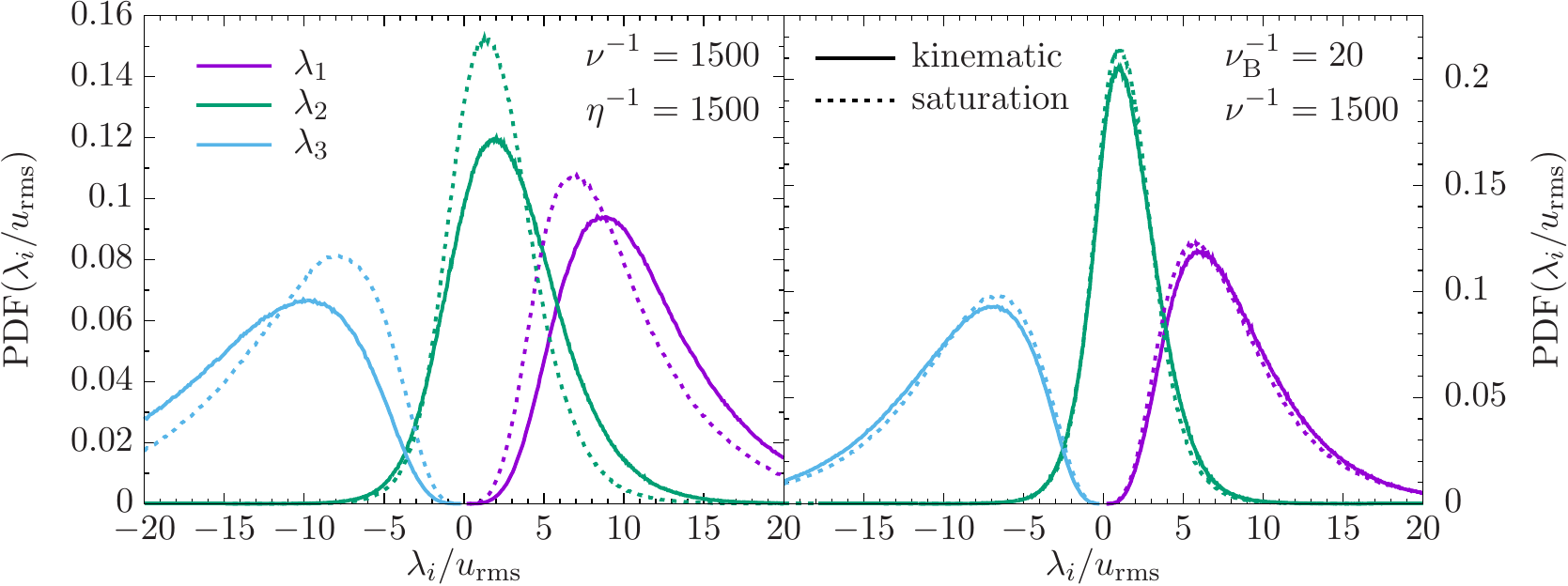}
    \caption[PDF of the rate of strain eigenvalues]{PDF of the symmetric rate-of-strain eigenvalues for simulations of (left) isotropic MHD and (right) unlimited Braginskii MHD in the kinematic stage (solid lines) and saturated state (dotted lines). Eigenvalues are ordered such that $\lambda_1 > \lambda_2 > \lambda_3$. }
    \label{fig:eigen_some}
\end{figure}

Finally, figure~\ref{fig:eigen_some} displays the PDFs of the eigenvalues $\lambda_1 > \lambda_2 > \lambda_3$ of the symmetric rate-of-strain tensor $\mathsfbi{S}$ for ${\rm Re}\gg{1}$ MHD (run MHD4; left) and for unlimited Braginskii MHD (run U1; right) in the kinematic (solid line) and saturated states (dashed line). As expected for turbulent systems, we find that $\langle \lambda_2 \rangle > 0$. The statistics of the rate-of-strain tensor change significantly between the two regimes in the isotropic-MHD simulation, while the change is much subtler in the unlimited Braginskii-MHD system. This signals once again that the unlimited Braginskii-MHD dynamo behaves much like that of saturated isotropic MHD.

\subsection{Magneto-immutability in the unlimited Braginskii dynamo}\label{sec:magnetoimmutability}

The minimization of $\ROS$ seen in our unlimited Braginskii-MHD simulations is reminiscent of recent studies of guide-field Braginskii-MHD turbulence \citep{Jono_magnetoimmutability}, in which the flow self-organized to minimize changes in magnetic-field strength and thus the production of pressure anisotropy---an effect named `magneto-immutability' by those authors. Given the above discussion on the anisotropization of the velocity field by the (unlimited) Braginskii viscosity, it is worth asking whether there is any physical relationship between the results in \S\,\ref{sec:anisotropization} and those presented by \citet{Jono_magnetoimmutability}.

\begin{figure}
    \centering
    \includegraphics[width=\textwidth]{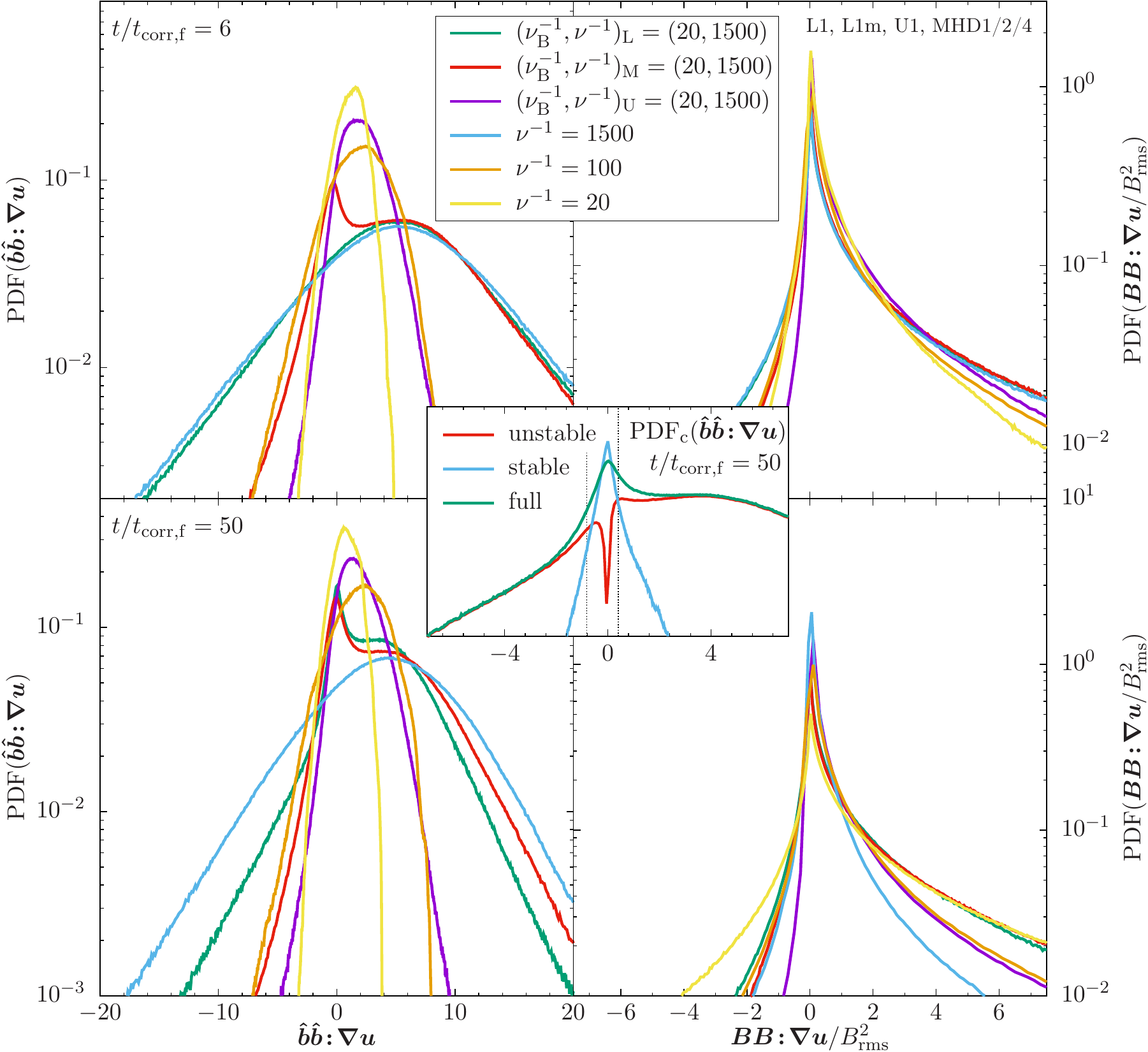}
    \caption[PDF of the bare and energy-weighted parallel rate of strain]{\label{fig:pdfROS} Probability distribution function of (left) the parallel rate of strain and (right) the energy-weighted parallel rate of strain (right) for various systems after six correlation times (top) and fifty correlation times (bottom), the latter pertaining to the exponential growth phase. 
    \emph{Inset:} PDF of the parallel rate of strain for a limited Braginskii-MHD simulation, conditioned on the stability criteria given by equation~\ref{stability} at $t/\tcorrf=50$. Vertical dashed lines indicate the \emph{approximate} thresholds for firehose and mirror instability given by equation~\eqref{stability} using $B_\mathrm{rms}$. }
\end{figure}

To address this question, we follow those authors in calculating the probability density functions (PDFs) of both the parallel rate of strain $\ROS$ and the energy-weighted parallel rate of strain $\boldsymbol{BB\!:\!\nabla u}/B^2_\mathrm{rms}$. The former quantity gives insight on how magneto-immutability affects the statistics of the parallel rate of strain, while the latter can be used to see how magneto-immutability affects the growth of the magnetic field on average.
The result is displayed in figure~\ref{fig:pdfROS}, with the upper panels showing these PDFs after six correlation times (corresponding to the start of the exponential growth phase) and the bottom panels showing these PDFs after $50$ correlation times (corresponding to the onset of saturation). Included are Braginskii-MHD simulations with hard-wall limiters (L1), without limiters (U1), and with a hard-wall limiter only on the mirror side (L1m). PDFs from runs MHD1--3 with varying $\visc$ are provided for comparison. At the initial stage of the dynamo, the PDF from the limited Braginskii-MHD run almost identically resembles that from the $\mathrm{Pm} = 1$, $\mathrm{Re} \gg 1$ MHD dynamo (MHD4). On the other hand, the unlimited Braginskii-MHD system more closely resembles the ${\rm Pm}\gg{1}$ MHD dynamo (MHD1), although the former's PDF is slightly broader as Braginskii viscosity is overall less efficient at damping motions than is an isotropic viscosity of similar magnitude. It was argued by~\citet{Malyshkin} that, for an isotropically tangled magnetic field, the anisotropic viscous stress with viscosity $\visc_\mr{B} = 1/20$ acts as an effective isotropic stress with viscosity $\visc = \visc_\mr{B}/5$.  Indeed, the PDFs for both the rate of strain and energy-weighted rate of strain for the unlimited simulation closely resembles the MHD simulation with $\visc = 1/100$ (MHD2). That being said, the magnetic-field growth rate is greatly reduced in the Braginskii system (compare the blue dashed and purple solid in the left panel of figure~\ref{fig:energy_unlim}). The reason for this is that, while the simulations exhibit similar stretching motions, the resistive dissipation is much higher in the Braginskii system due to the enhanced mixing motions, resulting in a slower dynamo (see \S\,\ref{sec:kazantsev} for more). This indicates that details of the sub-parallel-viscous range are important for the overall operation of the dynamo, and thus the closure advocated by~\citet{Malyshkin} is inappropriate for the case of large isotropic Reynolds number.

The Braginskii-MHD simulation employing a  pressure-anisotropy limiter only on the mirror side (run L1m) provides an interesting toy problem for the dynamo, where only the motions that act to decrease the magnetic-field strength ($\ROS < 0$) are targeted by the full Braginskii viscosity. One might expect that this scenario would lead to the fastest possible fluctuation dynamo for a given set of dissipation parameters. Indeed, the left panels of figure~\ref{fig:pdfROS} show that negative rate of strains are suppressed compared to the fully limited simulation, suggesting the potential for faster dynamo growth. However, these only amount to modest changes to low-probability regions in the PDF of the \emph{energy-weighted} rate of strain, whose average drives magnetic-field growth. Indeed, the evolution of the magnetic energy in this run is no different to those employing limiters on both regions of instability (see figure~\ref{fig:energy_lim}).

As the limited Braginskii-MHD dynamo evolves, a large notch appears in the PDF of the parallel rate of strain centered around $\ROS = 0$, signaling a relative preference in Braginskii MHD for motions with $\ROS \approx 0$. This notch, clearly seen at $t/\tcorrf = 50$ in the green and red curves, can be attributed to unlimited Braginskii viscosity acting on stable regions of the plasma. This becomes even clearer if one conditions the PDF to examine regions that either lie within or without of the stability region as defined by \eqref{stability}; this conditional PDF is displayed in the inset at the centre of figure~\ref{fig:pdfROS}. As portions of plasma enter the region of stability \eqref{stability}, unlimited viscous forces quickly damp the parallel rate of strain, condensing the PDF in that region near zero and forming a sharp peak that contrasts with the otherwise wide PDF in the unstable regions where the Braginskii viscosity is greatly reduced. This has the effect of rendering the PDF of $\ROS$ near the saturation for the limited simulation much thinner than its MHD counterpart, as can be seen in the bottom panels.

\begin{figure}
    \centering
    \includegraphics[scale=1.0]{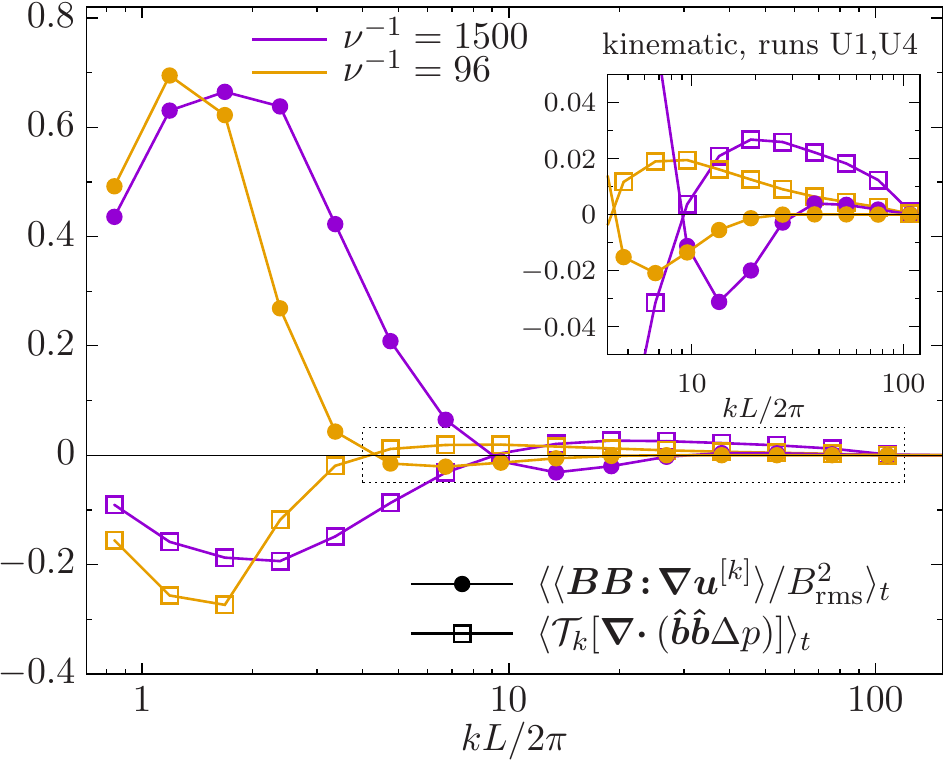}
    \caption[Shell-to-shell transfer between velocity and magnetic fields]{\label{fig:unlim_trans} Time-averaged shell-filtered energy transfer function of the pressure anisotropy $\mathcal{T}_k[\grad \bcdot (\eb \eb \rmDelta p)]$ along with the shell-filtered energy-weighted rate of strain $\langle\boldsymbol{BB\!:\!\nabla u}^{[k]} \rangle/B_\mathrm{rms}^2$, in $k$-space for runs U1 (purple) and U4 (orange). Bins have size $\sqrt{2}k$, where $k$ is center of the bin. \emph{Inset:} Enlargement of the region enclosed by the dotted rectangle. These quantities have been averaged over 60 correlation times during the kinematic stage of the unlimited Braginskii-MHD dynamo.}
\end{figure}

Further evidence of magneto-immutability can be seen in the $k$-space transfer of kinetic energy by the Braginskii viscous stress, which is shown in figure~\ref{fig:unlim_trans} for the kinematic stage of runs U1 (purple lines) and U4 (orange lines). While some of the energy extracted from the large-scale motions is damped away---note that $\mathcal{T}_k[\grad \bcdot (\eb \eb \rmDelta p)] < 0$ at large scales---a small portion of it is transferred to small-scale fluctuations.\footnote{The net effect of this term is damping, because
$3\visc_\mr{B}\langle\bb{u}\bcdot \grad \bcdot [\eb\eb(\ROS)] \rangle = - 3\visc_\mr{B} \langle|\ROS|^2\rangle$,
with the surface term disappearing in a triply periodic box.} These small-scale fluctuations attempt to nullify the net-positive parallel rate of strain by introducing \emph{negative} rates of strain, as highlighted in the inset of figure~\ref{fig:unlim_trans}. In both runs, these motions attempt to counteract the growth of magnetic energy driven by the large scales. In a system where the collisional relaxation of pressure anisotropy is governed by a constant $\visc_{\rm B}$ (as in our unlimited runs), the only means of regulating the pressure anisotropy and thereby avoiding strong damping of the large-scale motions is to re-organize the fields and flows to dynamically control $\ROS$.\footnote{Similar dynamical regulation occurs in the parallel firehose instability \citep{Scheko_2008,Rosin_2011}, where the contribution to $\ROS$ from the small-scale firehose fluctuations partially offsets the contribution to $\ROS$ from the large-scale motions to maintain the plasma at marginal stability. While this may occur in our simulations, the net pressure anisotropy is largely positive while regions that are firehose unstable are intermittent, and so such an effect may difficult to observe.} In kinetic systems, the pressure anisotropy, and thus the parallel viscous stress, can be reduced instead by anomalously increasing the collision frequency through the pitch-angle scattering of particles off firehose fluctuations. Indeed, this enhanced collisionality has been directly measured in hybrid-kinetic simulations of the firehose instability \citep{Kunz_2015}, and is what underpins the very idea of hall-wall limiters, in which $\visc_{\rm B}$ is effectively modified to maintain a kinetically stable plasma. No such collisional regulation can occur in our unlimited Braginskii-MHD runs, and so the motions responsible for driving $p_\perp\ne p_\parallel$ in the first place adjust. Interestingly, the high-Re isotropic MHD simulation in the saturated state also exhibits this behaviour, a feature not previously noted in the literature. Apparently, the unlimited Braginskii-MHD dynamo takes similar steps as the saturated MHD dynamo to reduce the overall stretching rate of the magnetic field.

 \begin{figure}
    \centering
    \includegraphics[width=\textwidth]{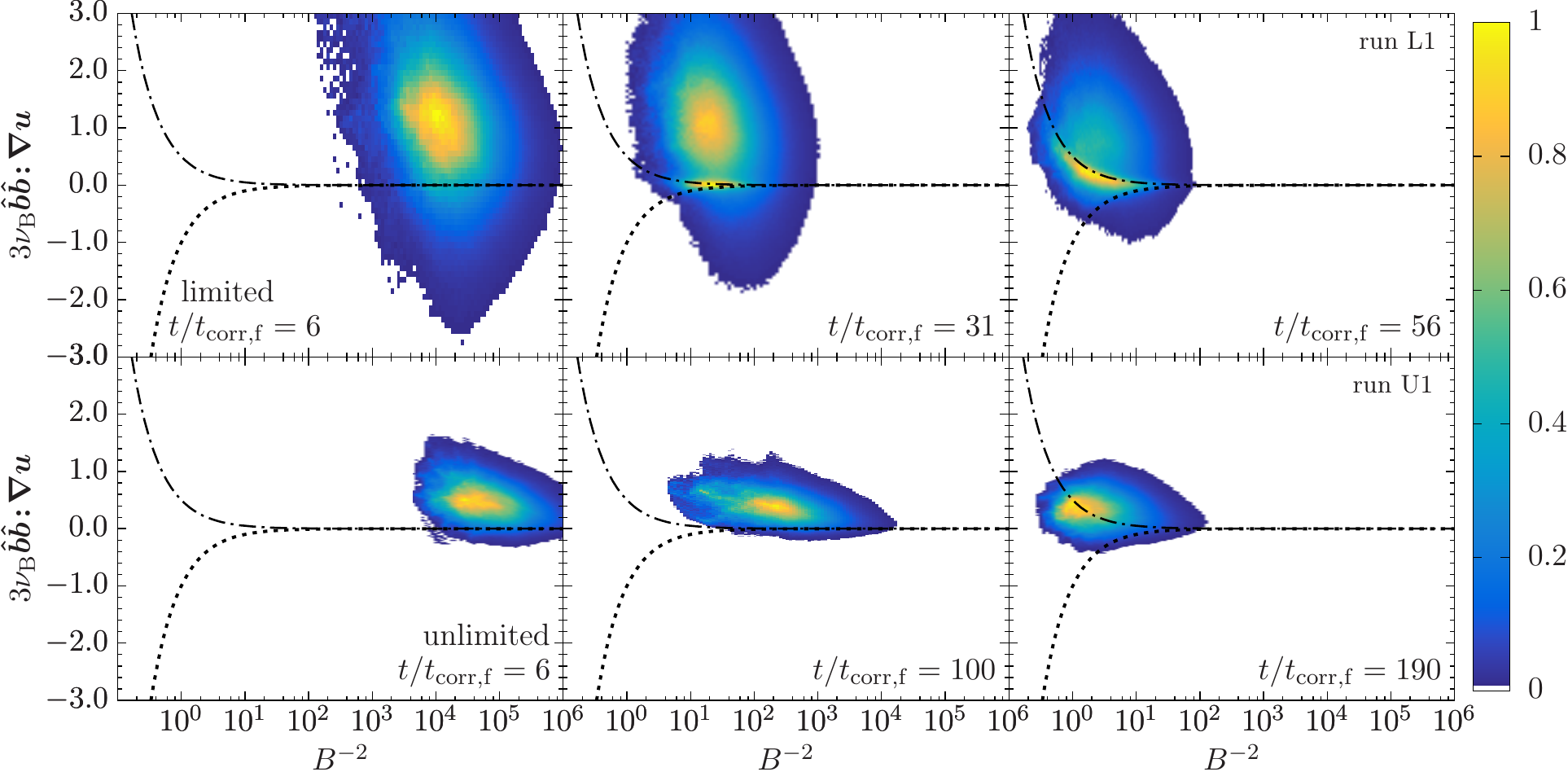}
    \caption[Brazil plot from limited and unlimited Braginskii-MHD]{ \label{fig:brazil} PDF of the parallel rate of strain with respect to $B^{-2}$ for the (top) hard-wall limited run L1 and (bottom)
    unlimited Braginskii-MHD run U1 at various times. Dashed-dotted (dotted) lines trace the mirror (firehose) instability thresholds given by equation~\eqref{stability}.}
\end{figure}
 
Finally, figure~\ref{fig:brazil} displays the PDF of the parallel rate of strain with respect to $B^{-2}$ (which serves as a proxy of $\betai$ as seen in similar histograms). Dashed-dotted (dashed) lines trace the mirror (firehose) instability thresholds given by \eqref{stability}, with the stable region lying to the left of these lines. In the limited case, the bulk of the plasma moves beyond the mirror instability threshold, migrating toward larger magnetic energies as the dynamo progresses. Because the viscous stress is limited, its ability to damp out the motions driving the PDF upwards in these panels is curbed. The portions of the plasma lying beyond the stability boundaries are then subject to a greatly reduced viscous stress comparable to the (small) Maxwell stress. However, as the magnetic energy grows, the viscous stress is able to regain its dynamical importance. An excess portion of the plasma accumulates in the stable region (as also seen in figure \ref{fig:pdfROS}), and evolves in a way that appears to respect the mirror boundary. No such behaviour is seen in the unlimited run (although the bulk of the plasma does coincidentally sit on the mirror threshold once saturation is reached). These are to be compared to the analogous plot for the magnetized kinetic regime computed using a collisionless hybrid-kinetic code and shown in figure~\ref{sim:aniso}, which displays a much stronger resemblance to the \emph{unlimited} run than the limited one. Again, with such small magnetic energy, the effective collision frequencies required to instantaneously pin the pressure anisotropy at its stability thresholds cannot be realized. In fact, the simulations in chapter~\ref{ch:simulation} have a box-averaged $\nu_\mathrm{eff}$ much less than the gyrofrequency, which further raises the question as to whether hard-wall limiters can truly serve as a panacea for dealing with excess pressure anisotropy in Braginskii-MHD simulations.

\subsection[Kazantsev--Kraichnan model for unlimited Braginskii dynamo]{Modified Kazantsev--Kraichnan model for unlimited Braginskii dynamo}\label{sec:kazantsev}

In light of the main conclusion from \S\S\,\ref{sec:unlimited}--\ref{sec:magnetoimmutability}---that the turbulence statistics during the `kinematic' phase of the Braginskii-MHD dynamo are strikingly similar to those found in the saturated state of the isotropic-MHD dynamo---we propose that the kinematic stage of the unlimited Braginskii dynamo can be described through a modified Kazantsev--Kraichnan model, originally developed to describe the saturated state of the MHD dynamo by \citet{Scheko_sim} and \citet{Scheko_saturated}. This model uses a modified form of the Kazantsev--Kraichnan \citep{Kazantsev,kraichnan} velocity correlation tensor
\begin{equation}\label{eqn:kazansev}
    \langle  u^i(t,\bb{x})u^j(t',\bb{x}')\rangle = \delta(t-t')\kappa^{ij}(\bb{x}-\bb{x}') 
\end{equation}
that features anisotropic statistics with respect to the magnetic-field direction:
\begin{equation}
    \kappa^{ij}(\bb{k}) = \kappa^{\rm (i)}(k,|\xi|) \bigl( \delta^{ij} - \hat{k}_i \hat{k}_j \bigr) + \kappa^{\rm (a)}(k,|\xi|) \bigl( \hat{b}^i \hat{b}^j + \xi^2 \hat{k}_i \hat{k}_j - \xi \hat{b}^i \hat{k}_j - \xi \hat{k}_i \hat{b}^j \bigr) ,
\end{equation}
where $\hat{\bb{k}} \doteq \bb{k}/k$ and $\xi \doteq \hat{\bb{k}}\bcdot\eb$. The $k$- and $|\xi|$-dependent amplitudes $\kappa^{\rm (i)}$ and $\kappa^{\rm (a)}$ quantify the sizes of the isotropic and field-anisotropic components of the correlation. Note that $\kappa^{ij}$ is trace-less as a result of incompressibility. When this model was put forward, the idea was that the dynamically important magnetic field feeds back on the velocity through the Lorentz force, biasing its statistics with respect to the field direction. Here, we ask whether  a modified version of this model might accurately describe the impact of the Braginskii viscosity on the turbulent flow during the {\em kinematic} stage.

\subsubsection{Formulation of the model}

To do so, we must first obtain equations for the magnetic-field fluctuations, the wavevector of these fluctuations, and the scale-dependent magnetic-field direction as functions of the fluctuating velocity field. If we only consider the straight regions of the folded magnetic-field structures, then spatial variations in $\eb$ are limited to changes of sign and, as $\eb$ arises in the momentum equation in pairs, cancels out. Thus, $\eb$ can be taken to depend only on time. This approximation is not as drastic as it may appear, as the bends of the folded structures occupy only a small fraction of the total volume. The evolution equations then follow straightforwardly from the non-ideal induction equation after adopting the {\it Ansatz}
\begin{equation}
    \bb{B}(t,\bb{x}) = \eb(t) \int \od^3\bb{k}_0 \, {B}(t,\bb{k}_0) \rme^{\imag \bs{x}\bscdot {\bs{k}}(t,\bs{k}_0)} ,
\end{equation}
where $\bb{k}$ is the wavevector that evolves in time from the initial value $\bb{k}_0$. Assuming statistical homogeneity and arbitrarily setting $\bb{x} = 0$ results in the closed set of equations
\begin{subequations}
\begin{gather}
    \partial_t {B} = \nrndb^i \nrndb^m u^i_{,m} {B} - \eta {k}^2 {B}, \\
    \partial_t {k}_m = - u^{i}_{,m} {k}_{i}, \\
\partial_t \nrndb^i = \nrndb^m u^{i}_{,m} - \nrndb^l \nrndb^m u^l_{,m} \nrndb^i,
    \end{gather}
\end{subequations}
where we have used the Einstein summation convention for repeated indices and indices appearing after commas denote derivatives in the dimension of that index (i.e.~$u_{,i} \doteq \partial_{x_i} u$). A closed equation for the probability density function
\begin{equation}
    \mathcal{P}({B},{\bb{k}},\eb) = \delta(|\eb|^2 -1)\delta(\eb\bcdot{\bb{k}})(4\upi^2 {k})^{-1}P({B},{k})
\end{equation}
can then be derived using (\ref{eqn:kazansev}). The calculation is detailed in appendix \ref{ap:kazantsev}. For $k\gg k_\visc$ (i.e.~at sub-viscous scales), the resulting evolution equation for the magnetic-energy spectrum $M(k) \doteq (1/2) \int_0^\infty \od {B}\, {B}^2 P({B},{k})$ is
\begin{equation}\label{eqn:mod_kazantsev}
    \pD{t}{M} = \frac{\gamma_\perp}{8} \pD{k}{} \biggl[ \bigl( 1 + 2\sigma_\parallel\bigr) k^2 \pD{k}{M} - \bigl( 1 + 4\sigma_\perp + 10\sigma_\parallel \bigr) kM \biggr] + 2 \bigl( \sigma_\perp + \sigma_\parallel\bigr) \gamma_\perp M - 2\eta k^2 M,
\end{equation}
where
\begin{subequations}
\begin{equation}\label{eqn:gamma_delta}
    \gamma_\perp = \int\frac{\rmd^3\bb{k}}{(2\upi)^3} \, k^2_\perp \kappa_\perp(\bb{k}) , \quad
    \sigma_\perp = \frac{1}{\gamma_\perp} \int\frac{\rmd^3\bb{k}}{(2\upi)^3} \, k^2_\parallel \kappa_\perp(\bb{k}) , \quad
    \sigma_\parallel = \frac{1}{\gamma_\perp} \int\frac{\rmd^3\bb{k}}{(2\upi)^3} \, k^2_\parallel \kappa_\parallel(\bb{k}) . \tag{\theequation {\it a,b,c}}
\end{equation}
\end{subequations}
Here $k_\parallel = k\xi$ and $k_\perp = k(1-\xi^2)^{1/2}$ define $\xi$, and 
\begin{align}
    \kappa_\perp(\bb{k}) &= \frac{1}{2} \bigl( \delta^{ij} - \hat{b}^i\hat{b}^j\bigr) \kappa^{ij}(\bb{k}) \nonumber\\*
    \mbox{} &= \frac{1}{2} \Bigl[ \bigl(1+\xi^2)\kappa^{\rm (i)}(k,|\xi|) + \xi^2\bigl(1-\xi^2\bigr) \kappa^{\rm (a)}(k,|\xi|) \Bigr] ,\label{eqn:kappa_prp} \\*
    \kappa_\parallel(\bb{k}) &= \frac{1}{2} \hat{b}^i \hat{b}^j \kappa^{ij}(\bb{k}) \nonumber\\*
    \mbox{} &= \frac{1}{2} \bigl(1-\xi^2\bigr) \Bigl[ \kappa^{\rm (i)}(k,|\xi|) + \bigl(1-\xi^2\bigr) \kappa^{\rm (a)}(k,|\xi|) \Bigr] \label{eqn:kappa_prl} 
\end{align}
are the correlations of velocities perpendicular and parallel to the magnetic field, respectively. The quantity $\gamma_\perp$ measures the strength of perpendicular variations of the perpendicular velocities, and thus gives the mixing rate of the magnetic field. It is on the related time scale that interchange-like motions shuffle the bring direction-reversing magnetic fields close enough together for them to annihilate resistively. The other two quantities, $\sigma_\perp$ and $\sigma_\parallel$, give the relative strength of the field-aligned stretching rates of the perpendicular and parallel velocities in terms of the mixing rate.\footnote{Notice that the rate of field-line slipping ${\propto}(k_\perp^2 \kappa_\parallel)^{1/2}$ does not appear in~\eqref{eqn:mod_kazantsev} (see also appendix \ref{ap:ROS}), which is a result of choosing to model regions of straight but alternating magnetic field. Thus, motions that may help unwind the magnetic field, which rely on some amount of magnetic field-curvature (such as those discussed in \citet{Malyshkin} and \citet{KulsrudZweibel}) are not captured by this model.} In the saturated state of the ${\rm Pm}\gg{1}$ MHD dynamo, $\sigma_\perp$ and $\sigma_\parallel$ are reduced until  weakened stretching of the magnetic field balances  two-dimensional mixing of the folded fields by the partially two-dimensionalized random flow \citep{Scheko_sim}. In order to compute these values for more physically relevant systems, equations~\eqref{eqn:gamma_delta} must be modified to account for the finite correlation time $\tau_\mathrm{c}$ of the driven velocity field. In this case, the Fourier-space velocity correlation tensor $I^{ij}(\bb{k}) \doteq \langle u^i(\bb{k})u^{j*}(\bb{k})\rangle$ is related to $\kappa^{ij}(\bb{k})$ via $I^{ij}(\bb{k}) = \tau^{-1}_\mathrm{c} \kappa^{ij}(\bb{k})$ (cf.~\eqref{eqn:kazansev}). We then take the correlation time to be associated with the `turnover' time of the motion in question, leading to
\begin{subequations}\label{eqn:kazantsev_param}
\begin{align}\label{eqn:kazantsev_gamma}
       \gamma_\perp &= \left[C\int\frac{\rmd^3\bb{k}}{(2\upi)^3} \, k^2_\perp  I_\perp\right]^{1/2}  = \left[C \langle | (\mathsfbi{I} - \eb\eb)\bcdot \grad\bb{u}\bcdot (\mathsfbi{I} - \eb\eb)|^2 \rangle \right]^{1/2} , \\
    \sigma_\perp &= \frac{1}{\gamma_\perp}\left[\frac{2}{3}C     \int\frac{\rmd^3\bb{k}}{(2\upi)^3} \,k^2_\parallel  I_\perp\right]^{1/2} = \frac{1}{\gamma_\perp}\left[\frac{2}{3}C\langle | \eb\bcdot \grad\bb{u}\bcdot (\mathsfbi{I} - \eb\eb)|^2 \rangle \right]^{1/2}, \label{eqn:kazantsev_sprp}\\
    \sigma_\parallel &= \frac{1}{\gamma_\perp} \left[\frac{1}{6}C\int\frac{\rmd^3\bb{k}}{(2\upi)^3} \, k^2_\parallel   I_\parallel\right]^{1/2} = \frac{1}{\gamma_\perp}\left[\frac{1}{6}C\langle | \ROS |^2 \rangle \right]^{1/2}, \label{eqn:kazantsev_sprl}
\end{align}
\end{subequations}
where $C$ is an adjustable coefficient that relates the eddy correlation time and $\gamma_\perp$; $I_\perp$ and $I_\parallel$ are defined in a similar way to~\eqref{eqn:kappa_prp} and \eqref{eqn:kappa_prl}, respectively. 

As $\eta\rightarrow 0$, the determination of whether or not a dynamo is viable in this model (i.e.~$\gamma>0$) depends only on the \emph{relative} size of the parallel (stretching) motions compared to the perpendicular (mixing) ones, rather than on their absolute magnitude. In particular, as $\gamma_\perp$ increases and the mixing motions become more important, the dynamo growth rate will decrease if the parallel motions do not increase commensurately. This outcome can be seen in figure~\ref{fig:energy_unlim}, where the unlimited Braginskii-MHD system with the largest isotropic viscosity (and thus the weakest mixing motions) reaches saturation first. In the limit $\eta\rightarrow 0$, equation~\eqref{eqn:mod_kazantsev} has the eigenvalue solution
\begin{equation}
 M(k) \approx k^s \rme^{\gamma t}K_0(k/k_\eta),
\end{equation}
where $K_0$ is the Macdonald function, $k_\eta = [(1+2\sigma_\parallel)\gamma_\perp/ 16 \eta]^{1/2}$, 
\begin{equation} \label{eqn:kazantsev_growth}
 \gamma = \frac{\gamma_\perp}{8}\frac{16(\sigma_\perp + \sigma_\parallel)(1+2\sigma_\parallel) - (1+2\sigma_\perp + 6 \sigma_\parallel)^2}{1+2\sigma_\parallel}   
\end{equation}
is the magnetic-energy growth rate, and 
\begin{equation}\label{eq:spectral_index}
    s = \frac{2(\sigma_\perp + 2 \sigma_\parallel)}{1+2 \sigma_\parallel}
\end{equation}
is the spectral index. For reference, the isotropic case with $\kappa^{\rm (i)} = \kappa^{\rm (i)}(k)$, $\kappa^{\rm (a)} = 0$, and $\delta$-correlated velocity statistics has $\sigma_\perp = 2/3$, $\sigma_\parallel = 1/6$, and $\gamma_\perp = (6/5)\overline{\gamma}$, where
\begin{equation}
    \overline{\gamma} \approx \frac{1}{3}\left[\int_{k_0}^\infty \od k \, k^2 E(k)\right]^{1/2}.
\end{equation}
 These values correspond to the classic \citet{Kazantsev} and \citet{Kulsrud1992} magnetic-energy spectrum: $M(k) \approx k^{3/2}\, \rme^{(3/4)\overline{\gamma} t} K_0(k\sqrt{10\eta/\overline{\gamma}})$.

\subsubsection{Solution of the model and comparison with simulation results}
 
\begin{figure}
    \centering
    \includegraphics[width=0.70\textwidth]{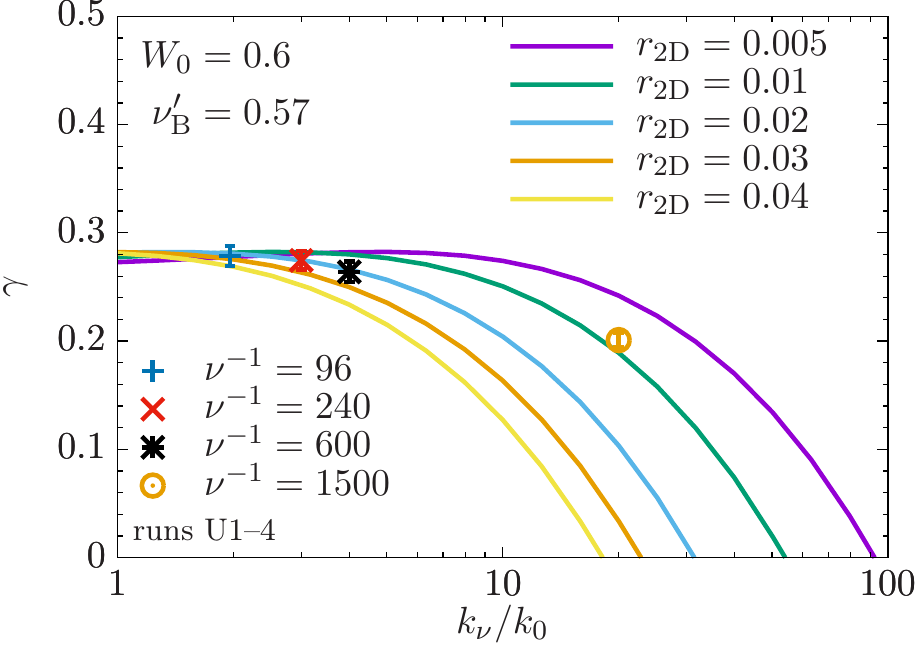}
    \caption[Growth rates of the modified Kazantsev--Kraichnan model]{\label{fig:kaz_growth} Growth rate of the modified Kazantsev--Kraichnan model given by \eqref{eqn:kaz_growth_spec} as a function of $k_\visc$ for $C=1$, $\alpha=5/3$, $\visc'_\mathrm{B}=0.57$, $W_0= 0.6$ and various values of $r_\mathrm{2D}$. Included are growth rates calculated from unlimited Braginskii-MHD simulations U1--4 in the exponential growth phase (cf. figure~\ref{fig:energy_unlim}).}
\end{figure}

Motivated by the tendency of the unlimited Braginskii-MHD dynamo to mimic the saturated statistics of the dynamo in standard MHD, we modify the arguments presented in~\citet{Scheko_saturated} in order to study how the Braginskii dynamo growth rate depends on the isotropic viscosity $\visc$ in the $\mathrm{Re}\rightarrow \infty$ limit.  The central crux of the original argument is that the magnetic field would disable the stretching motions below the scale $k_\mathrm{s}^{-1}$ at which the magnetic field is dynamically important. This scale was found by balancing the energy of the eddies below that scale with the total magnetic energy:
\begin{equation}
    \int_{k_\mathrm{s}}^{k_\visc} \od k\, E(k)  = \frac{1}{2}c_1\langle B^2\rangle ,
\end{equation}
where $c_1$ is an adjustable constant. In saturation, this gives $\langle U^2\rangle \sim \langle B^2\rangle$. To adapt this argument for Braginskii MHD, we posit that now the balance occurs between the hydrodynamic nonlinearity and the Braginskii viscous stress $\grad \bcdot (\eb \eb \Deltap)$:
\begin{equation} \label{eqn:model_kazantsev}
         \int_{k_0}^{k_\visc} \od k\, E(k)  = 3\visc_\mathrm{B}c_1\langle |\ROS|^2 \rangle^{1/2} \sim 3\visc'_\mathrm{B} (\sigma_\perp + \sigma_\parallel)\gamma_\perp,
\end{equation}
where $\visc'_\mathrm{B} \doteq c_1\visc_\mathrm{B}$. This balance is apparent at the large scales, which can be seen in figure~\ref{fig:trans_unlim} for the runs U1 and U4 in the kinematic regime. We adopt the same form of the Fourier-space velocity correlation tensor $I^{ij}(\bb{k})$ as formulated in~\citet{Scheko_saturated}, {\it viz.}, for $k_0 < k < k_\mathrm{s}$ the turbulence remains isotropic and
\begin{equation}
    I^{(\mathrm{i})}(k,|\xi|) = \frac{E(k)}{4\upi k^2}, \qquad I^{(\mathrm{a})}(k,|\xi|) =0,
\end{equation}
whereas the turbulence is anisotropized for $k_\mathrm{s} < k < k_\visc$ in such a way that the stretching motions ($\xi\ne{0}$) are disabled completely:
\begin{equation}
        I^{(\mathrm{i})}(k,|\xi|) = \frac{2r_\mathrm{2D}E(k)\delta(\xi)}{4\upi k^2}, \qquad I^{(\mathrm{a})}(k,|\xi|) =\frac{2\widetilde{E}(k)\delta(\xi)}{4\upi k^2}.
\end{equation}
Here, $r_\mathrm{2D}$ (${<}1$) parameterizes the efficiency of the mixing motions and $\widetilde{E}(k)$ is the anisotropic part of the sub-stretching mixing motions (the latter term does not figure into the following discussion as it is always accompanied by $\xi \delta(\xi) = 0$). Finally, the form of the energy spectrum will be assumed to be $E(k) = \beta k^{-\alpha}$ for $k\in[k_0,k_\visc]$ and zero otherwise, where $\beta$ is chosen such that $W_0 = \int_{k_0}^{k_\visc}\od k \, E(k)$ is the total kinetic energy.  Equations~\eqref{eqn:gamma_delta} can then be calculated explicitly:
\begin{align}
    \gamma_\perp &= \frac{6}{5}\overline{\gamma}\bigl[1 - (1+W_0/W_\visc)^{-\chi}\bigr]^{-1/2} \Gamma^{1/2},\\
    \sigma_\perp &= 4\sigma_\parallel = \frac{2}{3}\bigl[(1+W_\mathrm{s}/W_\visc)^{-\chi} - (1+W_0/W_\visc)^{-\chi}\bigr]^{1/2}\Gamma^{-1/2},\\
     \Gamma &\doteq (1+W_\mathrm{s}/W_\visc)^{-\chi} - (1+W_0/W_\visc)^{-\chi} + \frac{5}{4}r_\mathrm{2D}[1 -(1+W_\mathrm{s}/W_\visc)^{-\chi}],
\end{align}
where  $W_\mathrm{s} \doteq  \int_{k_\mathrm{s}}^{k_\visc} \od k \, E(k)$, $W_\visc \doteq \beta k_\visc^{1-\alpha}/(\alpha - 1)$, and $\chi \doteq (3-\alpha)/(\alpha-1)$. For Kolmorogov turbulence with $\alpha =5/3$, $\chi = 2$. The energy of the stretching motions, and thus the stretching wavenumber $k_\mathrm{s}$,  can be found using \eqref{eqn:model_kazantsev}: 
\begin{subequations}
\begin{equation}
W_\mathrm{s} + W_\visc =\frac{W_0 k_\mathrm{s}^{1-\alpha}}{k_0^{ 1-\alpha}-k_\visc^{1-\alpha}} =  \left[\frac{2}{5}\frac{W_0^2}{\visc'^2_\mathrm{B}}\frac{1}{QC} + (W_0+W_\visc)^{-\chi}\right]^{-1/\chi},
\end{equation}
\end{subequations}
where 
\begin{equation}
    Q \doteq \chi^{-1} \left(\frac{W_0}{k_0^{1-\alpha}-k_\visc^{1-\alpha}}\right)^{1+\chi}.
\end{equation}
It is a straightforward exercise to then calculate the resulting growth rate $\gamma$:
\begin{align}\label{eqn:kaz_growth_spec}
    \gamma = \frac{1}{8} \left( 35 \gamma_\perp -\frac{6W_0}{5\visc'_\mathrm{B}}  - \frac{540\visc'_\mathrm{B}\gamma_\perp^2}{15\visc'_\mathrm{B}\gamma_\perp+2W_0}\right),
\end{align}
where
\begin{align}\label{eqn:kaz_growth_spec_perp}
    \gamma_\perp &= \frac{6}{5}\left\{\frac{W_0^2}{9\visc'^2_\mathrm{B}}\left(1-\frac{5}{4}r_\mathrm{2D}\right) + \frac{25}{72}r_\mathrm{2D}QC\left[W_\visc^{-\chi} - (W_\visc + W_0)^{-\chi}\right]\right\}^{1/2}.
\end{align}
This model growth rate is plotted in figure~\ref{fig:kaz_growth} as a function of $k_\visc/k_0$, which serves as a measure of viscous dissipation and thus the isotropic (or perpendicular) Reynolds number. (Larger $k_\visc/k_0$ corresponds to larger Re.) For this figure, we have set $\alpha = 5/3$ and $C=1$; $W_0$ is chosen to be representative of the total kinetic energy computed during the kinematic stage in runs U1--4 (which are displayed in figure~\ref{fig:energy_unlim}). The scaled Braginskii viscosity $\visc'_\mathrm{B}$ (see (\ref{eqn:model_kazantsev})) is chosen so that (\ref{eqn:kaz_growth_spec}) returns the measured $\gamma$ in the kinematic stage of run U4. The value of $r_\mathrm{2D}$, which in the model effectively sets the value of Re at which $\gamma$ tends towards zero, is varied from $0.005$ to $0.04$. Overlaid on these curves are growth rates calculated from the unlimited Braginskii runs U1--4, with the corresponding value of $k_\visc/k_0$ chosen by measuring the maximum value of $k^2E(k)$ for each run.

While the above model is rather simplistic, it does capture the general trend seen in the simulation data, and only requires moderate tuning to achieve good agreement. The most striking feature of equation \eqref{eqn:kaz_growth_spec} and figure~\ref{fig:kaz_growth} is that, for sufficiently large $k_\visc/k_0$ (or isotropic Reynolds number), the growth rate of the magnetic energy becomes negative, and thus the dynamo ceases to operate. This is due to an enhancement of small-scale mixing, which tends to bring field lines closer together and encourages their resistive annihilation. This suggests that {\em there is no unlimited Braginskii-MHD dynamo in the limit $\visc\rightarrow 0$}, a somewhat paradoxical result whose physical origin is discussed at the end of this section. The value of $k_\visc/k_0$ at which this occurs can be calculated by solving~\eqref{eqn:kazantsev_growth} for $\gamma = 0$, which, for $\sigma_\perp = 4\sigma_\parallel$, leads to a quadratic equation in $\sigma_\parallel$, with the relevant solution $\sigma_\parallel = W_0/5\visc'_\mathrm{B}\gamma_\perp \approx 0.078\doteq \xi$. Using this expression in the definition of $\gamma_\perp$ given by \eqref{eqn:kaz_growth_spec_perp}, we find
\begin{equation}
     r_\mathrm{2D}  \left(\frac{W_\visc}{W_0}\right)^{-\chi}  \left[ 1-  ( 1 +W_0/ W_\visc)^{-\chi}\right] =   \frac{8}{25}\frac{W_0^{2+\chi}}{\visc'^2_\mathrm{B} QC } \left(\frac{1}{36\xi^2} +\frac{5}{4}r_\mathrm{2D}-1\right).
\end{equation}
When the sub-stretching mixing motions are disabled ($r_\mathrm{2D}=0$), this equation ceases to have a solution; in this case, we have the isotropic values $\sigma_\perp=2/3$ and $\sigma_\parallel = 1/6$ and the dynamo operates. For $r_\mathrm{2D}>0$, the limit $k_\visc /k_0 \gg 1$ and $\alpha < 3$ yields the approximate solution
\begin{equation}
 \left(\frac{k_\visc}{k_0}\right)^{3-\alpha} \approx \frac{8}{25}\frac{\chi W_0}{Ck_0^2\visc'^2_\mathrm{B}r_\mathrm{2D}}\left(  \frac{1}{36 \xi^2} + \frac{5}{4}r_\mathrm{2D} - 1\right),
\end{equation}
where we have used $W_0 = W_\visc [(k_\visc/k_0)^{\alpha-1}-1]$. Note that the right-hand side of this equation is always positive, implying a solution to this equation will exist (thus guaranteeing $\gamma=0$) if the right-hand side is larger than unity. Thus, the model predicts that the unlimited Braginskii dynamo will cease to operate if the isotropic Reynolds number is sufficiently large.

\begin{figure}
    \centering
    \includegraphics[width=\textwidth]{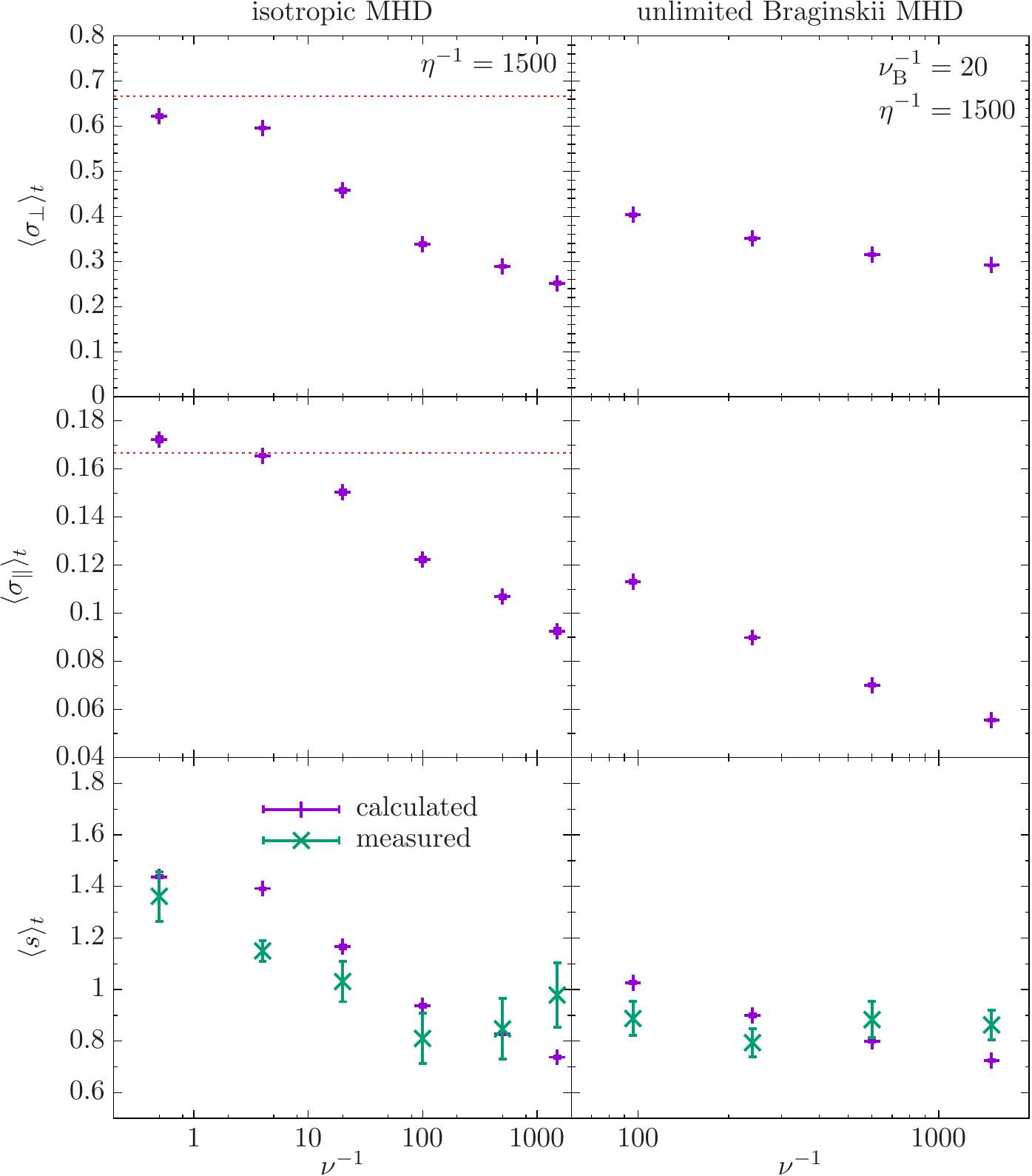}
    \caption[Average values of $\sigma_\perp$, $\sigma_\parallel$ and $s$]{Modified Kazantsev--Kraichnan model parameters $\sigma_\perp$ (top) and $\sigma_\parallel$ (center) calculated from simulations of MHD (left) and unlimited Braginskii MHD (right) using \eqref{eqn:kazantsev_param}({\it b},{\it c}). These values are averaged over the entire length of the simulation. The analytical values of $\sigma_\perp = 2/3$ and $\sigma_\parallel = 1/6$ obtained for isotropic, $\delta$-correlated velocity statistics are plotted in the left panels using red dotted lines. \emph{bottom:} Spectral index $s$ of the modified Kazantsev--Kraichnan model. Predicted values~\eqref{eq:spectral_index} are shown as purple plus symbols, while indices calculated from simulation spectra over the range $kL/2\upi \in [1,5]$ are plotted as green crosses.
    \label{fig:kazantsev}}
\end{figure}

To ascertain how the parameters $\sigma_\perp$ and $\sigma_\parallel$ depend on the isotropic and parallel viscosities, these quantities are calculated using the rate-of-strain tensor measured in our isotropic MHD and unlimited Braginskii-MHD simulations and are time averaged over the length of the simulation. The results are plotted in figure~\ref{fig:kazantsev}. As the viscosity for isotropic MHD is increased, the correlation time of the velocity statistics tends toward zero and the exact values of $\sigma_\perp = 2/3$ and $\sigma_\parallel = 1/6$ for isotropic $\delta$-correlated velocities are recovered. For both the isotropic-MHD and Braginskii-MHD systems, $\sigma_\parallel$ decreases as the isotropic viscosity is increased. This is due to the fact that the perpendicular motions, as measured by $\gamma_\perp$, tend to increase faster with $\visc$ than do the parallel ones. The dependence of $\sigma_\perp$ on $\visc$ is weaker, particularly in the unlimited Braginskii-MHD system, indicating that the assumption $\sigma_\perp = 4 \sigma_\parallel$ isn't strictly followed for all cases. However, the general trends are consistent with the model detailed above, which indicates the dynamo becomes less efficient as the isotropic viscosity decreases, conceivably shutting down entirely for a sufficiently large Reynolds number.

Finally, to reinforce the validity of the modified Kazantsev--Kraichnan model, the spectral indices as predicted by~\eqref{eq:spectral_index} are plotted on the bottom for figure~\ref{fig:kazantsev}, along with the spectral indices calculated using the first five wavenumbers from the magnetic spectra taken from simulation data. While the variance of the latter data set is quite large, the model does predict the general trend which shows a decreasing spectral exponent with decreasing isotropic viscosity $\visc$. Thus, the simplified model presented in this section overall gives sensible results.

\subsection[Is the unlimited Braginskii dynamo possible when $\Reprl < 1?$]{Stokes flow: Is the unlimited Braginskii dynamo possible when $\Reprl < 1?$}\label{sec:stokes}

In this section we ask whether the unlimited Braginskii-MHD dynamo exists in the `Stokes-flow' regime, i.e.~$\visc_\mathrm{B} \rightarrow \infty$ ($\Reprl \rightarrow 0$). In particular, it was predicted in \S\,\ref{sec:kazantsev} that the unlimited Braginskii-MHD dynamo ceases to operate in the limit $\mathrm{Re} \rightarrow \infty$, $\Reprl = \mathrm{const}$. Is the same true for the limit  $\mathrm{Re} = \mathrm{const}$,  $\Reprl \rightarrow 0$? 

In the isotropic-MHD case, the correlation time of the flow vanishes in the limit ${\rm Re}\rightarrow{0}$) and the dynamo is well described by the Kazantsev--Kraichnan model. Provided sufficient scale separation between the forcing and resistive scales, the dynamo will continue to operate in this regime regardless of $\mathrm{Re}$~\citep{Scheko_theory}. To investigate whether a similar result holds for the Braginskii system, we perform a set of low-$\mathrm{Re}$ isotropic-MHD and low-$\Reprl$ unlimited Braginskii-MHD simulations at reduced resolution ($N_\mathrm{cell} = 112^3$). For each run, the forcing amplitude $\varepsilon$ is adjusted so that $u_\mathrm{rms}\sim 1$ in steady-state. Details of these runs are given in the last block of table \ref{tab:runs}.

\begin{figure}
    \centering
    \includegraphics[width=\textwidth]{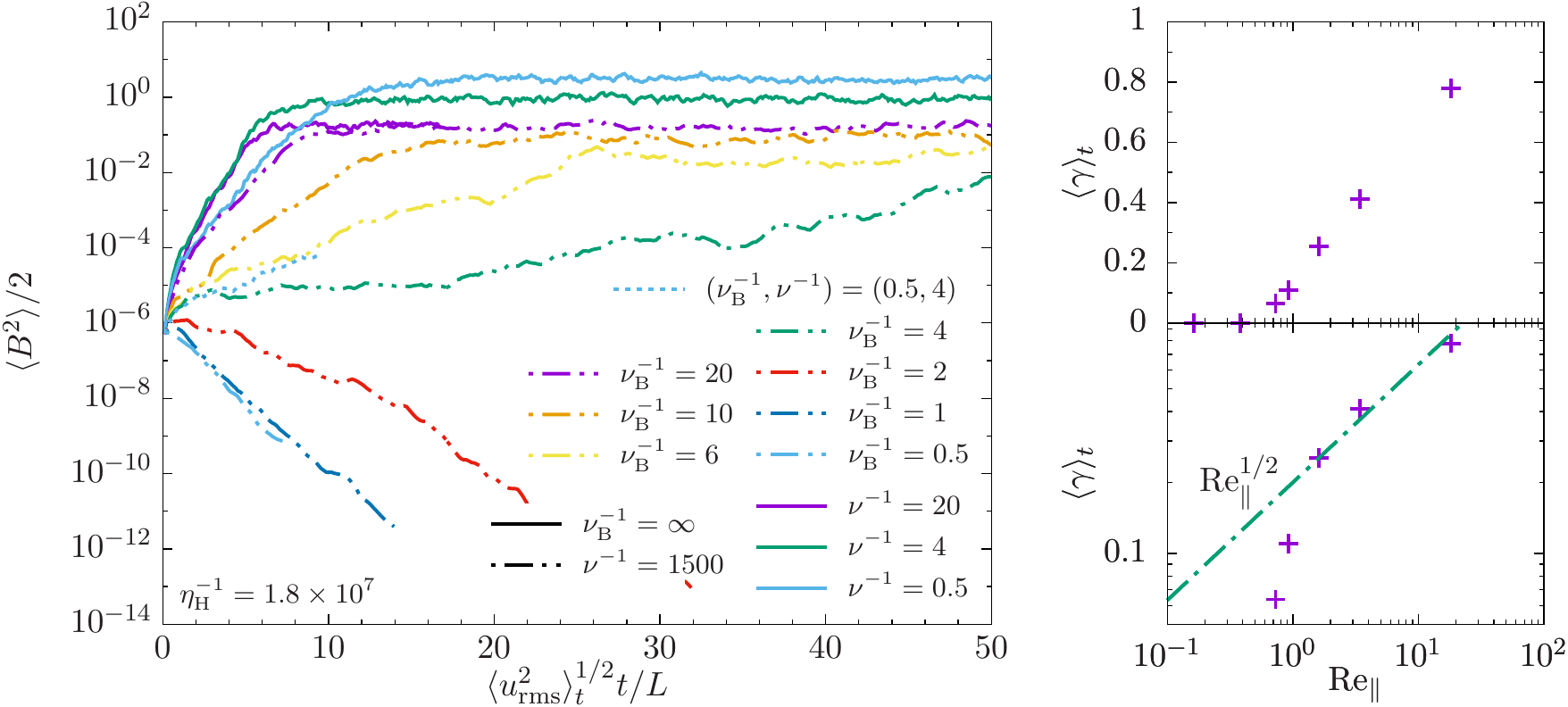}
    \caption[Magnetic energy evolution for Stokes flow simulations]{\emph{left:} Evolution of the magnetic energy for various $\mathrm{Re} < 1$ (MHD) and $\Reprl < 1$, $\mathrm{Re} \gg 1$ (unlimited Braginskii-MHD) simulations. For some values of $\Reprl$, the unlimited Braginskii-MHD simulations do not exhibit a dynamo. \emph{right:} Time-averaged growth rate $\gamma = \od \ln B^2_\mathrm{rms}/\od t$ of the magnetic energy during the kinematic stage as a function of  parallel Reynolds number for unlimited Braginskii in the Stokes regime on a linear (top) and logarithmic (bottom) scale. }
    \label{fig:stokes_en_growth}
\end{figure}

The evolution of the magnetic energy for simulations in the Stokes regime is plotted in figure~\ref{fig:stokes_en_growth}. Simulations using isotropic MHD are denoted using solid lines, while simulations using unlimited Braginskii MHD with $\visc^{-1}=1500$ are denoted with dotted lines; $\visc^{-1}_{\rm B}$ is varied from $20$ down to $0.5$. As predicted by the Kazantsev--Kraichnan model, the behaviour of the isotropic MHD simulations changes little, provided $u_\mathrm{rms}$ is kept constant between each run. All grow the magnetic energy and reach saturation. The unlimited Braginskii-MHD dynamo, on the other hand, operates only beyond a certain critical value of $\Reprl\gtrsim{1}$). Well above this cut-off, the scaling of the growth rate follows the expected \citet{Kolmogorov1941} scaling $\Reprl^{1/2}$ (figure \ref{fig:stokes_en_growth}, bottom right), while the cut-off itself is rather abrupt. These results, as well as those from the analytic model formulated in \S\,\ref{sec:kazantsev}, indicate that the dynamo is only viable for moderate values of the ratio $\Reprl/\mathrm{Re}$: too small a value results in a dynamo with too much mixing (relative to stretching) and thus excessive resistive dissipation. 

\begin{figure}
    \centering
    \includegraphics[width=\textwidth]{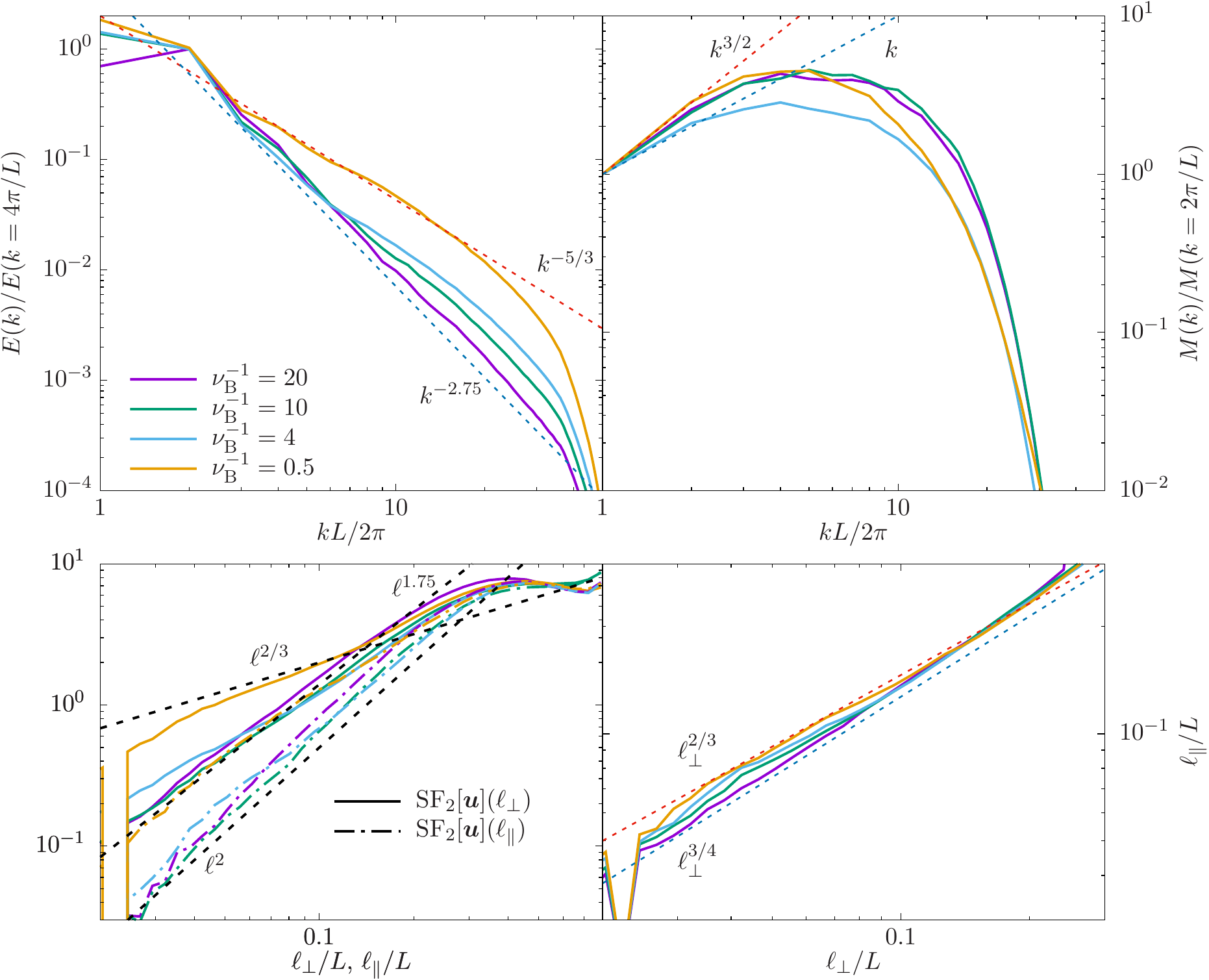}
    \caption[Energy spectra for the Stokes flow simulations]{\emph{top:} (left) Kinetic and (right) magnetic energy spectra for various unlimited Braginskii-MHD simulations with small and large parallel Reynolds number $\Reprl$. \emph{bottom:} (left) Perpendicular and parallel structure functions and (right) spectral anisotropy scalings (see \eqref{eqn:scale_aniso}) for the same runs).}
    \label{fig:stokes_spec}
\end{figure}

Both limits $\Reprl\rightarrow 0$, $\mathrm{Re}= \mathrm{const}$ and $\Reprl =\mathrm{const}$, $\mathrm{Re}\rightarrow \infty$ can be thought of as the result of a strong anisotropization of the underlying turbulence, leading to motions that are more two-dimensional than three-dimensional. (Recall that dynamo action cannot be sustained by a planar flow \citep{Zeldovich57}.) Interestingly, simulations that feature both $\Reprl$, $\mathrm{Re} < 1$ yet have $\Reprl/\mathrm{Re}\sim 1$ still experience a viable dynamo; compare dotted and dot-dashed light blue lines in figure~\ref{fig:stokes_en_growth}. This shows that it is the \emph{relative size of the stretching to mixing time scales}, rather than their absolute values, that lead to a viable dynamo (cf.~\S\,\ref{sec:kazantsev}). In some sense, having an isotropic viscosity comparable to an anisotropic viscosity allows the momentum to diffuse more isotropically, thwarting the anisotropic viscosity's tendency towards making the flow more two-dimesional. This is akin to the $\mathrm{Pm} < 1$ dynamo in the isotropic case, where mixing from motions at all scales makes magnetic energy amplification by motions at super-resistive scales more difficult. Conversely, the case with $\mathrm{Re} \sim \Reprl$ is more similar to the isotropic $\mathrm{Pm} \gtrsim 1$ dynamo, where the contributions of both stretching and mixing come from the same range of scales.

\begin{figure}
    \centering
    \includegraphics[width=\textwidth]{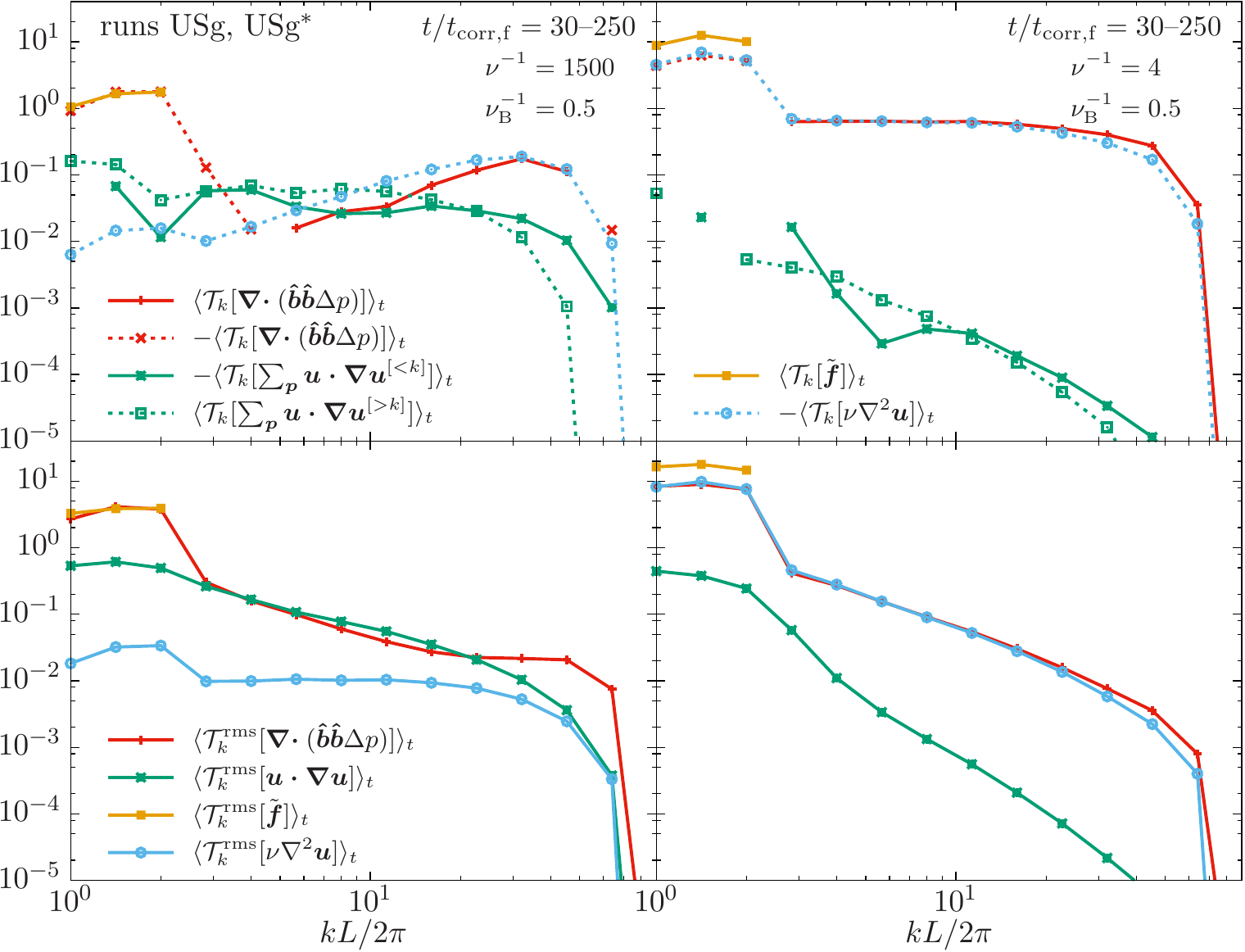}
    \caption[Shell-to-shell transfer for the Stokes flow simulations]{\emph{Top:} Shell-filtered kinetic-energy transfer function $\mathcal{T}_k$~\eqref{eq:shell_trans} for the unlimited Braginskii-MHD system in the Stokes flow regime employing $\visc_\mathrm{B}^{-1} = 0.5$ and $\visc^{-1} = 1500$ ($\visc^{-1} = 4$) in the left (right)  frame. The runs USg and USg$^*$ are chosen here as one corresponds to magnetic field decay (USg, left) while the other corresponds to a viable dynamo (USg$^*$, right).  Solid (dotted) lines denote energy flowing into (out of) the shell centered at mode $k$. \emph{Bottom:} root-mean-square shell-filtered transfer functions for the same runs. }
    \label{fig:stokes_trans}
\end{figure}

Despite not having a dynamo for $\Reprl\lesssim{1}$, the unlimited Braginskii-MHD system still features a cascade of turbulent energy to small scales. The kinetic energy spectra of various unlimited Braginskii simulations in the Stokes flow regime are displayed in figure~\ref{fig:stokes_spec}. As the parallel Reynolds is decreased, the spectral index of the energy cascade beyond the parallel viscous scale approaches 3/2 in the $\visc_\mathrm{B} = 0.5$ run. While this is reminiscent of isotropic Irishnikov--Kraichnan \citep{iroshnikov63,kraichnan65} or anisotropic \citet{Boldyrev2006} scalings for Alfv\'{e}nic turbulence, the scaling of the wavevector anisotropy is measured (see (\ref{eqn:scale_aniso})) to satisfy $\ell_\parallel \propto \ell_\perp^{2/3}$, incompatible with both of those theories. Computing the energy transfer functions [see \eqref{eq:shell_trans} and \eqref{eq:shell_trans_rms}], we find that the cascade satisfies a balance between the hydrodynamic nonlinearity and the (unlimited) Braginskii viscous stress, suggesting some form a critically balanced turbulence.

\section{Summary}\label{brag:sec:discussion}

For studying the fluctuation dynamo in the weakly collisional, magnetized regime, we used incompressible Braginskii-MHD simulations and analytic modeling. The pressure anisotropy, and thus the parallel viscous stress, was either hard-wall limited to lie within the firehose and mirror instability thresholds or allowed to venture beyond those thresholds. While the latter option has traditionally been considered unphysical, given the plethora of evidence---both theoretical and observational---that such kinetic instability thresholds are well respected in collisionless, magnetized plasmas, its study is important for (at least) three reasons. First, it offers an additional point of comparison to dynamo behavior in isotropic MHD and limited Braginskii MHD; in particular, our finding of its similarity to the saturated state of the MHD dynamo affords a better understanding of that more traditional case. Secondly, many aspects of its evolution are remarkably similar to those found in the hybrid-kinetic simulations performed in chapter~\ref{ch:simulation}; this is fortunate, as the unlimited Braginskii-MHD simulations provide a better controlled and more economical test bed with which to diagnose the field and flow statistics in this regime. Thirdly, we have argued that a significant period in the dynamo amplification of the intracluster magnetic field occurs at a time when the plasma $\betai$ is too large for kinetic instabilities to regulate the pressure anisotropy efficiently enough to pin it near its ${\sim}1/\betai$ stability boundaries. During this phase, a constant collision frequency that partially restrains the pressure anisotropy (as in the unlimited simulations presented herein) may in fact be the more realistic `closure'. Indeed, much of the evolution of the collisionless dynamo found in chapter~\ref{ch:simulation} using a hybrid-kinetic approach occurred during such a phase, with a suppressed parallel rate of strain, an anisotropization of the flow velocity, an imperfect regulation of the pressure anisotropy, and a Kolmogorov-like cascade of perpendicular kinetic energy---all of which are manifest in our unlimited Braginskii-MHD runs.

The main conclusions of this chapter are as follows:
\begin{enumerate}
    \item \label{bragcon1} The chaotic flow driven by large-scale forcing produces highly intermittent and structured magnetic fields, which are organized into folds and grow exponentially until the Lorentz tension force is strong enough to back-react dynamically on the velocity field. This folded structure, a hallmark of the ${\rm Pm}\gtrsim{1}$ fluctuation dynamo, persists into the saturated state. These results hold regardless of whether the plasma is described by isotropic MHD or Braginskii MHD with either limited or unlimited pressure anisotropy, so long as the Braginskii viscosity is not too large or the ratio of the Braginskii and isotropic viscosities is not too small (see~\ref{bragcon5} below). 
    \item \label{bragcon2} Hard-wall limiters on the parallel viscosity, intended to mimic the rapid regulation of pressure anisotropy by kinetic Larmor-scale instabilities otherwise not properly captured in Braginskii MHD, reduce the Braginskii dynamo to its ${\rm Re}\gg{1}$, ${\rm Pm}\sim{1}$ MHD counterpart. With the exception of some minor differences, such as a slight suppression of $\ROS$ in firehose/mirror-stable regions (figure \ref{fig:pdfROS}), the two are nearly indistinguishable. This conclusion is broadly consistent with the findings of \citet{SantosLima}. Regions of the plasma lying near or beyond the mirror instability threshold are subject to a magnetic tension that is effectively enhanced by a factor of only $3/2$ and therefore are largely unaffected by the positive, limited viscous stress. And regions of the plasma lying near or beyond the firehose instability threshold, at which the effective tension from the Maxwell and viscous stresses is zero, appear to be unimportant to the dynamics, most likely because such regions have a small volume-filling factor.
    \item \label{bragcon3} When the dynamical feedback of unbridled viscous dissipation on the field-stretching motions is allowed (the unlimited Braginskii model), the dynamo takes on a different character. Not only is the dynamo slower, but many characteristics of the flow and magnetic field change very little from the kinematic stage to the saturated state. Further, most of these characteristics bear a striking resemblance to those found in the saturated state of the ${\rm Re}\gg{1}$, ${\rm Pm}\gtrsim{1}$ isotropic-MHD dynamo. These include: the magnetic-energy spectrum (figure \ref{fig:unlim_spec}), the characteristic wavenumbers of the folded magnetic-field geometry (figure \ref{fig:wavenumbers_unlim}), the PDF of the magnetic-field-line curvature (figure \ref{fig:curvature}), and the PDF of the alignment angles between the magnetic field and the rate-of-strain tensor eigenvectors (figures \ref{fig:angle2d_all} and \ref{fig:eigen_some}). In addition, a scale-dependent anisotropy of the velocity field was found in the saturated state of the limited Braginskii dynamo and throughout the entire evolution of the unlimited Braginskii dynamo (specifically, $\ell_\parallel\propto\ell^{3/4}_\perp$; see figures \ref{fig:struct_lim} and \ref{fig:struct_unlim}). A similar anisotropy was found in the saturated state of the isotropic-MHD dynamo.
    \item \label{bragcon4} Motivated by this resemblance between the saturated-MHD and unlimited-Braginskii dynamos and by the structural similarity of the magnetic-tension and Braginskii-viscous stresses, we have constructed a theory for the unlimited Braginskii-MHD dynamo based on a similar framework developed by \citet{Scheko_saturated} to model the saturated sate of the MHD small-scale dynamo (\S\,\ref{sec:kazantsev}). This theory introduces a field-biased rate-of-strain tensor into the Kazantsev--Kraichnan model of the fluctuation dynamo that captures the partial two-dimensionalization of the velocity gradient statistics with respect to the local magnetic-field direction caused by the anisotropic viscosity. (In \citet{Scheko_saturated}, this partial two-dimensionalization is instead caused by the dynamically important Lorentz force in the saturated state.) This model predicts magnetic-energy spectra and dynamo growth rates in broad agreement with those found in our simulations.
    \item \label{bragcon5} Another prediction of our modified Kazantsev--Kraichnan model is that enhanced small-scale mixing and local two-dimensionalization of the flow as the isotropic viscosity $\visc\rightarrow{0}$ at fixed Braginskii viscosity $\visc_{\rm B}$ precludes the unlimited Braginskii-MHD dynamo in this regime. In the complementary limit of $\visc_{\rm B}\rightarrow\infty$ at constant $\visc$---what we have deemed the `Braginskii Stokes-flow' regime---the unlimited Braginskii-MHD dynamo fails for parallel Reynolds numbers $\Reprl\lesssim{1}$ (at fixed $u_{\rm rms}$). This is caused by excessive two-dimensionalization of the flow by the strong anisotropic viscosity. Isotropic viscosity comparable to the Braginskii viscosity saves the dynamo in this situation by diffusively bleeding momentum into the third direction; indeed, we find that the Stokes-flow dynamo works in isotropic MHD regardless of the value of ${\rm Re}$ so long as there is sufficient scale separation between the forcing and resistive scales. Combined with conclusion~\ref{bragcon4} above, this implies that the Braginskii dynamo is only viable for moderate values of the ratio $\Reprl/{\rm Re}$: too small a value results in too much field-line mixing by the perpendicular flows and gradients relative to field-line stretching and thus to excessive resistive dissipation of the magnetic field. This principle is quantified in our modified Kazantsev--Kraichnan model by the quantities $\sigma_\parallel$ and $\sigma_\perp$ [see \S\,\ref{sec:kazantsev} and, in particular, \eqref{eqn:gamma_delta}].
\end{enumerate}

\chapter{The Structure of Collisionless, High-$\betai$, Zero-net-flux Turbulence}\label{ch:structure}

\section{Motivation}

In this chapter, we reconsider the result presented in figure~\ref{sim:ROS_blowup} from our hybrid-kinetic simulations, which suggests that the growth rate of the collisionless plasma dynamo seems to grow without bound as both the magnetic Reynolds number and the number of grid points of the simulation $N_\mathrm{cell}$ are increased.  This is at odds with the observation that our kinetic simulations seem to more closely resemble the Braginskii-MHD system using unlimited, rather than limited, pressure anisotropy, which would suggest $\Reeff \sim 1$; indeed, the effective collisionalities calculated in \S\ref{sec:saturation} ($\nu_\mathrm{eff}/\Omegaio \approx \textrm{2--4} \ll S\betai$) support this conclusion. The idea here is a simple one: if the motions that drive the growth of the magnetic field are limited to the large scales by the parallel viscosity, then in the limit $\mathrm{Pm}\rightarrow \infty$ the dynamo growth rate should approach an asymptotic value  proportional to the inverse turnover time of the smallest eddies responsible for stretching the field.  This is not what is observed in our kinetic simulations. Keep in mind that the results presented in this chapter are \emph{preliminary} and have \emph{not been fully digested}. As a result, many questions remain unanswered and are left for future research. 

\begin{figure}
\centering
\includegraphics[scale=1.2]{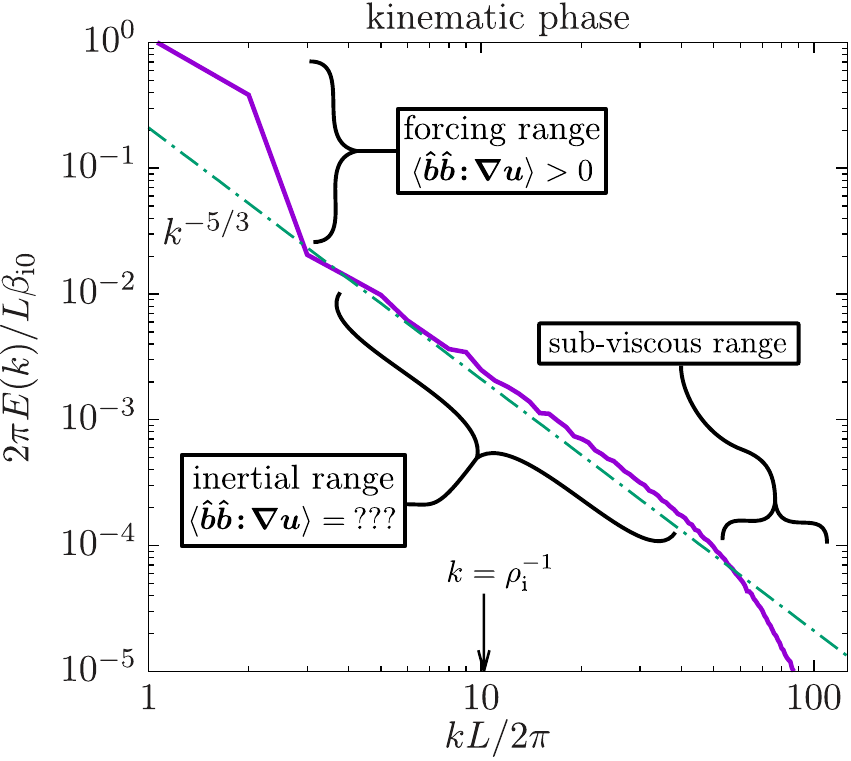}
\caption[Close-up of the kinematic kinetic energy spectra for $\betaio=10^6$]{\label{fig:kinspec_close} Time-averaged kinetic energy spectrum for the hybrid-kinetic run 1 (table~\ref{tab:sim:fiducial}) in the kinematic phase. This spectrum features three distinct regions: the forcing range, the inertial range, and the sub-viscous range. While motions in the forcing range grow the magnetic field, it is to be determined whether or not those in the inertial range do as well.}
\end{figure}

We will now examine more closely  the structure of the  turbulent velocity in our hybrid-kinetic simulations of the magnetized plasma dynamo.  To remind ourselves of what  this entails, a typical kinetic spectrum from the simulation is shown in figure~\ref{fig:kinspec_close}. Here, three distinct regions of the spectra are apparent: the forcing range ($kL/2\upi \in [1, 2]$), an inertial range ($ 2 < kL/2\upi \lesssim 50 $ in the figure) that exhibits a power law close to the Kolmogorov -5/3, and the sub-viscous region that exhibits an exponential cut-off due to dissipation.  We now wish to determine which velocity scales are responsible for growing the magnetic energy. Notice that there is an order of magnitude drop in the energy immediately beyond the forcing range, similar to what was observed with the unlimited Braginskii-MHD simulations in the Stokes flow regime (cf. figure~\ref{fig:stokes_spec} in \S\ref{sec:stokes}), suggesting that the hybrid-kinetic simulations are also operating in the Stokes flow regime.  If this were the case, then we would expect the characteristic scale of the parallel variation of the flow to be roughly the forcing scale (i.e. $k_\parallel \sim k_\mathrm{f}$), which results from a balance between energy injection and parallel viscous dissipation. Indeed, the measured growth rate of magnetic energy from the hybrid-kinetic simulations is roughly $\gamma \approx 0.3 u_\mathrm{rms}/\ell_0$. The remaining motions that make up the inertial range should na\"ively consist of perpendicular (Alfv\'enic) motions that only mix field lines, not stretch them.  Of course, these are isotropic spectra with $k^2 \sim k_\parallel^2 + k_\perp^2$, and so we may expect some amount of stretching in the inertial range. However, if $k_\parallel \sim k_\mathrm{f} = \mathrm{const}$, this leads to $k_\parallel u \sim k_\mathrm{f} k^{-2/3}$, and so the bulk of the stretching would come from motions in the forcing range.
Our goal in this chapter is to test this hypothesis. We shall see that, while the kinetic simulations share many features in common with the unlimited Braginskii-MHD system, the matter of \emph{increasing the magnetic energy} is quite different, and that the results of \S\ref{sec:kazantsev} and \S\ref{sec:stokes} may not actually pertain to collisionless plasmas.

\begin{figure}
\centering
\includegraphics[width=\textwidth]{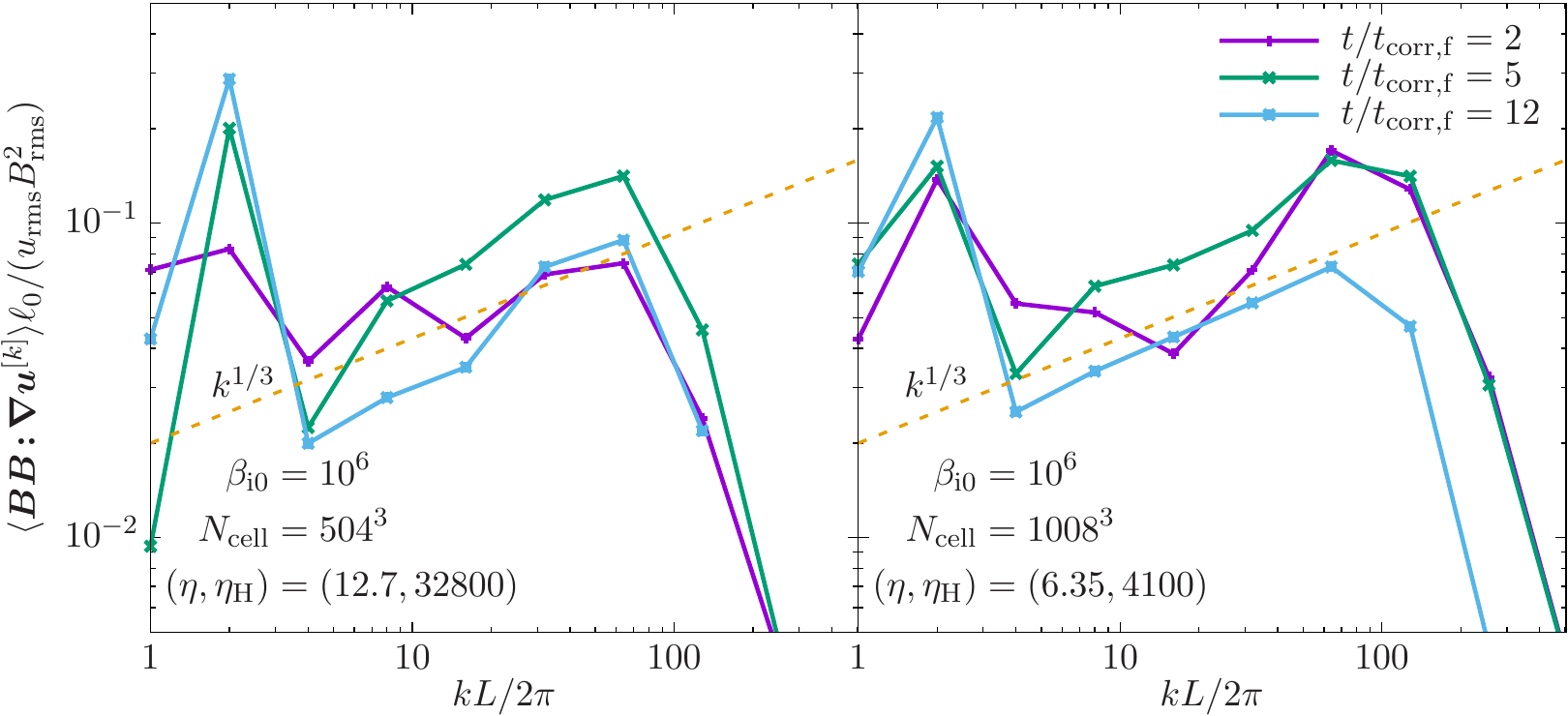}
\caption[Shell-filtered $\boldsymbol{BB\!:\!\nabla u}$ as a function of $k$ for $\betaio = 10^6$]{\label{fig:thesis_ros} Shell-filtered energy-weighted parallel rate of strain for hybrid kinetic simulations at various times. The discrete sum of all the points gives the growth rate of the magnetic field. A $k^{1/3}$ power law is displayed for reference.  }
\end{figure}

\section{Results}

Figure~\ref{fig:thesis_ros} displays the shell-filtered, energy-weighted parallel rate of strain as a function of scale at various times. These functions are binned logarithmically and displayed such that the discrete sum of the points gives the growth rate of the magnetic field. It is immediately apparent that all scales of the underlying velocity field contribute to the growth of the magnetic energy.  Even more surprising is that, in the inertial range, smaller scales have a larger contribution to this growth rate, ruling out the possibility that $k_\parallel \sim k_\mathrm{f}$. The power law of this trend is roughly $k^{1/3}$--$k^{1/2}$, which is shallower than the turnover-rate scaling of $k^{2/3}$ in the~\citet{Kolmogorov1941} phenomenology.
 As $\mathrm{Rm}$ and the number of cells are increased, the inertial range extends to even smaller scales, resulting in a larger growth rate.
 Interestingly, at $t/\tcorrf =5$  (right panel), the shell contributing the most to the growth rate is in the inertial range, rather than in the forcing range. 

\begin{figure}
\centering
\includegraphics{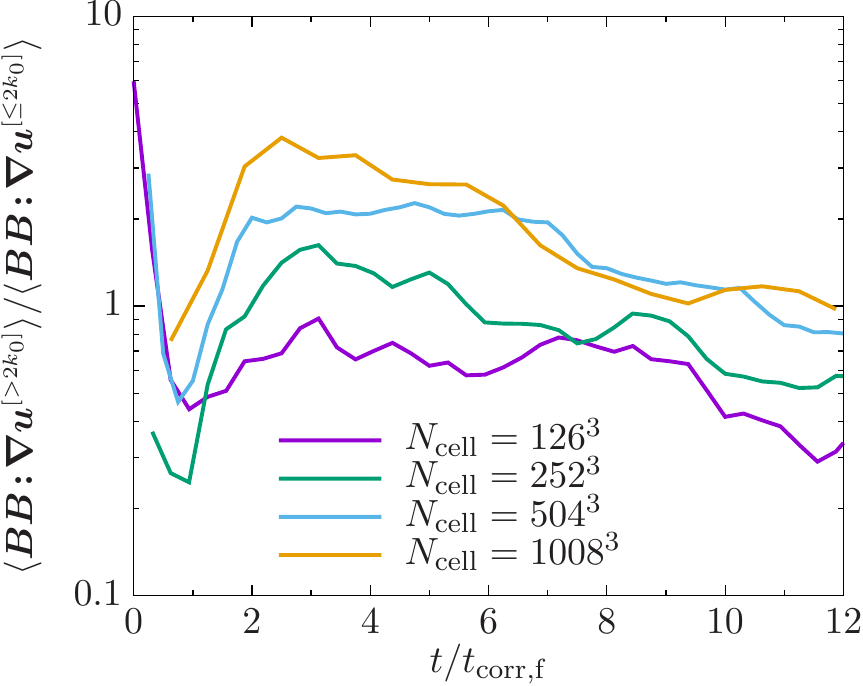}
\caption[Ratio of sub-forcing-scale and forcing-scale contributions to $\bb{BB\!:\! \nabla u}$]{\label{fig:urms_ratio} Ratio of the sub-forcing-scale and forcing-scale contributions to the energy-weighted parallel rate of strain.  As $\mathrm{Rm}$ and the number of collocation points are increased, the maximum value of this ratio increases as well. }
\end{figure}

Figure~\ref{fig:urms_ratio} displays the ratio of the sub-forcing-scale and forcing-scale contributions to the energy-weighted parallel rate of strain. For the simulation with the highest resistivity and lowest resolution, this ratio is less than unity throughout the simulation. However, as $\mathrm{Rm}$ and $N_\mathrm{cell}$ are increased, the ratio increases as well,  reaching values as high as 4 for the largest simulation.  As the simulation progresses and the magnetic energy continues to grow, %
this ratio decreases. The curves for the $N_\mathrm{cell}=504^3$ and $1008^3$ coincide at roughly $t/\tcorrf\approx 7$. At least for the simulation parameters used here, it does not seem that the maximum of this ratio approaches some asymptotic value as $\mathrm{Pm}$ and $N_\mathrm{cell}$ are increased. Thus without further information, it is difficult to predict the magnetic field growth rate in an \emph{a priori} way. At the very least, the growth rate for the simulations employing $N_\mathrm{cell} \ge 504^3$ is effectively doubled compared to $u_\mathrm{rms}/\ell_0$ throughout their runtimes. As such, the dynamo in a collisionless, weakly magnetized plasma seems to exhibit properties for both the $\mathrm{Re} \gg 1$ dynamo than the $\mathrm{Re} \sim 1$ dynamo, the latter being the expected result based on Coulomb collisions and weak regulation of the pressure anisotropy.

\begin{figure}
\centering
\includegraphics[width=\textwidth]{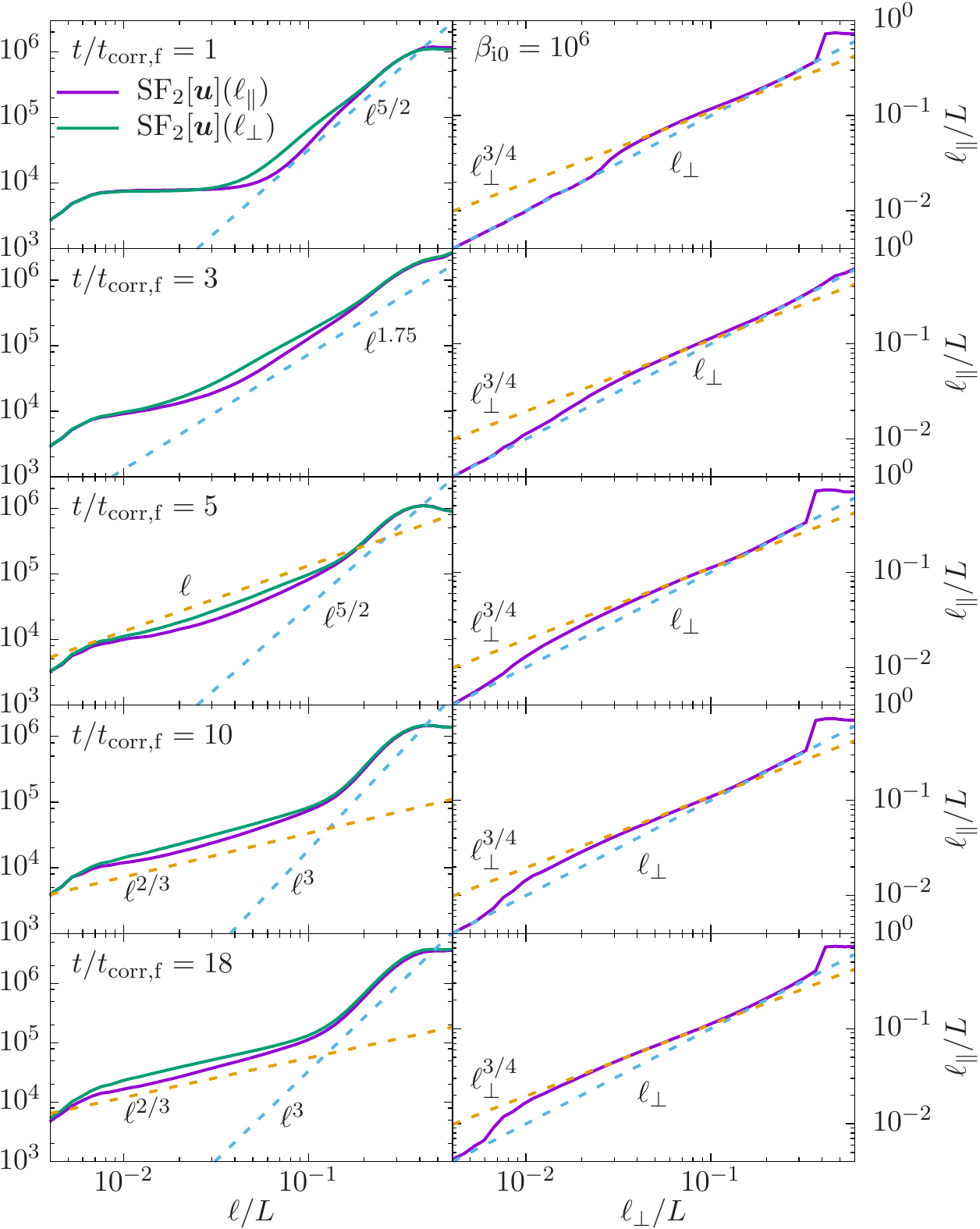}
\caption[Structure functions for the hybrid kinetic $\betaio=10^6$ simulation]{\label{fig:urms_structure} Perpendicular and parallel structure functions (left), and scale dependent anisotropy (right) for the $\betaio=10^6$ hybrid-kinetic simulation at various stages of the dynamo.}
\end{figure}

Figure~\ref{fig:urms_structure} displays the parallel and perpendicular structure functions (left) as well as the scale-dependent anisotropy (right) for the $\betaio=10^6$ simulation through various stages of the dynamo.  While the structure functions are approximately isotropic in the forcing range, they exhibit the anisotropy scaling $\ell_\parallel \sim \ell_\perp^{3/4}$ in the inertial range, similar to the unlimited Braginskii-MHD regime with $\visc_\mathrm{B}^{-1} = 20$, but steeper than the Stokes flow regime with $\visc_\mathrm{B}^{-1} = 0.5$ (see figures~\ref{fig:struct_unlim} and~\ref{fig:stokes_spec}). This is somewhat peculiar, as the spectral slope in the collisionless system, ${\approx}-5/3$, is closer to the latter case, which features very little sub-parallel-viscous motions that affect the magnetic field strength. More importantly, with an anisotropy scaling of $\ell-\perp^{3/4}$, any spectral index shallower than $9/4$ leads to a parallel rate of strain that \emph{increases} with wavenumber, provided $\bb{u}_\parallel(\ell_\parallel)$ scales similarly to $\bb{u}(\ell_\parallel)$. Thus, while the unlimited Braginskii-MHD simulations (with spectral index ${\approx}2.75$) led to a parallel rate of strain that decreased with wavenumber, the opposite is true for our hybrid-kinetic simulations: a spectral index of $5/3$ leads to $u_\parallel/\ell_\parallel \sim \ell_\perp^{0.4}$, consistent with what is observed in figure~\ref{fig:thesis_ros}.

\section{Where do these motions come from?}

\begin{figure}
\centering
\includegraphics{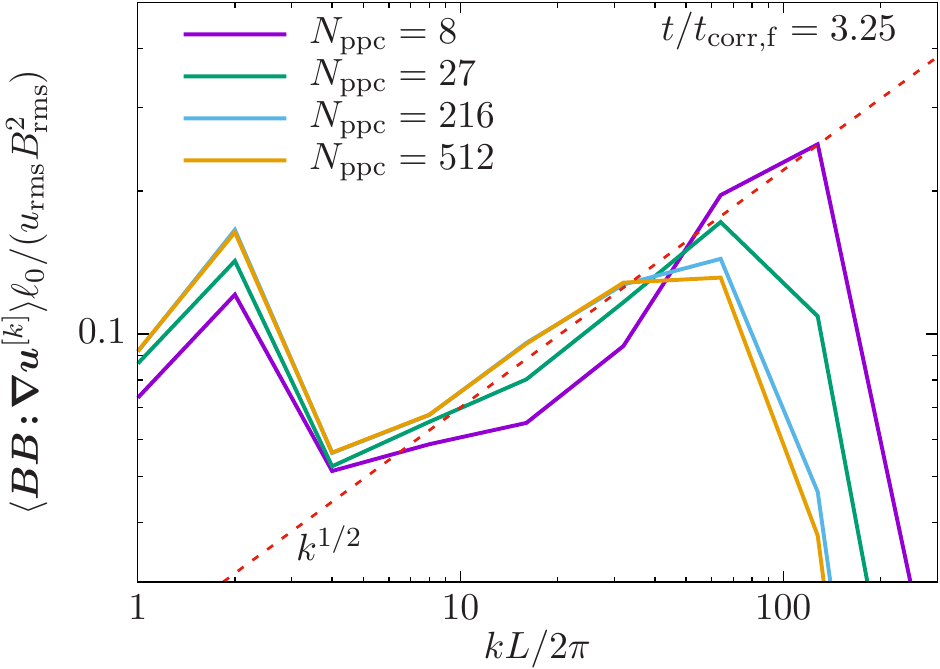}
\caption[Shell-filtered rate of strain for several values of $N_\mathrm{ppc}$]{\label{fig:urms_noise} The shell-filtered energy-weighted rate of strain for several values of the number of particles per cell for $\betaio=10^6$. These are the same simulations performed in \S\ref{sec:param_scan}. For $N_\mathrm{ppc} > 216$, the rate of strain appears to be resolved for the smallest wavenumber, and thus this effect appears to remain as $N_\mathrm{ppc}\rightarrow \infty$. }
\end{figure}

We now try to determine  the origin of these sub-forcing-scale motions that lead to growth of the magnetic energy. Remember that the unlimited Braginskii-MHD system also exhibited sub-parallel-viscous motions that would directly change the magnetic field strength (\emph{viz.} figure~\ref{fig:unlim_trans}). However, in this case these small-scale motions lead to \emph{damping} of the magnetic field, rather than growth.  In addition, these small-scale effects were lesser than the forcing-scale stretching by roughly an order of magnitude. Conversely, in the hybrid-kinetic simulations these range from being the same order to even larger than the forcing-scale stretching. Additionally, the cascade of perpendicular motions in the unlimited Braginskii-MHD system was found in \S\ref{sec:kazantsev} to be deleterious to the growth of magnetic energy, as small-scale mixing motions promoted resistive annihilation of the magnetic field.  It seems, at least according to figure~\ref{sim:ROS_blowup}, that in the $\mathrm{Pm}\rightarrow \infty$ limit, the issue of mixing versus stretching seems to be moot in the collisionless case.

One possible explanation for these motions is that they are a numerical artifact due to discrete particle noise.  Figure~\ref{fig:urms_noise} shows the shell-filtered energy-weighted rate of strain for the series of simulations scanning the particle-per-cell count $N_\mathrm{ppc}$ seen in \S\ref{sec:param_scan}. 
The close similarity of the curves for the simulations employing 216 particles per cell and 512 particles per cell indicate that this quantity 
 is numerically resolved, and that these sub-forcing-scale motions are not a product of discrete particle noise. Interestingly, while the curves for the under-resolved simulations would seem to suggest that particle noise is dominating the magnetic field growth, figure~\ref{sim:dynamo_PPC} reveals that such an incoherent effect only leads to minor differences in the growth rate of the magnetic energy. This difference mainly manifests itself as a slow secular growth of the magnetic energy at early times of the simulation, but as the kinetic energy grows this gives way to exponential growth that appears fairly consistent across all values of $N_\mathrm{ppc}$ that were studied.

One could argue that these motions are indicative of firehose and mirror instabilities, which can affect the turbulence in such a way as to bring the net pressure anisotropy closer to zero~\citep{Rosin_2011}. However, the parallel firehose instability, present in the unlimited Braginskii-MHD simulations, did not seem to bring about the type of motion seen here. The mirror instability could also explain these motions: if particles become trapped in mirror structures, then momentum transport, and thus viscous dissipation, along a field line is arrested, reducing the effective parallel viscosity of the plasma. Such trapping is difficult to observe in particle tracks when the fluid motions are turbulent, and the root-mean-square of the density fluctuation experiences a drop at $t/\tcorrf \approx 3$ for the simulation employing $N_\mathrm{cell}= 504^3$, suggesting that particles that may have been trapped by mirrors now become scattered. This is supported by inspection of the adiabatic invariant in the particle tracks, which suggest the scattering of particles begins around $t/\tcorrf \approx 1.5$.
 Finally, the fastest growing modes of the firehose and mirror instabilities have scales that are larger than the gyroradius by a factor of $\Deltai^{-1/2}$.  In the simulations studied here, the initial magnetization $L/\rhoio = 16$ is held fixed.  It is surprising then that, for simulations that only differ in resolution and resistivity, the growth rate increases substantially in the diffusion free regime $(t/\tcorrf \lesssim 3)$.

\begin{figure}
\centering
\includegraphics[width=0.7\textwidth]{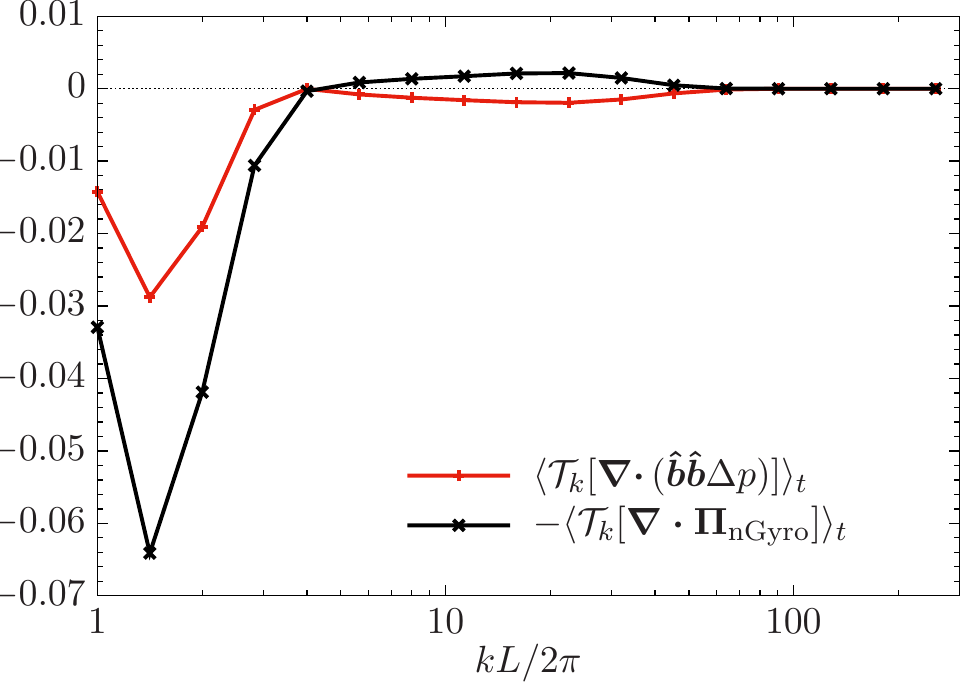}
\caption[Transfer functions of the gyrotropic and non-gyrotropic stress tensors]{\label{fig:urms_pressure} Shell-filtered transfer function of the gyrotropic (pressure anisotropy) and non-gyrotropic components of the deviatoric stress tensor. }
\end{figure}

We then must take a look at how the plasma viscosity is affecting the cascade of energy to smaller scales.  The plasma viscosity in the hybrid-kinetic system is contained in the deviatoric stress tensor $\boldsymbol{\mathit{\Pi}}$, defined as
\begin{equation}
\boldsymbol{\mathit{\Pi}} \doteq \mathsfbi{P} - p\mathsfbi{I},
\end{equation}
where $\mathsfbi{P} $ is the complete pressure tensor and $p = P_{ii}/3$ is the isotropic (ion) pressure.  We can further decompose the deviatoric stress tensor into gyrotropic  ($\boldsymbol{\mathit{\Pi}}_\parallel$) and non-gyrotropic components ($\boldsymbol{\mathit{\Pi}}_\mathrm{nGyro}$), the former being the usual pressure anisotropy given by equation~\eqref{eq:bragten} and the latter being equal to
\begin{equation}
\boldsymbol{\mathit{\Pi}}_{\mathrm{nGyro}} = \boldsymbol{\mathit{\Pi}} - \boldsymbol{\mathit{\Pi}}_\parallel.
\end{equation}
 The spectral energy transfer due to these two terms is plotted in figure~\ref{fig:urms_pressure}. Two distinct regions are manifest: the forcing range ($kL/2\upi \in [1,2]$) and the inertial range ($kL/2\upi > 2$). In the former, both terms (and thus the entire stress $\boldsymbol{\mathit{\Pi}}$) act to mostly balance the energy injection by the random forcing, with the remainder feeding into the inertial range cascade. In the simulations of Braginskii-MHD in chapter~\ref{ch:brag}, this balance is specifically between field-oriented dissipation and random forcing, allowing Alfv\'enic motion to proceed to smaller scales undamped. In our hybrid-kinetic system, the stress is more isotropic at large-scale, and as a result the flow should behave like an isotropic Stokes flow, similar to $\mathrm{Re}\sim 1$ MHD. However, in the inertial range these two contributions mostly cancel, which suggests that the parallel viscous stress may largely be nullified in this range, allowing a cascade of energy, both field-oriented and otherwise, to proceed. Here, the gyrotropic stress dissipates energy at all scales, while the non-gyrotropic stress acts to replenish it. This is in stark contrast to unlimited Braginskii-MHD presented in figure~\ref{fig:trans_unlim}: there, the Braginskii viscosity at small scales acts to inject energy, rather than dissipate it, whereas the isotropic viscosity always has a dissipative effect.

\section{Discussion}

  The suppression of viscous dissipation in the inertial range is reminiscent of the work by~\citet{Meyrand-echoes}, who observed that for some turbulent systems, inverse-Landau damping effected by stochastic echoes~\citep{Scheko_phasemixing} would allow the cascade of compressive fluctuations that  would normally be dissipated by linear Landau damping. While it is not clear that the results in this chapter are caused by the same phenomenon, the implications of such a process would have profound effects for the dynamo: if it were the case that $\mathrm{Re}_\parallel$ was a truly irrelevant parameter for collisionless systems undergoing dynamo, then arbitrarily fast fluctuation dynamo would be a generic feature of any collisionless, magnetized plasma.  
  
  The positivity of the non-gyrotropic component of the pressure-tensor energy transfer seen in figure~\ref{fig:urms_pressure} may be due to the gyroviscous stress, which causes a non-dissipative reorientation of the momentum relative to the direction of the magnetic field. While this term is typically a factor of $\nu_\mathrm{eff}/\Omegaio$ smaller than the gyrotropic viscous stress~\citep{Braginskii}, at the early stages of the magnetized plasma dynamo this factor may not be small, and the gyroviscosity may be important. Indeed, figure~\ref{fig:urms_ratio} would seem to suggest that as the magnetization of the plasma is increased, these motions become less important. Unfortunately this occurs at the same moment that the Lorentz force becomes dynamically significant at the smallest stretching scales. In order to separate these two processes, one would necessarily need more separation between the magnetic and kinetic energies, as well as the initial magnetization measured by $L/\rhoio$.
  
  Perhaps the main take-away from this chapter is that the determination of the viscosity for a collisionless plasma is an extremely complex undertaking, and that in order to so do, one requires scale-separation between several different quantities, a requirement difficult to achieve with current computational resources.

\chapter{ Explosive Dynamo Growth  Using `Soft-wall' Pressure-anisotropy Limiters}\label{ch:explosive}

\section{Motivation}

\begin{figure}
    \centering
    \includegraphics[scale=1.3]{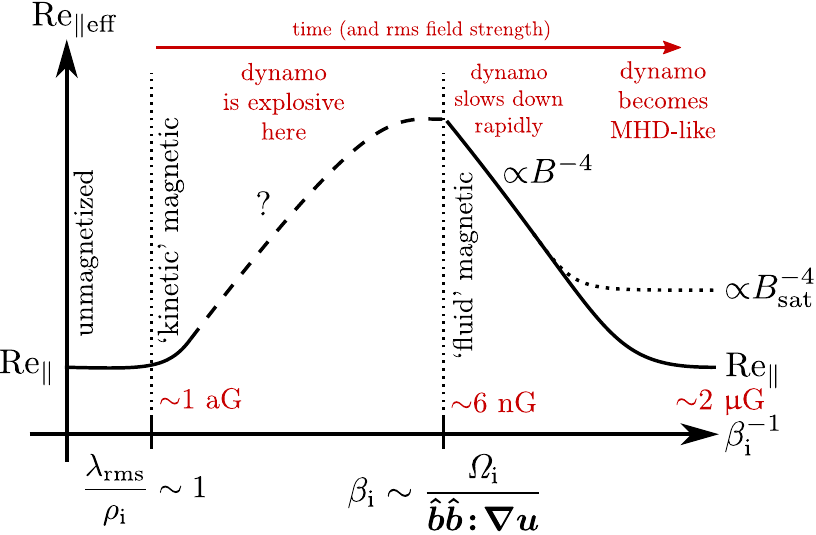}
    \caption[The evolution of the Reynolds number as a function of $\betai^{-1}$]{A qualitative illustration of the effective parallel Reynolds number $\Reeff$ vs.~$\betai^{-1}$ as the dynamo proceeds through the various collisionality regimes. The value of $\Reeff$ in the saturated state of the dynamo (i.e., when $\betai \sim M^{-2}$) is set by Coulomb collisions if the collisional mean free path  $\lambda_{\rm mfp,i} \lesssim \ell_0 \, M$; otherwise, it is given by \eqref{eqn:Re_eff} with $\betai \sim M^{-2}$, i.e., $\Reeff \sim 1$.}
    \label{fig:cartoon_brag}
\end{figure}

The potential for explosive growth of the magnetic energy in the magnetized `kinetic' regime of the plasma dynamo was described in \S\ref{sec:paniso}. There,  it was argued that, while the Reynolds number in the unmagnetized regime is ${\sim}100$ due to Coulomb collisions,  at the outset of the magnetized `fluid' regime the Reynolds number is ${\sim}\betai^2 M^4 \gg 1$ as a result of strong regulation of the pressure anisotropy by firehose and mirror instabilities that introduce an effective collisionality. There must be a period in the magnetized `kinetic' regime, then, during which the Reynolds number must increase in order to smoothly connect these two values. This scenario, which forms the basis of theories on the explosive growth of magnetic energy in collisionless plasma dynamos \citep{SchekoCowley06a,SchekoCowley06b},\footnote{While a paper by~\citet{Mogavero} also dealt with the possibility of explosive growth, only the magnetized `fluid' regime was considered, rather than the kinetic one. There, the most likely outcome is actually a \emph{slowing down} of the dynamo, see~\S\ref{sec:paniso}.}  is illustrated qualitatively in~\ref{fig:cartoon_brag}. When the dynamo process begins with an unmagnetized plasma, its Reynolds number $\mathrm{Re}_\parallel$ is determined by Coulomb collisions alone. As the plasma begins to become magnetized ($\lambda_\mathrm{mfp} \sim \rhoi$, roughly 1 aG in the ICM), the dynamo enters the `kinetic' magnetized regime. There, Larmor-scale instabilities develop which begin to scatter particles, leading to a decrease in the effective viscosity and thereby to an increase in the Reynolds number. The scattering from these instabilities becomes stronger as the plasma becomes more magnetized,  reaching a peak once these instabilitities can become well regulated ($\betai \sim \Omegai / \ROS$, roughly at 6 nG in the ICM), once this happens, the dynamo enters the `fluid' magnetized regime and, as the instabilities saturate, the collision frequency again returns to the Coulomb collision frequency.

\begin{table}
    \centering
    \small
    \begin{tabular}{ccccc ccccc}
    \hline 
    \hline
            Run &   $\visc_\mr{B}^{-1}$ & $\visc_\mathrm{H}^{-1}$  & $\eta^{-1}$ & $\!\!\langle u_\mathrm{rms}^2 \rangle^{1/2}_t\!\!\!\!$ & $\langle B_0^2\rangle^{1/2}\!\!$ &Re$_\parallel$ & Re & Rm & limiter   \\
    \hline
    SL1 &   20 & $1.8\times 10^{7}$ & $1.8\times 10^{7}$  & 1.29& $10^{-5}$ & 4.1 & 100 & 100 & $B_\mathrm{SL} = 0.002$\\
        SL2 &  20 &  $1.8\times 10^{7}$  & $1.8\times 10^{7}$ & 1.34& $10^{-5}$ & 4.3 & 100 & 100 & $B_\mathrm{SL} = 10^{-4}$ \\
    U1H & 20  & $1.8\times 10^{7}$  & $1.8\times 10^{7}$ & 1.21 & $10^{-3}$ & 3.9 & 100 & 300 & unlimited \\
  L1H &  20 &  $1.8\times 10^{7}$ & $1.8\times 10^{7}$  & 1.47 & $10^{-3}$ & 4.7 & 100 & 100 & hard-wall \\
    \hline
    \end{tabular}
    \caption[Index of runs for the soft-wall Braginskii-MHD dynamos.]{Index of runs for the Braginskii-MHD dynamo employing soft-wall limiters, SL1 and SL2.  Also included in this table are the Braginskii-MHD runs U1H and L1H (\emph{viz.} table~\ref{tab:runs}), which are displayed in figure~\ref{fig:wave_soft}.}
    \label{tab:soft}
\end{table}

   While it seems that the collision frequency of the hybrid-kinetic simulations in chapter~\ref{ch:simulation} plateaus before saturation is reached, it was shown in chapter~\ref{ch:structure} that the effective Reynolds number is larger than 
 what this collision frequency suggests, and so there is some mechanism present that alters the viscosity of the plasma, leading to faster dynamo.

 It is worthwhile, then, to see how the dynamo reacts and adjusts to a plasma whose viscosity is constantly in flux. So far this has been done for zero-dimensional models of the dynamo: what is needed is a study of this process using full three-dimensional geometry.   As such, we aim to probe the transition between the second and third regimes via the use a novel set of pressure-anisotropy limiters that are motivated by those models. While simulations of the dynamo employing  pressure anisotropy limiters have been performed in the past by~\citet{SantosLima},  a constant and uniform effective collision frequency was assumed, precluding any connection to earlier, somewhat speculative models of explosive magnetic-field growth in the plasma dynamo.  
 The pressure anisotropy limiters we employ in this section, so called `soft-wall' limiters,  have a collision frequency that depends on the local properties of the plasma and magnetic field, becoming more effective at regulating the pressure anisotropy as the system becomes more magnetized (as indicated by the dashed line in figure \ref{fig:cartoon_brag}). This results in a dynamo that is self-accelerating. 

\section{Description of the `soft'-wall pressure-anisotropy limiters}

The `soft-wall' limiters we propose in this section are distinct from the hard-wall limiters that are typically used in simulations of Braginskii-MHD (see \S\ref{ch:brag}). Rather than pinning the pressure anisotropy to the marginal threshold whenever the anisotropy ventures out of the stable region, these limiters reduce the magnitude of the pressure anisotropy by a fraction of what is needed to render it stable.
This fraction, which can be as large as unity, is controlled by the local properties of the plasma.

This is done by considering an effective collision frequency $\nu_\mr{eff}^\mr{SL}$ that is now inhomogeneous. The Braginskii viscosity in regions of firehose/mirror instability is  expressed as
\begin{equation}\label{eqn:nusoft1}
    \visc_\mr{B}^\mr{SL} = \frac{p}{\nu_\mr{i}+\nu_\mr{eff}^\mr{SL}} = \frac{\visc_\mr{B}}{1+\nu_\mr{eff}^\mr{SL}/\nu_\mr{i}}.
\end{equation}
While in principle $\nusl$ can depend non-trivially on various quantities, such as the structure and strength of the local magnetic field, the size of the local pressure anisotropy, etc., here we adopt the simple \emph{Ansatz} $\nusl = \alpha \Omegai$, where $\alpha$ is a constant of proportionality. 
With $\Omegai \propto B$, equation \eqref{eqn:nusoft1} may be written as
\begin{equation}\label{eqn:nusoft}
    \visc_\mr{B}^\mr{SL} = \frac{\visc_\mr{B}}{1+B/B_\mr{SL}}, 
\end{equation}
where $B_\mr{SL}$ parameterizes the ratio $\alpha\Omegai/\nu_\mr{i}$. This parameter is necessary because the gyrofrequency $\Omegai$ is ordered out of the Braginskii-MHD system given by equations~(\ref{eq:int:brag}a--d), and so $B_\mathrm{rm}$ serves as a reference magnetic-field magnitude above which the soft-wall limiter becomes relevant. 
Thus, these soft-wall limiters take the form
\begin{align}\label{eqn:mirror_soft}
    \rmDelta p = \left\{ \begin{array}{lc}
    3\visc_\mr{B}\ROS, &  3\visc_\mr{B}\ROS < B^2/8\upi \\
        \mr{max}[B^2/8\upi,3\visc_\mr{B}\ROS \,(1+B/B_\mr{SL})^{-1}], & 3\visc_\mr{B}\ROS \ge B^2/8\upi 
    \end{array}\right.
\end{align}
on the mirror ($\rmDelta p>0$) side and 
\begin{align}\label{eqn:firehose_soft}
    \rmDelta p = \left\{ \begin{array}{lc}
    3\visc_\mr{B}\ROS, &  3\visc_\mr{B}\ROS > -B^2/4\upi \\
       \mr{min}[-B^2/4\upi,3\visc_\mr{B}\ROS\,(1+B/B_\mr{SL})^{-1}], & 3\visc_\mr{B}\ROS \le -B^2/4\upi
    \end{array}\right.
\end{align}
on the firehose ($\rmDelta p<0)$ side [cf.~\eqref{eqn:mirror_hard} and \eqref{eqn:firehose_hard}]. These limiters work in a straightforward way: for regions that are mirror or firehose unstable, the pressure anisotropy is reduced by a factor of $(1+B/B_\mathrm{SL})^{-1}$ from what it would be if it were unlimited. If this fraction were to render it stable, then the pressure-anisotropy is simply pinned to the marginal threshold. Using these limiters, we hope to bridge the regimes of imperfect and perfect pressure-anisotropy regulation discussed in \S\,\ref{sec:paniso}.

\section{Soft-wall-limited simulations}

\begin{figure}
    \centering
    \includegraphics[width=0.9\textwidth]{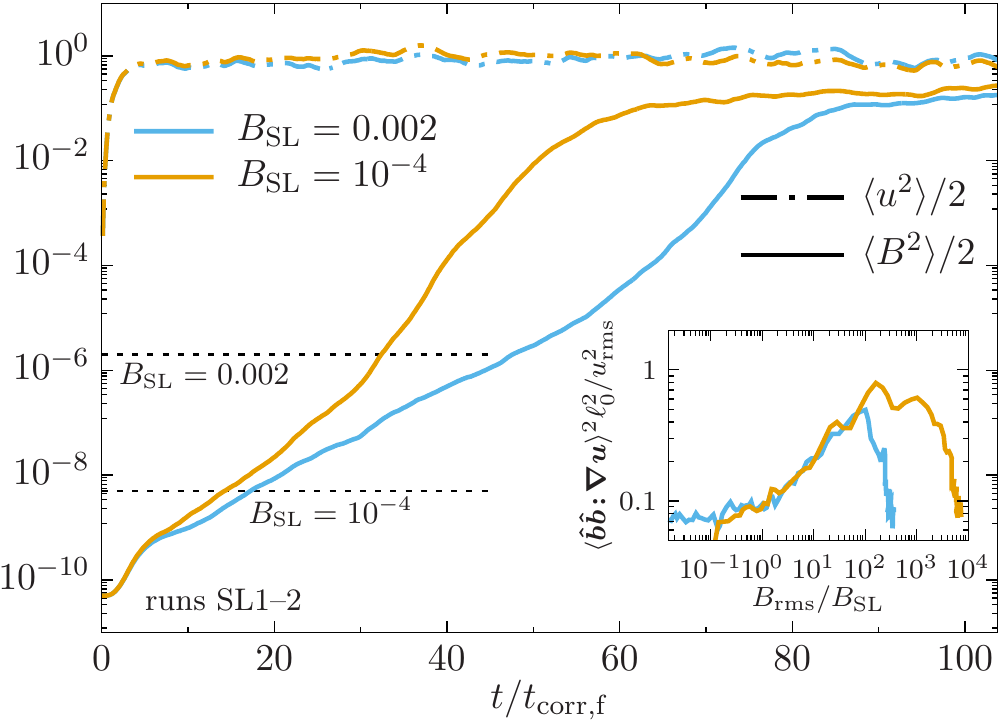}
    \caption[Energy evolution for `soft'-limited Braginskii-MHD]{Evolution of the kinetic and magnetic energies for the simulations employing soft-wall pressure-anisotropy limiters (equations ~\ref{eqn:mirror_soft} and~\ref{eqn:firehose_soft}). Dashed line indicates $B_\mathrm{SL}. $\emph{Inset:} Evolution of the squared mean parallel rate-of-strain  (${\propto}\Reeff$) as a function of $B(t)$.  C.f. figures 1 and 3 of~\citet{SchekoCowley06b}.}
    \label{fig:energy_soft}
\end{figure}

The simulations employing the soft-wall pressure-anisotropy limiters (equations~\ref{eqn:mirror_soft} and~\ref{eqn:firehose_soft}) have $B_\mathrm{SL} = 0.002$, $10^{-4}$ and an initial magnetic field strength given by $B_\mathrm{rms} = 10^{-5}$. Hyper-diffusivity is used in both simulations with $\visc_\mathrm{H}^{-1} = \eta_\mathrm{H}^{-1} = 1.8 \times 10^7$. The Braginskii viscosity is held fixed at $\visc_\mathrm{B}^{-1} = 20$. The parameters for these runs are recorded in table~\ref{tab:soft}. 

Figure~\ref{fig:energy_soft}  displays the evolution of the kinetic and magnetic energies for these simulations. In the early stage of the simulation with $B_\mathrm{SL} = 0.002$ ($t/t_\mathrm{corr} < 50$), the effective collision frequency remains small ($B_\mathrm{rms}/B_\mathrm{SL} \ll 1$) and the Braginskii viscosity is dominant, rendering it similar to the unlimited simulations.  Once $B_\mathrm{rms}/B_\mathrm{SL} \sim 1$ (at $t/t_\mathrm{corr} \approx 50$) the pressure-anisotropy begins to be regulated, and the parallel viscous stress diminishes, leading to a smaller parallel viscous scale. This in turn results in a dynamo which self-accelerates as the magnetic field gets stronger. This can be observed through the growth of $\Reeff$, which is shown in the inset of
figure~\ref{fig:energy_soft}. This behaviour appears much earlier in the simulation with lower $B_\mr{SL}$, with the self-acceleration being far more striking.
Once the pressure-anisotropy can be perfectly regulated ($t/t_\mathrm{corr} \approx 70$ for $B_\mathrm{SL}= 0.002$), the effective collision frequency needed to pin the anisotropy to the stability threshold diminishes with the field strength and the effective Reynolds number plummets (viz. equation~\ref{eqn:Re_eff}). At this point, the bulk of the plasma becomes stable and the Braginskii viscosity once again comes into play.  

\begin{figure}
    \centering
    \includegraphics[width=\textwidth]{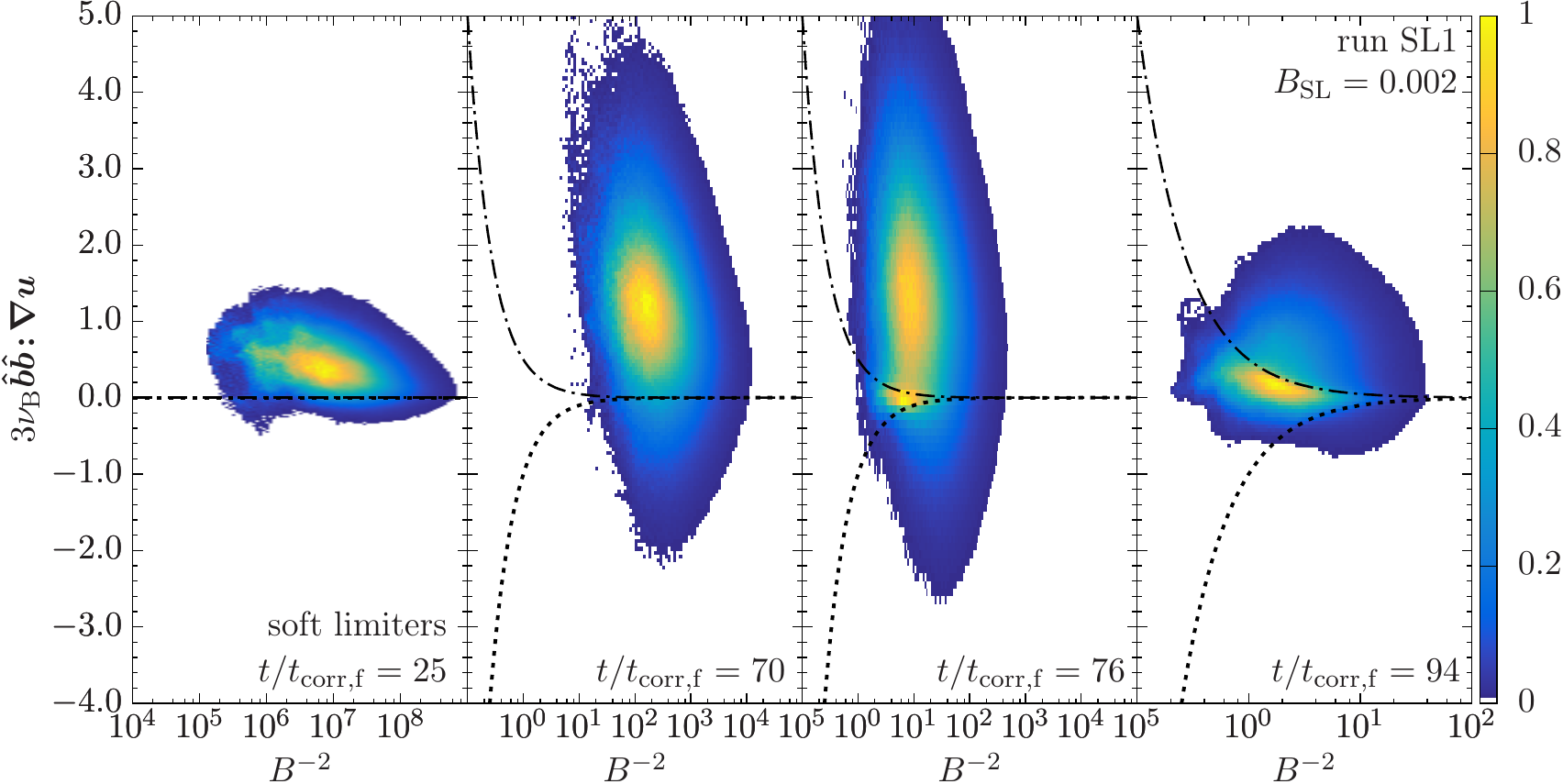}
    \caption[Brazil plots for soft-wall-limited Braginskii-MHD]{PDF of the parallel rate-of-strain  with respect to $B^{-2}$  for the simulation employing soft-wall pressure-anisotropy limiters [equations ~\eqref{eqn:mirror_soft} and~\eqref{eqn:firehose_soft}]. Dashed-dotted (dotted) lines denote mirror (firehose) stability thresholds given by equation~\ref{stability}.}
    \label{fig:aniso_soft}
\end{figure}

Figure~\ref{fig:aniso_soft} shows the PDF of the parallel rate-of-strain with respect to $B^{-2}$ for the soft-wall-limited simulation at various stages: a) the `unlimited' regime, b) the `self-acceleration' regime, c) the 'trans-saturation', and d) saturation. For the first regime, the PDF evolves very similarly to the unlimited regime (cf. figure~\ref{fig:brazil}). As the viscous stresses are reduced, the PDF is allowed to broaden and the dynamo growth rate increase.  The transition between the self-acceleration regime and saturation (in panel c) is rather abrupt.  While the system has saturated in panel d), the plasma still seems to exhibit knowledge of the stability threshold, contouring itself to the threshold much like the simulation with hard-wall limiters [figure~\ref{brag_MHD}(c)].

\begin{figure}
    \centering
    \includegraphics[width=\textwidth]{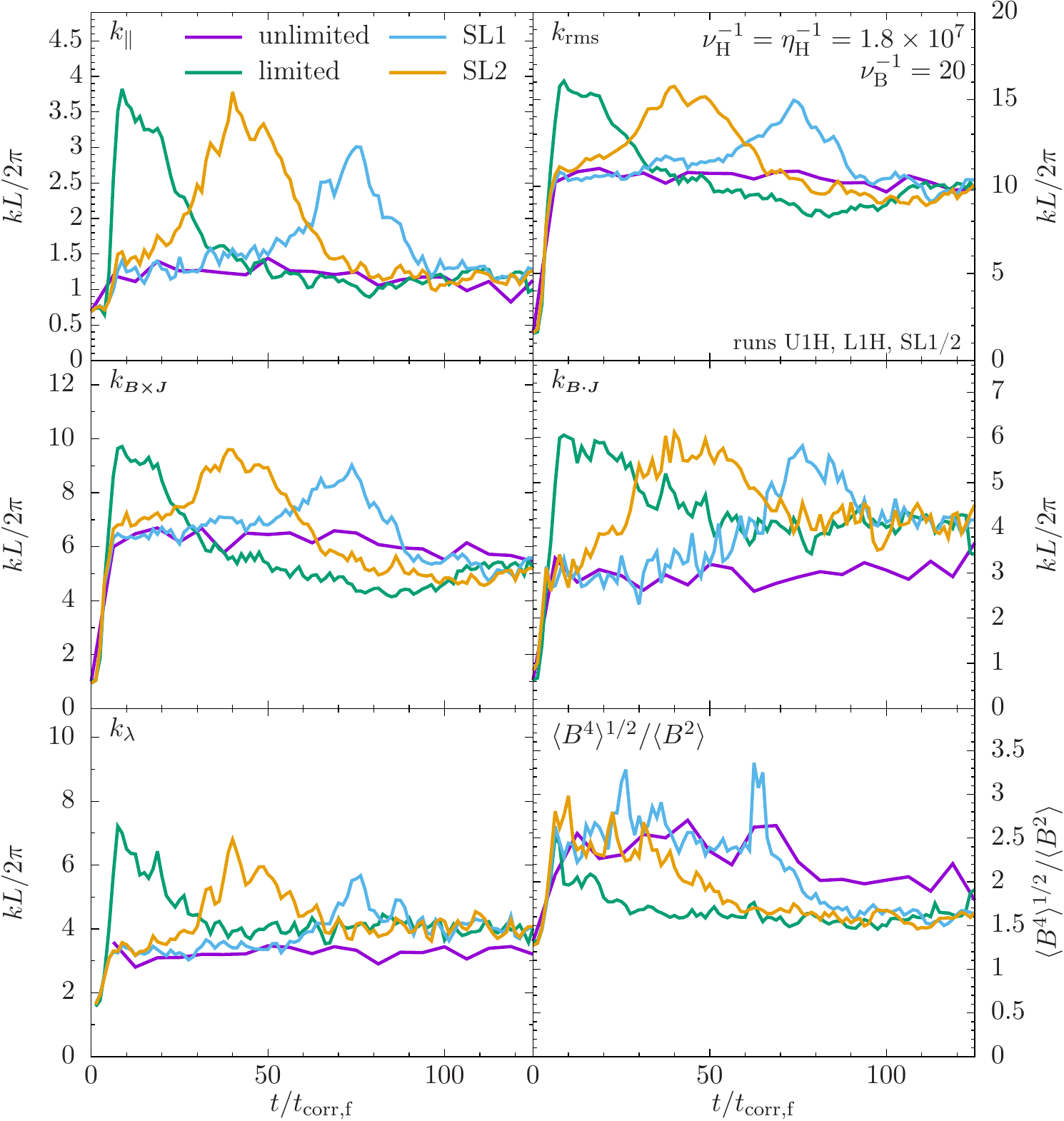}
   \caption[Characteristic wavenumbers for soft-wall-limited Braginskii-MHD dynamo]{Evolution of the characteristic wavenumbers [equations~\eqref{char-wavenumbers}] for the simulations employing soft-wall limiters with $B_\mathrm{SL} = 0.002$ (blue line) and $B_\mathrm{SL}=10^{-4}$ (yellow line), as well as for unlimited and hard-wall limited simulations employing identical diffusivities. Included are the characteristic wavenumber of the turbulence $k_\lambda \doteq \langle |\grad \bb{u}|^2\rangle^{1/2}/u_\mathrm{rms}$ and the square root of the magnetic field kurtosis $\ba{B^4}^{1/2}/B_\mathrm{rms}^2$.}
   \label{fig:wave_soft}
\end{figure}

The characteristic wavenumbers of the soft-wall-limited simulations are shown in figured~\ref{fig:wave_soft}, along with those of the unlimited and hard-wall limited Braginskii-MHD simulations employing identical diffusivities for comparison. Also included are the characteristic wavenumber of the velocity field $k_\lambda \doteq \langle |\grad \bb{u}|^2\rangle^{1/2}/u_\mathrm{rms}$ and the square root of the magnetic field kurtosis $\ba{B^4}^{1/2}/B_\mathrm{rms}^2$. The latter quantity gives a good estimate of the intermittency of the magnetic field, and whose inverse roughly gives the volume filling fraction of strong magnetic fields~\citep{Scheko_sim}. 
It is clear that the structure of both the magnetic and velocity fields transitions between an unlimited-like state to a state approaching that of hard-wall limited Braginskii-MHD.  As the explosive growth behavior is built into the model, this is perhaps not too surprising. The interesting result in this figure is that the saturated state of the Braginskii-MHD dynamo is largely insensitive to the specific details of the pressure anisotropy  regulation. In light of the results of chapter~\ref{ch:brag}, this result also may have been expected: the distinguishing feature of the unlimited Braginskii-MHD dynamo was its ability to mimic the saturated state of isotropic MHD, and thus of hard-wall limited Braginskii-MHD. Thus, if the two extreme cases of no regulation and perfect regulation exhibit similar saturated states, then the only quantities that matter in this state are the diffusivities themselves. While our choice of soft-wall limiters in this section, that based solely on the magnitude of the magnetic field, may have appeared overly simplistic, it would seem that \emph{any} choice of limiters will eventually yield the same saturated state.

\section{Discussion}

In this chapter, we proposed a set of soft-wall pressure anisotropy limiters that bridge the unlimited and hard-wall limited regimes of the Braginskii-MHD dynamo.
Figure~\ref{fig:energy_soft} should be compared to figures 1 and 3 of~\citet{SchekoCowley06b}, which qualitatively describes the same scenario. While the authors of that work only considered a zero-dimensional model, the dynamo using appropriate limiters leads to the same behavior even using full three-dimensional geometry. Thus, the soft-wall pressure-anisotropy limiters lead to magnetic field growth consistent with the explosive models proposed in~\citet{SchekoCowley06a,SchekoCowley06b}.

The soft-wall limiters proposed here were extremely simplistic, relying only on the local magnitude of the magnetic field.  In principle, this should depend on the \emph{structure} of the magnetic field as well, incorporating other features such as the fold separation length and local magnetic-field curvature. However, it was also shown in this chapter that regardless of the details of pressure anisotropy regulation, the dynamo always seems to reach a saturated state that resembles the $\mathrm{Pm} \gtrsim 1$ isotropic MHD dynamo. This puts further doubt on various proposed mechanisms that rely on the parallel Braginskii viscosity in order to generate large-scale magnetic fields, such as the idea of `field-line unfolding' put forward by~\citet{Malyshkin,KulsrudZweibel}. Clearly, some other physical effect must be present in order to yield a saturated state different than the typical isotropic MHD dynamo. One candidate could be magnetic reconnection, the description of which requires proper treatment of the electron scale dynamics. While this line of inquiry is outside the scope of this thesis, it may someday lead to a reconciliation of our current understanding of the dynamo with astrophysical observations.

\chapter{Conclusion}\label{ch:conclusion}

\section{Summary and discussion of the main results}

In this thesis we studied the fluctuation dynamo in both collisionless and weakly collisional, magnetized plasmas.  For the former, which was the subject of chapter~\ref{ch:simulation}, we used the hybrid-kinetic particle-in-cell code \textsc{Pegasus} \citep{Pegasus} to perform \emph{ab inito} simulations of the dynamo in the magnetized regime, eventually leading into the saturated state. We found in chapter~\ref{ch:simulation} that:
\begin{enumerate}

\item  The initialization and sustenance of the plasma dynamo rely heavily on the production and saturation of kinetic Larmor-scale instabilities, which  sever the adiabatic link between the thermal and magnetic pressures, effectively rendering the plasma weakly collisional by pitch-angle scattering particles. 

\item  This scattering causes much of the overall evolution of the plasma dynamo to resemble the large-Pm MHD dynamo, including an analogous `kinematic' phase during which the magnetic energy experiences steady exponential growth across several orders of magnitude. 

\item After an initial phase of rapid growth driven by these instabilities, the magnetic energy grows exponentially and exhibits a $k^{3/2}$ spectrum that peaks near the resistive scale, similar to the large-magnetic-Prandtl-number ($\Pm\gtrsim{1}$) MHD dynamo.

\item The magnetic field ultimately saturates at dynamical strengths, but without scale-by-scale equipartition with the kinetic energy. This feature, along with an anti-correlation of magnetic-field strength and field-line curvature and a gradual thinning of magnetic sheets into ribbons, resemble the saturated state of the large-$\Pm$ dynamo, the primary differences manifesting in firehose/mirror-unstable regions. 

\item Overall, it was found that the effective collisionality in saturation was sufficient to stabilize both the firehose and mirror instabilities  ($\nu_\mathrm{eff} \sim \beta \ROS$).

\end{enumerate}

For studying the fluctuation dynamo in the weakly collisional, magnetized regime, we used incompressible Braginskii-MHD simulations and analytic modeling.  In this study, which is the subject of chapter~\ref{ch:brag}, we arrive at three main conclusions:
\begin{enumerate}
    \item With hard-wall limiters on the pressure anisotropy that prevent $\rmDelta p$ from growing beyond its kinetically stable values, the Braginskii-MHD dynamo is in most respects identical to the standard high-${\rm Pm}$ MHD dynamo. This is to be expected, because the limited pressure anisotropy becomes dynamically important only once the Lorentz force does, {\it viz.}, as the dynamo starts saturating. Certain minor differences compared to isotropic MHD do indeed appear in the saturated state.
    \item When no pressure-anisotropy limiters are used (as relevant to regimes in which an effective collision frequency $\nu_{\rm eff} \gtrsim \Omegai$ would be required to keep $\rmDelta p$ near marginal stability), the Braginskii-MHD dynamo behaves quite differently to the MHD dynamo. Nearly all of these differences can be understood by noting that, in its growing phase, the structure and statistics of the magnetic field are remarkably similar to those found in the saturated state of the (high-${\rm Pm}$) MHD dynamo. This occurs because the form of the Braginskii-viscous stress is identical to that of the Lorentz force if one makes the substitution $B^2/4\upi \rightarrow \rmDelta p \sim \visc_\mr{B} \, \rmd\ln B/\rmd t$ (neglecting resistivity). 
    \item Without pressure anisotropy limiters, Braginskii MHD no longer supports a dynamo if the ratio of the Braginskii viscosity ($\visc_\mr{B}$) and the isotropic viscosity ($\visc$) is too large. This behaviour may be understood heuristically by noting that the Braginskii viscosity, by targeting only those fluid motions that stretch the magnetic field ($\ROS\ne{0}$), curbs the growth of that field while simultaneously promoting its resistive decay by allowing motions that mix the field lines. Finite isotropic viscosity moderates the mixing motions, thereby allowing the field to grow if the viscosity is sufficiently large. In the limit where the Braginskii viscosity is so strong that the outer-scale fluid motions become two-dimensionalised with respect to the magnetic-field direction, the dynamo shuts down unless the isotropic viscosity is large enough to diffuse velocity gradients into the field-perpendicular  direction, thus once again rendering the dynamics three-dimensional.
\end{enumerate}

However, while the unlimited Braginskii-MHD simulations exhibit many similarities to our hybrid-kinetic runs, in chapter~\ref{ch:structure} we showed that, unlike the unlimited Braginskii-MHD system, the appearance of a sub-parallel-viscous cascade is \emph{beneficial} to the growth of magnetic energy, rather than deleterious.  This is due to sub-parallel-viscous stretching that accompanies this cascade of energy and whose stretching rates increase at smaller scales. It was suggested that these motions may survive due to a cancellation of the parallel viscous stress by the non-gyrotropic component of the pressure tensor, which could potentially be caused by a reorientation of momentum by the gyroviscosity. Lack of scale separation, unfortunately, precluded any definitive answers.  

Finally, a novel set of pressure anisotropy limiters for Braginskii-MHD that are more appropriate for the magnetized `kinetic' regime were developed, which captures  the imperfect regulation of pressure anisotropy observed in simulations of weakly magnetized plasmas.  The efficiency of these limiters depend on local properties of the plasma and magnetic field,  resulting in the self-accelerating fluctuation dynamo that was originally proposed by~\citet{SchekoCowley06a,SchekoCowley06b}.

\section{Future work}

While the work presented in this thesis is a step forward in developing a better understanding of the fluctuation dynamo in the collisionless and weakly collisional regimes, many new questions have been raised while others go unanswered. In particular, the results of the hybrid-kinetic system do not fit squarely into either of the categories of unlimited or limited Braginskii-MHD; rather, it seems to share qualities from both regimes simultaneously. An effort should be made then to either develop a set of fluid equations with microphysical closures that can reconcile the differences between these systems, or to show that no set of fluid equations can correctly capture all the relevant pieces of  physics needed to model the plasma dynamo in a collisionless environment.  
Bear in mind that the parallel Braginskii viscosity is just one addition to the momentum equation that can arise when considering the dynamo in the weakly collisional, magnetized regime.  At the early stage of the magnetized plasma dynamo when the ion gyrofrequency is only somewhat larger than the ion collision frequency ($\Omegai \gtrsim \nu_\mathrm{i}$), the other components of the pressure tensor, such as the gyro-viscous contribution, may play an important role. Apart from the viscosity, heat fluxes in the temperature equation may also be important, as well as the effects of collisionless Landau damping and reconnection. Finally, any number of kinetic effects, such as the suppression of Landau damping caused by stochastic echoes~\citep{Scheko_phasemixing,Meyrand-echoes}, may possibly play a role. A complete theory of the dynamo in the weakly collisional regime will then hopefully take into account all these effects and specify their individual role, if any, in determining the details of the dynamo.   A promising  starting point would be to implement physics beyond the parallel Braginskii viscosity, such as implementing the full Braginskii viscosity, as well as incorporating higher fluid moments that include field-oriented transport of heat.

Clearly, efforts should focus on capturing the $\nueff\sim\Omegai\rightarrow{k}_\parallel\vthi\rightarrow{S}\betai$ transitions in the magnetized `fluid' regime before saturation occurs at $\betai{M}^2\sim{1}$. Sorting this out is all the more important in the context of determining the effective $\mr{Re}$ of the turbulent ICM \citep[e.g.,][]{Fabian05,ZuHone18}, which also plays a crucial role in viscous heating \citep[e.g.,][]{Lyutikov07,Kunz_2011a,Zweibel18} and the integrity of cold fronts \citep[e.g.,][]{ZuHone15} and rising bubbles \citep[e.g.,][]{Fabian03}.

One major shortcoming of the work presented in this thesis is the simplification of the electron dynamics in all systems. Indeed, in collisionless and weakly collisional plasmas the resistive scale is determined by either the electron gyroradius or the electron skin depth. If the former scenario pertains, then this implies that the resistive scale constantly shrinks as the dynamo progresses, and that the ion gyroradius is always greater than the fold separation! Additionally, the lack of proper electron dynamics hinders our ability to study the phenomenon of magnetic reconnection as the field strength increases and $v_\mathrm{A}\sim u_\mathrm{rms}$. This may have profound effects on the saturation properties of the dynamo, which has been given a semi-quantitative treatment in an upcoming review of MHD turbulence by~\citet{SchekoMHD}.  One avenue of inquiry would then be to perform simulations of the fluctuation dynamo in a fully kinetic system.  This could potentially be performed using reduced mass ratios, the extreme case consisting of a pure pair plasma.  One could then determine how the plasma creates folded magnetic fields with a self-consistently determined resistive scale, and whether or not reconnection substantially changes the statistics of the saturated state.

Another topic not mentioned in this thesis is the interplay between the mean-field dynamo and the fluctuation dynamo. While the e-folding time of the mean-field dynamo is controlled by some large-scale process, the growth rate of magnetic fluctuations $\ea{|B^2|} - |\ea{\bb{B}}|^2 $ as a result of the fluctuating dynamo is set by the fastest stretching motions, which in a turbulent medium are set by the smallest-scale eddies (see \S\ref{ch1:flucdyn}). As the growth rate of the latter can be much greater than the former, the fluctuating component of the magnetic field may attain dynamically important magnitudes before the mean field~\citep{Kulsrud1992}. This can lead to saturation of the dynamo before any appreciable increase in the net flux is experience, a scenario known as `catastrophic $\alpha$-quenching'~\citep{Gruzinov94}.  It has been observed, at least in some types of mean-field dynamos, that both dynamos can happily coexist, leading to growth of the net flux~\citep{Jono_shearcurrent}. Generally however, this problem is still unsettled. One can then study the role of kinetic effects and anisotropic plasma viscosity on how the fluctuation dynamo aids or deters the mean-field dynamo, and whether the mean-field dynamo can avoid catastrophic $\alpha$-quenching in a weakly collisional, turbulent environment.

Finally, we have not been able to assess definitively in our simulations whether anisotropic viscosity obviates the tendency for the small-scale dynamo to saturate with the majority of its magnetic energy residing near resistive scales. On the one hand, it is interesting that the magnetic spectra found in the unlimited Braginskii and hybrid-kinetic dynamos show little tendency to concentrate power on scales significantly smaller than the viscous scale, and, in the case of the unlimited dynamo, even less tendency to evolve in time in going from the kinematic stage to the saturation (see figure \ref{fig:unlim_spec}). On the other hand, because computational expense limits the maximum achievable scale separation between the viscous and resistive scales in our simulations, we cannot yet establish whether the peak of the spectrum is independent of ${\rm Rm}$, or whether the conjecture by \citet{Malyshkin}---that the interchange motions that are undamped by Braginskii viscosity unwrap the folded magnetic fields and thus promote their inverse cascade to larger scales---can be realized. Future numerical work, both fluid and kinetic, should maximize scale separation with the goal of definitively evaluating the ability of the fluctuation dynamo to generate saturated magnetic fields with large-scale coherence in weakly collisional plasmas such as the intracluster medium.

\appendix %
\chapter{Discrete Particle Noise}\label{ch:noise}
In this appendix, we calculate the kinetic energy spectrum due to discrete particle noise.  For any point in space with no net velocity, $N^{-1}_\mathrm{ppc}\sum_i^{N_\mathrm{ppc}} \bb{v}_i = 0 $ while  $N^{-1}_\mathrm{ppc}\sum_i^{N_\mathrm{ppc}}  v_{x,i}^2 = v_\mathrm{thi}^2/2 $ for any spatial index and $N_\mathrm{cell} \rightarrow \infty$, where $\bb{v}_i$ is the $i$th particle in the cell.  Thus, the standard error on the mean is $v_\mathrm{thi}/N_\mathrm{cell}^{1/2}$.

We now assume a Gaussian model for the discrete particle noise, with $u$ the velocity density and $d = 3$,
\begin{equation}
\ea{u_i u_j} = \frac{v_\mathrm{thi}^2}{2 N_\mathrm{ppc}}\frac{N_\mathrm{cell}^d}{L^d}\delta_{ij} \delta (\bb{x}_i - \bb{x}_j).
\end{equation}
What about the energy density spectrum? The total box-averaged kinetic energy is given by
\begin{align}
K &=  \frac{1}{2N_\mathrm{cell}^{2d}}\sum_\bb{k}\ea{\hat{u}_i^* \hat{u}_i} \nonumber
\\ &= \frac{1}{2N_\mathrm{cell}^{2d}}\sum_\bb{k}\int_0^L \od^3\bb{x}_i \int_0^L \od^3\bb{x}_j \, \ea{u_i u_i} e^{i\bb{k}\bcdot(\bb{x}_i - \bb{x}_j)}\nonumber
\\ &=  \frac{3v_\mathrm{thi}^2}{4 N_\mathrm{ppc}} \frac{1}{N_\mathrm{cell}^{d}L^{d}}\sum_\bb{k}\int_0^L \od^3\bb{x}_i  \nonumber
\\ &= \sum_\bb{k} \frac{3v_\mathrm{thi}^2}{4N_\mathrm{ppc}} \frac{1}{N_\mathrm{cell}^d}\nonumber
\\ &\rightarrow  \int \od k \,k^2 \frac{3v_\mathrm{thi}^2}{4N_\mathrm{ppc}} \frac{L^3}{N_\mathrm{cell}^d}\frac{4 \upi}{(2 \upi)^3},
\end{align}
where we gave passed to the continuum in the last step by making the substitution $\sum_\bb{k} \rightarrow (L/2\upi)^3\int \od^3 \bb{k}$.
The kinetic energy spectrum of the noise is then given as
\begin{align}
E(k)  &=  k^2 \frac{3v_\mathrm{thi}^2}{4N_\mathrm{ppc}} \frac{L^3}{N_\mathrm{cell}^d}\frac{4 \upi}{(2 \upi)^3}\nonumber
\\  &=  k^2 \frac{3v_\mathrm{thi}^2 L^3}{8 \upi^2 N_\mathrm{total}}.
\end{align}
This is what is observed in simulation.

\chapter{Energy Acceptance in a Collisionless Plasma}
\label{ap:forcing}

In this appendix we calculate the linear response of a collisionless plasma to a stochastic white-noise, incompressible and non-helical forcing $\vforce$ with the prescribed statistics
\begin{equation}
\ea{\force_i} = 0, \quad\quad \ea{\force_{i}(t,\bb{k})\force_{j}^*(t,\bb{k})} = \chi(k) \delta(t-t')(\delta_{ij}-\hat{k}_i\hat{k}_j).
\end{equation}
Here, $\chi(k)$ is the forcing correlator and $\hat{\bb{k}} = \bb{k}/k$. Noting that
\begin{align}
\int \od t \int \od t' \, \rme^{\imag\omega t - \imag \omega't' } \delta(t-t') =  \int \od t \, \rme^{ \imag t(\omega - \omega') } = 2\upi\delta(\omega ' - \omega),
\end{align}
we can Fourier transform the time coordinate to give
\begin{equation}
\ea{\force_{i}(\omega,\bb{k})\force_{j}^*(\omega',\bb{k})} = 2\upi \chi(k) \delta(\omega-\omega')(\delta_{ij}-\hat{k}_i\hat{k}_j).
\end{equation}

\section{Unmagnetized case ($\bb{B} = \bb{0}$)}

In this section we consider the electrostatic case with $\bb{B}= \bb{0}$, which was performed in the appendix of~\citet{Rincon_2016}.
We consider the hybrid-kinetic system of equations as described in~\S\ref{int:hybridkin} with $\bb{B}= 0$: 
\begin{subequations}
 \begin{gather}
  \frac{\partial f_\mathrm{i}}{\partial t} + \bb{v}\bcdot \grad f_\mathrm{i} + \frac{e}{m_\mathrm{i}} \bb{E} \bcdot \frac{\partial f_\mathrm{i} }{\partial \bb{v}} = 0, \\
  \bb{E}  = -\frac{T_\mathrm{e} \grad n}{e n}. 
   \end{gather}
 \end{subequations}
We first linearize these equations around a stationary background Maxwellian distribution, 
\begin{equation}
f_\mathrm{i0} = \frac{n_0}{v_\mr{thi}^3\upi^{3/2}}\rme^{-v^2/v_\mr{thi}^2},
\end{equation}
and Fourier transform, assuming $\bb{k} = k_z \ez$ without loss of generality.  This results in the linearized Ohm's law 
\begin{equation}
\bb{E}_1 = - \imag k_z \ez\frac{T_\ee  n_1 }{en_0 } = -\imag \bb{k} \frac{ T_\ee}{e }\frac{\bb{k}\bcdot \bb{u}_1}{\omega},
\end{equation}
where the continuity equation was used to obtain the last expression. 
With $\tau \doteq T_\ee / T_\ii$, the Vlasov equation becomes
\begin{gather}
(-\imag\omega + \imag k_zv_z)f_1 + \imag \tau f_\mathrm{i0}k_zv_z \frac{k_zu_{1z}}{\omega}-\frac{2f_\mathrm{i0}}{m_\ii v_\mr{thi}^2}\vforce\cdot \bb{v}=0,
\end{gather}
or
\begin{align}
f_1 &=   \tau \frac{k_z u_{1z}}{\omega}  \frac{f_\mathrm{i0}k_zv_z}{(\omega -  k_zv_z)}+ \frac{2\imag f_\mathrm{i0}}{m_\ii v_\mr{thi}^2}\frac{\vforce\cdot \bb{v}}{\omega -  k_zv_z} \nonumber
 \\ &=  - \tau \frac{k_z u_{1z}}{\omega} f_\mathrm{i0}\left[1 - \frac{\omega}{(\omega -  k_zv_z)}\right] + \frac{2 \imag f_\mathrm{i0}}{m_\ii v_\mr{thi}^2}\frac{\vforce\cdot \bb{v}}{\omega -  k_zv_z}.
\end{align}
Taking the first moment leads to
\begin{align}
\bb{u}_1 &=   \tau k_z u_{1z} \int \od \bb{v} \, f_\mathrm{i0}\frac{\bb{v}}{\omega -  k_z v_z}+ \int \od \bb{v}\frac{2\imag f_\mathrm{i0}}{m_\ii  v_\mr{thi}^2}\frac{\bb{v}(\vforce\cdot \bb{v})}{\omega -  k_z v_z} \nonumber
\\ &=  - \tau  u_{1z} \int \od \bb{v}\, f_\mathrm{i0}\frac{\bb{v}}{v_z - \omega/k_zv_\mr{thi}}+ \int \od \bb{v}\frac{2\imag f_\mathrm{i0}}{m_\ii  v_\mr{thi}^2}\frac{\bb{v}(\vforce\cdot \bb{v})}{\omega -  k_zv_z},
\end{align}
where $\bb{v}' \doteq \bb{v}/v_\mathrm{thi}$.
This can be rewritten in index notation as
\begin{equation}
M_i^ju_{1j} =   \int \od \bb{v} \, \frac{2\imag f_\mathrm{i0}}{m_\ii  v_\mr{thi}^2} \frac{v_i v_j\hat{f}^j }{\omega -  v_j k^j}
\end{equation}
where $M_{i}^j \doteq \delta_{ij} + \tau [1+\zeta Z(\zeta) ] \hat{k}_i\hat{k}_j$ is a diagonal tensor, $\zeta \doteq \omega / k_z v_\mathrm{thi}$ and $Z(\zeta)$ is the plasma dispersion function.
Defining $N_i^j = (M_i^j)^{-1}$, we have
\begin{align}
\ea{u_{1i}(t',\bb{k}) u_{1j}^*(t,\bb{k})}  &= \frac{1}{\upi^2}N^i_pN^j_q\int \od \omega \od \omega' \rme^{\imag\omega t - \imag \omega' t'}\int \frac{\od \bb{v} \od \bb{v}'}{m^2 v_\mr{thi}^4} \frac{ v_pv_m v'_qv'_n\ea{\force^m \force^{*n}}}{(\omega- v_l k^l)(\omega'- v'_l k^l)}f_\mathrm{i0}(\bb{v})f_\mathrm{i0}(\bb{v}'),
\end{align}
or, after performing the ensemble average,
\begin{align}
&\ea{u_{1i}(t',\bb{k}) u_{1j}^*(t,\bb{k})}  \nonumber
\\  &\qquad =   \frac{2}{\upi}\chi(k) N^i_pN^j_q(\delta_{mn}-\hat{k}_m\hat{k}_n)\int \od \omega \, \rme^{-\imag \omega( t- t')}\int \frac{\od \bb{v} \od \bb{v}'}{m^2 v_\mr{thi}^4} \frac{ v_pv_m v'_qv'_n f_\mathrm{i0}(\bb{v})f_\mathrm{i0}(\bb{v}')}{(\omega- v_l k^l)(\omega- v'_l k^l)}.
\end{align}
In order for the velocity integration to be non-zero, we require $p=m$ and $q=n$.  The projection operator  $\delta_{mn} - \hat{k}_m \hat{k}_n$ then ensures that the $\ez\ez$ component of  $\ea{u_{1i}(t',\bb{k}) u_{1j}^*(t,\bb{k})} $ is zero, and thus we can make the substitution $N^{i}_p \rightarrow \delta^i_p$ and $N^{j}_q \rightarrow \delta^j_q$. Performing the velocity integration over the $x$ and $y$ coordinates and switching to dimensionless integration variables leads to
\begin{align}
\ea{u_{1i}(t,\bb{k}) u_{1j}^*(t',\bb{k})} &=   \frac{\chi(k)}{2\upi k^2 m_\mr{i}^2 v_\mr{thi}^2} \left(\delta_{ij}-k_ik_j/k^2\right)\int \od \omega \, \rme^{-\imag\omega( t- t')}|Z(\zeta)|^2.
\end{align}
This agrees with the result in~\citet{Rincon_2016} up to a factor of $(2m_\mr{i}v_\mr{thi})^{-2}$, which is needed for dimensional correctness. Importantly, as the temperature of the plasma, and thus $v_\mathrm{thi}$, increases, the kinetic energy of the system \emph{decreases}. This is because the dominant dissipative process in a collisionless unmagnetized plasma is phase mixing, whose mixing frequency is proportional to $kv_\mathrm{thi}$. A hotter plasma features faster free-streaming particles and is therefore more viscous.

\begin{figure}
\centering
\includegraphics[scale=1.05]{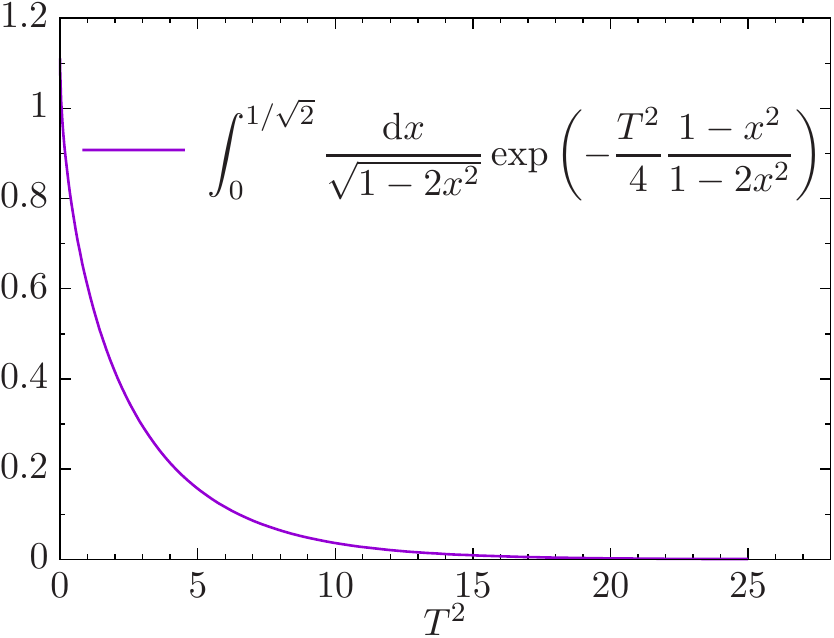}
\caption{\label{rincon}Numerical solution of $\int \od \zeta \rme^{-\imag \zeta T} |Z(\zeta)^2|$.}
\end{figure}

To get the saturated kinetic energy in the long-time limit, we look at the one-point correlation at $t=t'$. We need to calculate
\begin{align}
\frac{1}{k v_\mr{thi}}\int_{-\infty}^\infty \od \omega\, |Z(\zeta)|^2 &=\upi  \int_{-\infty}^{\infty} \od u\,  \rme^{-2u^2}+ 4\int_{-\infty}^{\infty} \od u\,\rme^{-2u^2}\int_{0}^u \int_0^u \rme^{x^2+ y^2} \od x \od y.
\end{align}
Going into cylindrical coordinates for the second term,
\begin{align}
\int_{-\infty}^{\infty} \od u \,\rme^{-2u^2}\int_{0}^u \int_0^u \rme^{x^2+ y^2} \od x \od y &= 2 \int_{-\infty}^{\infty}\od u \,\rme^{-2u^2}\int_0^{\upi/4} \int_{0}^{u/\cos\theta}  \rme^{r^2} r \od r  \od \theta  \nonumber
\\ \nonumber &= \int_{-\infty}^\infty\od u\int_0^{\upi/4} \left(\rme^{-u^2(2-\sec^2 \theta) }-\rme^{-2u^2}\right)\od \theta
\\ \nonumber &= \sqrt{\upi}\int_0^{\upi/4} \left(\frac{\cos \theta}{\sqrt{2\cos^2 \theta- 1}}-\sqrt{\frac{1}{2}}\right)\od \theta
\\ &= \frac{\upi^{3/2}}{4\sqrt{2}}.
\end{align}
So  $\int_{-\infty}^\infty \od \omega\, |Z(\zeta)|^2 =k v_\mr{thi} \sqrt{2}\upi^{3/2}$ and
\begin{equation}\label{eq:app:nobforce}
\ea{ u_{1i}(t,\bb{k})u_{1j}^*(t,\bb{k}) } = \sqrt{\frac{\upi}{2}}\frac{\chi(k)}{k m_\mr{i}^2 v_\mr{thi}}(\delta_{ij}-\hat{k}_i\hat{k}_j).
\end{equation}
This is what would be expected if one were to use a Landau-fluid closure~\citep{Hammett-Perkins}.
More generally, with $T = k v_\mr{thi}(t-t')$,
\begin{align}
\frac{1}{k v_\mr{thi}}\int_{-\infty}^\infty &\rme^{-\imag\zeta T}\od \omega\, |Z(\zeta)|^2   \nonumber
\\ &=\upi  \int_{-\infty}^{\infty} \od u\,  \rme^{-2u^2- \imag T \zeta}+ 8\int_{-\infty}^{\infty} \od u\,\rme^{-2u^2 - \imag uT}\int_0^{\upi/4} \int_{0}^{u/\cos\theta}  \rme^{r^2} r \od r  \od \theta \nonumber
\\ &= \frac{\upi^{3/2}\rme^{-T^2/8}}{\sqrt{2}} + 4\int_0^{\upi/4}\int_{-\infty}^{\infty} \od u\,\left(\rme^{-u^2(2-\sec^2\theta) -  \imag uT} - \rme^{-2u^2 - \imag uT}\right) \nonumber
\\ &= 4\sqrt{\upi}\int_0^{\upi/4}\od \theta\,  \frac{\cos \theta \rme^{-T^2/4(2-\sec^2\theta)}}{\sqrt{1 - 2\sin^2 \theta}} \nonumber
\\ &= 4\sqrt{\upi}\rme^{-T^2/8}\int_0^{1/\sqrt{2}} \frac{\od x}{\sqrt{1 - 2x^2 }}\exp \left(-\frac{T^2}{8}\frac{ 1}{1-2x^2}\right) \label{rinconInt}.
\end{align}
This is plotted in figure~\ref{rincon}.

\section{Magnetized case using the drift-kinetic equation}

\begin{figure}
\centering
\includegraphics[scale=1.15]{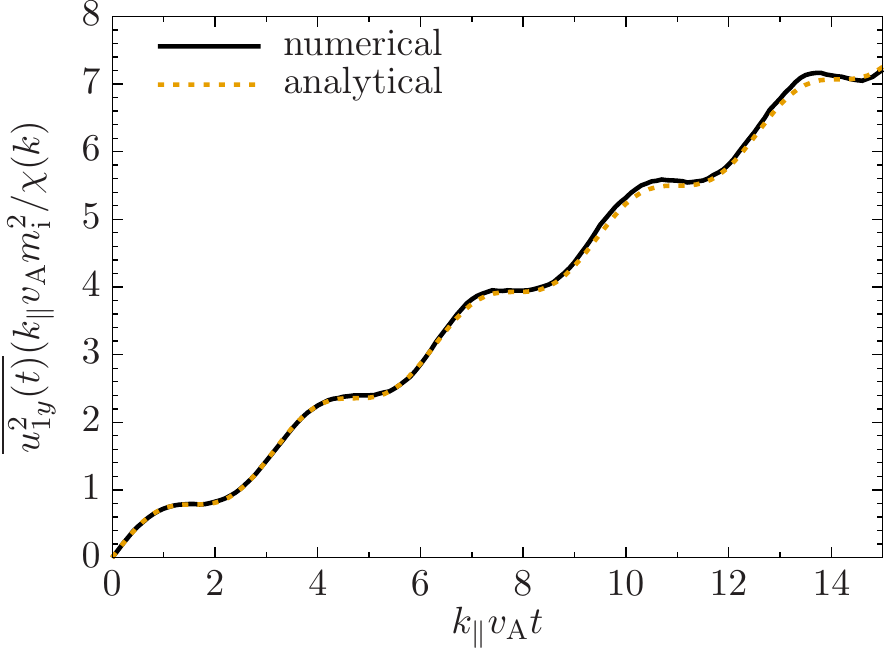}
\caption[Numerical and analytical solutions of equation~\eqref{eq:app_forcey}]{\label{DKE_forcey} Numerical solution  of equation~\eqref{eq:app_forcey} as well as the analytical prediction given by equation~\ref{eq:forcey_an} at $t'=t$.}
\end{figure}

For the magnetized case, we begin with the drift-kinetic equation~\citep[DKE, see \S\ref{sec:DKEder} and ][]{KulsrudMHD}:
\begin{subequations}
\begin{gather}
\frac{\oD f_0}{\oD t} + \frac{\oD\,\ln B}{\oD t} \frac{w_\perp}{2}\frac{\partial f_0}{\partial w_\perp} + \left( \frac{e E_\parallel}{m_\mathrm{i}}  + \frac{\force_\parallel}{m_\mathrm{i}} + \frac{w_\perp^2}{2}\grad\bcdot\bb{\hat{b}} - \frac{\oD \bb{u}_{\perp}}{\oD t}\bcdot\bb{\hat{b}}\right)\frac{\partial f_0}{\partial v_\parallel}= 0, \\
\left(\frac{\partial}{\partial t}  + \bb{u}\bcdot \grad\right) n = -n \grad \bcdot \bb{u}, \\
m_\mathrm{i}n\left(\frac{\partial}{\partial t}  + \bb{u}\bcdot \grad\right) \bb{u} =- \frac{T_\mathrm{e}\grad n}{en} - \grad \bcdot [p_\perp (\unitDyadic -\eb\eb) + p_\parallel \eb \eb] + \frac{1}{ c}\bb{J}\btimes \bb{B}+ n\vforce, \\
  \grad \btimes \bb{B} = \frac{4\upi }{c}\bb{J},\\
  \frac{\partial \bb{B}}{\partial t} = -c\grad \btimes \bb{E}, \\ 
  \bb{E} +\frac{1}{c}\bb{u}\btimes \bb{B} =   -\frac{T_\mathrm{e} \grad n}{e n},
\end{gather}
\end{subequations}
where we now have included a random Gaussian body force $\vforce$.
We now linearize around $\bb{B} = B_0 \bb{\hat{z}}$ and a bi-Maxwellian distribution
\begin{equation}
f_0(v_\parallel,w_\perp) = \frac{2n}{\upi^{1/2}v_{\mr{thi}\parallel} v_{\mr{thi}\perp}^2}\exp \left(- \frac{ w_\perp^2}{v_{\mr{thi}\perp}^2} - \frac{v_\parallel^2}{v_{\mr{thi}\parallel}^2}\right).
\end{equation}
After Fourier transforming in space and time, our linearized equations are
\begin{subequations}\label{eq:force:lin}
\begin{align}
\omega  n_1   &= n_0 \bb{k}\bcdot \bb{u}_1, \\
  \omega m n_0 \bb{u}_1 &= \bb{k} n_1 T_e  +\bb{k}p_{1\perp} + k_\parallel(  p_{1\parallel}-p_{1\perp})\bb{\hat{z}} + (  p_\parallel-p_\perp)\left[k_\parallel \bb{\hat{b}}_1 + (\bb{k}\bcdot\bb{\hat{b}}_1) \bb{\hat{z}}\right]  \nonumber\\
 &\qquad - \frac{ B_0}{4\upi}[(\bb{k}\btimes \bb{b}_1) \btimes \bb{\hat{z}}]  + \imag n_0\vforce,  \\
  \bb{b}_1 &= - \frac{B_0}{\omega}[\bb{u}_1 k_\parallel - \bb{\hat{z}}(\bb{k}\bcdot\bb{u}_1)],\\
    E_{1\parallel} &= -\imag k_\parallel \frac{T_e n_1}{e n_0}, \\
  f_1 &= \frac{b_{1\parallel}}{B_0}\frac{w_\perp^2}{v_{\mr{thi}\perp}^2} f_0  -  \frac{2}{v_{\mr{thi}\parallel}^2}\frac{v_\parallel f_0}{ k_\parallel v_\parallel -\omega  }\left(k_\parallel \frac{T_\mr{e} n_1 }{m_\mr{i}n_0} + \frac{\imag \force_\parallel}{m_\mr{i}} + k_\parallel\frac{b_{1\parallel}}{B_0} \frac{w_\perp^2}{2}\right).
\end{align}
\end{subequations}
Let $\bb{b}_1/B_0 \rightarrow \bb{b}'_1 $. Then
\begin{subequations}
\begin{align}
  \bb{b}'_1 &=   - \frac{1}{\omega}[k_\parallel\bb{u}_1  - (\bb{k}\bcdot\bb{u}_1)\bb{\hat{z}}],\\ 
  \omega m n_0 \bb{u}_1 &= \bb{k} n_1 T_e  + \bb{k}p_{1\perp} + k_\parallel(  p_{1\parallel}-p_{1\perp})\bb{\hat{z}}
\nonumber \\ &\qquad + (  p_\parallel-p_\perp)\left(k_\parallel \bb{b}'_1  - 2 k_\parallel b'_{1\parallel} \bb{\hat{z}}\right) - \frac{ B_0^2}{4\upi}(k_\parallel \bb{b}'_1 - b'_{1\parallel}\bb{k})  + \imag n_0\vforce.
\end{align}
\end{subequations}
We need $p_{1\perp}$ and $p_{1\parallel}$. Define $\zeta \doteq \omega / k_\parallel v_{\mr{thi}\parallel}$. The following identities will be useful:
\begin{subequations}
\begin{align}
\frac{1}{\sqrt{\upi}}\int \od v \frac{v \rme^{-v^2}}{v -\zeta }   &= 1 + \zeta Z(\zeta), \\
\frac{1}{\sqrt{\upi}}\int \od v \frac{v^2 \rme^{-v^2}}{v -\zeta }   &= \zeta + \zeta^2 Z(\zeta), \\
\frac{1}{\sqrt{\upi}}\int \od v \frac{v^3 \rme^{-v^2}}{v -\zeta }  &= \frac{1}{2} +\zeta^2 + \zeta^3 Z(\zeta).
\end{align}
\end{subequations}
Defining $g_0 \doteq \int \od w_\perp\, w_\perp f_0$, the second perpendicular moment of the DKE gives
\begin{gather}
\int \od w_\perp \, w_\perp^3 f_1 = 2 v_{\mr{thi}\perp}^2 \frac{b_{1\parallel}}{B_0}  g_0  -  \frac{2v_{\mr{thi}\perp}^2}{v_{\mr{thi}\parallel}^2}\frac{v_\parallel f_0}{ k_\parallel v_\parallel -\omega  }\left(k_\parallel \frac{T_\mr{e} n_1 }{m_\mr{i}n_0} + \frac{\imag \force_\parallel}{m_\mr{i}} + k_\parallel v_{\mr{thi}\perp}^2\frac{b_{1\parallel}}{B_0}\right).
\end{gather}
Then, with $v' = v_\parallel / v_{\mr{thi}\parallel}$,
\begin{align}
\frac{p_{1\perp}}{p_\perp}&= 2 \frac{b_{1\parallel}}{B_0}  -  \frac{2}{k_\parallel v_{\mr{thi}\parallel}^2}\left(k_\parallel \frac{T_\mr{e} n_1 }{m_\mr{i}n_0} + \frac{\imag \force_\parallel}{m_\mr{i}} + k_\parallel v_{\mr{thi}\perp}^2\frac{b_{1\parallel}}{B_0}\right)\left(\frac{1}{\sqrt{\upi}}\int \od v' \frac{v' \rme^{-v'^2}}{v' - \zeta}\right) \nonumber
\\ &= 2 \frac{b_{1\parallel}}{B_0}  - [1 + \zeta Z(\zeta)]\left(\frac{T_e}{T_\parallel}\frac{n_1}{n_0} + \frac{\imag \force_\parallel}{k_\parallel T_\parallel} + 2\frac{T_\perp}{T_\parallel}\frac{b_{1\parallel}}{B_0}\right). \label{ap:eq:pprp}
\end{align}
For the parallel pressure, we will need
\begin{align*}
\int \od w_\perp \, w_\perp f_1 &=  \frac{b_{1\parallel}}{B_0}  g_0  -  \frac{2}{v_{\mr{thi}\parallel}^2}\frac{v_\parallel f_0}{ k_\parallel v_\parallel -\omega  }\left(k_\parallel \frac{T_\mr{e} n_1 }{m_\mr{i}n_0} + \frac{\imag \force_\parallel}{m_\mr{i}} + \frac{1}{2}k_\parallel v_{\mr{thi}\perp}^2\frac{b_{1\parallel}}{B_0}\right),
\end{align*}
leading to
\begin{align}
\frac{p_{1\parallel}}{p_\parallel} &=  \frac{b_{1\parallel}}{B_0}   -  \int \od v_\parallel \frac{1}{v_{\mr{thi}\parallel}^2}\frac{v_\parallel^3 f_0}{ k_\parallel v_\parallel -\omega  }\left(k_\parallel \frac{T_\mr{e} n_1 }{m_\mr{i} v_{\mr{thi}\parallel}^2 n_0} + \frac{\imag \force_\parallel}{m_\mr{i}v_{\mr{thi}\parallel}^2} + \frac{1}{2}k_\parallel \frac{v_{\mr{thi}\perp}^2}{v_{\mr{thi}\parallel}^2}\frac{b_{1\parallel}}{B_0}\right) \nonumber
\\ &=  \frac{b_{1\parallel}}{B_0}   -  \left[1 + 2\zeta^2 +2\zeta^3 Z(\zeta)\right]\left( \frac{T_e}{T_\parallel} \frac{ n_1 }{n_0} + \frac{\imag \force_\parallel}{k_\parallel T_\parallel} +  \frac{T_\perp}{T_\parallel}\frac{b_{1\parallel}}{B_0}\right). \label{ap:eq:pprl}
\end{align}
From here on we assume background pressure isotropy for simplicity. 
Thus
\begin{align}
\frac{p_{1\perp} - p_{1\parallel}}{p} = -\zeta Z(\zeta )\frac{b_{1\parallel}}{B_0} - [ 2\zeta^2 +2\zeta^3 + \zeta Z(\zeta)]\left(\tau\frac{n_1}{n_0} + \frac{\imag \force_\parallel}{k_\parallel T_\parallel} + \frac{b_{1\parallel}}{B_0}\right). 
\end{align}
We now have all the equations needed to solve for the linear response of $\bb{u}$ to the forcing.  Take $\bb{k} = k_x \bb{\hat{x}} + k_\parallel \bb{\hat{z}}=k (\sin \theta\bb{\hat{x}} + \cos \theta \bb{\hat{z}})$, leading to
\begin{subequations}
\begin{align}
  \omega m n_0 u_{1x} &=  k_x \frac{k_x u_{1x} + k_\parallel u_{1\parallel}}{\omega} n_0T_e   + \frac{ k^2 B_0^2}{4\upi}\frac{u_{1x}}{\omega}  - \imag n_0 \force_x \nonumber
  \\   &\quad + k_x p\Bigg[\frac{2 k_xu_{1x}}{\omega } 
 - [1 + \zeta Z(\zeta)]\Bigg(\tau\frac{k_x u_{1x} + k_\parallel u_{1\parallel} }{\omega} + \frac{\imag \force_\parallel}{k_\parallel T_\parallel} + 2\frac{k_xu_{1x}}{\omega}\Bigg)\Bigg] ,   \\ 
 \omega m n_0 u_{1\parallel} &=  k_\parallel  \frac{k_x u_{1x} + k_\parallel u_{1\parallel}}{\omega}n_0 T_e   - \imag n_0 \force_\parallel  \nonumber 
\\  &\quad+ k_\parallel  p \Bigg[\frac{k_x u_{1x} }{\omega}  -  [1 +  2\zeta^2 +2\zeta^3 Z(\zeta)]\left( \tau \frac{k_x u_{1x} + k_\parallel u_{1\parallel}  }{\omega} + \frac{\imag \force_\parallel}{k_\parallel T_\parallel} + \frac{k_x u_{1x}}{\omega}\right)\Bigg],   \\ 
 \omega m n_0 u_{1y} &= \frac{ B_0^2}{4\upi} \frac{k_\parallel^2 u_{1y}}{\omega}   -\imag n_0\force_y.
\end{align}
\end{subequations}
The $y$ component immediately gives
\begin{gather}\label{eq:app_forcey}
 u_{1y} =  \frac{\imag\omega}{\omega^2 - k_\parallel^2 v_\mr{A}^2}\frac{\force_y}{m},
\end{gather}
whose solution exhibits secular growth  due to the forcing being in resonance with the shear Alfv\'en wave.
To see this, we obtain the solution by calculating the Green's function,
\begin{align}
G(t) = \int_{-\infty}^\infty  \od \omega \frac{\imag \omega \rme^{\imag \omega t}}{(\omega - k_\parallel v_\mr{A})(\omega + k_\parallel v_\mr{A})} = \cos k_\parallel v_\mr{A} t.
\end{align}
So
\begin{align}
u_{1y}(t) = \frac{1}{m_\mr{i}}\int_{0}^t \od s \cos k_\parallel v_\mr{A} (t-s) \force_y,
\end{align}
and, with  $t > t'$,
\begin{align}
\ea{ u_{1y}(t)u^*_{1y}(t')} &= \frac{1}{m_\mr{i}^2}\int_{0}^t \od s \int_{0}^{t'} \od s' \cos [k_\parallel v_\mr{A} (t-s)]\cos [k_\parallel v_\mr{A} (t'-s')] \ea{ \force_y(s)\force_y(s')} \nonumber
\\ &= \frac{\chi(k)}{m_\mr{i}^2}\int_{0}^t \od s \cos [k_\parallel v_\mr{A} (t-s)]\cos [k_\parallel v_\mr{A} (t'-s)]  \nonumber
\\ &= \frac{\chi(k)}{2k_\parallel v_\mr{A} m_\mr{i}^2}\left[ k_\parallel v_\mr{A}\mr{Max}(t,t') \cos k_\parallel v_\mr{A}(t-t')+ \cos k_\parallel v_\mr{A}t \sin k_\parallel v_\mr{A}t'\right] \label{eq:forcey_an}.
\end{align}
 This has been verified numerically by integrating a one-dimensional Langevin equation that represents the forced shear Alfv\'en waves, which is shown in figure~\ref{DKE_forcey}.

 The other two components now form a 2 by 2 matrix that must be inverted. The  $x$ component gives
\begin{align}
u_{1x}\Big[-  \omega^2   +   k^2 v_\mr{A}^2 - & (1+\tau/2)k_x^2 v_\mr{thi}^2  \zeta Z(\zeta)\Big]   \nonumber
 \\ &=  \frac{\tau}{2} k_x k_\parallel  v_\mr{thi}^2 \zeta Z(\zeta)  u_{1\parallel}   + \frac{k_x}{k_\parallel}   [1 + \zeta Z(\zeta)]\frac{\imag\omega \force_\parallel}{m} - \frac{\imag\omega \force_x}{m}, 
\end{align}
or
\begin{align}
 u_{1x}\Big[\beta^{-1} -&  \zeta^2  \cos^2 \theta      - (1+\tau/2)  \zeta Z(\zeta) \sin^2\theta\Big]    \nonumber
 \\ &=  \frac{\tau}{2}   \zeta Z(\zeta) \sin \theta \cos \theta  u_{1\parallel}   +   [1 + \zeta Z(\zeta)]\sin \theta \frac{\imag\zeta \force_\parallel}{k m v_\mr{thi}} - \cos\theta\frac{\imag\zeta \force_x}{k m v_\mr{thi}}. 
\end{align}
Similarly for the $z$ component,
\begin{align}
- \cos^2 \theta  \Big[ \zeta^2  &  + \tau  \left(\zeta^2 +\zeta^3 Z(\zeta)\right)\Big]u_{1\parallel} \nonumber
\\ &=   \sin\theta \cos \theta (\tau+1)\left[ \zeta^2 +\zeta^3 Z(\zeta)\right]  u_{1x}    +  \left[ \zeta^2 +\zeta^3 Z(\zeta)\right] \cos \theta\frac{ 2 \imag \zeta \force_\parallel}{k m v_\mr{thi}}.
\end{align}
After  rescaling the forces $\force' = \force/ k m v_\mr{thi} $, we have the solution
\begin{equation}
\left(\begin{matrix} u_{1x} \\ u_{1\parallel}\end{matrix}\right) = \frac{\imag \zeta}{\mathit{\Theta}}\left(\begin{matrix} \mathit{\Gamma}_{\perp x} & \mathit{\Gamma}_{\perp z}\\ \mathit{\Gamma}_{\perp x} & \mathit{\Gamma}_{\parallel z}    \end{matrix}\right)
\left(\begin{matrix}\force'_x \\ \force'_z\end{matrix}\right),
\end{equation}
where the determinant $\mathit{\Theta}$ is given by
\begin{align}
\mathit{\Theta} &\doteq  \cos^2 \theta  \left[1    + \tau  \left(1 +\zeta Z(\zeta)\right)\right]\left[  \zeta^2  \cos^2 \theta       + (1+\tau/2)  \zeta Z(\zeta) \sin^2\theta- \beta^{-1}\right] 
\nonumber \\ & \qquad  - \frac{1}{2} \tau(\tau+1) \sin^2 \theta \cos^2 \theta   \zeta Z(\zeta)  \left[ 1 +\zeta Z(\zeta)\right],  
\end{align}
while the forcing matrix coefficients are 
\begin{subequations}
\begin{align}
\mathit{\Gamma}_{\perp x} &=   \left[ 1    + \tau  \left(1 +\zeta Z(\zeta)\right)\right] \cos^3 \theta, \\
\mathit{\Gamma}_{\perp z} &=\mathit{\Gamma}_{\parallel x}   = -    (\tau+1)\left[ 1 +\zeta Z(\zeta)\right]  \sin\theta \cos^2 \theta,   \\
\mathit{\Gamma}_{\parallel z} &=  \left[2\beta^{-1} - 2 \zeta^2  \cos^2 \theta   +   \left(\tau + 1 -  \zeta Z(\zeta) \right)\sin^2\theta \right]   \left[ 1+\zeta Z(\zeta)\right] \cos \theta. 
\end{align}
\end{subequations}
The projection operator in the forcing due to incompressibility leads to $\ea{ \force_x \force_x^*} \propto \cos^2 \theta$, $\ea{ \force_\parallel \force_\parallel^*} \propto \sin^2 \theta$, $\ea{ \force_x \force_\parallel^*} \propto \sin \theta \cos \theta$. Also note that $\ea{ \force_i \force_j^* }^* = \ea{  \force_i \force_j^* }$, and thus the velocity correlations are
\begin{subequations}
\begin{align}
\ea{ u_{1x}u^*_{1x} } &= \frac{\zeta^* \zeta}{\mathit{\Theta}^*\mathit{\Theta}}\ea{ \left(\mathit{\Gamma}_{\perp x}  \force'_x + \mathit{\Gamma}_{\perp z}\force'_z\right)\left(\mathit{\Gamma}^*_{\perp x}  \force'^*_x + \mathit{\Gamma}^*_{\perp z}\force'^*_z\right)} \nonumber
\\  &\propto \frac{\zeta^* \zeta}{\mathit{\Theta}^*\mathit{\Theta}}\left[ |\mathit{\Gamma}_{\perp x}|^2\cos^2 \theta   +  |\mathit{\Gamma}_{\perp z}|^2\sin^2 \theta + 2 \Re(\mathit{\Gamma}_{\perp x}\mathit{\Gamma}_{\perp z}^*) \sin \theta \cos \theta\right] \label{eq:app:bforce1}, \\
\ea{ u_{1x}u^*_{1\parallel} } &= \frac{\zeta^* \zeta}{\mathit{\Theta}^*\mathit{\Theta}}\ea{ \left(\mathit{\Gamma}_{\perp x}  \force'_x + \mathit{\Gamma}_{\perp z}\force'_z\right)\left(\mathit{\Gamma}^*_{\parallel x}  \force'^*_x + \mathit{\Gamma}^*_{\parallel z}\force'^*_z\right)}  \nonumber
 \\ & = \frac{\zeta^* \zeta}{\mathit{\Theta}^*\mathit{\Theta}}\left[\mathit{\Gamma}_{\perp x}  \mathit{\Gamma}^*_{\parallel x} \cos^2\theta+ \left( \mathit{\Gamma}_{\perp x}   \mathit{\Gamma}^*_{\parallel z}+ \mathit{\Gamma}_{\perp z}\mathit{\Gamma}^*_{\parallel x}\right) \sin\theta\cos\theta+   \mathit{\Gamma}_{\perp z}\mathit{\Gamma}^*_{\parallel z}\sin^2\theta \right], \\
\ea{ u_{1\parallel}u^*_{1\parallel} } &= \frac{\zeta^* \zeta}{\mathit{\Theta}^*\mathit{\Theta}}\ea{ \left(\mathit{\Gamma}_{\parallel x}  \force'_x + \mathit{\Gamma}_{\parallel z}\force'_z\right)\left(\mathit{\Gamma}^*_{\parallel x}  \force'^*_x + \mathit{\Gamma}^*_{\parallel z}\force'^*_z\right)}  \nonumber
\\ &\propto \frac{\zeta^* \zeta}{\mathit{\Theta}^*\mathit{\Theta}}\left[|\mathit{\Gamma}_{\parallel x}| ^2\cos^2 \theta  +  |\mathit{\Gamma}_{\parallel z}|^2 \sin^2 \theta+ 2\Re(\mathit{\Gamma}_{\parallel x}\mathit{\Gamma}_{\parallel z}^*)\sin \theta \cos \theta \right]. \label{eq:app:bforce2}
\end{align}
\end{subequations}
 To proceed, we consider the asymptotic small- and large-angle long-time limits. For $k_x \rightarrow 0$ ($ \theta \rightarrow 0$) and $\theta \ll \beta^{-1} \ll 1$, to lowest order in $\theta$ the determinant becomes
\begin{align}
\mathit{\Theta} &\approx   \left[ 1    + \tau  \left(1 +\zeta Z(\zeta)\right)\right]\left[  \zeta^2       + (1+\tau/2)  \zeta Z(\zeta) \theta^2 - \beta^{-1}\right]  \nonumber
\\ & \qquad    - \frac{1}{2}\tau(\tau+1) \theta^2  \zeta Z(\zeta)  \left[1 +\zeta Z(\zeta)\right] .
\end{align}
For small $\theta$, the dominant pole appears for $\zeta^2 \approx \beta^{-1}$, so  the plasma dispersion function is expanded in the small argument ($Z(\zeta) \approx \imag \sqrt{\upi} - 2 \zeta + (2/3)\zeta^3 -  \ldots$),
\begin{align}
\mathit{\Theta} &\approx   \left[1 + \tau    + \tau(\imag\sqrt{\upi} \zeta - 2\zeta^2)  \right]\left[  \zeta^2       + \imag \sqrt{\upi}(1+\tau/2)  \zeta  \theta^2 - \beta^{-1}\right]    \nonumber \\ & \qquad - \frac{1}{2}\tau(\tau+1) \theta^2  (\imag \sqrt{\upi}\zeta - 2 \zeta^2)  \left[1 + \imag \sqrt{\upi}\zeta \right]  .
\end{align}
Using the \emph{Ansatz} $\zeta^2 \approx \beta^{-1} + \imag a \theta^2$, we find after some algebra $a = \sqrt{\upi \beta^{-1}}$.
To lowest order in $\theta$ and $\beta^{-1}$, 
\begin{subequations}
\begin{align}
\mathit{\Gamma}_{\perp x} &\approx  1 + \tau,\\
\mathit{\Gamma}_{\perp z } &\approx \mathit{\Gamma}_{\parallel x} \approx  - (1+\tau) \theta, \\
\mathit{\Gamma}_{\parallel z} &\approx (1+ \tau)\theta^2,
\end{align}
\end{subequations}
and so we  have near the dominant pole
\begin{align}
&\ea{ u_{1\parallel}(t,\bb{k}) u_{1\parallel}^* (t,\bb{k})}  \nonumber
\\ &\qquad \approx  \frac{\chi(k)}{2\upi k m_\mr{i}^2 v_\mr{thi}}  \int \od \zeta \,  \frac{  \theta^2 \beta^{-1}}{  (\zeta^2 - \beta^{-1})^2 + \upi \beta^{-1}  \theta^4} \nonumber
\\ &\qquad =  \frac{\chi(k)}{2\upi k m_\mr{i}^2 v_\mr{thi}} \int \od \zeta \,   \frac{\sqrt{\beta^{-1}}}{2\imag \sqrt{\upi}}\left(\frac{  1 }{ \zeta^2 - \beta^{-1}  -\imag  \theta^2\sqrt{\upi \beta^{-1}}  } - \frac{  1 }{ \zeta^2 - \beta^{-1} + \imag  \theta^2\sqrt{\upi \beta^{-1}}  }\right) \nonumber
\\ &\qquad =  \frac{\chi(k)}{2\upi^{1/2} k m_\mr{i}^2 v_\mr{thi}}.
\end{align}
Similarly for $u_{1x}$,
\begin{align}
\ea{ u_{1x}(t,\bb{k}) u_{1x}^* (t,\bb{k})} &\approx  \frac{\chi(k)}{2\upi^{1/2} \theta^2 k m_\mr{i}^2 v_\mr{thi}}.
\end{align}
This indicates that parallel variation of $u_{1\parallel}$ is strongly damped and saturates at  low amplitude, similar to the unmagnetized case. However, the parallel variation of $u_{1x}$ is not strongly damped, and as a result it experiences secular growth without saturation in the linear regime.

For the opposite limit  $\theta \rightarrow \upi/2$, we define $\phi \doteq (\upi/2) - \theta$. Then the determinant is approximately
\begin{align}
\mathit{\Theta} &\approx  \phi^2   \left[1    + \tau  \left(1 +\zeta Z(\zeta)\right)\right]\left[  \zeta^2  \phi^2       + (1+\tau/2)  \zeta Z(\zeta) - \beta^{-1}\right] 
\nonumber \\ & \qquad  - \frac{1}{2} \tau(\tau+1) \phi^2 \zeta Z(\zeta)  \left[ 1 +\zeta Z(\zeta)\right].
\end{align}
 For $\beta \ll 1$, the dominant pole now appears around $\zeta^2 \approx -(1+\tau/2) \zeta Z(\zeta)/\phi^2$.  The consistent limit for $\phi \rightarrow 0$ is $\zeta \rightarrow \infty$ and so $\zeta^2 \approx (1+\tau/2)/\phi^2$. Using the plasma dispersion function expanded in the large argument with the  assumption that $\Im(\zeta) \ll |  \Re(\zeta)^{-1}|$,  i.e. $Z(\zeta) \approx  \imag \upi^{1/2} \exp (-\zeta^2) - \zeta^{-1}(1 + 1/2\zeta^2 + \ldots)$, we arrive at 
\begin{align}
\mathit{\Theta} &\approx  \phi^2   \left[1    + \tau  \left(\imag \alpha \zeta + 1/2\zeta^2\right)\right]\left[  \zeta^2  \phi^2       - (1+\tau/2)  +  \imag \alpha (1+\tau/2)  \zeta  - \beta^{-1}\right] 
\nonumber \\ & \qquad  - \frac{1}{2} \tau(\tau+1) \phi^2 (\imag \alpha \zeta - 1)  \left[ \imag\alpha \zeta  - 1/2\zeta^2\right] \nonumber
\\ &\approx \phi^4 [\zeta^2  - (1+\tau/2)/\phi^2 + \imag \alpha \phi^{-3} (1+\tau/2)^{3/2}],
\end{align}
where $\alpha \doteq \upi^{1/2}\exp(-\zeta^2) \lll 1$ for $\theta \rightarrow \upi/2$. Then
\begin{subequations}
\begin{align}
\mathit{\Gamma}_{\perp x} &\approx  (1+\tau) \phi^3, \\
\mathit{\Gamma}_{\perp z } &\approx \mathit{\Gamma}_{\parallel x} \approx   - \frac{1+\tau}{2+\tau} \phi^4, \\
\mathit{\Gamma}_{\parallel z} &\approx    - \frac{\phi^5}{(2+\tau)^2}.
\end{align}
\end{subequations}
\begin{figure}
\centering
\includegraphics[scale=1.15]{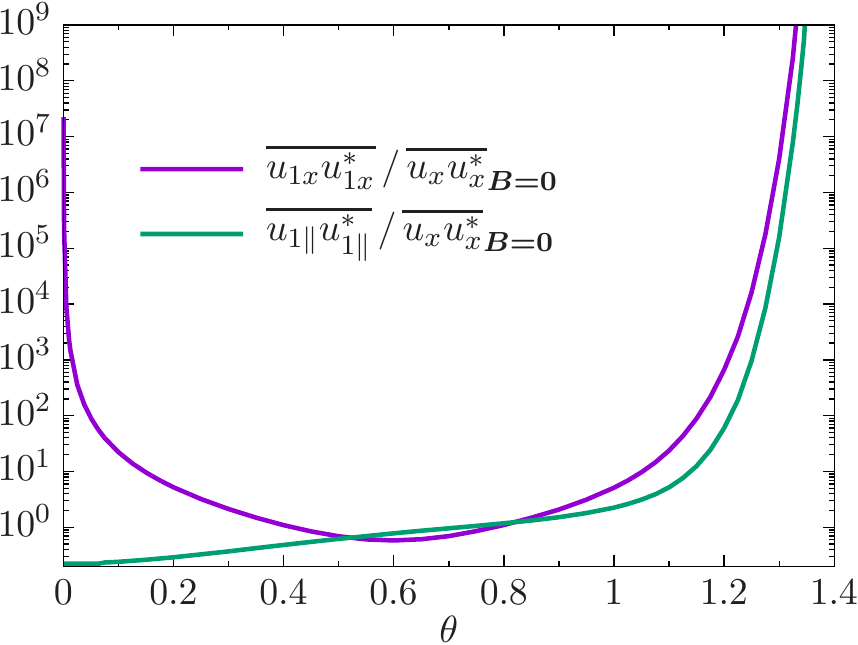}
\caption[Numerical solution of equations~\eqref{eq:app:bforce1} and~\eqref{eq:app:bforce1}]{\label{DKE_force} Numerical solution of equations~\eqref{eq:app:bforce1} and~\eqref{eq:app:bforce1} at $t' = t$, normalized by the unmagnetized result given by equation~\eqref{eq:app:nobforce}.}
\end{figure}
\noindent This leads to
\begin{align}
&\ea{ u_{1\parallel}(t,\bb{k})  u_{1\parallel}^* (t,\bb{k})} \nonumber
\\ & \qquad\approx   \frac{\chi(k)}{4\upi k m_\mr{i}^2 v_\mr{thi}}  \int \od \zeta\, \frac{(1+3\tau +\tau^2)^2}{(2+\tau)^3} \phi  \frac{1}{[\zeta^2 - (1+\tau/2)/\phi^2]^2 + \alpha^2(1+\tau/2)^3/\phi^6} \nonumber
\\ & \qquad = \frac{\chi(k)}{4\upi k m_\mr{i}^2 v_\mr{thi}}  \frac{(1+3\tau +\tau^2)^2}{(2+\tau)^3} \phi \frac{\upi \phi^3}{ \alpha (1+\tau/2)^{3/2}}\frac{\phi}{(1+\tau/2)^{1/2}}
\nonumber \\ &\qquad= \frac{\chi(k)}{\upi^{1/2} k m_\mr{i}^2 v_\mr{thi}}   \frac{(1+3\tau +\tau^2)^2}{(2+\tau)^5}\phi^5\exp [\phi^{-2}(1+\tau/2)]. 
\end{align}
Likewise for $u_{1x}$,
\begin{align}
&\ea{ u_{1x}(t,\bb{k}) u_{1x}^* (t,\bb{k})} 
\nonumber
\\  &  \qquad \approx   \frac{\chi(k)}{4\upi k m_\mr{i}^2 v_\mr{thi}}  \int \od \zeta\, \frac{(1+\tau)^3(3+\tau)}{2+\tau}\frac{1}{\phi} \frac{1}{[\zeta^2 - (1+\tau/2)/\phi^2]^2 + \alpha^2(1+\tau/2)^3/\phi^6} \nonumber
\\ &\qquad = \frac{\chi(k)}{4\upi k m_\mr{i}^2 v_\mr{thi}}   \frac{(1+\tau)^3(3+\tau)}{2+\tau}\frac{1}{\phi} \frac{\upi \phi^3}{ \alpha (1+\tau/2)^{3/2}}\frac{\phi}{(1+\tau/2)^{1/2}}
\nonumber \\ &  \qquad= \frac{\chi(k)}{\upi^{1/2} k m_\mr{i}^2 v_\mr{thi}} \frac{(1+\tau)^3(3+\tau)}{(2+\tau)^3} \phi^3\exp [\phi^{-2}(1+\tau/2)]. 
\end{align}
In both of these cases, Landau damping becomes vanishingly small as $\theta \rightarrow \upi/2$, and thus the saturated velocities  grow strongly with $\theta$.  Equations~\eqref{eq:app:bforce1} and~\eqref{eq:app:bforce1} are evaluated numerically at $t=t'$ in figure~\ref{DKE_force}, which demonstrates the behaviour quantified above.

\chapter{Linear Stability Analysis of the Firehose and Mirror Instabilities}\label{ap:linear}

In this Appendix, we compute the linear theory for both the unlimited, incompressible, Braginskii--MHD system of equations (\ref{brag_MHD}a--d), and the hybrid-kinetic system of equations~(\ref{intro:hybkin}a--b). In doing so, we demonstrate that while Braginskii-MHD system exhibits the proper parallel firehose instability, it does not correctly capture the mirror instability

\section{Hybrid kinetics}

We can quickly perform the stability analysis of the hybrid-kinetic equations by starting with the results from appendix~\ref{ap:forcing} and setting $\vforce =0 $. Specifically, using equations~\eqref{ap:eq:pprp} and~\eqref{ap:eq:pprl} along with~\eqref{eq:force:lin}, we have
\begin{subequations}
\begin{align}
  \omega m n_0 u_{1x} &= k_x n_1 T_e  + k_x p_{0\perp} \left[\frac{2 k_xu_{1x}}{\omega }  - [1 + \zeta Z(\zeta)]\left(\frac{T_\mr{e}}{T_{\parallel}}\frac{k_x u_{1x} + k_\parallel u_{1\parallel} }{\omega}  + 2\frac{T_\perp}{T_\parallel}\frac{k_xu_{1x}}{\omega}\right)\right]  \nonumber
\\  &\qquad - (  p_{0\parallel}-p_{0\perp})k_\parallel^2 \frac{u_{1x}}{\omega} + \frac{ k^2 B^2_0}{4\upi}\frac{u_{1x}}{\omega}, \\
 \omega m n_0 u_{1\parallel} &= k_\parallel n_1 T_e   + k_\parallel  p_{0\parallel}\left[ \frac{k_x u_{1x} }{\omega}   -  \left[1 + 2\zeta^2 +2\zeta^3 Z(\zeta)\right]\left( \frac{T_\mr{e}}{T_\parallel} \frac{k_x u_{1x} + k_\parallel u_{1\parallel}  }{\omega}  + \frac{T_\perp}{T_\parallel} \frac{k_x u_{1x}}{\omega}\right)\right]  \nonumber
\\ &\qquad - (  p_{0\parallel} - p_{0\perp}) k_\parallel k_x \frac{u_{1x}}{\omega}, 
\end{align}
\end{subequations}
or
\begin{subequations}
\begin{align}
  \omega^2 m n_0 u_{1x} &=  k_x n_0 T_e  (k_x u_{1x} + k_\parallel u_{1\parallel})  - (  p_{0\parallel} - p_{0\perp})k_\parallel^2 u_{1x} +\frac{ k^2 B^2_0}{4\upi}u_{1x}  \nonumber
  \\  &+k_x p_{0\perp} \left[2 k_xu_{1x}  - [1 + \zeta Z(\zeta)]\left(\frac{T_\mr{e}}{T_{\parallel}}\frac{k_x u_{1x} + k_\parallel u_{1\parallel} }{\omega}  + 2\frac{T_\perp}{T_\parallel}k_xu_{1x}\right)\right] , \\
-   k_\parallel u_{1\parallel} &=       \left[ 1 +\zeta Z(\zeta)\right]\left[ \frac{T_\mr{e}}{T_\parallel} (k_x u_{1x} + k_\parallel u_{1\parallel})  + \frac{T_\perp}{T_\parallel} k_x u_{1x}\right].
\end{align}
\end{subequations}
At marginal stability, $\omega \rightarrow 0$, so
\begin{subequations}
\begin{align}
0 &=  k_x n_0 T_e  (k_x u_{1x} + k_\parallel u_{1\parallel}) + k_x p_{0\perp} \left[2 k_xu_{1x}  -\left(\frac{T_\mr{e}}{T_{\parallel}}(k_x u_{1x} + k_\parallel u_{1\parallel})  + 2\frac{T_\perp}{T_\parallel}k_xu_{1x}\right)\right] \nonumber \\ &\qquad -(  p_{0\parallel} - p_{0\perp})k_\parallel^2 u_{1x} +\frac{ k^2 B^2_0}{4\upi}u_{1x}, \\
-   k_\parallel u_{1\parallel} &=       \frac{T_\mr{e}}{T_\parallel} (k_x u_{1x} + k_\parallel u_{1\parallel})  + \frac{T_\perp}{T_\parallel} k_x u_{1x}. 
\end{align}
\end{subequations}
After defining $\tau_\parallel \doteq T_\mathrm{e}/T_\parallel$ and $\tau_\perp \doteq T_\mathrm{e}/T_\perp$, we arrive at
\begin{subequations}
\begin{gather}
  \left( (\tau_\perp -\tau_\parallel) k^2_x  + 2 k_x^2  - 2\frac{p_{0\perp}}{p_{0\parallel}}k_x^2\right) u_{1x}  + (\tau_\perp   -\tau_\parallel) k_x k_\parallel u_{1\parallel}   =   -\left(\frac{  p_{0\perp} - p_{0\parallel}}{p_{0\perp}}k_\parallel^2  + \frac{ k^2 B^2_0}{4\upi p_{0\perp}}\right)u_{1x}, \\
   - (1+\tau_\parallel)  k_\parallel u_{1\parallel} =       \left(\tau_\parallel + \frac{p_{0\perp}}{p_{0\parallel}}\right)  k_x u_{1x}. 
\end{gather}
\end{subequations}
Combining these two leads to the stability condition
\begin{gather}
 \frac{1}{2}\frac{\tau_\perp}{1+\tau_\perp}\left(\frac{p_{0\perp}-p_{0\parallel}}{p_{0\parallel}}\right)^2 +   \frac{ 1}{\beta_{\mr{i}0\perp}}  -   \frac{p_{0\perp} - p_{0\parallel}}{p_{0\parallel}} +\frac{k_\parallel^2}{2k_\perp^2} \left(\frac{  p_{0\perp} - p_{0\parallel}}{p_{0\perp}}  + \frac{ 2}{\beta_{\mr{i}0\perp}}\right) >  0.
\end{gather}
In the limit $k_\parallel/k_\perp \gg 1$, we recover the firehose instability threshold $(p_{0\perp} -p_{0\parallel})/p_{0\perp} -  > - 2/\beta_{\mr{i}0\perp}$.  In the opposite limit, we  have the mirror stability threshold $(p_{0\perp} - p_{0\parallel})/p_{0\parallel} \lesssim 1/\beta_{\mr{i}0\perp}$, now with a correction stemming from a warm electron contribution. If the pressure anisotropy is small ($\Delta \ll 1$), then so is this electron correction.

\section{Braginskii-MHD}
For the linear analysis of the Braginskii-MHD system, we assume an unforced, homogeneous, stationary equilibrium state with a uniform pressure anisotropy $\Deltap \ne 0$ and $\bb{B}_0 = B_0\ez$, subject to small-amplitude perturbations (subscripted with a `1') of the form $\exp(\gamma t + \imag\bb{k}\bcdot\bb{x})$.
 While this equilibrium is formally incompatible with the Braginskii closure $\Deltap = 3\visc_\mr{B} \ROS$, we consider perturbations with sufficiently high frequencies ($\omega \gg | \grad \bb{u}|$) and small scales ($k \gg k_\visc$) that $\Deltap$ may be considered constant in space and time (a similar analysis using a $\Deltap$ obtained from the drift kinetic equation was carried out by \citealt{Scheko_2005}).

Without loss of generality, let $\bb{k} = k_\perp \ex + k_\parallel \ez$. The equations, written to first order in the perturbation amplitudes and assuming Laplacian diffusion, are
\begin{align}
( \gamma + \visc k^2 ) \bb{u}_1 &= - \imag\bb{k} p_1  +  \imag k_\parallel  B_0^2 \bb{b}_1 + \imag k_\parallel \Deltap \left(\bb{b}_1 - 2b_{1z}\ez\right) - \visc_\mr{B} k^2_\parallel u_{1z}\ez,\label{eqn:linmom}
\\ \bb{b}_1 &= \frac{\imag k_\parallel }{\gamma + k^2 \eta} \bb{u}_1,  \label{eqn:A2}
\end{align}
where $\bb{b}_1 \doteq \bb{B}_1/B_0$. If one were to consider other types of diffusion, $\visc$ and $\eta$ may be simply redefined (e.g.~$\visc \rightarrow \visc_\mr{H}k^2$ for hyper-diffusion). Incompressibility ($\bb{k}\bcdot\bb{u}_1$ = 0) is used to determine $p_1$:
\begin{equation}\label{eqn:p1}
p_1 = ( \imag k_\parallel \visc_\mr{B} u_{1z} - 2b_{1z}\Deltap  )\frac{k_\parallel^2}{k^2} ,
\end{equation}
where we have used the solenoidality constraint on the perturbed magnetic field, \emph{viz}.,  $\bb{k}\bcdot\bb{b}_1=0$. Substituting \eqref{eqn:p1} back into \eqref{eqn:linmom} leads to 
\begin{align} \label{eqn:A4}
(\gamma + \visc k^2) \bb{u}_1 &=  \imag k_\parallel B_0^2 \bb{b}_1 + \imag k_\parallel \Deltap \Bigl[\bb{b}_1 - 2b_{1z}\ez\bcdot(\msb{I}-\bb{\hat{k}\hat{k}})\Bigr] - \visc_\mr{B} k_\parallel^2 u_{1z}\ez \bcdot(\msb{I}-\bb{\hat{k}\hat{k}}).
\end{align}
Combining \eqref{eqn:A2} and \eqref{eqn:A4} yields the matrix equation
\begin{align}\label{eqn:dr_matrix}
\bigl( \gamma + \eta k^2 \bigr) &\bigl[ (\gamma + \visc k^2 )\msb{I} + \visc_\mr{B}k_\parallel^2  (\msb{I}-\bb{\hat{k}\hat{k}})\bcdot\ez\ez  \bigr]\bcdot \bb{b}_1 =  
\nonumber\\* \mbox{} &- k_\parallel^2 B_0^2 \bb{b}_1 -  k_\parallel^2 \Deltap \, \bb{b}_1\bcdot\bigl[\msb{I} - 2\ez\ez\bcdot(\msb{I}-\bb{\hat{k}\hat{k}})\bigr].
\end{align}
Solenoidality requires that $b_{1z} = -(k_\perp/k_\parallel) b_{1x}$, breaking the dispersion relation into two branches:
\begin{align}
\bigl(\gamma + \eta k^2 \bigr) \bigl(\gamma + \visc k^2 + \visc_\mr{B}  k_\parallel^2k_\perp^2/k^2 \bigr)  &=  - k_\parallel^2 \bigl[B_0^2 +   \Deltap (k_\parallel^2 -k_\perp^2  )/k^2 \bigr], \label{ap:branch1} \\*
\bigl(\gamma + \eta k^2 \bigr) \bigl(\gamma + \visc k^2\bigr)  &=  - k_\parallel^2  \bigl(B_0^2 + \Deltap \bigr). \label{ap:branch2}
\end{align}
 The solution of~\eqref{ap:branch2} is
\begin{equation}
    \gamma =  -D_+ \pm \sqrt{ D_-^2 - k_\parallel^2(B_0^2+\Deltap)},
\end{equation}
where $D_\pm \doteq (\visc \pm \eta) k^2/2$. The dissipationless limit $\visc = \eta =0$ returns the firehose stability threshold $\Deltap > -B_0^2  $ (see equation \eqref{stability}). 

Now we turn to~\eqref{ap:branch1} and consider it in the limit of $\visc_\mr{B} k_\parallel^2 \gg \gamma$ and $\visc = \eta = 0$. This leads to 
\begin{align}
 \gamma  \approx \frac{1}{\visc_\mr{B}} \frac{k^2}{k_\perp^2} \biggl( - B_0^2 + \Deltap\frac{k_\perp^2-k_\parallel^2}{k^2} \biggr),
\end{align}
with a stability threshold given by $\Deltap < B_0^2$ and a scale-independent growth rate. This illustrates the inability of the Braginskii-MHD system to capture correctly the mirror instability, which in the kinetic calculation should exhibit a growth rate that scales with $|k_\parallel|$ (until ion-Larmor scales) and has a stability threshold given by $\Deltap < B_0^2/2$. Instead, the Braginskii mirror instability grows at a rate proportional to $\Delta p/\visc_\mathrm{B}$, i.e., comparable to the rate of strain of the field-stretching motions.

\chapter{Alignment of the Rate-of-Strain Eigenvectors  with the Magnetic Field} \label{ap:ROS}

Here we analyze how the eigenvectors of the symmetric rate-of-strain tensor $\mathsfbi{S} \doteq (\grad \bb{u} + \grad \bb{u}^{\mathrm{T}})/2$ might align themselves with the magnetic field in the saturated state. The presentation follows A.~B.~Iskakov \& A.~A.~Schekochihin (2008, unpublished), and is relevant to the interpretation of figures \ref{fig:angle_some} and \ref{fig:pdfROS}.

As in the main text, the eigenvalues $\lambda_i$ of $\mathsf{S}_{ij}$ are ordered so that $\lambda_1 \ge \lambda_2 \ge \lambda_3$; the corresponding eigenvectors are denoted $\eig_1$, $\eig_2$, and $\eig_3$. The rate-of-strain tensor can then be written as $\mathsf{S}_{ij} = \mathsf{R}_{il}{\Lambda}_{lm}\mathsf{R}_{mj}$, where $\mathsf{R}_{ij}$ is the tensor whose columns  are the eigenvectors and $\Lambda_{ij}$ is the diagonal matrix composed of the eigenvalues. Here we have assumed distinct eigenvalues for simplicity. The anti-symmetric portion of the rate-of-strain tensor encompasses vortical motion, and the full tensor can be expressed in index notation as
\begin{equation}\label{eqn:duij}
    \nabla_i u_j = \mathsf{S}_{ij} + \frac{1}{2}\epsilon_{ijk}\omega_k,
\end{equation}
where $\epsilon_{ijk}$ is the Levi-Civita symbol, $\nabla_i = \partial/\partial x_i$, and $\omega_i = \epsilon_{ijk}\nabla_j u_k$ is the vorticity of the fluid motion.

The quantity that must be constrained in order to achieve saturation is the parallel rate of strain, which is given by
\begin{equation}\label{eqn:appROS}
    \ROS = b_ib_j \mathsf{S}_{ij} = \lambda_1 \cos^2 \theta_1 + \lambda_2 \cos^2 \theta_2 + \lambda_3 \cos^2 \theta_3,
\end{equation}
where $\theta_i$ is the angle between $\eig_i$ and $\eb$.  The  other components of the rate of strain tensor are
\begin{subequations}\label{eqn:Gu}
\begin{align}
    \grad_\parallel \bb{u}_\perp &\doteq \eb\eb \bcdot \grad \bb{u} \bcdot(\unitDyadic - \eb\eb), \\*
    \grad_\perp\bb{u}_\parallel &\doteq (\unitDyadic - \eb\eb) \bcdot \grad \bb{u} \bcdot \eb\eb,\\*
    \grad_\perp \bb{u}_\perp &\doteq (\unitDyadic - \eb\eb) \bcdot \grad \bb{u} \bcdot(\unitDyadic - \eb\eb) \nonumber
     \\ &= \grad \bb{u} -\grad_\parallel \bb{u} -\grad \bb{u}_\parallel   + \eb\eb (\ROS),\label{eqn:Gpup}
\end{align}
\end{subequations}
which respectively represent the shearing, slipping, and squeezing (or mixing) of magnetic-field lines. In \eqref{eqn:Gpup} and for the remainder of this appendix, the parallel subscript is to be taken \emph{outside} the derivative; i.e., the gradient works only on the velocity $\bb{u}$ and not the unit vector $\eb$.

The combination $\grad_\perp\bb{u}_\perp$ \eqref{eqn:Gpup} is of particular interest, as it results in field lines coming closer together. This results in resistive annihilation and so this component of the rate-of-strain tensor is deleterious to the growth of magnetic energy; it should thus also be minimized alongside the parallel rate of strain. To do so, we start by computing the square magnitude of the mixing motions,
\begin{equation}\label{eqn:Gpupsq}
    |\grad_\perp\bb{u}_\perp|^2 = |\grad \bb{u}|^2  - |\grad \bb{u}_\parallel|^2 - |\grad_\parallel \bb{u}|^2 +  |\ROS|^2.
\end{equation}
First, the square magnitude of \eqref{eqn:duij} is
\begin{align}
    |\grad \bb{u}|^2 &= \left(\mathsf{S}_{ij} +  \frac{1}{2}\epsilon_{ijk}\omega_k\right)\left(\mathsf{S}_{ij}+ \frac{1}{2}\epsilon_{ijl}\omega_l\right)
   \nonumber  \\* 
    \mbox{} &= \mathsf{S}_{ij}\mathsf{S}_{ij} +  \mathsf{S}_{ij}\epsilon_{ijk}\omega_k + \frac{1}{4}\epsilon_{ijk}\epsilon_{ijl}\omega_k\omega_l
    \nonumber \\*
    \mbox{} &= \sum_i \lambda_i^2 + \frac{1}{2} \omega^2,
\end{align}
where we have used the fact that $\mathsf{R}_{ij}\mathsf{R}_{jk} = \delta_{ik}$,  $\mathsf{S}_{ij}\epsilon_{ijk} = 0$ by symmetry, and   $\epsilon_{ijk}\epsilon_{ijl} = 2 \delta_{kl}$. The next term on the right-hand side of~\ref{eqn:Gpupsq} is
\begin{align}
    |\grad \bb{u}_\parallel|^2 &= ( \nrndb_m \nabla_i u_m ) ( \nrndb_n \nabla_i u_n)  \nonumber \\* 
    \mbox{} &= \left(\mathsf{S}_{im} \nrndb_m + \frac{1}{2}\epsilon_{imk}\omega_k \nrndb_m\right)\left(\mathsf{S}_{in} \nrndb_n + \frac{1}{2}\epsilon_{inl}\omega_l \nrndb_n\right) \nonumber \\* 
    \mbox{} &= \mathsf{S}_{im}\mathsf{S}_{in}\nrndb_m \nrndb_n + \mathsf{S}_{im}\epsilon_{inl}\omega_l \nrndb_n \nrndb_m + \frac{1}{4}\epsilon_{imk}\epsilon_{inl} \omega_k \omega_l \nrndb_m \nrndb_n.\label{eqn:Gpu}
\end{align}
The first term in \eqref{eqn:Gpu} may be simplified as follows:
\begin{align}
    \mathsf{S}_{im}\mathsf{S}_{in}\nrndb_m \nrndb_n &= (\mathsf{R}_{ip}\Lambda_{pq}\mathsf{R}_{mq})( \mathsf{R}_{ir}\Lambda_{rs}\mathsf{R}_{ns})\nrndb_m \nrndb_n \nonumber\\*
    \mbox{} &= \Lambda_{pq} \Lambda_{ps}\nrndb_m\mathsf{R}_{mq}\nrndb_n\mathsf{R}_{ns} \nonumber \\*
    \mbox{} &= \sum_i \lambda_i^2 \cos^2\theta_i .
\end{align}
The second and third terms in \eqref{eqn:Gpu} are simplified using the contracted epsilon identity $\epsilon_{lmi}\epsilon_{ijk} = \delta_{jl}\delta_{km} - \delta_{kl}\delta_{jm}$, yielding
\begin{align}
    S_{im}\epsilon_{inl}\omega_l \nrndb_n \nrndb_m  &= \frac{1}{2}(\nabla_i u_m + \nabla_m u_i)\epsilon_{inl}\epsilon_{lpq} \nabla_p u_q \nrndb_n \nrndb_m \nonumber\\
     &= \frac{1}{2}(\nrndb_m\nabla_i u_m + \nrndb_m \nabla_m u_i)(\nrndb_n \nabla_i u_n - \nrndb_n \nabla_n u_i)
     \nonumber\\ &= \frac{1}{2}(|\grad \bb{u}_\parallel|^2 - |\grad_\parallel \bb{u}|^2)
\end{align}
and
\begin{equation}
   \frac{1}{4} \epsilon_{imk}\epsilon_{inl} \omega_k \omega_l \nrndb_m \nrndb_n =  \frac{1}{4}(\delta_{kl}\delta_{mn} - \delta_{kn}\delta_{lm}) \omega_k \omega_l \nrndb_m \nrndb_n = \frac{1}{4}\omega_\perp^2 ,
\end{equation}
respectively. Therefore,
\begin{equation}\label{eqn:GupGpu}
    |\grad \bb{u}_\parallel|^2 + |\grad_\parallel \bb{u}|^2 = 2\sum_i \lambda_i^2 \cos^2\theta + \frac{1}{2} \omega_\perp^2 .
\end{equation}
Thus, with the use of \eqref{eqn:GupGpu}, equation \eqref{eqn:Gpupsq} becomes
\begin{align}
       |\grad_\perp\bb{u}_\perp|^2 &= \sum_i \lambda_i^2 + \frac{1}{2} \omega^2 - 2\sum_i \lambda_i^2 \cos^2\theta - \frac{1}{2}\omega_\perp^2 + \left(\sum_i \lambda_i \cos^2\theta_i\right)^2
      \nonumber \\&= \sum_i \lambda_i^2 (1-2\cos^2 \theta_i) + \left(\sum_i \lambda_i \cos^2\theta_i\right)^2 + \frac{1}{2}\omega_\parallel^2.\label{eqn:Gpupsq_2}
\end{align}

We wish to minimize \eqref{eqn:Gpupsq_2} subject to some constraint on the stretching motions given by equation~\eqref{eqn:appROS}. To simplify this procedure, we define $c_i \doteq \cos^2 \theta_i$ and $\varsigma \doteq \lambda_2/ \lambda_1$. Using $\sum_i \lambda_i = 0$ (incompressibility) and $\sum_i c_i = 1 $ (orthogonality), equations~\eqref{eqn:appROS} and \eqref{eqn:Gpupsq_2} may be written as
\begin{subequations}
\begin{align}
    \frac{\ROS}{\lambda_1}  &= x+y-1-\varsigma,   \label{eq:apAlexROS2}\\
    \frac{|\grad_\perp \bb{u}_\perp|^2}{\lambda_1^2} - \frac{1}{2}\frac{\omega_\parallel^2}{\lambda_1^2} &= (x+y-1)^2 + 2y(1-\varsigma) + \varsigma^2 ,\label{eq:apAlexROS1}
\end{align}
\end{subequations}
where $x \doteq (2+\varsigma)c_1$, $y \doteq (1+2\varsigma) c_2$, and $\varsigma \doteq \lambda_2/\lambda_1$. We then have two free parameters that describe the alignment of the magnetic field, $c_1$ and $c_2$; these can be adjusted in order to minimize~\eqref{eq:apAlexROS1} subject to some specified constraint on~\eqref{eq:apAlexROS2}. We consider two constraints on the latter that can potentially lead to saturation of the dynamo: (i) the parallel rate of strain tends towards zero ($\ROS \approx 0$); or (ii) the magnetic field aligns itself in such a way that $\ROS \propto \lambda_2$. Scenario (i) obtains for $x+y-1=\varsigma$. Substituting this solution into \eqref{eq:apAlexROS1}, we find that, so long as $-1/2 < \varsigma < 1$, the resulting expression for $|\grad_\perp \bb{u}_\perp|^2$ is minimized at $c_2 = 0$ ($\theta_2=90^\circ$). In this case, $c_1 = (1+\varsigma)/(2+\varsigma)$ and $|\grad_\perp\bb{u}_\perp|^2 = 2\lambda^2_2 + \omega^2_\parallel/2$. In the limit $\varsigma \rightarrow 0$, the magnetic field aligns equidistant in angle between the stretching and compression directions, i.e., $\theta_1 = \theta_3 = 45^\circ$. Scenario (ii) obtains when all the terms not multiplying $\varsigma$ in \eqref{eq:apAlexROS2} vanish, {\it viz.}, when $2c_1+c_2-1=0$. In this case, $\nabla_\parallel u_\parallel = \lambda_2 P_2(\cos\theta_2)$, where $P_2$ is the second Legendre polynomial, and \eqref{eq:apAlexROS1} becomes
\begin{equation}
    \frac{|\grad_\perp \bb{u}_\perp|^2}{\lambda_1^2} - \frac{1}{2}\frac{\omega_\parallel^2}{\lambda_1^2} = 2 c_2 (1+\varsigma) + \frac{\varsigma^2}{4}[5 + c_2(9 c_2 - 10)].
\end{equation}
For $\varsigma \ll 1$, the mixing is minimized when $c_2 = 0$, $c_1 = 1/2$. Saturation is then achieved when $\theta_2 = 90^\circ$ and $\theta_1 = \theta_3 = 45^\circ$, in which case $\nabla_\parallel u_\parallel = -\lambda_2/2$ and $|\grad_\perp\bb{u}_\perp|^2 = (5/4)\lambda^2_2 + \omega^2_\parallel/2$. Note that these two scenarios coincide in the limit $\varsigma\ll{1}$, with 45$^\circ$ alignment between $\eone$ and $\ethree$.

\chapter{Derivation of the Anisotropic Kazantsev--Kraichnan Model}\label{ap:kazantsev}

For completeness, we provide here the full derivation of \eqref{eqn:mod_kazantsev}, the anisotropic Kazantsev--Kraichnan model presented and utilized in \S\,\ref{sec:kazantsev}. This derivation was omitted in~\citet{Scheko_saturated} for lack of space and never came to be published anywhere, but we feel that it is important to spell out all the steps and assumptions that went into it in order for the reader to be able to judge the level of plausibility of our arguments in \S \ref{sec:kazantsev}.

We start with the evolution equations
\begin{subequations}
\begin{gather}
    \partial_t \rndB = \rndb^i\rndb^m u^i_{,m} \rndB - \eta \rndk^2 \rndB, \\*
    \partial_t \rndk_m = - \rndu^{i}_{,m} \rndk_{i}, \\*
\partial_t \rndb^i = \rndb^m \rndu^{i}_{,m} - \rndb^l \rndb^m \rndu^l_{,m} \rndb^i,
\end{gather}
\end{subequations}
where tildes denote random variables and indices appearing after a comma denote a spatial derivative, i.e. $\rndu_{i,j} \doteq \partial \rndu_i / \partial x^{j}$. We assume that $\tilde{\bb{u}}$ is approximately linear in space,\footnote{Such a Taylor expansion of the flow is a good approximation when both the anisotropic (i.e.~Braginskii) and isotropic magnetic Prandtl numbers are large.} white in time, and anisotropic with respect to the local magnetic-field direction:
\begin{equation}\label{eqn:ap:velo}
    \rndu^i(t,\bb{x}) = \rnds^i_m (t) x^m,
\end{equation}   
where the rate-of-strain tensor $\rnds^i_m \doteq \rndu^i_{,m}$ satisfies the two-time correlation
\begin{equation}\label{eqn:ttcorr}
    \ea{\rnds^i_m(t)\rnds^j_n(t')} = \Gamma^{ij}_{mn}\delta(t-t') ,
\end{equation}
and
\begin{align}\label{eqn:Gammaijmn}
        {\Gamma}^{ij}_{mn} &= \kappa_2 \Bigl[\delta^{ij}\delta_{mn} + a(\delta^i_m \delta^j_n + \delta^i_n \delta^j_m) + \chi_1 \delta^{ij}\nrndb^m\nrndb^n + \chi_2 \nrndb^i\nrndb^j \delta_{mn} + \chi_3 \nrndb^i\nrndb^j \nrndb^m \nrndb^n \nonumber\\*
        \mbox{} &\quad +\chi_4(\delta^i_m \nrndb^n \nrndb^j + \delta^i_n \nrndb^m \nrndb^j + \nrndb^i\nrndb^m \delta^j_n + \nrndb^i\nrndb^n \delta^j_m)\Bigr]
\end{align}
is the general fourth-rank tensor that is anisotropic with respect to the magnetic-field direction and symmetric under interchange of $i$, $j$ and $m$, $n$. Here, $\kappa_2$ is the second-order coefficient in the Taylor expansion of the field-anisotropic rate-of-strain tensor
\begin{align}
    u_{i,j}(\bb{y}) &= \kappa_0 \delta^{ij} - \frac{1}{2} \kappa_2 \Bigl[ y^2 \delta^{ij} + 2a y^i y^j + \chi_1 \delta^{ij} (\bb{y}\bcdot\eb)^2 + \chi_2 \nrndb^i \nrndb^j y^2 + \chi_3 \nrndb^i \nrndb^j (\bb{y}\bcdot\eb)^2
    \nonumber\\*
    \mbox{} &+ 2\chi_4 (\bb{y}\bcdot\eb) \bigl( \nrndb^j y^i + \nrndb^i y^j \bigr) \Bigr] + \dots 
\end{align}
The constants $\chi_i$ and $a$  parameterize the rate of strain. Two of them can be fixed by assuming an incompressible flow: $\delta^i_m\Gamma^{ij}_{mn}=0$ and $\delta^j_n\Gamma^{ij}_{mn} = 0$, so
\begin{equation}\label{eqn:kappa_incomp}
    a = -\frac{1+\chi_4}{1+d} , \quad \chi_1+\chi_2+\chi_3=-(d+2)\chi_4,
\end{equation}
where $d$ is the dimensionality of the system. The isotropic case is recovered when $\chi_i = 0$ for all $i$. 

Already, a major simplification has been made in performing the ensemble average in equation~\eqref{eqn:ttcorr}. While the magnetic field unit vector is a random variable, it acted as a \emph{non-random} variable when the ensemble average was calculated, and so statistical correlations between the velocity and $\eb$ were neglected.  In the striped region of a magnetic fold, only the \emph{sign} of $\eb$ changes significantly in space, and so the dyad $\eb\eb$, a quadratic quantity, can be approximated as a constant and non-random, provided that it changes on a much longer time scale than either of the correlation times of the underlying velocity field and magnetic-field strength.  Having noted this shortcoming of the model, we now proceed forward.

The evolution equation for the magnetic-energy spectrum $M(k)$ can be obtained by first deriving an evolution equation for the joint probability density function
\begin{equation}
   \mathcal{P}(t;B,\bb{k},\eb) = \ea{\tilde{\mathcal{P}}} = \ea{\delta(B-\rndB(t))\delta(k_m - \rndk_m(t))\delta(\nrndb^i - \rndb^i(t))}.
\end{equation} 
Here and throughout this appendix, angular brackets denote ensemble averages. 
Then
\begin{align}
\pD{t}{\mc{P}} &= \ea{\pD{t}{\tilde{\mc{P}}}} = \ea{\left[-\pD{t}{\rndB(t)} \pD{B}{} - \pD{t}{\rndk_m(t)} \pD{k_m}{} - \pD{t}{\rndb^i(t)} \pD{\nrndb^i}{} \right]\tilde{\mc{P}}}
\nonumber\\*
\mbox{} &= -\ea{\bigg[ \rndb^i(t) \rndb^m(t) \rnds^i_m(t) \rndB(t) \pD{B}{}  - \eta \rndk^2(t)\rndB(t) \pD{B}{}  -   \rnds^i_m (t)\rndk_i(t) \frac{\partial}{\partial k_m} }
\nonumber\\*
\mbox{} & \quad
\ea{+ \rnds^i_m(t) \rndb^m(t) \frac{\partial}{\partial \nrndb^i} - \rnds^l_m(t) \rndb^l(t) \rndb^m(t) \rndb^i(t) \frac{\partial}{\partial \nrndb^i}\bigg]\tilde{\mathcal{P}}}
\nonumber\\*
\mbox{} &= -\ea{\bigg[\frac{\partial}{\partial B} B \nrndb^i \nrndb^m  - \frac{\partial}{\partial k_m}k_i + \frac{\partial}{\partial \nrndb^i}b^m - \frac{\partial}{\partial \nrndb^l}\nrndb^l \nrndb^i \nrndb^m\bigg] \rnds^i_m(t) \tilde{\mathcal{P}} }
+ \eta k^2 \frac{\partial}{\partial B} B \mathcal{P}.\label{eqn:dPdt}
\end{align}
To arrive at the final line of this, the identity $a\,\delta(a-b) = b\,\delta(a-b)$ was used. Note that everything in the square brackets in the final line is non-random. If we define the differential operator 
\begin{equation}
\hat{L}^m_i \doteq \pD{B}{} B \nrndb^i \nrndb^m  - \pD{k_m}{} k_i + \pD{\nrndb^i}{} b^m - \pD{\nrndb^l}{} \nrndb^l \nrndb^i \nrndb^m,
\end{equation}
equation \eqref{eqn:dPdt} can be succinctly written  as
\begin{equation}
\frac{\partial \mathcal{P}}{\partial t} =  - \hat{L}_i^m \ea{\rnds^i_m(t) \tilde{\mathcal{P}}} + \eta k^2 \pD{B}{}B \mathcal{P}.
\end{equation}
To perform the ensemble average, we make use of the Furutsu--Novikov formula~\citep{furutsu,Novikov}, which generalizes Gaussian splitting to functions:
\begin{equation}\label{eq:furutsu-novikov}
\ea{\rnds_m^i(t) \tilde{\mathcal{P}}(t)} = \int_0^t \od t' \ea{\rnds^i_m(t) \rnds^j_n(t')}\ea{\frac{\delta \tilde{\mathcal{P}}(t) }{\delta \rnds^j_n(t')}},
\end{equation}
where $\delta / \delta \rnds^j_n(t')$ is the functional derivative with respect to $\rnds^j_n(t')$ and can be calculated by formally integrating $\partial \tilde{\mathcal{P}}(t)/\partial t$ with respect to time:
\begin{align}
\frac{\delta \tilde{\mathcal{P}}(t)}{\delta \rnds^j_n(t')} &=- \hat{L}^m_i\int_0^{t} \od t'' \, \frac{\delta \rnds^i_m(t'')}{\delta \rnds^j_n(t')} \tilde{\mathcal{P}}(t'') - \int_0^{t} \od t'' \,  \left(\hat{L}^m_i \rnds^i_m(t'') - \eta k^2 \pD{B}{} B\right)\!\cancel{\frac{\delta\tilde{\mathcal{P}}(t'')}{\delta \rnds^j_n(t')} }_{\mathrm{causality}}
\nonumber\\* 
\mbox{} &=- \hat{L}^n_j\int_0^{t} \od t'' \,\delta(t'-t'')\tilde{\mathcal{P}}(t'') = -\hat{L}^n_j\tilde{\mathcal{P}}(t') .
\end{align}
Using this in \eqref{eq:furutsu-novikov} alongside the expression for the two-time correlation \eqref{eqn:ttcorr}, we integrate in time over half of the resulting delta function to find
\begin{equation}
\ea{\rnds_m^i(t) \tilde{\mathcal{P}}(t)} = -\frac{1}{2} \hat{L}^n_j {\Gamma}_{mn}^{ij}(t)  \ea{\tilde{\mathcal{P}}(t)}.
\end{equation}

The result of these manipulations is a closed equation for the joint probability density function:
\begin{equation}\label{eqn:dPdt_averaged}
    \pD{t}{\mc{P}} = \frac{1}{2} \hat{L}^m_i \hat{L}^n_j {\Gamma}^{ij}_{mn}(t) \mc{P} + \eta k^2 \pD{B}{} B\mc{P} .
\end{equation}
To make further progress, we use the chain rule to write
\begin{equation}
\hat{L}_j^n = -k_j \frac{\partial}{\partial k_n} + \left(\nrndb^n \frac{\partial}{\partial \nrndb^j} -\nrndb^j \nrndb^n \nrndb^q \frac{\partial}{\partial \nrndb^q} \right) + \nrndb^j \nrndb^n \left(\frac{\partial}{\partial B}B - d - 2\right),
\end{equation}
where $d$ is the dimensionality of the system, and then calculate the combination $\hat{L}^n_j \Gamma^{ij}_{mn}\mathcal{P}(B,\bb{k},\eb) = \hat{L}^n_j \delta(|\eb|^2 -1) \delta(\eb\bcdot \bb{k})  \Gamma^{ij}_{mn} P(B,k)$. Note that $\mathcal{P}(B,\bb{k},\eb)$ must obey this factorization: the  delta functions result from having $\eb$ be a unit vector and from  solenoidality ($\eb \bcdot \bb{k} = 0$), respectively. Finally, as  the statistics are homogeneous and the relatively alignment of $\eb$ and $\bb{k}$ is fixed, the remaining factor $P$ must be gyrotropic and thus can only depend on the magnitudes of $\bb{B}$ and $\bb{k}$.
For a test function $f$,
\begin{align}
\hat{L}^n_j \delta(\eb\bcdot \bb{k})f &=\delta(\eb\bcdot \bb{k})\hat{L}^n_j  f  - k_j \nrndb^n \delta'(\eb\bcdot\bb{k}) f+ (\nrndb^n k_j - \nrndb^j \nrndb^n \eb \bcdot \bb{k})\delta'(\eb\bcdot\bb{k})f
\nonumber\\*
\mbox{} &= \delta(\eb\bcdot\bb{k}) \hat{L}_j^n f- \nrndb^j \nrndb^n (\eb\bcdot\bb{k})\delta'(\eb\bcdot \bb{k})f
\nonumber\\*
\mbox{} &= \delta(\eb \bcdot \bb{k})(\hat{L}^n_j + \nrndb^n \nrndb^j)f,\label{eqn:Lop1}
\end{align}
where we have used $x \delta'(x) = -\delta(x)$ to obtain the final equality. Similarly,
\begin{equation}
\hat{L}^n_j \delta(|\eb|^2 -1)f = \delta(|\eb|^2-1)(\hat{L}^n_j + 2 \nrndb^n \nrndb^j)f.\label{eqn:Lop2}
\end{equation}
Combining \eqref{eqn:Lop1} and \eqref{eqn:Lop2} leads to
\begin{align}
\hat{L}^n_j\Gamma^{ij}_{mn}\mathcal{P}  &= \delta(\eb\bcdot\bb{k})\delta(|\eb|^2-1)(\hat{L}^n_j + 3\nrndb^n\nrndb^j)\Gamma^{ij}_{mn}P(B,k)
\nonumber\\*
\mbox{} &= \delta(\eb\bcdot\bb{k})\delta(|\eb|^2-1)\bigg[-\frac{k_jk_n}{k^2} \Gamma_{mn}^{ij}k \pD{k}{} + \nrndb^j \nrndb^n \Gamma^{ij}_{mn}\left(\pD{B}{} B - d +1\right) 
\nonumber\\*
\mbox{}&\quad + \nrndb^n \pD{\nrndb^j}{\Gamma^{ij}_{mn}} - \nrndb^j\nrndb^n\nrndb^q \pD{\nrndb^q}{\Gamma^{ij}_{mn}} \bigg]P. \label{eqn:Lop3}
\end{align}
Taking into account the solenoidality constraint $\eb\bcdot \bb{k} = 0$ imposed by the prefactor in~\eqref{eqn:Lop3}, the following combinations in \eqref{eqn:Lop3} may be expressly calculated:
\begin{subequations}
\begin{align}
\frac{k_jk_n}{k^2}\Gamma^{ij}_{mn} &=\kappa_2 \left[a\delta^i_m + (a+1) \frac{k_ik_m}{k^2} + \chi_4 \nrndb^i \nrndb^m \right], 
\\*
\nrndb^j \nrndb^n \Gamma^{ij}_{mn} &= \kappa_2 \bigl[(a+\chi_4)\delta^i_m + (1+a+\chi_1+\chi_2+\chi_3 +3\chi_4)\nrndb^i\nrndb^m \bigr], 
\\*
\nrndb^n \frac{\partial}{\partial \nrndb^j}\Gamma^{ij}_{mn} &= \kappa_2 \bigl[\delta^i_m (\chi_1 + (d+2)\chi_4) 
\nonumber \\*
\mbox{} &\quad + \nrndb^i\nrndb^m(\chi_1+ (d+1)\chi_2 + (d+3)\chi_3 + (d+4)\chi_4) \bigr], 
\\*
\nrndb^j \nrndb^n \nrndb^q \frac{\partial}{\partial \nrndb^q} \Gamma^{ij}_{mn} &= \kappa_2 \bigl[2\chi_4 \delta^i_m + 2(\chi_1+\chi_2 + 2\chi_3 + 3\chi_4)\nrndb^i\nrndb^m \bigr].
\end{align}
\end{subequations}
Using these formulae in \eqref{eqn:Lop3} gives the result
\begin{align}
\hat{L}^n_j\Gamma^{ij}_{mn}\mathcal{P}  &= \kappa_2 \delta(\eb\bcdot\bb{k})\delta(|\eb|^2-1)\bigg\{  -\bigg[a\delta^i_m + (a+1)\frac{k_ik_m}{k^2} + \chi_4 \nrndb^i \nrndb^m\bigg] k \frac{\partial}{\partial k}
\nonumber\\*
\mbox{} &\quad + \bigl[(a+\chi_4)\delta^i_m + (1+a+\chi_1+\chi_2+\chi_3+3\chi_4)\nrndb^i\nrndb^m \bigr]\frac{\partial}{\partial B}B 
\nonumber\\*
\mbox{} &\quad + \bigl[-a(d-1) + \chi_1 + \chi_4\bigr]\delta^i_m 
\nonumber\\
\mbox{} &\quad + \bigl[-(1+a)(d-1) -d \chi_1 - (2d-1)\chi_4\bigr] \nrndb^i \nrndb^m \bigg\}P(B,k).
\end{align}
Further applying the operator $\hat{L}_i^m$ and expending much effort gives the desired expression appearing in \eqref{eqn:dPdt_averaged}:
\begin{align}\label{eqn:Lop4}
\hat{L}^n_j\Gamma^{ij}_{mn}\mathcal{P} 
&= \bigg\{(2a+1)k^2 \frac{\partial^2}{\partial k^2} + (1+2a+\chi_1 + \chi_2 + \chi_3 + 4\chi_4) \frac{\partial}{\partial B}B\frac{\partial}{\partial B}B 
\nonumber\\*
\mbox{} &\quad -2(a+\chi_4) \frac{\partial}{\partial B}B k \frac{\partial}{\partial k} + \bigl[d+(3d-1)a-\chi_1 - \chi_4\bigr]k \frac{\partial}{\partial k}
\nonumber\\*
\mbox{} &\quad - (1+3a+\chi_1 + 3\chi_4)(d-1)\frac{\partial}{\partial B}B + \bigl[(d-1)a-\chi_1-\chi_4\bigr](d-1)  \bigg\}P.
\end{align}
Normalization of the PDF requires that
\begin{align}
1 &= \int \od^d\eb \, \delta(|\eb|^2 -1) \int \od^d \bb{k}\, \delta(\eb\bcdot \bb{k}) \int \od B \, P(B,k)
\nonumber\\*
\mbox{} &= \frac{S_{d-1}S_{d-2}}{2}\int_0^\infty \od k \, k^{d-2}\int \od B \, P(B,k),
\end{align}
where $S_n$ is the surface area of unit $n$-sphere (e.g.~$S_1 = 2\upi$, $S_2 = 4\upi$). Taking this normalization into consideration, we make the substitution $k^{d-2}P(B,k) \rightarrow P(B,k)$ in \eqref{eqn:Lop4} and use
\begin{subequations}
\begin{align}
k^{d-2}k\frac{\partial}{\partial k} \frac{1}{k^{d-2}}P &= \left[k\frac{\partial}{\partial k} - (d-2)\right]P, 
\\*
k^{2-d}k^2 \frac{\partial^2}{\partial k^2} \frac{1}{k^{d-2}} P &= \left[k^2 \frac{\partial^2}{\partial k^2} - 2(d-2) k \frac{\partial}{\partial k} + (d-2)(d-1)\right]P.
\end{align}
\end{subequations}
Thus,
\begin{align}\frac{\partial P }{\partial t} &= \frac{1}{2}\kappa_2 \bigg \{(1+2a)\frac{\partial}{\partial k} k^2 \frac{\partial}{\partial k} + ( 1+2a  + \chi_1 + \chi_2 + \chi_3 + 4 \chi_4 ) \frac{\partial}{\partial B}B \frac{\partial}{\partial B}B 
\nonumber \\*
\mbox{} &\quad - 2(a+\chi_4) \frac{\partial}{\partial B}B \frac{\partial}{\partial k}k - \bigl[(d-2) + (d-3)a + \chi_1 + \chi_4\bigr] \frac{\partial}{\partial k}k 
\nonumber \\*
\mbox{} &\quad - (1+a+\chi_1 +\chi_4 )(d-1)\frac{\partial}{\partial B}B\bigg\} P +  \eta k^2 \pD{B}{} B P .\label{eq:app_almost}
\end{align}
Up to this point, we have kept the model as general as possible. We now enforce incompressibility. Substituting the expressions \eqref{eqn:kappa_incomp} into \eqref{eq:app_almost} leads to the final expression for the evolution equation of the joint PDF:
\begin{align}\label{eq:app_there}
\frac{\partial P}{\partial t} &= \frac{\kappa_2}{2(d+1)}\bigg\{ (d-1-2\chi_4)\frac{\partial}{\partial k}k^2 \frac{\partial}{\partial k} + (d-1)(1-d\chi_4)\frac{\partial}{\partial B} B \frac{\partial}{\partial B} 
 \nonumber \\*
 \mbox{} &\quad  +2(1+d\chi_4) \frac{\partial}{\partial B} B \frac{\partial}{\partial k} k - \bigl[(d-1)^2 +4 \chi_4 + (d+1)\chi_1\bigr]\frac{\partial}{\partial k}k 
\nonumber \\*
\mbox{} &\quad -(d-1)\bigl[d(1+\chi_4) + (d+1)\chi_1\bigr] \frac{\partial}{\partial B}B   \bigg\} P +  \eta k^2 \pD{B}{} B P.
\end{align}
The magnetic energy spectrum is defined as $M(k) = \int_0^\infty \od B\, B^2 P(B,k)$. Taking the $B^2$ moment of \eqref{eq:app_there} leads in three dimensions ($d=3$) to
\begin{align}\label{eq:app_final}
\frac{\partial M}{\partial t} &= \frac{1}{4}\kappa_2 \left[ (1-\chi_4) \frac{\partial}{\partial k}k^2 \frac{\partial}{\partial k} - 2(2+\chi_1 -2 \chi_4) \frac{\partial}{\partial k}k + 2(5+4\chi_1-3\chi_4 )\right] M
\nonumber \\*
\mbox{} &- 2\eta k^2 M.
\end{align}
Equation \eqref{eqn:mod_kazantsev} for the time evolution of the magnetic spectrum written as a function of $\gamma_\perp$, $\sigma_\perp$, and $\sigma_\parallel$ follows from \eqref{eq:app_final} after noting that
\begin{subequations}
\begin{align}
    \gamma_\perp &\doteq \int \frac{\od^3\bb{k}}{(2\upi)^3} \, k_\perp^2 \kappa_\perp = \frac{3-\chi_4}{2}\kappa_2, \\*
    \sigma_\perp &\doteq \frac{1}{\gamma_\perp}\int \frac{\od^3 \,\bb{k}}{(2\upi)^3} k_\parallel^2 \kappa_\perp = \frac{2(1+\chi_1)}{3-\chi_4},\\*
    \sigma_\parallel &\doteq \frac{1}{\gamma_\perp}\int \frac{\od^3\bb{k}}{(2\upi)^3} \,k_\parallel^2 \kappa_\parallel = \frac{1-3\chi_4}{2(3-\chi_4)}.
\end{align}
\end{subequations}
The isotropic case, equation~\eqref{FK}, is recovered by noting $\gamma_\perp = 3\kappa_2/2$, $\sigma_\perp = 2/3$ and $\sigma_\parallel = 1/6$ when $\chi_i = 0$.
\singlespacing
\bibliographystyle{plainnat2}

\cleardoublepage
\ifdefined\phantomsection
  \phantomsection  %
\else
\fi
\addcontentsline{toc}{chapter}{Bibliography}

\bibliography{refs}

\end{document}